\documentclass{IEEEtran}
\pdfoutput=1

\newif\ifanon
\anonfalse

\newif\iffull
\fulltrue

\usepackage[utf8]{inputenc}
\usepackage[T1]{fontenc}
\usepackage{amssymb}
\usepackage{amsmath}
\usepackage{ntheorem}
\usepackage{stmaryrd}
\usepackage{amsfonts}
\usepackage{mathtools}
\usepackage{mathpar}
\usepackage{float}
\usepackage{array}
\usepackage[shortlabels]{enumitem}
\usepackage{balance}
\usepackage{url}
\usepackage{multirow}
\usepackage{xspace}
\usepackage{tikz}
\usetikzlibrary{decorations.pathreplacing}
\usepackage{framed}
\usepackage{placeins}
\usepackage{wasysym}
\usepackage{graphicx}
\usepackage{listings}
\usepackage{wrapfig}

\usepackage[pass]{geometry}

\definecolor{javared}{rgb}{0.6,0,0} 
\definecolor{javagreen}{rgb}{0.25,0.5,0.35} 
\definecolor{javapurple}{rgb}{0.5,0,0.35} 
\definecolor{javadocblue}{rgb}{0.25,0.35,0.75} 
 
\newtheorem{theorem}{Theorem}
\newtheorem{proposition}{Proposition}
\newtheorem{lemma}{Lemma}

\theoremstyle{definition}
\newtheorem{assumption}{Assumption}
\newtheorem{definition}{Definition}

\theoremstyle{remark}
\newtheorem{remark}{Remark}
\newtheorem{example}{Example}

\newcounter{ncomm}

\newcommand{\numberthis}{\addtocounter{equation}{1}\tag{\theequation}}

\newcommand{\dom}{\textit{dom}}
\newcommand{\irule}[1]{({\sc #1})}
\newcommand{\sem}{{$\mu\text{-Dalvik}_{A}$}\xspace}

\newcommand{\tool}{{HornDroid}\xspace}
\newcommand{\flowsensitive}{flow-sensitive\xspace}
\newcommand{\flowinsensitive}{flow-insensitive\xspace}

\newcommand{\absdom}{\hat{D}}
\newcommand{\abspo}{\mathbin{{\sqsubseteq}}}
\newcommand{\absnpo}{\mathbin{{\not\sqsubseteq}}}
\newcommand{\absjoin}{\mathbin{{\sqcup}}}
\newcommand{\absmeet}{\mathbin{{\sqcap}}}

\newcommand{\taintcup}{\mathbin{\sqcup^{\textsf{t}}}}
\newcommand{\taintpo}{\mathbin{\sqsubseteq^{\textsf{t}}}}

\newcommand{\class}{\mathit{cls}}
\newcommand{\cls}[5]{\texttt{cls}\ #1 \subtype #2\ \texttt{imp}\ #3\ \{#4; #5\}}
\newcommand{\field}{\mathit{fld}}
\newcommand{\method}{\mathit{mtd}}

\newcommand{\stm}{\mathit{st}}
\newcommand{\prim}{\mathit{prim}}
\newcommand{\rhs}{\mathit{rhs}}
\newcommand{\lhs}{\mathit{lhs}}
\newcommand{\goto}[1]{\texttt{goto}\ #1}
\newcommand{\move}[2]{\texttt{move}\ #1\ #2}
\newcommand{\new}[2]{\texttt{new}\ #1\ #2}
\newcommand{\newarray}[3]{\texttt{newarray}\ #1\ #2\ #3}
\newcommand{\checkcast}[2]{\texttt{checkcast}\ #1\ #2}
\newcommand{\instanceof}[3]{\texttt{instof}\ #1\ #2\ #3}
\newcommand{\invoke}[3]{\texttt{invoke}\ #1\ #2\ #3}
\newcommand{\sinvoke}[3]{\texttt{sinvoke}\ #1\ #2\ #3}
\newcommand{\return}{\texttt{return}}
\newcommand{\ifbr}[3]{\texttt{if}_{\comp}\ #1\ #2\ \texttt{then}\ #3}
\newcommand{\comp}{\varolessthan}
\newcommand{\pc}{\mathit{pc}}
\newcommand{\spc}{\mathsf{pc}}
\newcommand{\define}{::=}
\newcommand{\unop}[2]{\texttt{unop}_{\odot}\ #1\ #2}
\newcommand{\arrtype}[1]{\texttt{array}[#1]}
\newcommand{\type}[1]{\texttt{#1}}
\newcommand{\primtype}{\mathit{\tau_{prim}}}
\newcommand{\binop}[3]{\texttt{binop}_{\oplus}\ #1\ #2\ #3}

\makeatletter
\newcommand\oo@ocircle[5]{\mathrel{\ooalign{\hfil\raisebox{#1\height}{\scalebox{#2}{$#4#5$}}\hfil\cr$#4#3$}}}
\newcommand\oeq{\mathpalette{\oo@ocircle{0.025}{0.95}{\ocircle}}{=}}
\newcommand\oneq{\mathpalette{\oo@ocircle{0.025}{0.95}{\oslash}}{=}}
\newcommand\olt{\mathpalette{\oo@ocircle{0.06}{0.87}{\ocircle}}{\mkern-2mu<}}
\newcommand\onlt{\mathpalette{\oo@ocircle{0.06}{0.87}{\oslash}}{\mkern-2mu<}}
\makeatother

\newcommand{\subtype}{\leq}
\newcommand{\callstack}{\alpha}
\newcommand{\ind}{\mathit{idx}}
\newcommand{\regval}[1]{\Sigma \llbracket #1 \rrbracket}
\newcommand{\aregval}[1]{\Sigma_A \llbracket #1 \rrbracket}
\newcommand{\meth}[3]{#1: #2\ \{#3\}}
\newcommand{\true}{\mathit{true}}
\newcommand{\false}{\mathit{false}}
\newcommand{\obj}[2]{\{\!| #1; #2 |\!\}}
\newcommand{\arr}[2]{#1[#2]}
\newcommand{\super}{\textit{super}}
\newcommand{\interfaces}{\textit{inter}}
\newcommand{\getst}[1]{\textit{get-stm} (#1)}
\newcommand{\gettype}[2]{\textit{type}_{#1} (#2)}
\newcommand{\lookup}{\textit{lookup}}
\newcommand{\loc}{\mathit{loc}}
\newcommand{\locstate}[4]{\langle #1 \cdot #4 \cdot #2 \cdot #3 \rangle}
\newcommand{\size}{\mathit{len}}
\newcommand{\intent}[2]{\{\!|@ #1; #2 |\!\}}
\newcommand{\defvalue}{\mathbf{0}}

\newcommand{\newintent}[2]{\texttt{newintent}\ #1\ #2}
\newcommand{\result}{\textsf{result}}

\newcommand{\sign}{\textit{sign}}
\newcommand{\methsign}[3]{#1 \xrightarrow{#3} #2}
\renewcommand{\pointer}[2]{#1_{#2}}

\newcommand{\actstack}{\Omega}
\newcommand{\heap}{H}
\newcommand{\sheap}{S}

\newcommand{\ocallstack}{\overline{\callstack}}
\newcommand{\actstate}[1]{\textit{#1}}
\newcommand{\finished}{\textsf{finished}}
\newcommand{\parent}{\textsf{parent}}
\newcommand{\getcb}[2]{\callstack_{#1.#2}}
\newcommand{\lifecycle}{\textit{Lifecycle}}
\newcommand{\actstates}{\textit{ActStates}}
\newcommand{\putextra}[3]{\texttt{put-extra}\ #1\ #2\ #3}
\newcommand{\getextra}[3]{\texttt{get-extra}\ #1\ #2\ #3}
\newcommand{\startact}[1]{\texttt{start-act}\ #1}

\newcommand{\handlers}{\textit{handlers}}
\newcommand{\cb}{\mathit{cb}}
\newcommand{\fintent}{\textsf{intent}}
\newcommand{\serval}[1]{\textit{ser}_{\textit{Val}}^{#1}}
\newcommand{\serblock}[1]{\textit{ser}_{\textit{Blk}}^{#1}}
\newcommand{\serialized}{\Gamma}
\newcommand{\newpointer}[1]{\nu(#1)}

\newcommand{\fact}{\mathsf{f}}
\newcommand{\absual}{\hat{u}}
\newcommand{\absval}{\hat{v}}
\newcommand{\absheap}{\mathsf{H}}
\newcommand{\abssheap}{\mathsf{S}}
\newcommand{\absloc}{\ann}
\newcommand{\apc}{\mathsf{c},\mathsf{m},\mathsf{pc}}
\newcommand{\apcn}{\mathsf{c},\mathsf{m},\mathsf{pc+1}}
\newcommand{\apcp}{\mathsf{c},\mathsf{m},\mathsf{pc'}}
\newcommand{\ainst}[1]{(\!| #1 |\!)_{\pp}}
\newcommand{\ainstb}[1]{(\!| #1 |\!)_{c,m,\pc}}
\newcommand{\acomp}{\mathbin{\hat{\comp}}}
\newcommand{\ancomp}{\mathbin{\hat{\onlt}}}
\newcommand{\abinop}{\mathbin{\hat{\oplus}}}
\newcommand{\aunop}{\hat{\odot}}
\newcommand{\earhs}[1]{\langle\!\langle #1 \rangle\!\rangle_{c,m,\pc}}
\newcommand{\arhs}[1]{\langle\!\langle #1 \rangle\!\rangle_{\pp}}
\newcommand{\prhs}[1]{\mathsf{RHS}_{\spp} (#1)}
\newcommand{\rhsname}{\mathsf{RHS}}
\newcommand{\eprhs}[1]{\mathsf{RHS}_{\apc} (#1)}
\newcommand{\absprim}{\widehat{\prim}}
\newcommand{\absobj}[2]{\{\!| #1; #2 |\!\}}
\newcommand{\absarray}[2]{#1[#2]}
\newcommand{\absintent}[2]{\{\!| @#1; #2 |\!\}}
\newcommand{\absblock}{\hat{b}}
\newcommand{\absprog}{\Delta}
\newcommand{\adefvalue}{\hat{\mathbf{0}}}

\newcommand{\absgettype}{\widehat{\textit{get-type}}}
\newcommand{\translate}[1]{(\!| #1 |\!)}

\newcommand{\rulename}[1]{\textit{#1}}
\newcommand{\absdispatch}{\mathsf{I}}
\newcommand{\abslookup}{\widehat{\textit{lookup}}}
\newcommand{\const}[1]{\mathsf{#1}}

\newcommand{\taint}{\mathit{t}}
\newcommand{\ataint}{\hat{\taint}}
\newcommand{\taintf}[1]{\mathsf{taint}_{#1}}
\newcommand{\ataintfname}{\mathsf{Taint}}
\newcommand{\ataintf}[3]{\ataintfname(#1,#2,#3)}
\newcommand{\public}{\mathsf{public}}
\newcommand{\secret}{\mathsf{secret}\xspace}

\newcommand{\sinks}{\textit{Sinks}}
\newcommand{\sources}{\textit{Sources}}
\newcommand{\ann}{\lambda}
\newcommand{\astart}[1]{in(#1)}
\newcommand{\pp}{\mathit{pp}}
\newcommand{\spp}{\mathsf{pp}}

\definecolor{goodgreen}{rgb}{0.20,0.43,0.09}

\newcommand{\abslh}{\hat{h}}
\newcommand{\absfi}{\hat{k}}
\newcommand{\lfilter}{\mathsf{lk}}
\newcommand{\abslab}{\hat{\absloc}}
\newcommand{\absl}[1]{\mathsf{FS}(#1)}
\newcommand{\absg}[1]{\mathsf{NFS}(#1)}
\newcommand{\bool}{\textit{bb}}

\newcommand{\mmax}[2]{\texttt{max}(#1,#2)}

\newcommand{\cfiltername}{\mathsf{Reach}}
\newcommand{\cfilter}[3]{\cfiltername(#1; #2; #3)}
\newcommand{\lhlift}[2]{\mathsf{hlift}(#1; #2)}
\newcommand{\lift}[2]{\mathsf{lift}(#1; #2)}
\newcommand{\liftlh}[2]{\mathsf{LiftHeap}(#1; #2)}

\newcommand{\absregname}[1]{\mathsf{LState}_{#1}}
\newcommand{\absreg}[5]{\absregname{#1}(#2; #3; #4; #5)}

\newcommand{\absresultname}{\mathsf{Res}}
\newcommand{\absresult}[5]{\absresultname_{#1}(#2; #3; #4; #5)}
\newcommand{\absinvoke}[5]{\mathsf{Inv}_{#1}^{#2}(#3; #4; #5)}

\newcommand{\absuncaughtname}{\mathsf{Uncaught}}
\newcommand{\absuncaught}[5]{\absuncaughtname_{#1}(#2 ; #3 ; #4; #5)}
\newcommand{\absabnormalname}{\mathsf{AState}}
\newcommand{\absabnormal}[5]{\absabnormalname_{#1}(#2 ; #3 ; #4; #5)}

\newcommand{\rfprim}{\beta_{\textit{Prim}}}
\newcommand{\rfloc}[1]{\beta_{\textit{Loc}}^{#1}}
\newcommand{\rffilter}{\beta_{\textit{Filter}}}
\newcommand{\rflab}{\beta_{\textit{Lab}}}
\newcommand{\rflval}[1]{\beta_{\textit{LocVal}}^{#1}}
\newcommand{\rfval}{\beta_{\textit{Val}}}
\newcommand{\rfblock}{\beta_{\textit{Blk}}}
\newcommand{\rflblock}[1]{\beta_{\textit{LocBlk}}^{#1}}
\newcommand{\rflheap}[1]{\beta_{\textit{LHeap}}^{#1}}
\newcommand{\rfcall}[1]{\beta_{\textit{Call}}^{#1}}
\newcommand{\rfheap}[1]{\beta_{\textit{Heap}}^{#1}}
\newcommand{\rfstat}{\beta_{\textit{Stat}}}
\newcommand{\rflconf}{\beta_{\textit{Lcnf}}}
\newcommand{\rfconf}{\beta_{\textit{Cnf}}}
\newcommand{\rfastk}[1]{\beta_{\textit{Stk}}^{#1}}
\newcommand{\rflocstate}[1]{\beta_{\textit{Lst}}^{#1}}
\newcommand{\rfalocstate}[1]{\beta_{\textit{ALst}}^{#1}}
\newcommand{\rfinvoke}[1]{\beta_{\textit{LstInv}}^{#1}}
\newcommand{\rfframe}[1]{\beta_{\textit{Frm}}^{#1}}
\newcommand{\rfdispatch}[1]{\beta_{\textit{Pact}}^{#1}}
\newcommand{\rftdispatch}[1]{\beta_{\textit{Pthr}}^{#1}}

\newcommand{\dcall}{D_{\textit{Call}}}
\newcommand{\dheap}{D_{\textit{Heap}}}
\newcommand{\dstat}{D_{\textit{Stat}}}
\newcommand{\ddispatch}{D_{\textit{Pact}}}
\newcommand{\dtdispatch}{D_{\textit{Pthr}}}

\newcommand{\delcall}{\Delta_{\textit{Call}}}
\newcommand{\delheap}{\Delta_{\textit{Heap}}}
\newcommand{\delstat}{\Delta_{\textit{Stat}}}
\newcommand{\deldispatch}{\Delta_{\textit{Pact}}}
\newcommand{\deltdispatch}{\Delta_{\textit{Pthr}}}

\newcommand{\polab}{\mathbin{\sqsubseteq_{\textit{Loc}}}}
\newcommand{\poval}{\mathbin{\abspo^{\textsf{nfs}}}}
\newcommand{\povalp}{\mathbin{\abspo}}
\newcommand{\poseq}{\mathbin{\sqsubseteq_{\textit{Seq}}^{\textsf{nfs}}}}
\newcommand{\poseqp}{\mathbin{\sqsubseteq_{\textit{Seq}}}}
\newcommand{\poblk}{\mathbin{\sqsubseteq_{\textit{Blk}}^{\textsf{nfs}}}}
\newcommand{\polblk}{\mathbin{\sqsubseteq_{\textit{Blk}}}}
\newcommand{\poreg}{\mathbin{\sqsubseteq_{\textit{R}}}}
\newcommand{\poabnormal}{\mathbin{\sqsubseteq_{\textit{A}}}}
\newcommand{\poinvoke}[1]{\mathbin{\sqsubseteq_{\textit{Inv}}^{#1}}}
\newcommand{\pofilter}{\mathbin{\sqsubseteq_{\textit{Filter}}}}

\newcommand{\funion}{\sqcup_f}
\newcommand{\lfunion}{\sqcup^{\mathsf{loc}}}
\newcommand{\afunion}{\mathbin{\hat{\sqcup}}}

\newcommand{\lheap}{K}
\newcommand{\gheap}{G}
\newcommand{\heapto}{\rightarrow_{\mathsf{ref}}}
\newcommand{\lheapd}{\ensuremath{\gheap,(\lheap_i)_{i}}\xspace}
\newcommand{\lheapdp}{\ensuremath{\gheap',(\lheap_i')_{i}}\xspace}

\newcommand{\flist}{(\lfilter^j)_j}
\newcommand{\flistp}{(\lfilter'^j)_j}
\newcommand{\fhist}{(\lheap_a,\flist)}
\newcommand{\fhistp}{(\lheap'_a,\flistp)}
\newcommand{\fhistget}[2]{\Gamma^{#2}{#1}}

\newcommand{\lheapdh}{\ensuremath{(\gheap,(\lheap_i)_i, \lheap,(\lfilter^j)_j)}\xspace}
\newcommand{\lheapdhp}{\ensuremath{(\gheap',(\lheap_i')_i,\lheap',(\lfilter'^j)_j)}\xspace}

\newcommand{\cheapd}{\ensuremath{(\gheap,(\lheap_i,(\lfilter^{i,j})_j)_{i})}\xspace}
\newcommand{\cheapdp}{\ensuremath{(\gheap',(\lheap'_i,(\lfilter'^{i,j})_j)_{i})}\xspace}

\newcommand{\threadobj}[2]{\obj{#1}{#2}}
\newcommand{\threadstack}{\gamma}
\newcommand{\thread}{\textsf{Thread}}
\newcommand{\void}{\mathsf{Void}}
\newcommand{\activity}{\mathsf{Activity}}
\newcommand{\threadrun}{\mathsf{run}\xspace}

\newcommand{\tmethconf}[6]{#6 \cdot #1 \cdot #2 \cdot #3 \cdot #4 \cdot #5}
\newcommand{\startthread}[1]{\texttt{start-thread}\ #1}
\newcommand{\startthreadbf}[1]{\textbf{\texttt{start-thread}}\ #1}
\newcommand{\jointhread}[1]{\texttt{join}\ #1}
\newcommand{\jointhreadbf}[1]{\textbf{\texttt{join}}\ #1}
\newcommand{\interruptthread}[1]{\texttt{interrupt}\ #1}
\newcommand{\interruptthreadbf}[1]{\textbf{\texttt{interrupt}}\ #1}
\newcommand{\interruptedthread}[1]{\texttt{interrupted}\ #1}
\newcommand{\interruptedthreadbf}[1]{\textbf{\texttt{interrupted}}\ #1}
\newcommand{\isinterruptedthread}[1]{\texttt{is-interrupted}\ #1}
\newcommand{\isinterruptedthreadbf}[1]{\textbf{\texttt{is-interrupted}}\ #1}
\newcommand{\threadpool}{\Xi}
\newcommand{\tactframe}[5]{\langle #1, #2, #3, #4, #5 \rangle}
\newcommand{\threadframe}[5]{\llangle #1, #2, #3, #4, #5 \rrangle}
\newcommand{\tuactframe}[5]{\underline{\tactframe{#1}{#2}{#3}{#4}{#5}}}
\newcommand{\tactconf}[4]{#1 \cdot #2 \cdot #3 \cdot #4}

\newcommand{\abstdispatch}{\mathsf{T}}
\newcommand{\absthreadobj}[2]{\obj{#1}{#2}}
\newcommand{\absthread}{\abslab_t}

\newcommand{\res}{\mathsf{res}}
\newcommand{\abswal}{\hat{w}}
\newcommand{\callinv}[4]{\mathsf{Call}_{#1}^{#2}(#3;#4)}

\newcommand{\rlookupname}[1]{\mathsf{GetBlk}_{#1}}
\newcommand{\rlookup}[5]{\rlookupname{#1}(#2;#3;#4;#5)}

\newcommand{\monitorenter}[1]{\texttt{monitor-enter}\ #1}
\newcommand{\monitorenterbf}[1]{\textbf{\texttt{monitor-enter}}\ #1}
\newcommand{\monitorexit}[1]{\texttt{monitor-exit}\ #1}
\newcommand{\monitorexitbf}[1]{\textbf{\texttt{monitor-exit}}\ #1}

\newcommand{\monitor}{\ensuremath{\mathsf{acquired}}\xspace}
\newcommand{\monitorcounter}{\ensuremath{\mathsf{m}\text{-}\mathsf{cnt}}\xspace}
\newcommand{\interrupted}{\ensuremath{\textsf{inte}}\xspace}
\newcommand{\startwait}[1]{\texttt{wait}\ #1}
\newcommand{\startwaitbf}[1]{\textbf{\texttt{wait}}\ #1}
\newcommand{\waiting}[2]{\textsf{waiting}(#1,#2)}
\newcommand{\throw}[1]{\texttt{throw}\ #1}
\newcommand{\throwbf}[1]{\textbf{\texttt{throw}}\ #1}
\newcommand{\moveexcpt}[1]{\texttt{move-except}\ #1}
\newcommand{\moveexcptbf}[1]{\textbf{\texttt{move-except}}\ #1}

\newcommand{\abnormal}[1]{\texttt{AbNormal}(#1)}
\newcommand{\throwable}{\textsf{Throwable}}
\newcommand{\excpt}{\textsf{excpt}}
\newcommand{\excpttablename}{\textsf{ExcptTable}}
\newcommand{\excpttable}[2]{\excpttablename(#1,#2)}
\newcommand{\interruptedexception}{\mathsf{IntExcpt}}

\makeatletter
\DeclareFontFamily{OMX}{MnSymbolE}{}
\DeclareSymbolFont{MnLargeSymbols}{OMX}{MnSymbolE}{m}{n}
\SetSymbolFont{MnLargeSymbols}{bold}{OMX}{MnSymbolE}{b}{n}
\DeclareFontShape{OMX}{MnSymbolE}{m}{n}{
    <-6>  MnSymbolE5
   <6-7>  MnSymbolE6
   <7-8>  MnSymbolE7
   <8-9>  MnSymbolE8
   <9-10> MnSymbolE9
  <10-12> MnSymbolE10
  <12->   MnSymbolE12
}{}
\DeclareFontShape{OMX}{MnSymbolE}{b}{n}{
    <-6>  MnSymbolE-Bold5
   <6-7>  MnSymbolE-Bold6
   <7-8>  MnSymbolE-Bold7
   <8-9>  MnSymbolE-Bold8
   <9-10> MnSymbolE-Bold9
  <10-12> MnSymbolE-Bold10
  <12->   MnSymbolE-Bold12
}{}

\let\llangle\@undefined
\let\rrangle\@undefined
\DeclareMathDelimiter{\llangle}{\mathopen}%
                     {MnLargeSymbols}{'164}{MnLargeSymbols}{'164}
\DeclareMathDelimiter{\rrangle}{\mathclose}%
                     {MnLargeSymbols}{'171}{MnLargeSymbols}{'171}
\makeatother

\title{A Sound Flow-Sensitive Heap Abstraction for the Static Analysis of Android Applications}

\ifanon

\author{\IEEEauthorblockN{Anonymous submission}
\IEEEauthorblockA{ }
\IEEEauthorblockA{ }
\IEEEauthorblockA{ }}

\else 

\author{\IEEEauthorblockN{
	Stefano Calzavara\IEEEauthorrefmark{1},
	Ilya Grishchenko\IEEEauthorrefmark{2},
	Adrien Koutsos\IEEEauthorrefmark{3},
	Matteo Maffei\IEEEauthorrefmark{2}
}

\IEEEauthorblockA{\IEEEauthorrefmark{1}
Universit\`{a} Ca' Foscari Venezia}
\IEEEauthorblockA{\IEEEauthorrefmark{2}
TU Wien},
\IEEEauthorblockA{\IEEEauthorrefmark{3}
LSV, CNRS, ENS Paris-Saclay}
}

\fi

\begin{document}

\maketitle

\begin{abstract}
The present paper proposes the first static analysis for Android applications which is both flow-sensitive on the heap abstraction and provably sound with respect to a rich formal model of the Android platform. We formulate the analysis as a set of Horn clauses defining a sound over-approximation of the semantics of the Android application to analyse, borrowing ideas from recency abstraction and extending them to our concurrent setting. Moreover, we implement the analysis in HornDroid, a state-of-the-art information flow analyser for Android applications. Our extension allows HornDroid to perform strong updates on heap-allocated data structures, thus significantly increasing its precision, without sacrificing its soundness guarantees. We test our implementation on DroidBench, a popular benchmark of Android applications developed by the research community, and we show that our changes to HornDroid lead to an improvement in the precision of the tool, while having only a moderate cost in terms of efficiency. Finally, we assess the scalability of our tool to the analysis of real applications.
\end{abstract}

\section{Introduction}
Android is today the most popular operating system for mobile phones and tablets, and it boasts the largest application market among all its competitors. Though the huge number of available applications is arguably one of the main reasons for the success of Android, it also poses an important security challenge: there are way too many applications to ensure that they go through a timely and thorough security vetting before their publication on the market. Automated analysis tools thus play a critical role in ensuring that security
verification does not fall behind with respect to the release of malicious (or buggy) applications. 

There are many relevant security concerns for Android applications, e.g., privilege escalation~\cite{FeltWMHC11,BugliesiCS13} and component hijacking~\cite{LuLWLJ12}, but the most important challenge in the area is arguably \emph{information flow control}, since Android applications are routinely granted access to personal information and other sensitive data stored on the device where they are installed. To counter the threats posed by malicious applications, the research community has proposed a plethora of increasingly sophisticated (static) information flow control
frameworks for Android~\cite{YangY12,ZhaoO12,MannS12,GiblerCEC12,Kim12,ArztRFBBKTOM14,WeiROR14,GordonKPGNR15,CalzavaraGM16}. Despite all this progress, however, none of these static analysis tools is able to properly reconcile soundness and precision in its treatment of heap-allocated data structures.

\subsection{Soundness vs. Precision in Android Analyses}
Designing a static analysis for Android applications which is both sound and precise on the heap abstraction is very challenging, most notably because the Android ecosystem is highly concurrent, featuring multiple components running in the same application at the same time and sharing part of the heap. More complications come from the scheduling of these components, which is user-driven, e.g., via button clicks, and thus statically unknown. This means that it is hard to devise precise \emph{flow-sensitive} heap abstractions for Android applications without breaking their soundness. Indeed, most existing static analysers for Android applications turn out to be unsound and miss malicious information leaks ingeniously hidden in the control flow: for instance, Table~\ref{tab:unsound} shows a leaky code snippet that cannot be detected by FlowDroid~\cite{ArztRFBBKTOM14}, a state-of-the-art taint tracker for Android applications\footnote{Android applications are written in Java and compiled to bytecode run by a register-based virtual machine (Dalvik). Most static analysis tools for Android analyse Dalvik bytecode, but we present our examples using a Java-like language to improve readability.}.

\begin{table}[htb]
\begin{lstlisting}[language=Java]
public class Leaky extends Activity {
  Storage st = new Storage();
  Storage st2 = new Storage(); 
  onRestart() { st2 = st; }
  onResume() { st2.s = getDeviceId(); }
  onPause() { send(st.s, "http://www.myapp.com/"); }
}
\end{lstlisting}
\caption{\label{tab:unsound} A Subtle Information Leak}
\end{table}

Assume that the \texttt{Storage} class has only one field \texttt{s} of type \texttt{String}, populated with the empty string by its default constructor. The activity class \texttt{Leaky} has two fields \texttt{st} and \texttt{st2} of type \texttt{Storage}. A leak of the device id may be performed in three steps. First, the activity is stopped and then restarted: after the execution of the \texttt{onRestart()} callback, \texttt{st2} becomes an alias of \texttt{st}. Then, the activity is paused and resumed. As a result,
the execution of the \texttt{onPause()} callback communicates the empty string over the Internet, while the \texttt{onResume()} callback stores the device id in \texttt{st2} and thus in \texttt{st} due to aliasing. Finally, the activity is paused again and the device id is leaked by \texttt{onPause()}.

HornDroid~\cite{CalzavaraGM16} is the only provably sound static analyser for Android applications to date and, as such, it correctly deals with the code snippet in Table~\ref{tab:unsound}. In order to retain soundness, however, HornDroid is quite conservative on the prediction of the control flow of Android applications and implements a \emph{flow-insensitive} heap abstraction by computing just one static over-approximation of the heap, which is proved to be correct at all reachable program points. This is a significant limitation of the tool, since it prevents \emph{strong updates}~\cite{Lhotak:2011:PAE:1925844.1926389} on heap-allocated data structures and thus negatively affects the precision of the analysis. Concretely, to understand the practical import of this limitation, consider the Java code snippet in Table~\ref{tab:anonymous}.   

\begin{table}[htb]
\begin{lstlisting}[language=Java]
public class Anon extends Activity {
  Contact[] m = new Contact[]();  
  onStart() {
    for (int i = 0; i < contacts.length(); i++) { 
      Contact c = contacts.getContact(i);
      c.phone = anonymise(c.phone);
      m[i] = c;
    }
    send(m, "http://www.cool-apps.com/");
  }
}
\end{lstlisting}
\caption{\label{tab:anonymous} Anonymizing Contact Information}
\end{table}

This code reads the contacts stored on the phone, but then calls the \texttt{anonymise} method at line 6 to erase any sensitive information (like phone numbers) before sending the collected data on the Internet. Though this code is benign, HornDroid raises a false alarm, since the field \texttt{c.phone} stores sensitive information after line 5 and strong updates of object fields are not allowed by the static analysis implemented in the tool.

\subsection{Contributions}
In the present paper we make the following contributions:
\begin{enumerate}
\item we extend an operational semantics for a core fragment of the Android ecosystem~\cite{CalzavaraGM16} with multi-threading and exception handling, in order to provide a more accurate representation of the control flow of Android applications;

\item we present the first static analysis for Android applications which is both flow-sensitive on the heap abstraction and provably sound with respect to the model above. Our proposal borrows ideas from \emph{recency abstraction}~\cite{Balakrishnan:2006:RHS:2090874.2090894} in order to hit a sweet spot between precision and efficiency, extending it for the first time to a concurrent setting;

\item we implement our analysis as an extension of HornDroid~\cite{CalzavaraGM16}. This extension allows HornDroid to perform strong updates on heap-allocated data structures, thus significantly increasing the precision of the tool;

\item we test our extension of HornDroid against DroidBench, a popular benchmark proposed by the research community~\cite{ArztRFBBKTOM14}. We show that our changes to HornDroid lead to an improvement in the precision of the tool, while having only a moderate cost in terms of efficiency. We also discuss analysis results for 64 real applications to demonstrate the scalability of our approach. Our tool and more details on the experiments are available online~\cite{website}.
\end{enumerate}



\section{Design and Key Ideas}
\label{sec:design}
%
%

\subsection{Our Proposal}
Our proposal starts from the pragmatic observation that statically predicting the control flow of an Android application is daunting and error-prone~\cite{GordonKPGNR15}. For this reason, our analysis simply assumes that all the activities, threads and callbacks of the application to analyse are concurrently executed under an interleaving semantics\footnote{We are aware of the fact that the Java Memory Model allows more behaviours than an interleaving semantics (see~\cite{Lochbihler:2014:MJM:2560142.2518191} for a formalisation), but since its connections with Dalvik depend on the Android version and its definition is very complicated, in this work we just consider an interleaving semantics for simplicity.}. (In the following paragraphs, we just refer to threads for brevity.)

The key observation to recover precision despite this conservative assumption is that the runtime behaviour of a given thread can only invalidate the static approximation of the heap of another thread whenever the two threads share memory. This means that the heap of each thread can be soundly analysed in a flow-sensitive fashion, as long as the thread runs isolated from all other threads. Our proposal refines this intuition and achieves a much higher level of precision by using two separate static approximations of the heap: a \emph{flow-sensitive abstract heap} and a \emph{flow-insensitive abstract heap}. 

Abstract objects on the flow-sensitive abstract heap approximate concrete objects which are guaranteed to be local to a single thread (not shared). Moreover, these abstract objects always approximate exactly one concrete object, hence it is sound to perform \emph{strong updates} on them. Abstract objects on the flow-insensitive abstract heap, instead, approximate either (1) one concrete object which may be shared between multiple threads, or (2) multiple concrete objects, e.g., produced by a loop. Thus, abstract objects on the flow-insensitive abstract heap only support \emph{weak updates} to preserve soundness. In case (1), this is a consequence of the analysis conservatively assuming the concurrent execution of all the threads and the corresponding loss of precision on the control flow. In case (2), this follows from the observation that only one of the multiple concrete objects represented by the abstract object is updated at runtime, but the updated abstraction should remain sound for all the concrete objects, including those which are not updated. The analysis moves abstract objects from the flow-sensitive abstract heap to its flow-insensitive counterpart when one of the two invariants of the flow-sensitive abstract heap may be violated: this mechanism is called \emph{lifting}.

Technically, the analysis identifies heap-allocated data structures using their allocation site, like most traditional abstractions~\cite{DBLP:conf/cgo/PereiraB09,Hardekopf:2007:AGF:1273442.1250767,Lhotak:2011:PAE:1925844.1926389,Kastrinis:2013:HCP:2499370.2462191}.  Unlike these, however, each allocation site $\absloc$ is bound to \emph{two} distinct abstract locations: $\absl{\absloc}$ and $\absg{\absloc}$. We use $\absl{\absloc}$ to access the flow-sensitive abstract heap and $\absg{\absloc}$ to access the flow-insensitive abstract heap. The abstract location $\absl{\absloc}$ contains the abstraction of the \emph{most-recently-allocated} object created at $\absloc$, provided that this object is \emph{local} to the creating thread. Conversely, the abstract location $\absg{\absloc}$ contains a sound abstraction of all the other objects created at $\absloc$. 

Similar ideas have been proposed in \emph{recency abstraction}~\cite{Balakrishnan:2006:RHS:2090874.2090894}, but standard recency abstraction only applies to sequential programs, where it is always sound to perform strong updates on the abstraction of the most-recently-allocated object. Our analysis, instead, operates in a concurrent setting and assumes that all the threads are concurrently executed under an interleaving semantics. As we anticipated, this means that, if a pointer may be shared between different threads, performing strong updates on the abstraction of the object indexed by the pointer would be unsound. Our analysis allows strong updates without sacrificing soundness by statically keeping track of a set of pointers which are known to be local to a single thread: only the abstractions of the most-recently-allocated objects indexed by these pointers are amenable for strong updates.

\subsection{Examples}
By being conservative on the execution order of callbacks, our analysis is able to soundly analyse the leaky example of Table~\ref{tab:unsound}. We recall it in Table~\ref{tab:unsound-analysed}, where we annotate it with a simplified version of the facts generated by the analysis: the heap fact $\absheap$ provides a \emph{flow-insensitive} heap abstraction, while the $\mathsf{Sink}$ fact denotes communication to a sink. We use line numbers to identify allocation sites and to index the heap abstractions.

\begin{table}[htb]
\begin{lstlisting}[language=Java]
public class Leaky extends Activity {%*\\\textcolor{blue}{$\absheap(1,\absobj{\texttt{Leaky}}{\texttt{st} \mapsto \absg{2},\texttt{st2} \mapsto \absg{3}})$ \\// flow-insensitivity on activity object}*)
  Storage st = new Storage();%*\\\textcolor{blue}{$\absheap(2,\absobj{\texttt{Storage}}{\texttt{s} \mapsto \texttt{""}})$ // after the constructor}*)
  Storage st2 = new Storage();%*\\\textcolor{blue}{$\absheap(3,\absobj{\texttt{Storage}}{\texttt{s} \mapsto \texttt{""}})$ // after the constructor}*)
  onRestart() { st2 = st; }%*\\\textcolor{blue}{$\absheap(1,\absobj{\texttt{Leaky}}{\texttt{st} \mapsto \absg{2},\texttt{st2} \mapsto \absg{2}})$ // aliasing}*)
  onResume() { st2.s = getDeviceId(); }%*\\\textcolor{blue}{$\absheap(2,\absobj{\texttt{Storage}}{\texttt{s} \mapsto \texttt{id}}) \wedge \absheap(3,\absobj{\texttt{Storage}}{\texttt{s} \mapsto \texttt{id}})$ \\// due to flow-insensitivity on activity object}*)
  onPause() { send(st.s, "http://www.myapp.com/");%*\\\textcolor{blue}{$\textsf{Sink}(\texttt{""}) \wedge \textsf{Sink}(\texttt{id})$ // the leak is detected}*)
  }
}
\end{lstlisting}
\caption{\label{tab:unsound-analysed} A Subtle Information Leak (Detected)}
\end{table}

In our analysis, activity objects are always abstracted in a flow-insensitive way, which is crucial for soundness, since we do not predict the execution order of their callbacks. When the activity is created, an abstract flow-insensitive heap fact
$\absheap(1,\absobj{\texttt{Leaky}}{\texttt{st} \mapsto \absg{2},\texttt{st2} \mapsto \absg{3}})$ is introduced, and two facts $\absheap(2,\absobj{\texttt{Storage}}{\texttt{s} \mapsto \texttt{""}})$ and
$\absheap(3,\absobj{\texttt{Storage}}{\texttt{s} \mapsto \texttt{""}})$ abstract the objects pointed by the activity fields \texttt{st} and \texttt{st2}. Then the life-cycle events are abstracted: the \texttt{onRestart} method performs a weak update on the activity object, adding a fact $\absheap(1,\absobj{\texttt{Leaky}}{\texttt{st} \mapsto \absg{2},\texttt{st2} \mapsto \absg{2}})$ which tracks aliasing; after the \texttt{onResume} method, \texttt{st} can thus point to two possible objects, as reflected by the abstract flow-insensitive heap facts generated at line 2 and at line 5. Since the latter fact tracks a sensitive value in the field \texttt{s}, the leak is caught in \texttt{onPause}. 

Our analysis can also precisely deal with the benign example of Table~\ref{tab:anonymous} thanks to recency abstraction. We show a simplified version of the facts generated by the analysis in Table~\ref{tab:anonymous-analysed}. If our static analysis only used a traditional allocation-site abstraction, the benefits of flow-sensitivity would be voided by the presence of the ``for'' loop in the code. Indeed, the allocation site of \texttt{c} would need to identify all the concrete objects allocated therein, hence a traditional static analysis could not perform strong updates on \texttt{c.phone} without breaking soundness and would raise a false alarm on the code.

\begin{table}[htb]
\begin{lstlisting}[language=Java]
public class Anon extends Activity { %*\\\textcolor{blue}{$\absheap(1,\absobj{\texttt{Anon}}{\texttt{m} \mapsto \absg{2}})$ \\// flow-insensitivity on activity object}*)
  Contact[] m = new Contact[](); %*\\\textcolor{blue}{$\absheap(2,[])$ // new empty array is created}*)
  onStart() {%*\\\textcolor{blue}{$ \absregname{3}(\texttt{c}\mapsto \texttt{null};5 \mapsto \bot)$ \\// no allocated contact at location 5 yet}*)
    for (int i = 0; i < contacts.length(); i++) { %*\\\textcolor{blue}{$\absregname{4}(\texttt{c} \mapsto \texttt{null};5 \mapsto \bot) \wedge \absregname{4}(\texttt{c} \mapsto \absg{5};5 \mapsto \bot)$ \\// loop invariant (see below)}*)
      Contact c = contacts.getContact(i);%*\\\textcolor{blue}{$\absregname{5}(\texttt{c} \mapsto \absl{5};5 \mapsto o_\texttt{c})$ // flow-sensitivity}*)
      c.phone = anonymise(c.phone);%*\\\textcolor{blue}{$\absregname{6}(\texttt{c} \mapsto \absl{5};5 \mapsto o_\texttt{c}\{\texttt{phone}\mapsto \texttt{""}\})$ // strong update}*)
      m[i] = c;%*\\\textcolor{blue}{$\absregname{7}(\texttt{c} \mapsto \absg{5};5 \mapsto \bot) \wedge \absheap(5,o_\texttt{c}\{\texttt{phone}\mapsto \texttt{""}\}) \wedge \absheap(2,[\absg{5}])$ // lifting is performed}*)
    }
    send(m, "http://www.cool-apps.com/");%*\\\textcolor{blue}{$\mathsf{Sink}([o_c\{\texttt{phone}\mapsto \texttt{""}\}])$ // no leak is detected}*)
  }
}
\end{lstlisting}
\caption{\label{tab:anonymous-analysed} Anonymizing Contact Information (Allowed)}
\end{table}

The local state fact $\absregname{\spp}$ provides a flow-sensitive abstraction of the state of the registers and the heap at program point $\pp$. Recall that activity objects are always abstracted in a flow-insensitive fashion, therefore the \texttt{Contact} array \texttt{m} is also abstracted by a flow-insensitive heap fact $\absheap(2,[])$. At each loop ite\-ration, our static analysis abstracts the most-recently-allocated \texttt{Contact} object at line 5 in a flow-sensitive fashion. This is done by putting the abstract flow-sensitive location $\absl{5}$ in \texttt{c} and by storing the abstraction of the \texttt{Contact} object $o_c$ in the flow-sensitive local state abstraction $\absregname{5}$, using its allocation site $5$ as a key. This allows us to perform a strong update on the \texttt{c.phone} field at line 6, overwriting the private information with a public one. At line 7 the program stores the public object in the array \texttt{m}, which is abstracted by a flow-insensitive heap fact: to preserve soundness, the flow-sensitive abstraction of $o_c$ is \emph{lifted} (downgraded) to a flow-insensitive abstraction by generating a flow-insensitive heap fact $\absheap(5,o_c[\texttt{phone}\mapsto \texttt{""}])$ and by changing the abstraction of \texttt{c} from $\absl{5}$ to $\absg{5}$. We then perform a weak update on the array stored in \texttt{m} by generating a flow-insensitive heap fact $\absheap(2,[\absg{5}])$. Thanks to the previous strong update, however, the end result is that \texttt{m} only stores public information at the end of the loop and no leak is detected.


\section{Concrete Semantics}
\label{sec:activity}
Our static analysis is defined on top of an extension of \sem{}, a formal model of a core fragment of the Android ecosystem~\cite{CalzavaraGM16}. It includes the main bytecode instructions of Dalvik, the register-based virtual machine running Android applications, and a few important API methods. Moreover, it captures the life-cycle of the most common and complex application components (\emph{activities}), as well as inter-component communication based on asynchronous messages (\emph{intents}, with a dictionary-like structure). Our extension of \sem{} adds two more ingredients to the model: \emph{multi-threading} and \emph{exceptions}, which are useful to get a full account of the control flow of Android applications. For space reasons, the presentation focuses on a relatively high-level overview of our extensions: the formal details, including the full operational semantics, 
\iffull
are provided in Appendix~\ref{sec:concrete}.
\else
can be found in the long version of the paper~\cite{full-version}.
\fi

\subsection{Basic Syntax}
We write $(r_i)^{i \leq n}$ to denote the sequence $r_1,\ldots,r_n$. When the length of the sequence is unimportant, we simply write $r^*$. Given a sequence $r^*$, $r_j$ stands for its $j$-th element and $r^*[j \mapsto r']$ denotes the sequence obtained from $r^*$ by substituting its  $j$-th element with $r'$. We let $k_i \mapsto v_i$ denote a key-value binding and we represent partial maps using a sequence of key-value bindings $(k_i \mapsto v_i)^*$, where all the keys $k_i$ are pairwise distinct; the order of
the keys in a partial map is immaterial.

We introduce in Table~\ref{tab:dalvik} a few basic syntactic categories. A program $P$ is a sequence of classes. A class $\cls{c}{c'}{c^*}{\field^*}{\method^*}$ consists of a name $c$, a super-class $c'$, a sequence of implemented interfaces $c^*$, a sequence of fields $\field^*$, and a sequence of methods $\method^*$. A method $\meth{m}{\methsign{\tau^*}{\tau}{n}}{\stm^*}$ consists of a name $m$, the type of its arguments $\tau^*$, the return type $\tau$, and a sequence of statements $\stm^*$ defining the method body; the syntax of statements is explained below. The integer $n$ on top of the arrow declares how many registers are used by the method. Observe that field declarations $f: \tau$ include the type of the field. A left-hand side $\lhs$ is either a register $r$, an array cell $r_1[r_2]$, an object field $r.f$, or a static field $c.f$, while a right-hand side  $\rhs$ is either a left-hand side $\lhs$ or a primitive value $\prim$.
\begin{table}[htb]
\[
\begin{array}{lcl}
P         & \define & \class^*                                     \\
\class    & \define & \cls{c}{c'}{c^*}{\field^*}{\method^*}        \\
\primtype & \define & \type{bool} ~|~ \type{int} ~|~ \dots         \\
\tau      & \define & c ~|~ \primtype ~|~ \arrtype{\tau}           \\
\field    & \define & f: \tau                                      \\
\method   & \define & \meth{m}{\methsign{\tau^*}{\tau}{n}}{\stm^*} \\
\lhs      & \define & r  ~|~ r[r] ~|~ r.f ~|~ c.f                  \\
\prim  & \define & \true ~|~ \false ~|~ \dots \\
\rhs   & \define & \lhs ~|~ \prim             \\
\end{array}
\]
\caption{\label{tab:dalvik} Basic Syntactic Categories}
\end{table}

Table~\ref{tab:dalvik-stm} reports the syntax of selected statements, along with a brief intuitive explanation of their semantics. Observe that statements do not operate directly on values, but rather on the content of the registers of the Dalvik virtual machine. The extensions with respect to~\cite{CalzavaraGM16} are in bold and are discussed in more detail in the following. Some of the next definitions are dependent on a program $P$, but we do not make this dependency explicit to keep the notation more concise.

\begin{table*}[htb]
\begin{mathpar}
\begin{array}{llll}
\stm \define\\
 \goto{\pc}                & \text{unconditionally jump to program counter $\pc$}                                                &  \invoke{r_o}{m}{r^*}      & \text{invoke method $m$ of the object in $r_o$ with args $r^*$}\\
 \ifbr{r_1}{r_2}{\pc}      & \text{jump to program counter $\pc$ if $r_1 \comp r_2$}                & \return & \text{get the value of the special return register $r_\res$}\\
 \move{\lhs}{\rhs}         & \text{move $\rhs$ into $\lhs$}                                      &  \newintent{r_i}{c}        & \text{put a pointer to a new intent for class $c$ in $r_i$}\\
 \unop{r_d}{r_s}           & \text{compute  $\odot r_s$ and put the result in $r_d$}             &  \putextra{r_i}{r_k}{r_v}  & \text{bind the value of $r_v$ to key $r_k$ of the intent in $r_i$}\\
 \binop{r_d}{r_1}{r_2}     & \text{compute  $r_1 \oplus r_2$ and put the result in $r_d$}        &  \getextra{r_i}{r_k}{\tau} & \text{get the $\tau$-value bound to  key $r_k$ of the intent in $r_i$ } \\
 \new{r_d}{c}              & \text{put a pointer to a new object of class $c$ in $r_d$}          &  \startact{r_i}            & \text{start a new activity by sending the intent in $r_i$}\\
 \newarray{r_d}{r_l}{\tau} & \text{put a pointer to a new $\tau$-array of length $r_l$ in $r_d$} &  \startthreadbf{r_t}         & \text{start the thread in $r_t$}\\
\throwbf{r_e}                & \text{throw the exception stored in $r_e$}                          &  \interruptthreadbf{r_t}     & \text{interrupt the thread in $r_t$}\\
\moveexcptbf{r_e}            & \text{store a pointer to the last thrown exception in $r_e$}        &  \jointhreadbf{r_t}          & \text{join the current thread with the thread in $r_t$}\\
\end{array}
\end{mathpar}
\caption{\label{tab:dalvik-stm} Syntax and Informal Semantics of Selected Statements}
\end{table*}

\subsection{Local Reduction}
\label{sec:local}

\begin{table}[htb]
\begin{mathpar}
\begin{array}{lllcl}
\text{Pointers} & & p & \in & \textit{Pointers} \\
\text{Program counters} & & \pc & \in & \mathbb{N} \\
\text{Program points} & & \pp & \define & c,m,\pc \\
\text{Annotations} & & \ann & \define & \pp ~|~ c ~|~ \astart{c} \\
\text{Locations} & & \ell & \define & \pointer{p}{\ann} \\
\text{Values} & & u,v & \define & \prim ~|~ \ell \\
\text{Register states} & & R & \define & (r \mapsto v)^* \\
\text{Local states} & & L & \define & \locstate{\pp}{\stm^*}{R}{u^*} \\
\text{Local state lists} & & L^\# &\define & \varepsilon ~|~ L :: L^\#\\
\text{Call stacks} & & \callstack & \define & L^\# ~|~ \abnormal{L^\#}\\
\text{Objects} & & o & \define & \obj{c}{(f_{\tau} \mapsto v)^*} \\
\text{Arrays} & & a & \define & \arr{\tau}{v^*} \\
\text{Intents} & & i & \define & \intent{c}{(k \mapsto v)^*} \\
\text{Memory blocks} & & b & \define & o ~|~ a ~|~ i\\
\text{Heaps} & & \heap & \define & (\ell \mapsto b)^* \\
\text{Static heaps} & & \sheap & \define & (c.f \mapsto v)^* \\
\text{Pending activity stacks} & & \pi & \define & \varepsilon ~|~ i :: \pi \\
\text{Pending thread stacks} & & \threadstack & \define & \varepsilon ~|~ \ell :: \threadstack\\
\text{Local configurations} & & \Sigma & \define & \tmethconf{\callstack}{\pi}{\threadstack}{\heap}{\sheap}{\ell}
\end{array}
\end{mathpar}
\caption{\label{tab:dalvik-domains} Semantic Domains for Local Reduction}
\end{table}

\paragraph{Notation} 
Table~\ref{tab:dalvik-domains} shows the main semantic domains used in the present section. We let $p$ range over pointers from a countable set \textit{Pointers}. A program point $\pp$ is a triple $c,m,\pc$ including a class name $c$, a method name $m$ and a program counter $\pc$ (a natural number identifying a specific statement of the method). Annotations $\ann$ are auxiliary information with no semantic import, their use in the static analysis is discussed in Section~\ref{sec:analysis}. A location $\ell$ is an annotated pointer $\pointer{p}{\ann}$ and a value $v$ is either a primitive value or a location.

A \emph{local state} $L = \locstate{\pp}{\stm^*}{R}{u^*}$ stores the state information of an invoked method, run by a given thread or activity. It is composed of a program point $\pp$, identifying the currently executed statement; the method calling context $u^*$, which keeps track of the method arguments and is only used in the static analysis; the method body $\stm^*$, defining the method implementation; and a register state $R$, mapping registers to their content. Registers are local to a given method invocation. 

A \emph{local state list} $L^\#$ is a list of local states. It is used to keep track of the state information of all the methods invoked by a given thread or activity. The \emph{call stack} $\callstack$ is modeled as a local state list $L^\#$, possibly qualified by the $\abnormal{\cdot}$ modifier if the thread or activity is recovering from an exception.

Coming to memory, we define the \emph{heap} $H$ as a partial map from locations to \emph{memory blocks}. There are three types of memory blocks in the formalism: objects, arrays and intents. An \emph{object}
$o = \obj{c}{(f_{\tau} \mapsto v)^*}$ stores its class $c$ and a mapping between fields and values. Fields are annotated with their type, which is typically omitted when unneeded. An \emph{array} $a = \arr{\tau}{v^*}$ contains the type $\tau$ of its elements and the sequence of the values $v^*$ stored into it. An \emph{intent} $i = \intent{c}{(k \mapsto v)^*}$ is composed by a class name $c$, identifying the intent recipient, and a sequence of key-value bindings $(k \mapsto v)^*$, defining the intent payload (a dictionary). The \emph{static heap} $S$ is a partial map from static fields to values.

Finally, we have \emph{local configurations} $\Sigma = \tmethconf{\callstack}{\pi}{\threadstack}{\heap}{\sheap}{\ell}$, representing the full state of a specific activity or thread. They include a location $\ell$, pointing to the corresponding activity or thread object; a call stack $\callstack$; a pending activity stack $\pi$, which is a list of intents keeping track of all the activities that have been started; a pending thread stack $\threadstack$, which is a list of pointers to the threads which have been started; a heap $\heap$, storing memory blocks; and a static heap $\sheap$, storing the values of static fields.

We use several substitution notations in the reduction rules, with an obvious meaning. The only non-standard notations are $\Sigma^+$, which stands for $\Sigma$ where the value of $\pc$ is replaced by $\pc + 1$ in the top-most local state of the call stack, and the substitution of registers $\Sigma[r_d \mapsto u]$, which sets the value of the register $r_d$ to $u$ in the top-most local state of the call stack. This reflects the idea that the computation is performed on the local state of the last invoked method.

\paragraph{Local Reduction Relation}
The \emph{local reduction} relation $\Sigma \rightsquigarrow \Sigma'$ models the evolution of a local configuration $\Sigma$ into a new local configuration $\Sigma'$ as the result of a computation step. The definition of the local reduction relation uses two auxiliary relations:
\begin{itemize}
\item $\regval{\rhs}$, which evaluates a right-hand side expression $\rhs$ in the local configuration $\Sigma$;
\item $\Sigma,\stm \Downarrow \Sigma'$, which executes the statement $\stm$ on the local configuration $\Sigma$ to produce $\Sigma'$.
\end{itemize}
The simplest rule defining a local reduction step $\Sigma \rightsquigarrow \Sigma'$ just fetches the next statement $\stm$ to run and performs a look-up on the auxiliary relation $\Sigma,\stm \Downarrow \Sigma'$. Formally, assuming a function $\getst{\Sigma}$ fetching the next statement based on the program counter of the top-most local state in $\Sigma$, we have:
\[
\small
\inferrule[(R-NextStm)]
{\Sigma, \getst{\Sigma} \Downarrow \Sigma'}
{\Sigma \rightsquigarrow \Sigma'}\]
We show a subset of the new local reduction rules added to \sem{} in Table~\ref{tab:small-sem-excerpt} and we explain them below.

\begin{table*}[htb]
\begin{mathpar}
\inferrule*[width=20em,lab=(R-Throw)]
{\ell = \regval{r_e} \\ 
\heap(\ell) = \obj{c'}{(f \mapsto v)^*}
}
{\Sigma, \throw{r_e} \Downarrow \Sigma[\callstack \mapsto \abnormal{\callstack}][r_\excpt \mapsto \ell]}

\inferrule*[width=35em,lab=(R-Caught)]
{\ell = \aregval{r_\excpt} \\ 
\heap(\ell) = \obj{c'}{(f \mapsto v)^*}\\
\excpttable{c,m,\pc}{c'} = \pc'\\
\callstack_c = \locstate{c,m,\pc'}{\stm^*}{R}{u^*} :: \callstack'}
{\Sigma_A \rightsquigarrow \Sigma_A[\callstack_A \mapsto \callstack_c]}

\inferrule*[width=24em,lab=(R-UnCaught)]
{\ell = \aregval{r_\excpt} \\
\heap(\ell) = \obj{c'}{(f \mapsto v)^*}\\
\excpttable{c,m,\pc}{c'} = \bot}
{\Sigma_A \rightsquigarrow \Sigma_A[\callstack_A \mapsto \abnormal{\callstack'}][r_\excpt \mapsto \ell]}

\inferrule*[width=15em,lab=(R-MoveException)]
{\ell = \regval{r_\excpt} \\
}
{\Sigma, \moveexcpt{r_e} \Downarrow \Sigma^+[r_e \mapsto \ell]}

\inferrule*[width=23em,lab=(R-StartThread)]
{\ell = \regval{r_t} \\ \heap(\ell) = \threadobj{c'}{(f \mapsto v)^*} \\
\threadstack' = \ell :: \threadstack }
{\Sigma, \startthread{r_t} \Downarrow \Sigma^+[\threadstack \mapsto \threadstack']}

\inferrule*[width=30em,lab=(R-InterruptThread)]
{\ell = \regval{r_t} \\ \heap(\ell) = \threadobj{c'}{(f \mapsto v)^*,\interrupted \mapsto \_} \\
\heap' = \heap[\ell \mapsto \threadobj{c'}{(f \mapsto v)^*,\interrupted \mapsto \true}]
}
{\Sigma, \interruptthread{r_t} \Downarrow \Sigma^+[\heap \mapsto \heap']}

\inferrule*[width=30em,lab=(R-JoinThread)]
{\heap(\ell_r) = \obj{c_r}{(f_r \mapsto v_r)^*,\interrupted \mapsto \false}\\
\ell = \regval{r_t} \\ 
\heap(\ell) = \threadobj{c'}{(f \mapsto v)^*,\finished \mapsto \true}
}
{\Sigma, \jointhread{r_t} \Downarrow \Sigma^+}

\inferrule*[width=35em,lab=(R-InterruptJoin)]
{\heap(\ell_r) = \obj{c_r}{(f_r \mapsto v_r)^*,\interrupted \mapsto \true} \\
o = \obj{c_r}{(f_r \mapsto v_r)^*,\interrupted \mapsto \false} \\
\pointer{p}{c,m,\pc} \not\in \dom(H) \\
\heap' = \heap, \pointer{p}{c,m,\pc} \mapsto \obj{\interruptedexception}{} \\
\callstack_c = \abnormal{\callstack[r_\excpt \mapsto \pointer{p}{c,m,\pc}]}}
{\Sigma, \jointhread{r_t} \Downarrow \Sigma[\callstack \mapsto \callstack_c, \heap \mapsto \heap'[\ell_r \mapsto o]]}
\end{mathpar}
\textbf{Convention:} let $\Sigma  = \tmethconf{\callstack}{\pi}{\threadstack}{\heap}{\sheap}{\ell_r}$ with $\callstack = \locstate{c,m,\pc}{\stm^*}{R}{u^*} :: \callstack'$ and $\Sigma_A = \tmethconf{\callstack_A}{\pi}{\threadstack}{\heap}{\sheap}{\ell_{r}}$ with $\callstack_A = \abnormal{\locstate{c,m,\pc}{\stm^*}{R}{u^*} :: \callstack'}$. \\
\caption{\label{tab:small-sem-excerpt} Small step semantics of extended \sem{} - Excerpt}
\end{table*}

\paragraph{Exception Rules}
In Dalvik, method bodies can contain special annotations for exception handling, specifying which exceptions are caught and where, as well as the program counter of the corresponding exception handler (handlers are part of the method body). In our formalism, we assume the existence of a partial map $\excpttable{\pp}{c} = \pc$ which provides, for all program points $\pp$ where exceptions can be thrown and for all classes $c$ extending the $\throwable$ interface, the program counter $\pc$ of the corresponding exception handler. If no handler exists, then $\excpttable{\pp}{c} = \bot$. Moreover, all local states contain a special register $r_\excpt$ that is only accessed by the exception handling rules: this stores the location of the last thrown exception. 

An exception object stored in $r_e$ can be thrown by the statement $\throw{r_e}$ using rule \irule{R-Throw}: it checks that $r_e$ contains the location of a (throwable) object, stores this location into the register $r_\excpt$ and moves the local configuration into an abnormal state. After entering an abnormal state, there are two possibilities:  if there exists an handler for the thrown exception, we exit the abnormal state and jump to the program counter of the exception handler using rule \irule{R-Caught}; otherwise, the exception is thrown back to the method caller using rule \irule{R-UnCaught}. Finally, the location of the last thrown exception object can be copied from the register $r_\excpt$ into the register $r_e$ by the statement $\moveexcpt{r_e}$, as formalized by rule \irule{R-MoveException}

\paragraph{Thread Rules} 
Our formalism covers the core methods of the Java Thread API~\cite{JavaThread}: they enable thread spawning and thread communication by means of interruptions and synchronizations. Rule \irule{R-StartThread} models the statement $\startthread{r_t}$: it allows a thread to be started by simply pushing the location of the thread object stored in $r_t$ on the pending thread stack. The actual execution of the thread is left to the virtual machine, which will spawn it at an unpredictable point in time, as we discuss in the next section. The statement $\interruptthread{r_t}$ sets the interrupt field (named $\interrupted$) of the thread object whose location is stored in $r_t$ to $\true$, as formalized by rule \irule{R-InterruptThread}. We now describe the semantics of thread synchronizations. If the thread $t'$ calling $\jointhread{r_t}$ was not interrupted at some point, rule \irule{R-JoinThread} checks whether the thread whose location is stored in $r_t$ has finished; if this is the case, it resumes the execution of $t'$, otherwise $t'$ remains stuck. If instead $t'$ was interrupted before calling $\jointhread{r_t}$, rule \irule{R-InterruptJoin} performs the following operations: the $\interrupted$ field of $t'$ is reset to $\false$, an $\interruptedexception$ exception is thrown (this creates a new exception object) and the local configuration enters an abnormal state.

\subsection{Global Reduction}

\paragraph{Notation}
Table~\ref{tab:ext-dalvik} introduces the main semantic domains used in the present section. First, we assume the existence of a set of activity states $\actstates$, which is used to model the Android activity life-cycle (see~\cite{PayetS14}). Then we have two kinds of \emph{frames}, modeling running processes. An \emph{activity frame} $\varphi = \tactframe{\ell}{s}{\pi}{\threadstack}{\callstack}$ describes the state of an activity: it includes a location $\ell$, pointing to the activity object; the activity state $s$; a pending activity stack $\pi$, representing other activities started by the activity; a pending thread stack $\threadstack$, representing threads spawned by the activity; and a call stack $\callstack$. A \emph{thread frame} $\psi = \threadframe{\ell}{\ell'}{\pi}{\threadstack}{\callstack}$ describes a running thread: it includes a location $\ell$, pointing to the activity object that started the thread; a location $\ell'$ pointing to the thread object; a pending activity stack $\pi$, representing activities started by the thread; a pending thread stack $\threadstack$, representing other threads spawned by the thread; and a call stack $\callstack$.

Activity frames are organized in an \emph{activity stack} $\actstack$, containing all the running activities; one of the activities may be singled out as \emph{active}, represented by an underline, and it is scheduled for execution. We assume that each $\actstack$ contains at most one underlined activity frame. Thread frames, instead, are organized in a \emph{thread pool} $\threadpool$, containing all the running threads. A \emph{configuration} $\Psi = \tactconf{\actstack}{\threadpool}{\heap}{\sheap}$ includes an activity stack $\actstack$, a thread pool $\threadpool$, a heap $\heap$ and a static heap $\sheap$. It represents the full state of an Android application.

\begin{table}[htb]
\[
\begin{array}{lllcl}
\text{Activity states} & \quad\quad & s & \in & \actstates \\
\text{Activity frames} & & \varphi & \define & \tactframe{\ell}{s}{\pi}{\threadstack}{\callstack} ~|~ \tuactframe{\ell}{s}{\pi}{\threadstack}{\callstack} \\
\text{Activity stacks} & & \actstack & \define & \varphi ~|~ \varphi :: \actstack \\
\text{Thread frames} & & \psi & \define & \threadframe{\ell}{\ell'}{\pi}{\threadstack}{\callstack}\\
\text{Thread pools} & & \threadpool & \define & \emptyset ~|~ \psi :: \threadpool \\
\text{Configurations} & & \Psi & \define & \tactconf{\actstack}{\threadpool}{\heap}{\sheap}
\end{array}
\]
\caption{\label{tab:ext-dalvik} Semantic Domains for Global Reduction}
\end{table}

\paragraph{Global Reduction Relation}
The \emph{global reduction} relation $\Psi \Rightarrow \Psi'$ models the evolution of a configuration $\Psi$ into a new configuration $\Psi'$, either by executing a statement in a thread or activity according to the local reduction rules, or as the result of processing life-cycle events of the Android platform, including user inputs, system callbacks, inter-component communication, etc.

Before presenting the global reduction rules, we define a few auxiliary notions. First, we let $\lookup$ be the function such that $\lookup(c,m) = (c',\stm^*)$ iff $c'$ is the class obtained when performing dispatch resolution of the method $m$ on an object of type $c$ and $\stm^*$ is the corresponding method body. Then, we assume a function $\sign$ such that $\sign(c,m) = \methsign{\tau^*}{\tau}{n}$ iff there exists a class $\class_i$ such that $\class_i = \cls{c}{c'}{c^*}{\field^*}{\method^*, \meth{m}{\methsign{\tau^*}{\tau}{n}}{\stm^*}}$. Finally, we let a \emph{successful} call stack be the call stack of an activity or thread which has completed its computation, as formalized by the following definition.

\begin{definition}
A call stack $\callstack$ is \emph{successful} if and only if $\callstack = \locstate{\pp}{\return}{R}{u^*} :: \varepsilon$ for some $\pp$, $u^*$ and $R$. We let $\ocallstack$ range over successful call stacks.
\end{definition}

The core of the global reduction rules are taken from~\cite{CalzavaraGM16}, extended with a few simple rules used, e.g., to manage the thread pool. 
The main new rules are given in Table~\ref{tab:activity-tsemantics-body} and the full set can be found in
\iffull
 Appendix~\ref{sec:concrete}.
\else
 the long version~\cite{full-version}.
\fi
We start by describing rule \irule{A-ThreadStart}, which models the starting of a new thread by some activity. Let $\ell'$ be a pointer to a pending thread spawned by an activity identified by the pointer $\ell$, the rule instantiates a new thread frame $\psi = \threadframe{\ell}{\ell'}{\varepsilon}{\varepsilon}{\callstack'}$ with empty pending activity stack and empty pending thread stack, executing the $\threadrun$ method of the thread object referenced by $\ell'$. We then have two other rules: rule \irule{T-Reduce} allows the reduction of any thread in the thread pool, using the reduction relation for local configurations; rule \irule{T-Kill} allows the system to remove a thread which has finished its computations, by checking that its call stack is successful.

\begin{table*}[h]
\begin{mathpar}
\inferrule[(A-ThreadStart)]
{\varphi = \tuactframe{\ell}{s}{\pi}{\threadstack :: \ell' :: \threadstack'}{\callstack}\\
\varphi' =\tuactframe{\ell}{s}{\pi}{\threadstack :: \threadstack'}{\callstack}\\
\psi =\threadframe{\ell}{\ell'}{\varepsilon}{\varepsilon}{\callstack'}\\
\heap(\ell') = \threadobj{c'}{(f\mapsto v)^*}\\
\lookup(c',\threadrun) = (c'',\stm^*)\\
\sign(c'',\threadrun) = \methsign{\thread}{\void}{\loc}\\
\callstack' = \locstate{c'',\threadrun,0}{\stm^*}{(r_k \mapsto \defvalue)^{k \leq loc},r_{loc + 1} \mapsto \ell'}{\ell'}}
{\tactconf{\actstack :: \varphi:: \actstack'}{\threadpool}{\heap}{\sheap} \Rightarrow \tactconf{\actstack :: \varphi' :: \actstack'}{\psi :: \threadpool}{\heap}{\sheap}}

\inferrule[(T-Reduce)]
{\tmethconf{\callstack}{\pi}{\threadstack}{\heap}{\sheap}{\ell_t} \rightsquigarrow \tmethconf{\callstack'}{\pi'}{\threadstack'}{\heap'}{\sheap'}{\ell_t}}
{\tactconf{\actstack}{\threadpool :: \threadframe{\ell}{\ell_t}{\pi}{\threadstack}{\callstack} :: \threadpool'}{\heap}{\sheap} \Rightarrow \tactconf{\actstack}{\threadpool :: \threadframe{\ell}{\ell_t}{\pi'}{\threadstack'}{\callstack'} :: \threadpool'}{\heap'}{\sheap'}}

\inferrule[(T-Kill)]
{ \heap(\ell') = \obj{c}{(f\mapsto v)^*,\finished \mapsto \_}\\
\heap'= \ \heap[\ell' \mapsto \obj{c}{(f\mapsto v)^*,\finished \mapsto \true}]
}
{\tactconf{\actstack}{\threadpool :: \threadframe{\ell}{\ell'}{\varepsilon}{\varepsilon}{\ocallstack} :: \threadpool'}{\heap}{\sheap} \Rightarrow \tactconf{\actstack}{\threadpool :: \threadpool'}{\heap'}{\sheap} }


\end{mathpar}
\caption{\label{tab:activity-tsemantics-body}New Global Reduction Rules - Excerpt}
\end{table*}


\section{Abstract Semantics}
\label{sec:analysis}
Our analysis takes as input a program $P$ and generates a set of Horn clauses $\translate{P}$ that over-approximate the concrete semantics of $P$. We can then use an automated theorem prover such as Z3~\cite{MouraB08} to show that $\translate{P}$, together with a set of facts $\absprog$ over-approximating the initial state of the program, does not entail a formula $\phi$ representing the reachability of some undesirable program state (e.g., leaking sensitive information). By the over-approximation, the unsatisfiability of the formula ensures that also $P$ does not reach such a program state.

\subsection{Syntax of Terms}
We assume two disjoint countable sets of variables $\textit{Vars}$ and $\textit{BVars}$. The syntax of the \emph{terms} of the abstract semantics is defined in Table~\ref{tab:absdoms} and described below. 

\begin{table}[htb]
\[
\begin{array}{lllclllcl}
\text{Boolean variables} & \quad\quad & x_b & \in & \textit{BVars} \\
\text{Variables} & & x & \in & \textit{Vars} \\
\text{Abstract elements} & & \hat{d} & \in &\absdom\\
\text{Booleans} & & \bool & ::= & 0 ~|~ 1 ~|~ x_b \\
\text{Abstract locations} & & \abslab & \define & {\absl{\absloc}} \,|\, {\absg{\absloc}}\\
\text{Abstract values} & & \absual,\absval & \define & \hat{d} ~|~ x ~|~  f(\absval^*)\\
\text{Abstract objects} & & \hat{o} & \define & \absobj{c}{(f_{\tau} \mapsto \absval)^*} \\
\text{Abstract arrays} & & \hat{a} & \define & \absarray{\tau}{\absval} \\
\text{Abstract intents} & & \hat{i} & \define & \absintent{c}{\absval} \\
\text{Abstract blocks} & & \absblock & \define & \hat{o} ~|~ \hat{a} ~|~ \hat{i} \\
\text{Abstract flow-sensitive blocks} & & \hat{l} & \define & \hat{b}  ~|~ \bot\\
\text{Abstract flow-sensitive heap} & & \abslh & \define & (\pp \mapsto \hat{l})^* \\
\text{Abstract filter} & & \absfi & \define & (\pp \mapsto \bool)^*
\end{array}
\]
\caption{\label{tab:absdoms} Syntax of Terms}
\end{table}

Each location $\pointer{p}{\absloc}$ is abstracted by an \emph{abstract location} $\abslab$, which is either an abstract \flowsensitive location  $\absl{\absloc}$ or an abstract \flowinsensitive location $\absg{\absloc}$. Recall the syntax of annotations: in the concrete semantics, $\absloc = c$ means that $\pointer{p}{\absloc}$ stores an activity of class $c$; $\absloc = \astart{c}$ means that $\pointer{p}{\absloc}$ stores an intent received by an activity of class $c$; and $\absloc = \pp$ means that $\pointer{p}{\absloc}$ stores a memory block (object, array or intent) created at program point $\pp$. Only the latter elements are amenable for a sound flow-sensitive analysis, since activity objects are shared by all the activity callbacks and received intents are shared between at least two activities, but the analysis assumes the concurrent execution of all callbacks and activities.

The analysis assumes a bounded lattice $(\absdom,\abspo,\absjoin,\absmeet,\top,\bot)$ for approximating concrete values such that the abstract domain $\absdom$ contains at least all the abstract locations $\abslab$ and the abstractions $\absprim$ of any primitive value $\prim$. We also assume a set of interpreted functions $f$, containing at least sound over-approximations $\hat \odot,\hat \oplus, \hat \comp$ of the unary, binary and comparison operators $\odot,\oplus,\comp$. Abstract values $\absval$ are elements $\hat{d}$ of the abstract domain $\absdom$, variables $x$ from $\textit{Vars}$ or function applications of the form $f(\absval^*)$. 

The abstraction of objects $\hat{o}$ is field-sensitive, while the abstraction of arrays $\hat{a}$ and intents $\hat{i}$ is field-insensitive. The reason is that the structure of objects is statically known thanks to their type, while array lengths and intent fields (strings) may only be known at runtime. It would clearly be possible to use appropriate abstract domains to have a more precise representation of array lengths and intent fields, but we do not do it for the sake of simplicity. An \emph{abstract block} $\hat{b}$ can be an abstract object $\hat{o}$, an abstract array $\hat{a}$ or an abstract intent $\hat{i}$. An abstract \emph{flow-sensitive} heap $\abslh$ is a total mapping from the set of allocation sites $\pp$ to abstract memory blocks $\hat{b}$ or the symbol $\bot$, representing the lack of a flow-sensitive abstraction of the memory blocks created at $\pp$.

There is just one syntactic element in Table~\ref{tab:absdoms} which we did not discuss yet: \emph{abstract filters}. Abstract filters $\absfi$ are total mappings from the set of allocation sites $\pp$ to boolean flags $\bool$. They are technically needed to keep track of the allocation sites whose memory blocks must be downgraded to a flow-insensitive analysis when returning from a method call. The downgrading mechanism, called \emph{lifting} of an allocation site, is explained in Section~\ref{sec:lifting}.

\subsection{Ingredients of the Analysis}

\paragraph{Overview} 
Our analysis is \emph{context-sensitive}, which means that the abstraction of the elements in the call stack keeps track of a representation of their calling context. In this work, contexts are defined as tuples $(\absthread,\absual^*)$, where $\absthread$ is an abstraction of the location storing the thread or activity which called the method, while $\absual^*$ is an abstraction of the method arguments. Abstracting the calling thread or activity increases the precision of the analysis, in particular when dealing with the $\jointhread{r_t}$ statement for thread synchronization.

Moreover, our analysis is \emph{flow-sensitive} and computes a different over-approximation $\abslh$ of the state of the heap at each reachable program point, satisfying the following invariant: for each allocation site $\pp$, if $\abslh(\pp) = \hat{b}$, then $\hat{b}$ is an over-approximation of the most-recently allocated memory block at $\pp$ and this memory block is local to the allocating thread or activity. Otherwise, $\abslh(\pp) = \bot$ and the memory blocks allocated at $\pp$, if any, do not admit a flow-sensitive analysis. These memory blocks are then abstracted by an abstract \emph{flow-insensitive} heap, defining an over-approximation of the state of the heap which is valid at all reachable program points. As such, the abstract flow-insensitive heap is not indexed by a program point.

For space reasons, we just present selected excerpts of the analysis in the remaining of this section: the full analysis specification 
\iffull
is given in Appendix~\ref{sec:abs-sem}.
\else
can be found in~\cite{full-version}.
\fi

\paragraph{Analysis Facts}
The syntax of the analysis \emph{facts} $\fact$ is defined in Table~\ref{tab:absfacts}. The fact \(\absreg{\apc}{(\absthread,\absual^*)}{\absval^*}{\abslh}{\absfi}\) is used to abstract local states: it denotes that, if the method $m$ of the class $c$ is invoked in the context $(\absthread,\absual^*)$, the state of the registers at the $\pc$-th statement is over-approximated by $\absval^*$, while $\abslh$ provides a flow-sensitive abstraction of the state of the heap and $\absfi$ tracks the set of the allocation sites which must be lifted after returning from the method. The fact $\absabnormal{\apc}{(\absthread,\absual^*)}{\absval^*}{\abslh}{\absfi}$ has an analogous meaning, but it abstracts local states trying to recover from an exception. The fact $\absresult{\mathsf{c},\mathsf{m}}{(\absthread,\absual^*)}{\absval}{\abslh}{\absfi}$ states that, if the method $m$ of the class $c$ is invoked in the context $(\absthread,\absual^*)$, its return value is over-approximated by $\absval$; the information $\abslh$ and $\absfi$ has the same meaning as before and it is used to update the abstract state of the caller after returning from the method $m$. The fact $\absuncaught{\apc}{(\absthread,\absual^*)}{\absval}{\abslh}{\absfi}$ ensures that, if the method $m$ of the class $c$ is invoked in the context $(\absthread,\absual^*)$, it throws an uncaught exception at the $\pc$-th statement and the location of the exception object is over-approximated by $\absval$; here, $\abslh$ and $\absfi$ are needed to update the abstract state of the caller of $m$, which becomes in charge of handling the uncaught exception. The fact $\prhs{\absval}$ states that $\absval$ over-approximates the right-hand side of a $\move{\lhs}{\rhs}$ statement at program point $\pp$.

\begin{table}[t]
\[
\begin{array}{ll}
\fact \define\\
 \absreg{\spp}{(\abslab,\absval^*)}{\absval^*}{\abslh}{\absfi}  & \text{Abstract local state} \\
 \absabnormal{\spp}{(\abslab,\absval^*)}{\absval^*}{\abslh}{\absfi} & \text{Abstract abnormal state} \\
 \absresult{\mathsf{c},\mathsf{m}}{(\abslab,\absval^*)}{\absval}{\abslh}{\absfi} & \text{Abstract result of method call} \\
 \absuncaught{\spp}{(\abslab,\absval^*)}{\absval}{\abslh}{\absfi} & \text{Abstract uncaught exception} \\
 \prhs{\absval} & \text{Abstract value of right-hand side} \\
 \liftlh{\abslh}{\absfi} & \text{Abstract heap lifting} \\
 \cfilter{\absval}{\abslh}{\absfi} & \text{Abstract heap reachability} \\
 \rlookup{i}{\absval^*}{\abslh}{\abslab}{\absblock} & \text{Abstract heap look-up} \\
 \absheap(\absloc,\absblock) & \text{Abstract flow-insensitive heap entry} \\
 \abssheap_{\mathsf{c},\mathsf{f}}(\absval) & \text{Abstract static field} \\
 \absdispatch_{\mathsf{c}}(\hat{i}) & \text{Abstract pending activity} \\
 \abstdispatch(\absloc,\hat{o}) & \text{Abstract pending thread} \\
 \absual \abspo \absval & \text{Partial ordering on abstract values} \\
 \tau \subtype \tau' & \text{Subtyping fact} \\
\end{array}
\]
\caption{\label{tab:absfacts} Analysis Facts}
\end{table}

We then have a few facts used to abstract the heap and lift the allocation sites. The facts $\liftlh{\abslh}{\absfi}$, $\cfilter{\absval}{\abslh}{\absfi}$ and $\rlookup{i}{\absval^*}{\abslh}{\abslab}{\absblock}$ are the most complicated and peculiar, so they are explained in detail later on. The fact $\absheap(\ann,\absblock)$ models the abstract flow-insensitive heap: it states that the location $\pointer{p}{\ann}$ stores a memory block over-approximated by $\absblock$ at some point of the program execution. The fact $\abssheap_{\mathsf{c},\mathsf{f}}(\absval)$ states that the static field $f$ of class $c$ contains a value over-approximated by $\absval$ at some point of the program execution.

Finally, the fact $\absdispatch_{\mathsf{c}}(\hat{i})$ tracks that an activity of class $c$ has sent an intent over-approximated by $\hat{i}$. The fact $\abstdispatch(\ann,\hat{o})$ tracks that an activity or thread has started a new thread stored at some location $\pointer{p}{\ann}$ and over-approximated by $\hat{o}$. We then have standard partial order facts $\absual \abspo \absval$ and subtyping facts $\tau \subtype \tau'$. 

\paragraph{Horn Clauses}
We define \emph{Horn clauses} as logical formulas of the form $\forall x_1,\ldots,\forall x_m.\fact_1 \wedge \ldots \wedge \fact_n \implies \fact$
without free variables. In order to improve readability, we always omit the universal quantifiers in front of Horn clauses and we distinguish constants from universally quantified variables by using a $\mathsf{sans\ serif}$ font for constants, e.g., we write $\const{c}$ to denote some specific class $c$. When an element in a Horn clause is unimportant, we just replace it with an underscore ($\_$). Also, we write $\forall x_1,\ldots,\forall x_m.\fact_1 \wedge \ldots \wedge \fact_n \implies \fact_1' \wedge \ldots \wedge \fact_k'$ for the set $\{\forall x_1,\ldots,\forall x_m.\fact_1 \wedge \ldots \wedge \fact_n \implies \fact_i' ~|~ i \in [1,k]\}$.

\paragraph{Abstract Programs}
We define \emph{abstract programs} $\Delta$ as sets of facts and Horn clauses, where facts over-approximate program states, while Horn clauses over-approximate the concrete semantics of the analysed program.

\subsection{The Lifting Mechanism}
\label{sec:lifting}
The \emph{lifting} mechanism is the central technical contribution of the static analysis. It is convenient to abstract for a moment from the technical details and explain it in terms of three separate sequential steps, even though in practice these steps are interleaved together upon Horn clause resolution.

\paragraph{Computing the Abstract Filter}
Let $\pp_a$ be the allocation site to lift, i.e., assume that the most-recently-allocated memory block $b$ at $\pp_a$ must be downgraded to a flow-insensitive analysis, for example because it was shared with another activity or thread. Hence, all the memory blocks which can be reached by fol\-lo\-wing a chain of locations (pointers) starting from any location in $b$ must also be downgraded for soundness. In the analysis, we over-approximate this set of locations with facts of the form $\cfilter{\absval}{\abslh}{\absfi}$, meaning that the abstract filter $\absfi$ represents a subset of the flow-sensitive abstract locations which are reachable along $\abslh$ from any flow-sensitive abstract location over-approximated by $\absval$. The Horn clauses deriving $\cfilter{\absval}{\abslh}{\absfi}$ are in Table~\ref{tab:abs-reach} and should be read as a recursive computation, whose goal is to find the set of all the abstract flow-sensitive locations reachable from $\absval$ and hence a sound over-approximation of the set of the allocation sites which need to be lifted. The definition uses the function $\absfi \afunion \absfi'$, computing the point-wise maximum between $\absfi$ and $\absfi'$.

\begin{table*}[t]
\begin{mathpar}
 \cfilter{\absprim}{\abslh}{0^*}

 \cfilter{\absg{\absloc}}{\abslh}{0^*}

 \cfilter{\absl{\spp}}{\abslh}{0^*[\spp \mapsto 1]}

\cfilter{\absual}{\abslh}{\absfi} \wedge \absual \abspo \absval \implies \cfilter{\absval}{\abslh}{\absfi}

\cfilter{\absval}{\abslh}{\absfi} \wedge \cfilter{\absval}{\abslh}{\absfi'} \implies \cfilter{\absval}{\abslh}{\absfi \afunion \absfi'}

\left.
\begin{array}{r}
\abslh(\spp) = \absobj{c}{\_, f \mapsto \absval} \\
\abslh(\spp) = \absarray{\tau}{\absval}\\
\abslh(\spp) = \absintent{c}{\absval}
\end{array} \right\}
 \wedge \cfilter{\absval}{\abslh}{\absfi}
 \implies \cfilter{\absl{\spp}}{\abslh}{\absfi}
\end{mathpar}
\caption{\label{tab:abs-reach} Horn Clauses Used to Derive the Predicate $\cfilter{\absval}{\abslh}{\absfi}$}
\end{table*}

\paragraph{Performing the Lifting}
Once $\cfilter{\absl{\pp_a}}{\abslh}{\absfi}$ has been recursively computed, the analysis introduces a fact $\liftlh{\abslh}{\absfi}$ to force the lifting of the allocation sites $\pp$ such that $\absfi(\pp) = 1$, moving their abstract blocks from the abstract flow-sensitive heap $\abslh$ to the abstract flow-insensitive heap. The lifting is formalized by the following Horn clause:
\[
\liftlh{\abslh}{\absfi} \wedge \absfi(\spp) = 1 \wedge  \abslh(\spp) = \absblock \implies \absheap(\spp;{\absblock})
\]

\paragraph{Housekeeping}
Finally, we need to update the data structures used by the analysis to reflect the lifting, using the computed abstract filter $\absfi$ to update: 
\begin{enumerate}
\item the current abstraction of the registers $\absval^*$. This is done by using a function $\lift{\absval^*}{\absfi}$, which updates $\absval^*$ so that all the abstract flow-sensitive locations $\absl{\pp}$ such that $\absfi(\pp) = 1$ are changed to $\absg{\pp}$. This ensures that the next abstract heap accesses via the register abstractions perform a look-up on the abstract flow-insensitive heap for lifted allocation sites. Formally, we require the $\mathsf{lift}$ function to satisfy the axioms in Table~\ref{tab:axioms-lift};

\begin{table}[t]
\begin{mathpar}
\inferrule
{\absfi(\pp) = 0}
{\lift{\absl{\pp}}{\absfi} = \absl{\pp}}

\inferrule
{\absfi(\pp) = 1}
{\lift{\absl{\pp}}{\absfi} = \absg{\pp}}

\inferrule
{}
{\lift{\absg{\absloc}}{\absfi} =  \absg{\absloc}}

\inferrule
{}
{\lift{\absprim}{\absfi} =  \absprim}

\inferrule
{\absual \abspo \absval }
{\lift{\absual}{\absfi} \abspo \lift{\absval}{\absfi}}

\inferrule
{\forall i: \lift{\absval_i}{\absfi}) = \absual_i}
{\lift{\absval^*}{\absfi} = \absual^*}
\end{mathpar}
\caption{\label{tab:axioms-lift} Axioms Required on the Function $\lift{\absval^*}{\absfi}$}
\end{table}

\item the current abstract \flowsensitive heap $\abslh$. This is done by the function $\lhlift{\abslh}{\absfi}$, which replaces all the entries of the form $\pp \mapsto \absblock$ in $\abslh$ with $\pp \mapsto \bot$ if $\absfi(\pp) = 1$, thus invalidating their flow-sensitive abstraction. If $\absfi(\pp) = 0$, instead, the function calls $\lift{\absval}{\absfi}$ on all the abstract values $\absval$ occurring in $\absblock$, so that $\absblock$ itself is still analysed in a flow-sensitive fashion, but it is correctly updated to reflect the lifting of its sub-components;

\item the current abstract filter $\absfi'$. This is done by the function $\absfi \afunion \absfi'$, computing the point-wise maximum between $\absfi$ and $\absfi'$. This tracks the allocation sites which must be lifted upon returning from the current method call, so that also the caller can correctly update the abstraction of its registers by using the $\mathsf{lift}$ function.
\end{enumerate}
For simplicity, we just say that we lift some abstract value $\absval$ when we lift all the allocation sites $\pp$ such that $\absl{\pp} \abspo \absval$. 

\paragraph{Example}
Assume integers are abstracted by their sign and consider the following abstract flow-sensitive heap:
\[
\begin{array}{ll}
\abslh =  &\pp_1 \mapsto \absarray{\tau}{\absl{\pp_2}}, \pp_2 \mapsto \absobj{c}{g \mapsto \absl{\pp_1},g' \mapsto +} \\
 &\pp_3 \mapsto \absobj{c'}{f \mapsto \absg{\pp_2},f' \mapsto\absl{\pp_4}} \\
 &\pp_4 \mapsto \absobj{c'}{f \mapsto \absl{\pp_1},f' \mapsto\absl{\pp_3}}
\end{array}
\]
Assume we want to lift the allocation site $\pp_1$, the computation of the abstract filter gives: $\absfi = \pp_1 \mapsto 1, \pp_2 \mapsto 1, \pp_3 \mapsto 0, \pp_4 \mapsto 0$. The result of the lifting is then the following:
\[
\begin{array}{lcl}
\lhlift{\abslh}{\absfi} & = & \pp_1 \mapsto \bot, \pp_2 \mapsto \bot, \\
& & \pp_3 \mapsto \absobj{c'}{f \mapsto \absg{\pp_2},f' \mapsto\absl{\pp_4}} \\
& & \pp_4 \mapsto \absobj{c'}{f \mapsto \absg{\pp_1},f' \mapsto\absl{\pp_3}}
\end{array}
\]

\subsection{Abstracting Local Reduction}

\paragraph{Accessing the Abstract Heaps}
We observe that in the concrete semantics one often needs to read a location stored in a register and then access the contents of that location on the heap. In the abstract semantics we rely on a similar mechanism, adapted to read from the correct abstract heap. The fact $\rlookup{i}{\absval^*}{\abslh}{\abslab}{\absblock}$ states that if $\absval^*$ is an over-approximation of the content of the registers and $\abslh$ is an abstract \flowsensitive heap, then $\abslab$ is an abstract location over-approximated by $\absval_i$ and $\absblock$ is an abstract block over-approximating the memory block that re\-gi\-ster $i$ is pointing to. Formally, this fact can be proved by the two Horn clauses below, discriminating on the flow-sensitivity of $\abslab$:
\[
\begin{array}{lcl}
\absl{\absloc} \abspo \absval_i \wedge \abslh(\absloc) = \absblock & \implies & \rlookup{i}{\absval^*}{\abslh}{\absl{\absloc}}{\absblock} \\
\absg{\absloc} \abspo \absval_i \wedge \absheap(\absloc,\absblock) & \implies & \rlookup{i}{\absval^*}{\abslh}{\absg{\absloc}}{\absblock}
\end{array}
\]

\paragraph{Evaluation of Right-Hand Sides}
The abstract semantics needs to be able to over-approximate the evaluation of right-hand sides. This is done via a translation $\arhs{\rhs}$ generating a set of Horn clauses, which over-approximate the value of $\rhs$ at program point $\pp$. For example, the following translation rule generates one Horn clause which approximates the content of the register $r_i$ at $\pp$, based on the information stored in the corresponding local state abstraction:
\[
\arhs{r_i} = \{\absreg{\spp}{ \_ }{\absval^*}{\_}{\_} \implies \prhs{\absval_i }\}
\]

\paragraph{Standard Statements} 
The abstract semantics defines, for each possible form of statement $\stm$, a translation $\ainst{\stm}$ into a set of Horn clauses which over-approximate the semantics of $\stm$ at program point $\pp$. We start by discussing the top part of Table~\ref{tab:abs-statements-excerpt}, presenting the abstract semantics of some statements considered in the original HornDroid paper~\cite{CalzavaraGM16}. We focus in particular on the main additions needed to generalize their abstraction to implement a flow-sensitive heap analysis: 

\begin{table*}[t]
\begin{itemize}
\item $\ainstb{\new{r_d}{c'}} =$\\
$ \{ \absreg{\apc}{\_}{\absval^*}{\abslh}{\absfi} \wedge \cfilter{\absl{\apc}}{\abslh}{\absfi'} \\
\implies \liftlh{\abslh}{\absfi'}\, \wedge \absreg{\apcn}{\_}{\lift{\absval^*}{\absfi'}[d \mapsto \absl{\apc}]}{\lhlift{\abslh}{\absfi'}[\apc \mapsto \absobj{\mathsf{c'}}{(f \mapsto \adefvalue_{\tau})^*}]}{\absfi \afunion \absfi'}\}$

\item $\ainstb{\move{r_o.f}{\rhs}}  =$\\
$ \earhs{\rhs} \cup \{\eprhs{\absval''}  \wedge \absreg{\apc}{ \_} {\absval^*}{\abslh}{\absfi}\, \wedge 
\rlookup{o}{\absval^*}{\abslh}{\absl{\absloc}}{\absobj{c'}{(f' \mapsto \absual')^*, f \mapsto \absval'}} \implies \\
\absreg{\apcn}{ \_ }{\absval^*}{\abslh[\absloc \mapsto \absobj{c'}{(f' \mapsto \absual')^*, f \mapsto \absval''}}{\absfi}\}\, \cup\\
\{\eprhs{\absval''} \wedge \absreg{\apc}{ \_} {\absval^*}{\abslh}{\absfi} \wedge \rlookup{o}{\absval^*}{\abslh}{\absg{\absloc}}{\absobj{c'}{(f' \mapsto \absual')^*, f \mapsto \absval'}}\, \wedge \cfilter{\absval''}{\abslh}{\absfi'} \implies \\
\absheap(\absloc,\absobj{\mathsf{c'}}{(f' \mapsto \absual')^*, f \mapsto {\absval''})}) \wedge \liftlh{\abslh}{\absfi'}\, \wedge 
\absreg{\apcn}{ \_ }{\lift{\absval^*}{\absfi'}}{\lhlift{\abslh}{\absfi'}}{\absfi \afunion \absfi'}\} $

\item $\ainstb{\return} =$
$ \{\absreg{\apc}{(\absthread,\absval^*_{call})}{\absval^*}{\abslh}{\absfi} \implies \absresult{\mathsf{c},\mathsf{m}}{(\absthread,\absval^*_{call})}{\absval_{\res}}{\abslh}{\absfi}\} $

\item $\ainstb{\invoke{r_o}{m'}{(r_{i_j})^{j \leq n}}}  =$\\
$ \{\absreg{\apc}{(\absthread, \_ )}{\absval^*}{ \abslh }{\absfi}\, \wedge 
\rlookup{o}{\absval^*}{\abslh}{\_}{\absobj{c'}{(f \mapsto \absual)^*}} \wedge c' \subtype \const{c''} \implies \\
\absreg{\const{c''},\mathsf{m'},\mathsf{0}}{(\absthread,(\absval_{i_j})^{j \leq n})}{(\adefvalue_k)^{k \leq \loc}, (\absval_{i_j})^{j \leq n}}{ \abslh }{0^*}  ~|~ \const{c''} \in \abslookup(\mathsf{m'})\, \wedge 
\sign(\const{c''},\mathsf{m'}) = \methsign{(\tau_j)^{j \leq n}}{\tau}{\loc}\}\, \cup \hfill\mathbf{(1)}\\
\{ \absreg{\apc}{ (\absthread, \_ )}{\absval^*}{ \abslh }{\absfi} \wedge \rlookup{o}{\absval^*}{\abslh}{\_}{\absobj{c'}{(f \mapsto \absual)^*}}  \wedge c' \subtype \const{c''}\, \wedge \absresult{\const{c''},\mathsf{m'}}{(\absthread',\abswal^*)}{\absval'_{\res}}{\abslh_{{\res}}}{\absfi_\res} \\
\wedge \absthread = \absthread' \wedge \left(\bigwedge_{j \le n} \absval_{i_j} \absmeet \abswal_j \absnpo\bot\right) \implies \absreg{\apcn}{ (\absthread, \_ )} {\lift{\absval^*}{\absfi_\res}[\res \mapsto \absval'_{\res}]}{ \abslh_{{\res}} }{\absfi \afunion \absfi_\res} ~|~ \const{c''} \in \abslookup(\mathsf{m'})\}\, \cup \hfill\mathbf{(2)}\\
\{\absreg{\apc}{ (\absthread,\_) }{\absval^*}{ \abslh }{\absfi} \wedge \rlookup{o}{\absval^*}{\abslh}{\_}{\absobj{c'}{(f \mapsto \absual)^*}}  \wedge c' \subtype \const{c''}\, \wedge \absuncaught{\const{c''},\mathsf{m'}}{(\absthread',\abswal^*))}{\absval'_\excpt}{\abslh_{{\res}}}{\absfi_\res} \\
 \wedge \absthread = \absthread' \wedge \left(\bigwedge_{j \le n} \absval_{i_j} \absmeet \abswal_j \absnpo \bot\right) \implies 
\absabnormal{\apc}{ (\absthread, \_ )} {\lift{\absval^*}{\absfi_\res}[\excpt \mapsto \absval'_{\excpt}]}{ \abslh_{{\res}} }{\absfi \afunion \absfi_\res} ~|~ \const{c''} \in \abslookup(\mathsf{m'})\} \hfill\mathbf{(3)}$

\item $\ainstb{\throw{r_i}} =$
$\{ \absreg{\apc}{\_ }{\absval^*}{\abslh}{\absfi} \implies \absabnormal{\apc}{\_}{\absval^*[\excpt \mapsto \absval_i]}{\abslh}{\absfi}\}$

\item $\ainstb{\startthread{r_i}}  =$\\
$ \{\absreg{\apc}{\_}{\absval^*}{\abslh}{\absfi} \wedge \rlookup{i}{\absval^*}{\abslh}{\absg{\absloc}}{\absthreadobj{c'}{(f \mapsto \absual)^*}}\, \wedge c' \le \thread \\
\implies \abstdispatch(\absloc,\absthreadobj{c'}{(f \mapsto \absual)^*}) \wedge  \absreg{\apcn}{\_}{\absval^*}{\abslh}{\absfi}\}\, \cup\\ 
\{\absreg{\apc}{\_}{\absval^*}{\abslh}{\absfi} \wedge\rlookup{i}{\absval^*}{\abslh}{\absl{\absloc}}{\absthreadobj{c'}{(f \mapsto \absual)^*}} \wedge c' \le \thread\, \wedge \cfilter{\absl{\absloc}}{\abslh}{\absfi'}\\
 \implies \abstdispatch(\absloc,\absthreadobj{c'}{(f \mapsto \absual)^*})\wedge \liftlh{\abslh}{\absfi'}\, \wedge \absreg{\apcn}{\_}{\lift{\absval^*}{\absfi'}}{\lhlift{\abslh}{\absfi'}}{\absfi \afunion \absfi'}\}$

\item $\ainstb{\jointhread{r_i}}  =$\\
$\{ \absreg{\apc}{(\absg{\absloc_t},\_)}{\absval^*}{\abslh}{\absfi}   \wedge \absheap(\absloc_t,\absthreadobj{c'}{(f \mapsto \absual)^*,\interrupted \mapsto \absval'})\, \wedge 
\widehat{\false} \abspo \absval' \implies \absreg{\apcn}{(\absg{\absloc_t},\_)}{\absval^*}{\abslh}{\absfi}\}\, \cup\\
\{\absreg{\apc}{(\absg{\absloc_t},\_)}{\absval^*}{\abslh}{\absfi} \wedge\absheap(\absloc_t,\absthreadobj{c'}{(f \mapsto \absual)^*,\interrupted \mapsto \absval'})  \wedge \widehat{\true} \abspo \absval'  \implies\\
\absheap(\apc;\absobj{\interruptedexception}{})\wedge \absabnormal{\apc}{(\absg{\absloc_t},\_)}{\absval^*[\excpt \mapsto \absg{\apc}]}{\abslh}{\absfi}\, \wedge 
\absheap(\absloc_t,\absthreadobj{c'}{(f \mapsto \absual)^*,\interrupted \mapsto \widehat{\false}})\}$
\end{itemize}
\caption{\label{tab:abs-statements-excerpt} Abstract Semantics of Statements - Excerpt}
\end{table*}

\begin{itemize}
\item $\ainst{\new{r_d}{c'}}$: When allocating a new object at $\pp$, the abstraction of the object that was the most-recently allocated one before the new allocation, if any, must be downgraded to a flow-insensitive analysis. Therefore, we lift the allocation site $\pp$ by computing an abstract filter $\absfi'$ via the $\cfiltername$ predicate and using it to perform the lifting as described in Section~\ref{sec:lifting}. We then put in the resulting abstract \flowsensitive heap a new abstract object $\absobj{c'}{(f \mapsto \adefvalue_{\tau})^*}$ initialized to default values ($\adefvalue_{\tau}$ represents the abstraction of the default value used to populate fields of type $\tau$). The abstraction of the register $r_d$ is set to the abstract \flowsensitive location $\absl{\pp}$ to enable a flow-sensitive analysis of the new most-recently-allocated object;

\item $\ainst{\move{r_o.f}{\rhs}}$: We first use $\arhs{\rhs}$ to generate the Horn clauses over-approximating the value of $\rhs$ at program point $\pp$. Assume then we have the over-approximation $\absval''$ in a $\rhsname$ fact. We have two possibilities, based on the abstract value $\absval_o$ over-approximating the content of the register $r_o$. If $\rlookupname{o}$ returns an abstract flow-sensitive location $\absl{\absloc}$, then we perform a strong update on the corresponding element of the abstract \flowsensitive heap. If $\rlookupname{o}$ returns an abstract flow-insensitive location $\absg{\absloc}$, we use $\absloc$ to get an abstract heap fact $\absheap(\absloc,\absobj{c'}{(f' \mapsto \absual')^*, f \mapsto \absval'})$ and we update the field $f$ of this object in a new heap fact: this implements a weak update, since the old fact is still valid. The abstract value $\absval''$ moved to the \flowinsensitive heap fact may contain abstract \flowsensitive locations, which must be downgraded by lifting  $\absval''$ when propagating the local state abstraction to the next program point;

\item $\ainst{\return}$: The callee generates a return fact $\absresultname$ containing the calling context $(\absthread,\absval^*_{call})$, the abstract value $\absval_{\res}$ over-approximating the return value, its abstract \flowsensitive heap $\abslh$ and its abstract filter $\absfi$ recording which allocation sites were lifted during its computation. All this information is propagated to the analysis of the caller, as we explain in the next item; 

\item $\ainst{\invoke{r_o}{m'}{(r_{i_j})^{j \leq n}}}$: We statically know the name $m'$ of the invoked method, but not the class of the receiver object in the register $r_o$. In part $\mathbf{(1)}$ we over-approximate dynamic dispatching as follows: we collect all the abstract objects accessible via the abstraction $\absval_o$ of the content of the register $r_o$, but we only consider as possible receivers the ones whose type is a subtype of a class $c'' \in \abslookup(m')$, where $\abslookup(m')$ just returns the set of classes which define or inherit a method named $m'$. For all of them, we introduce an abstract local state fact $\absregname{}$ over-approximating the local state of the invoked method, instantiating it with the calling context, the abstract \flowsensitive heap of the caller and an empty abstract filter.

Part $\mathbf{(2)}$ handles the propagation of the abstraction of the return value from the callee to the caller. This is done by using the $\mathsf{Res}$ fact generated by the $\return$ statement of the callee: the caller matches appropriate callees by checking the context of the $\mathsf{Res}$ fact. Specifically, the caller checks that: $(i)$ its own abstraction $\absthread$ matches the abstraction $\absthread'$ in the context of the callee, and $(ii)$ that the meet of its arguments $\absval_{i_j}$ and the context arguments $\abswal_j$ is not $\bot$. This prevents a callee from returning to a caller that could not have invoked it, in case $(i)$ because caller and callee are being executed by different threads, and in case $(ii)$ because the over-approximation of the arguments used by the caller and the over-approximation of the arguments supplied to the callee are disjoint. We then instantiate the abstract local state of the next program point by inheriting the abstract \flowsensitive heap of the callee $\abslh_{{\res}}$, lifting the abstraction of the caller registers, joining the caller abstract filter $\absfi$ with the callee abstract filter $\absfi_{\res}$, and storing the abstraction of the returned value $\absval_{\res}'$ in the abstraction of the return register.

Finally, part $\mathbf{(3)}$ of the rule is used to handle the propagation of uncaught exceptions from the callee to the caller. It uses an abstract uncaught exception fact $\absuncaughtname$, generated by the exception rules explained below: it tries to throw back the exceptions to an appropriate caller, by matching the context of the $\absuncaughtname$ fact with the abstract local state of the caller.
\end{itemize}

\paragraph{Exceptions and Threads}
The bottom part of Table~\ref{tab:abs-statements-excerpt} presents the abstract semantics of some selected new statements of the concrete semantics:
\begin{itemize}
\item $\ainst{\throw{r_i}}$: We generate an abstract \emph{abnormal} local state fact $\absabnormalname$ from the abstract local state throwing the exception, and we set the abstraction of the special exception register accordingly;

\item $\ainst{\startthread{r_i}}$: We create an abstract pending thread fact $\abstdispatch$, tracking that a new thread was started. The actual instantiation of the abstract thread object is done by the abstract counterpart of the global reduction rules, which we discuss later. Observe that, if the abstract location pointing to the abstract thread object has the form $\absl{\absloc}$, then $\absloc$ is lifted, since the parent thread can access the state of the new thread, but the two threads are concurrently executed;

\item $\ainst{\jointhread{r_i}}$: We just check whether the $\interrupted$ field of the abstract object over-approximating the running thread or activity is over-approximating $\widehat{\true}$, in which case an abstract abnormal local state throwing an $\interruptedexception$ exception is generated, or $\widehat{\false}$, in which case the abstract local state is propagated to the next program point.
\end{itemize}

\paragraph{Example}
We show in Table~\ref{tab:unsound-bytecode-excerpt} a (simplified) bytecode program corresponding to the code snippet in Table~\ref{tab:unsound}. A few comments about the bytecode: the activity constructor \texttt{<init>} is explicitly defined; by convention, the first register after the local registers of a method is used to store a pointer to the activity object and the register \texttt{ret} is used to store the result of the last invoked method. 

We assume that the class \texttt{Leaky} extends \texttt{Activity} and implements at least the methods \texttt{send} and \texttt{getDeviceId}, whose code is not shown here. We also use line numbers to refer to program points, which makes the notation lighter. Notice that there are only two allocation points, lines $7$ and $9$, therefore the abstract \flowsensitive heap will contain only two entries and have the form $7 \mapsto \hat{l}_1, 9 \mapsto \hat{l}_2$.

We selected three bytecode instructions and we give for each of them the Horn clauses generated by our analysis. We briefly comment on the clauses: the $\texttt{new}$ instruction at line $7$ computes all the abstract \flowsensitive locations reachable from $\absl{7}$ with the predicate $\cfiltername$: $bb'_1$ (resp. $bb'_2$) is set to 1 iff the location $7$ (resp. $9$) needs to be lifted. These abstract \flowsensitive locations are then lifted, if needed, using:
\[
\liftlh{7 \mapsto \hat{l}_1,9 \mapsto \hat{l}_2}{7 \mapsto bb'_1,9 \mapsto bb'_2},
\] 
and the abstract \flowsensitive heap is updated by putting a fresh $\texttt{Storage}$ object in $7$ and by lifting $9$, if needed: 
\[
  7 \mapsto \absobj{\texttt{Storage}}{\texttt{s} \mapsto ""},9 \mapsto \lhlift{\hat{l}_2}{7 \mapsto bb'_1,9 \mapsto bb'_2}.
\]

The \texttt{invoke} instruction at line $18$ has two clauses: the first clause retrieves the callee's class $c'$ and performs an abstract virtual method dispatch (here there is only one class implementing $\texttt{getDeviceId}$, hence this step is trivial); the second clause gets the result from the called method and returns it to the caller, checking that the caller's abstract thread pointer $\absthread$ and supplied argument $\absval$ match the callee's context $(\absthread',\absval')$ with the constraint \(\absthread = \absthread' \wedge \absval \absmeet \absval' \absnpo\bot\). We removed the exception handling clauses, as they are not relevant here.

Finally, the \texttt{move} instruction at line $20$ is abstracted by four Horn clauses: the first one evaluates the right-hand side of the \texttt{move}; the two subsequent clauses execute the move in case the left-hand side is the field \texttt{s} of, respectively, the abstract \flowsensitive location $7$ or $9$; finally, the last clause is used if the left-hand side is the field $\texttt{s}$ of an abstract flow-insensitive location, in which case a new abstract flow-insensitive heap entry is created.

\begin{table*}[htb]
\textbf{Bytecode Example:}\\
\begin{minipage}[t]{0.33\linewidth}
\begin{lstlisting}[escapeinside={\%*}{*)}, firstnumber=1]
.class public Leaky
.super Activity
.field st:Storage
.field st2:Storage%*\\[-0.65em]*)
.method constructor <init>()
.1 local register
    new r0 Storage
    move r1.st r0
    new r0 Storage
    move r1.st2 r0
.end method
\end{lstlisting}
\end{minipage}
\begin{minipage}[t]{0.33\linewidth}
\begin{lstlisting}[escapeinside={\%*}{*)}, firstnumber=12]
.method onRestart()
.1 local register
    move r1.st2 r1.st
.end method%*\\[-0.65em]*)
.method onResume()
.1 local register
    invoke r1 getDeviceId()
    move r0 r1.st2
    move r0.s ret
.end method
\end{lstlisting}
\end{minipage}
\begin{minipage}[t]{0.33\linewidth}
\begin{lstlisting}[escapeinside={\%*}{*)}, firstnumber=22]
.method onPause()
.2 local registers
    move r0 r2.st
    move r1 r0.s
    move r0 "http://myapp.com/"
    invoke r2 send() r1 r0
.end method
\end{lstlisting}
\end{minipage}
\vspace{0.5em}

\(\begin{array}{ll}
&\textbf{Generated Horn Clauses for Line 7:}\\
\bullet\;&\absreg{7}{\_}{r_0 \mapsto \absual, r_1 \mapsto \absval}{7 \mapsto \hat{l}_1,9 \mapsto \hat{l}_2}{7 \mapsto bb_1,9 \mapsto bb_2} \wedge 
\cfilter{\absl{7}}{7 \mapsto \hat{l}_1,9 \mapsto \hat{l}_2}{7 \mapsto bb'_1,9 \mapsto bb'_2} \implies\\
&\null\hfill\liftlh{7 \mapsto \hat{l}_1,9 \mapsto \hat{l}_2}{7 \mapsto bb'_1,9 \mapsto bb'_2} \wedge 
\absreg{8}{\_}{r_0 \mapsto \absl{7}, r_1 \mapsto \lift{\absual}{7 \mapsto bb'_1,9 \mapsto bb'_2}}{\\&\null\hfill 7 \mapsto \absobj{\texttt{Storage}}{\texttt{s} \mapsto ""},9 \mapsto \lhlift{\hat{l}_2}{7 \mapsto bb'_1,9 \mapsto bb'_2}}{7 \mapsto bb_1 \afunion bb_1',9 \mapsto bb_2 \afunion bb_2'}\\[0.5em]
&\textbf{Generated Horn Clauses for Line 18:}\\
\bullet&\absreg{18}{(\absthread,\_)}{r_0 \mapsto \absual, r_1 \mapsto \absval, \texttt{ret} \mapsto \abswal}{7 \mapsto \hat{l}_1,9 \mapsto \hat{l}_2}{7 \mapsto bb_1,9 \mapsto bb_2} \wedge\\
&\rlookup{1}{r_0 \mapsto \absual, r_1 \mapsto \absval, \texttt{ret} \mapsto \abswal}{7 \mapsto \hat{l}_1,9 \mapsto \hat{l}_2}{\_}{\absobj{c'}{\_}} \wedge c' \subtype \texttt{Leaky} \implies \\
&\null\hfill\absreg{0}{(\absthread,\absval)}{r_0 \mapsto \absval}{7 \mapsto \hat{l}_1,9 \mapsto \hat{l}_2}{7 \mapsto 0,9 \mapsto 0}\\
\bullet&\absreg{18}{(\absthread,\_)}{r_0 \mapsto \absual, r_1 \mapsto \absval, \texttt{ret} \mapsto \abswal}{7 \mapsto \hat{l}_1,9 \mapsto \hat{l}_2}{7 \mapsto bb_1,9 \mapsto bb_2} \wedge\\
&\rlookup{1}{r_0 \mapsto \absual, r_1 \mapsto \absval, \texttt{ret} \mapsto \abswal}{7 \mapsto \hat{l}_1,9 \mapsto \hat{l}_2}{\_}{\absobj{c'}{\_}} \wedge c' \subtype \texttt{Leaky} \wedge\\
&\absresult{\texttt{getDeviceId}}{(\absthread',\absval')}{\absual'_{\res}}{7 \mapsto \hat{l}'_1,9 \mapsto \hat{l}'_2}{7 \mapsto bb'_1,9 \mapsto bb'_2}
\wedge \absthread = \absthread' \wedge \absval \absmeet \absval' \absnpo\bot \implies \\
&\null\hfill\absreg{19}{(\absthread,\_)}{r_0 \mapsto \absual, r_1 \mapsto \absval, \texttt{ret} \mapsto \absual'_{\res}}{7 \mapsto \hat{l}'_1,9 \mapsto \hat{l}'_2}{7 \mapsto bb_1 \afunion bb_1',9 \mapsto bb_2 \afunion bb_2'}\\[0.5em]
&\textbf{Generated Horn Clauses for Line 20:}\\
\bullet&\absreg{20}{\_}{r_0 \mapsto \absual, r_1 \mapsto \absval, \texttt{ret} \mapsto \abswal}{7 \mapsto \hat{l}_1,9 \mapsto \hat{l}_2}{7 \mapsto bb_1,9 \mapsto bb_2} \implies \rhsname_{20}(\abswal) \\
\bullet&\absreg{20}{\_}{r_0 \mapsto \absual, r_1 \mapsto \absval, \texttt{ret} \mapsto \abswal}{7 \mapsto \hat{l}_1,9 \mapsto \hat{l}_2}{7 \mapsto bb_1,9 \mapsto bb_2} \wedge\\
&\rhsname_{20}(\absual') \wedge \rlookup{0}{r_0 \mapsto \absual, r_1 \mapsto \absval, \texttt{ret} \mapsto \abswal}{7 \mapsto \hat{l}_1,9 \mapsto \hat{l}_2}{\absl{7}}{\absobj{\texttt{Storage}}{\texttt{s} \mapsto \absval'}} \implies \\
&\null\hfill\absreg{21}{ \_ }{r_0 \mapsto \absual, r_1 \mapsto \absval, \texttt{ret} \mapsto \abswal}{7 \mapsto \absobj{\texttt{Storage}}{\texttt{s} \mapsto \absual'},9 \mapsto \hat{l}_2}{7 \mapsto bb_1,9 \mapsto bb_2}\\[0.5em]
\bullet&\absreg{20}{\_}{r_0 \mapsto \absual, r_1 \mapsto \absval, \texttt{ret} \mapsto \abswal}{7 \mapsto \hat{l}_1,9 \mapsto \hat{l}_2}{7 \mapsto bb_1,9 \mapsto bb_2} \wedge\\
&\rhsname_{20}(\absual') \wedge 
\rlookup{0}{r_0 \mapsto \absual, r_1 \mapsto \absval, \texttt{ret} \mapsto \abswal}{7 \mapsto \hat{l}_1,9 \mapsto \hat{l}_2}{\absl{9}}{\absobj{\texttt{Storage}}{\texttt{s} \mapsto \absval'}} \implies \\
&\null\hfill\absreg{21}{ \_ }{r_0 \mapsto \absual, r_1 \mapsto \absval, \texttt{ret} \mapsto \abswal}{7 \mapsto \hat{l}_1,9 \mapsto \absobj{\texttt{Storage}}{\texttt{s} \mapsto \absual'}}{7 \mapsto bb_1,9 \mapsto bb_2}\\[0.5em]
\bullet&\absreg{20}{\_}{r_0 \mapsto \absual, r_1 \mapsto \absval, \texttt{ret} \mapsto \abswal}{7 \mapsto \hat{l}_1,9 \mapsto \hat{l}_2}{7 \mapsto bb_1,9 \mapsto bb_2} \wedge\rhsname_{20}(\absual') \wedge\\& \rlookup{0}{r_0 \mapsto \absual, r_1 \mapsto \absval, \texttt{ret} \mapsto \abswal}{7 \mapsto \hat{l}_1,9 \mapsto \hat{l}_2}{\absg{\spp}}{\absobj{\texttt{Storage}}{\texttt{s} \mapsto \absval'}} \wedge
\cfilter{\absual'}{7 \mapsto \hat{l}_1,9 \mapsto \hat{l}_2}{7 \mapsto bb'_1,9 \mapsto bb'_2} \implies\\
&\null\hfill\liftlh{7 \mapsto \hat{l}_1,9 \mapsto \hat{l}_2}{7 \mapsto bb'_1,9 \mapsto bb'_2} \wedge 
\absheap(\spp,\absobj{\texttt{Storage}}{\texttt{s} \mapsto \absual'}) \wedge \\
&\absreg{21}{\_}{r_0 \mapsto \lift{\absual}{7 \mapsto bb'_1,9 \mapsto bb'_2}, r_1 \mapsto \lift{\absval}{7 \mapsto bb'_1,9 \mapsto bb'_2}, \texttt{ret} \mapsto \lift{\abswal}{7 \mapsto bb'_1,9 \mapsto bb'_2}}{\\&\null\hfill 7 \mapsto \lhlift{\hat{l}_1}{7 \mapsto bb'_1,9 \mapsto bb'_2},9 \mapsto \lhlift{\hat{l}_2}{7 \mapsto bb'_1,9 \mapsto bb'_2}}{7 \mapsto bb_1 \afunion bb_1',9 \mapsto bb_2 \afunion bb_2'}\\[0.5em]
\end{array}\)
\caption{\label{tab:unsound-bytecode-excerpt} Example of Dalvik Bytecode and Excerpt of the Corresponding Horn Clauses}
\end{table*}

\subsection{Abstracting Global Reduction}

\begin{table*}[htb]
\[
\begin{array}{lcl}
\rulename{Tstart} & = & \{\abstdispatch(\absloc,\absobj{c}{(f \mapsto \_)^*}) \wedge c \subtype \mathsf{c'} \wedge c \le \thread \implies\\
&&  \absreg{\mathsf{c'},\threadrun,\mathsf{0}}{(\absg{\absloc},\absg{\absloc})}{(\adefvalue_k)^{k \leq \loc},\absg{\absloc}}{(\bot)^*}{0^*}  \mid\mathsf{c'} \in \abslookup(\threadrun) \wedge \sign(\mathsf{c'},\threadrun) = \methsign{\thread}{\void}{\loc}\}\\
\rulename{AbState} & = & \{ \absabnormal{\apc}{\_ }{\absval^*}{\abslh}{\absfi}  \wedge \rlookup{\excpt}{\absval^*}{\abslh}{\_}{\absobj{\mathsf{c'}}{\_}} \wedge \mathsf{c'} \le \throwable  \implies\\
&& \absreg{\apc'}{\_}{\absval^*}{\abslh}{\absfi}~|~ \excpttable{\apc}{\mathsf{c'}} = \spc'\} \cup \hfill\mathbf{(A)}\\
&& \{ \absabnormal{\apc}{\_ }{\absval^*}{\abslh}{\absfi}  \wedge \rlookup{\excpt}{\absval^*}{\abslh}{\_}{\absobj{\mathsf{c'}}{\_}}\wedge \mathsf{c'} \le \throwable  \implies\\
&& \absuncaught{\mathsf{c},\mathsf{m}}{\_}{\absval_\excpt}{\abslh}{\absfi}~|~ \excpttable{\apc}{\mathsf{c'}} = \bot\}\hfill\mathbf{(B)}
\end{array}
\]
\caption{\label{tab:global-rules-excerpt} Global Rules of the Abstract Semantics - Excerpt}
\end{table*}

The abstract counterpart of the global reduction rules is a set of Horn clauses over-approximating system events and the Android activity life-cycle. We extended the original rules of HornDroid~\cite{CalzavaraGM16} with some new rules needed to support our richer concrete semantics including threads and exceptions. Table~\ref{tab:global-rules-excerpt} shows two of these rules to exemplify, the other rules 
\iffull
are in Appendix~\ref{sec:abs-sem}. 
\else
can be found in~\cite{full-version}.
\fi
Rule \rulename{Tstart} over-approximates the spawning of new threads by generating an abstract local state executing the $\threadrun$ method of the corresponding thread object. Rule \rulename{AbState} abstracts the mechanism by which a method recovers from an exception: part $\mathbf{(A)}$ turns an abstract abnormal state into an abstract local state if the abstraction of the exception register contains the abstract location of an object of class $c$ extending the $\throwable$ interface and if there exists an appropriate entry for exception handling in the exception table; part $\mathbf{(B)}$ is triggered if no such entry exists, and generates an abstract uncaught exception fact, which is then used in the abstract semantics of the method invocation performed by the caller.

Let $\mathcal{R}$ denote the set of all the Horn clauses defining the auxiliary facts, like $\rlookupname{i}$, plus the Horn clauses abstracting system events and the activity life-cycle. We define the translation of a program $P$ into Horn clauses, noted as $\translate{P}$, by adding to $\mathcal{R}$ the translation of the individual statements of $P$.


\subsection{Formal Results}
The soundness of the analysis is proved by using \emph{representation functions}~\cite{NielsonNH99}: we define a function $\rfconf$ mapping each concrete configuration $\Psi$ to a set of abstract configurations over-approximating it. We then define a partial order $<:$ between abstract configurations, where $\absprog <: \absprog'$ should be interpreted as: $\absprog$ is no coarser than $\absprog'$. The soundness theorem can be stated as follows; its proof 
\iffull
is given in Appendix~\ref{sec:proof}.
\else
can be found in~\cite{full-version}.
\fi

\begin{theorem}[Global Preservation]
\label{thm:preservation}
If $\Psi \Rightarrow^* \Psi'$ under a given program $P$, then for any $\absprog_1 \in \rfconf(\Psi)$ and $\absprog_2 :> \absprog_1$ there exist $\absprog_1' \in \rfconf(\Psi')$ and $\absprog_2' :> \absprog_1'$ s.t. $\translate{P} \cup \absprog_2 \vdash \absprog_2'$.
\end{theorem}

We now discuss how a sound static taint analysis can be implemented on top of our formal result. First, we extend the syntax of concrete values as follows:
\[
\begin{array}{llcl}
\text{Taint} & \taint &::=& \public \mid \secret\\
\text{Values} & u,v & ::=& \prim^\taint \mid \ell
\end{array}
\]
The set of taints is a two-valued lattice, and we use $\taintpo$ and $\taintcup$ to denote respectively the standard ordering on taints (where $\public \taintpo \secret$) and their join. When performing unary and binary operations, taints are propagated by having the taint of the result be the join of the taints of the arguments.

\iffull
We then define the taint extraction function $\taintf{\Psi}$ which satisfies the following relations:
  \begin{multline*}
    \taintf{\Psi}(v) = \\
    \begin{cases}
      \taintcup_{i}\; \taintf{\Psi}(v_i) & \text{if } v = \ell \wedge H(\ell) = \obj{c}{(f_i \mapsto v_i)^*} \\
      \taintcup_{i}\; \taintf{\Psi}(v_i) & \text{if } v = \ell \wedge H(\ell) = \arr{\tau}{v^*} \\
      \taintcup_{i}\; \taintf{\Psi}(v_i) & \text{if } v = \ell \wedge H(\ell) = \intent{c}{(k_i \mapsto v_i)^*} \\
      \taint & \text{if } v = \prim^\taint
    \end{cases}
  \end{multline*}
Informally, given a value $v$, it extracts its taint by doing a recursive computation: if $v$ is a primitive value this is straightforward; if $v$ is a pointer it recursively computes the join of all the taint accessible from $v$ in the heap of $\Psi$. 

We describe in Table~\ref{tab:abstract-taint} the abstract counter-part of $\taintf{\Psi}$: intuitively $\ataintf{\absval}{\abslh}{\ataint}$ holds when $\absval$ has taint $\ataint$ in the abstract local heap $\abslh$. The rules defining $\ataintfname$ are similar to the rules defining $\cfiltername$, since both predicate need to perform a fix-point computation in the abstract heap.

\begin{table*}[htp]
  \begin{mathpar}
    \ataintf{\widehat{\prim^\taint}}{\abslh}{\taint}
        
    \ataintf{\absual}{\abslh}{\ataint} \wedge \absual \abspo \absval \implies \ataintf{\absval}{\abslh}{\ataint}
    
    \ataintf{\absval}{\abslh}{\ataint} \wedge \ataintf{\absval}{\abslh}{\ataint'} \implies \ataintf{\absval}{\abslh}{\ataint \taintcup \ataint'}

    \rlookup{0}{\absual}{\abslh}{\_}{\absblock} \wedge
    \left\{
      \begin{array}{r}
        \absblock = \absobj{c}{\_, f \mapsto \absval} \\
        \absblock = \absarray{\tau}{\absval}\\
        \absblock = \absintent{c}{\absval}
      \end{array} \right\}
    \wedge \ataintf{\absval}{\abslh}{\ataint}
    \implies \ataintf{\absual}{\abslh}{\ataint}
  \end{mathpar}
  \caption{\label{tab:abstract-taint} Horn Clauses Rules used to Derive $\ataintf{\absval}{\abslh}{\ataint}$.}
\end{table*}

\else

We then define a taint extraction function $\taintf{\Psi}$: informally it is a function that, given a value $v$, extracts its taint by doing a recursive computation: if $v$ is a primitive value, the function just returns the taint of the value; if $v$ is a location, the function recursively computes the join of all the taints accessible from $v$ along the heap of $\Psi$.

We also define the abstract counter-part $\ataintfname$ of $\taintf{\Psi}$: the analysis fact $\ataintf{\absval}{\abslh}{\ataint}$ holds when $\absval$ has taint $\ataint$ in the abstract \flowsensitive heap $\abslh$. The rules defining $\ataintfname$ are similar to the rules defining $\cfiltername$, since both predicates need to perform a recursive computation on the abstract heap. The formal definitions underlying this intuitive description can be found in the long version~\cite{full-version}.
\fi

Finally, we assume two sets \sinks{} and \sources, where \sinks{} (resp. \sources) contains a pair ($c$, $m$) if and only if a method $m$ of a class $c$ is a sink (resp. a source). We assume that when a source returns a value, it always has the $\secret$ taint.

\begin{definition}
\label{def:leak}
A program $P$ \emph{leaks} starting from a configuration $\Psi$ if there exists $(c,m) \in \sinks$ such that $\Psi \Rightarrow^* \tactconf{\actstack}{\threadpool}{\heap}{\sheap}$ and there exists $\tactframe{\ell}{s}{\pi}{\threadstack}{\callstack} \in \actstack$ or $\threadframe{\ell}{\ell'}{\pi}{\threadstack}{\callstack} \in \threadpool$ such that $\callstack = \locstate{c,m,0}{\stm^*}{R}{u^*} :: \callstack'$, $R(r_k) = v$ and $\taintf{\Psi}(v) = \secret$ for some $r_k$ and $v$. 
\end{definition}

We then state the soundness of our taint tracking analysis in the following lemma:
\iffull
 its proof can be found in Section~\ref{subsec:taint-tracking}.
\else
 its proof is rather simple and can be found in~\cite{full-version}.
\fi

\begin{lemma}
\label{lem:no-leak}
If for all sinks $(c, m) \in \sinks{}$, $\absprog \in \rfconf(\Psi)$:
\[
 \translate{P} \cup \absprog \vdash  \absreg{{c,m,0}}{\_}{\absval^*}{\abslh}{\absfi} \wedge \ataintf{\absval_i}{\abslh}{\secret}
\] 
is unsatisfiable for each $i$, then $P$ does not leak from $\Psi$.
\end{lemma}

%


\section{Experiments}
\label{sec:experiments}
We implemented a prototype of our flow-sensitive analysis as an extension of an existing taint tracker, \tool~\cite{CalzavaraGM16}. Our tool encodes the application to analyse as a set of Horn clauses, as we detailed in the previous section, and then uses the SMT solver Z3~\cite{MouraB08} to statically detect information leaks. More specifically, the tool automatically generates a set of queries for the analysed application based on a public database of Android sources and sinks~\cite{RasthoferAB14}; if no query is satisfiable according to Z3, no information leak may occur by the soundness results of our analysis.

\subsection{Testing on DroidBench}
We tested our flow-sensitive extension of HornDroid (called fsHornDroid) against DroidBench~\cite{ArztRFBBKTOM14}, a common benchmark of 115 small applications proposed by the research community to test information flow analysers for Android\footnote{We removed from DroidBench 4 applications testing implicit information flows, since none of the available tools aims at supporting them.}. In our experiments we compared with the most popular and advanced static taint trackers for Android applications: FlowDroid~\cite{ArztRFBBKTOM14}, AmanDroid~\cite{WeiROR14}, DroidSafe~\cite{GordonKPGNR15} and the original version of HornDroid~\cite{CalzavaraGM16}. For all the tools, we computed standard validity measures (sensitivity for soundness and specificity for precision) and we tracked the analysis times on the 115 applications included in DroidBench: the experimental results are summarised in Table~\ref{tab:spec-sens}.

\begin{table*}[htb]
\textbf{Validity Measures on DroidBench:}
\begin{center}
\begin{tabular}{c|c|c|c|c||c|}
\cline{2-6}
& FlowDroid & AmanDroid & DroidSafe & HornDroid & fsHornDroid \\
\cline{1-6}
\multicolumn{1}{|c|}{\emph{Sensitivity}} & 0.67 & 0.74 & 0.92 & 1 &
  1 \\ 
\cline{1-6}
\multicolumn{1}{|c|}{\emph{Specificity}} & 0.58 & 0.74 & 0.47 & 0.68 &
  0.79\\
\cline{1-6}
\multicolumn{1}{|c|}{\emph{F-Measure}} & 0.62 & 0.74 & 0.62 & 0.81 & 0.88 \\
\cline{1-6}
\end{tabular}
\end{center}
\hspace*{31pt} \emph{Sensitivity} = $tp / (tp + fn)$ $\sim$ Soundness \\
\hspace*{31pt} \emph{Specificity} = $tn / (tn + fp)$ $\sim$ Precision \\
\hspace*{31pt} \emph{F-Measure} = $2 * (sens * spec) / (sens + spec)$ $\sim$ Aggregate \\

\textbf{Analysis Times on DroidBench:}
\begin{center}
\begin{tabular}{c|c|c|c|c||c|}
\cline{2-6}
& FlowDroid & AmanDroid & DroidSafe & HornDroid & fsHornDroid \\
\cline{1-6}
\multicolumn{1}{|c|}{\emph{Average}} & 22s & 11s & 2m92s & 1s & 14s \\ 
\cline{1-6}
\multicolumn{1}{|c|}{\emph{1st Quartile}} & 13s  & 9s  & 2m38s  & 1s  & 1s  \\
\cline{1-6}
\multicolumn{1}{|c|}{\emph{2nd Quartile}} & 14s & 10s & 3m1s & 1s & 2s \\
\cline{1-6}
\multicolumn{1}{|c|}{\emph{3rd Quartile}} & 15s  & 11s & 3m26s  & 1s  & 5s  \\
\cline{1-6}
\end{tabular}
\end{center}
\caption[Validity Measures and Analysis Times on DroidBench]{Validity Measures and Analysis Times on DroidBench}
\label{tab:spec-sens}
\end{table*}

Like the original version of HornDroid, fsHornDroid detects all the information leaks in DroidBench, since its sensitivity is 1. However, fsHornDroid turns out to be the most precise static analysis tool to date, with a value of specificity which is strictly higher than the one of all its competitors. In particular, fsHornDroid produces only 4 false positives on DroidBench: a leak inside an exception that is never thrown; a leak inside an unregistered callback which cannot be triggered; a leak inside an undeclared activity which cannot be started; and a leak of a public element of a list which contains also a confidential element. The last two cases should be easy to fix: the former by parsing the application manifest and the latter by implementing field-sensitivity for lists.

We also evaluated the analysis times of the applications in DroidBench for the different tools. In terms of performances, the original version of HornDroid is better than fsHornDroid as expected. However, the performances of fsHornDroid are satisfying: the median analysis time does not change too much with respect to HornDroid, which is the fastest tool, while the average analysis time is comparable with other flow-sensitive analysers like FlowDroid and AmanDroid.

\subsection{Testing on Real Applications}
In order to test the scalability of fsHornDroid, we picked the top 4 applications from 16 categories in a publicly available snapshot of the Google Play market~\cite{AndroidApps}. For each application, we run fsHornDroid setting a timeout of 3 hours for finding the first information leak. In the end, we managed to get the analysis results within the timeout for 62 applications, whose average and median sizes were 7.4 Mb and 5 Mb respectively. The tool reported 47 applications as leaky and found no direct information leaks for 15 applications. Unfortunately, the absence of a ground truth makes it hard to evaluate the validity of the reported leaks, which we plan to manually investigate in the future. To preliminarily assess the improvement in precision due to flow-sensitivity, however, we sampled 3 of the potentially leaky applications and we checked all their possible information leaks. On these applications, fsHornDroid eliminated 17 false positives with respect to HornDroid, which amount to the 18\% of all the checked flows.

In terms of performances, fsHornDroid spent 17 minutes on average to perform the analysis, with a median analysis time of 2 minutes on an Intel Xeon E5-4650L 2.60 GHz. The constantly updated experimental evaluation is available online, along with the web version of the tool and its sources~\cite{website}. Our results demonstrate that fsHornDroid scales to real applications, despite the increased performance overhead with respect to the original HornDroid.  

\subsection{Limitations} 
Our implementation of fsHornDroid does not aim at solving a few important limitations of HornDroid. First, a comprehensive implementation of \emph{analysis stubs} for unknown methods is missing: this issue was thoroughly discussed by the authors of DroidSafe~\cite{GordonKPGNR15} and we think their research may be very helpful to improve on this. Moreover, the analysis does not capture \emph{implicit} information flows, but only direct information leaks, and it does not cover native code, but only Dalvik bytecode. Finally, the analysis has no way of being less conservative on \emph{intended} information flows: implementing declassification mechanisms would be important to analyse real applications without raising a high number of false alarms.


\section{Related Work}
\label{sec:related}
There are several static information flow analysers for Android applications (see, e.g.,~\cite{YangY12,ZhaoO12,MannS12,GiblerCEC12,Kim12,ArztRFBBKTOM14,WeiROR14,GordonKPGNR15,CalzavaraGM16}). We thoroughly compared with the current state of the art in the rest of the paper, so we focus here on other related works.

\paragraph{Sound Analysis of Android Applications}
The first paper proposing a formally sound static analysis of Android applications is a seminal work by Chaudhuri~\cite{Chaudhuri09}. The paper presented a type-based analysis to reason on the data-flow security properties of Android applications modeled in an idealised calculus. A variant of the analysis was implemented in a prototype tool, SCanDroid~\cite{FuchsCF09}. Unfortunately, SCanDroid is in an early prototype phase and it cannot analyse the applications in DroidBench~\cite{ArztRFBBKTOM14}. 

Sound type systems for Android applications have also been proposed in~\cite{LortzMSBSW14} to prove non-interference and in~\cite{BugliesiCS13} to prevent privilege escalation attacks. In both cases, the considered formal models are significantly less detailed than ours and the purpose of the static analyses is different. Though the framework in~\cite{LortzMSBSW14} can be used to prevent implicit information flows, unlike our approach, the analysis proposed there is not fully automatic, it does not approximate runtime value, thus sacrificing precision, and it was not experimentally evaluated. 

Julia is a static analysis tool based on abstract interpretation, first developed for Java and recently extended to Android~\cite{PayetS12}. It is a commercial product and supports many useful features, including class analysis, nullness analysis and termination analysis for Android applications, but it does not track information flows. Moreover, Julia does not handle multi-threading and we are not aware of the existence of a soundness proof for its extension to Android.

\paragraph{Pointer Analysis}
Pointer analysis aims at over-approximating the set of objects that a program variable can refer to, and it is a well-established and rich research field~\cite{DBLP:journals/corr/KanvarK14,Sridharan:2013:AAO:2554511.2554523,Smaragdakis:2015:PA:2802194.2802195}. The  most prominent techniques in pointer analysis are variants of the classical Andersen algorithm~\cite{Andersen94programanalysis}, including flow-insensitive analyses~\cite{Das:2000:UPA:358438.349309,DBLP:conf/cgo/PereiraB09,Hardekopf:2007:AGF:1273442.1250767,Kastrinis:2013:HCP:2499370.2462191} and flow-sensitive analyses~\cite{Choi:1993:EFI:158511.158639,Emami:1994:CIP:773473.178264,Kahlon:2008:BTS:1379022.1375613,Lhotak:2011:PAE:1925844.1926389}; light-weight analyses in the flavor of the unification-based Steensgaard analysis~\cite{Steensgaard:1996:PAA:237721.237727}, which are flow-insensitive and very efficient; and shape analysis techniques~\cite{Sagiv:1999:PSA:292540.292552}, which can be used to prove complex properties about the heap, often at the price of efficiency. 

Although pointer analysis of sequential programs is well-studied, much less attention has been paid to pointer analysis of concurrent programs. Most flow-insensitive analyses for sequential programs remain sound for concurrent programs~\cite{Rugina:1999:PAM:301631.301645}, because flow-insensitivity forces a sound analysis to consider all the possible interleavings of reads and writes to the heap. Designing a sound flow-sensitive pointer analysis for concurrent programs is more complicated and most flow-sensitive analyses for sequential programs cannot be easily adapted to concurrent programs. Still, flow-sensitive sound analyses for concurrent programs exist. The approach of Rugina and Rinard~{\cite{Rugina:1999:PAM:301631.301645} handles concurrent programs with an unbounded number of threads, recursion and dynamic allocations, but it does not allow strong updates on dynamically allocated heap objects. Gotsman \textit{et al.}~\cite{export:73587} proposed a framework to prove complex properties about programs with dynamic allocations by using shape analysis and separation logic, but their approach requires  users or external tools to provide annotations, and it is restricted to a bounded number of threads.


\section{Conclusion}
We presented the first static analysis for Android applications which is both flow-sensitive on the heap abstraction and provably sound with respect to a rich formal model of the Android ecosystem. Designing a sound yet precise analysis in this setting is particularly challenging, due to the complexity of the control flow of Android applications. In this work, we adapted ideas from \emph{recency abstraction}~\cite{Balakrishnan:2006:RHS:2090874.2090894} to hit a sweet spot in the analysis design space: our proposal is sound, precise, and efficient in practice. We substantiated these claims by implementing the analysis in HornDroid~\cite{CalzavaraGM16}, a state-of-the-art static information flow analyser for Android applications, and by performing an experimental evaluation of our extension. Our work takes HornDroid one step further towards the sound information flow analysis of real Android applications. 


\paragraph*{Acknowledgements}
This work has been partially supported by the MIUR project ADAPT, by the CINI Cybersecurity National Laboratory within the project FilieraSicura: Securing the Supply Chain of Domestic Critical Infrastructures from Cyber Attacks (www.filierasicura.it) funded by CISCO Systems Inc. and Leonardo SpA, and by the German Federal Ministry of Education and Research (BMBF) through the Center for IT-Security, Privacy and Accountability (CISPA). This work also acknowledges support by the FWF project W1255-N23 and the DAAD-MIUR Joint Mobility Program ``Client-side Security Enforcement for Mobile and Web Applications''.

\bibliographystyle{splncs03}
\bibliography{local}

\iffull
\clearpage
\newgeometry{lmargin=2cm,rmargin=2cm,tmargin=2.5cm,bmargin=2.5cm}
\onecolumn
\appendices

\paragraph{Appendix outline:} In Section~\ref{sec:concrete} we give the small-step semantics of the local states reduction for the Dalvik bytecode, as well as the reduction rules for activities and threads; in Section~\ref{sec:abs-sem} we give the full abstract semantics\iffull; in Section~\ref{sec:proof} we give the soundness proof\fi.

\section{Concrete Semantics}
\label{sec:concrete}
As in~\cite{CalzavaraGM16}, we require that Dalvik programs are \emph{well-formed}.

\begin{definition}[Well-formed Program~\cite{CalzavaraGM16}]
A program $P$ is \emph{well-formed} iff all its class names are pairwise distinct and, for each of its classes, all the field names and the method names are pairwise distinct.
\end{definition}

From now on, we always consider a fixed well-formed program $P = \class^*$. We give in Table~\ref{table:add-stm} the syntax and an informal explanation of the Dalvik statements that were omitted in the body. The extensions with respect to \cite{CalzavaraGM16} are in bold.

\subsection{Extensions : Waiting Sets and Monitors}

\begin{table}[b]
\[\begin{array}{ll}
 \sinvoke{c}{m}{r^*} & \text{invoke the static method $m$ of the class $c$ with args $r^*$} \\
 \checkcast{r_s}{\tau} & \text{jump to the next statement if the value of $r_s$ has type $\tau$}\\
 \instanceof{r_d}{r_s}{\tau} & \text{put $\true$ in $r_d$ iff the value of $r_s$ has type $\tau$}\\
 \interruptedthreadbf{r_t}& \text{read and reset the interrupt field of the thread in $r_t$}\\
 \isinterruptedthreadbf{r_t}& \text{read the interrupt field of the thread in $r_t$}\\
 \monitorenterbf{r_o}& \text{acquire the monitor of the object in $r_o$}\\
 \monitorexitbf{r_o}& \text{release the monitor of the object in $r_o$}\\
 \startwaitbf{r_o}& \text{enter the waiting set of the object in $r_o$}
\end{array}\]
\caption{\label{table:add-stm} Syntax and Informal Semantics of Additional Statements}
\end{table}

In order to give a full account of Java concurrency we extended our model to include waiting sets and monitors~\cite{JavaObject}, as well as two other interrupting methods of the Java $\thread$ API. We start by extending the concrete semantics to handle the $\startwait{}$ statement: we introduce a new semantic domain for waiting states and extend the local state lists domain: we use a special type of state, called \emph{waiting state} and denoted by $\omega = \waiting{j}{\ell}$, to model that the thread running the method is currently waiting on some object stored at location $\ell$; the integer parameter $j$ stores how many times the object monitor was acquired prior to entering the waiting state. A \emph{local state list} $L^\#$ is now a list of local states and waiting states. Since a thread entering a waiting state is paused until it is ready to resume its execution, we assume that a local state list never contains more than one waiting state. Moreover, we assume this waiting state is always the head of the local state list (if present). 
\[\begin{array}{lllcl}
\text{Waiting states} & & \omega &\define & \waiting{\ell}{j}\\
\text{Local state lists} & & L^\# &\define & \varepsilon ~|~ L :: L^\# ~|~  \omega:: L^\#\\
\end{array}\]

\paragraph{Statements Description}  A monitor is a synchronization construct attached to an object, which can be acquired and released by threads, but cannot be acquired by more than one thread at once. Any thread holding an object monitor can start waiting on the object: this makes the thread enter the object waiting set, release the monitor, and pause until it is woken-up, notified or interrupted by another thread. Since we do not model timing aspects in our formalism and \emph{spurious} wake-ups may happen in practice, we make the conservative assumption that waiting threads can non-deterministically wake up at any time. Moreover, we assume that all objects contain two special fields: the $\monitor$ field storing the location of the thread currently holding the object monitor, and the $\monitorcounter$ field counting the number of monitor acquisitions. These fields can only be accessed by the monitor and wait rules.

When $\monitorenter{r_o}$ is called, there are two possibilities. If the $\monitorcounter$ field of the monitor of the object whose location is stored in $r_o$ is set to 0, it is immediately set to 1 and the corresponding $\monitor$ field is set to the location of the acquiring thread. Otherwise, we check that the $\monitor$ field points to the location of the acquiring thread: if this is the case, the $\monitorcounter$ field is incremented by 1 to reflect the presence of multiple acquisitions. A monitor is released only when all its acquisitions have been released via the statement $\monitorexit{r_o}$, which checks that the running thread holds the monitor of the object whose location is stored in $r_o$ and decrements the monitor counter $\monitorcounter$ by 1.

The statement $\startwait{r_o}$ checks that the running thread holds the monitor of the object $o$ whose location is stored in $r_o$, releases the monitor and pushes on the call stack a waiting state $\waiting{\ell}{j}$, where $\ell$ is the location of $o$ and $j$ tracks how many times the released monitor was acquired before calling $\startwait{r_o}$. An uninterrupted thread can exit a waiting state and reacquire back the released monitor $j$ times, provided that the monitor is not held by another thread. If a thread in a waiting state gets interrupted, an $\interruptedexception$ exception is thrown, the thread wakes up and starts recovering from the exception.

Finally   $\interruptedthread{r_t}$ and $\isinterruptedthread{r_t}$ are simple write or read operations on the interrupt field ($\interrupted$) of the thread object whose location is stored in $r_t$.

\subsection{Local Reduction Relation}

\subsubsection{Type System}

Local registers are untyped in Dalvik, and have default value $\defvalue$. We also assume that for all type $\tau$, there exists a default value $\defvalue_{\tau}$ that will be used for field initialization. Before giving the concrete semantics of the Dalvik bytecode, we need some definitions. 
First we define a function $\gettype{\heap}{v}$ that retrieve from the heap $\heap$ the type of the memory block $v$ is pointing to.
\begin{definition}
Given a heap $\heap$, we let the partial function $\gettype{\heap}{v}$ be defined as follows:
\[
\gettype{\heap}{v} =
\begin{cases}
c & \text{if } v = \ell \wedge H(\ell) = \obj{c}{(f \mapsto v)^*} \\
\arrtype{\tau} & \text{if } v = \ell \wedge H(\ell) = \arr{\tau}{v^*} \\
\texttt{Intent} & \text{if } v = \ell \wedge H(\ell) = \intent{c}{(k \mapsto v)^*} \\
\primtype & \text{if } v = \prim 
\end{cases}
\]
where $\primtype$ is the type of the primitive value $\prim$.
\end{definition}

Given a class name $c$, we let $\super(c) = c'$ if there exists a class $\class_i$ such that $\class_i = \cls{c}{c'}{c^*}{\field^*}{\method^*}$, and $\interfaces(c) = \{c^*\}$ iff there exists a class $\class_i$ such that $\class_i = \cls{c}{c'}{c^*}{\field^*}{\method^*}$. The subtyping relation is quite simple: a class $c$ is a subclass of its super class $\super(c)$ and of the interfaces $\interfaces(c)$ it implements (plus reflexive and transitive closure). There is also a co-variant subtyping rule for array, which is unsound in presence of side-effects (types are checked dynamically at run-time to avoid errors). The typing rules are summarized below.
\begin{mathpar}
\inferrule[(Sub-Refl)]
{ }
{\tau \subtype \tau}

\inferrule[(Sub-Trans)]
{\tau \subtype \tau' \\ \tau' \subtype \tau''}
{\tau \subtype \tau''}

\inferrule[(Sub-Ext)]
{ }
{c \subtype \super(c)}

\inferrule[(Sub-Impl)]
{c' \in \interfaces(c)}
{c \subtype c'}

\inferrule[(Sub-Array)]
{\tau \subtype \tau'}
{\arrtype{\tau} \subtype \arrtype{\tau'}}
\end{mathpar}

\subsubsection{Right-Hand Side Evaluation}

Let $a[i] = v_i$ whenever $a = \arr{\tau}{v^*}$ and $o.f = v$ whenever $o = \obj{c}{(f_i \mapsto v_i)^*,f \mapsto v}$. We define in Table~\ref{tab:rhs-eval} the relation $\regval{\rhs}$ that evaluates a right-hand side expression in a given local configuration $\Sigma$.
\begin{table}[htb]
\begin{mathpar}
\inferrule*[lab=(Rhs-Register)]
{ }
{\regval{r} = R(r)}

\inferrule*[lab=(Rhs-Array),width=10em]
{\ell = \regval{r_a} \\ a = \heap(\ell) \\ 
j = \regval{r_{\ind}}}
{\regval{r_a[r_{\ind}]} = a[j]}

\inferrule*[lab=(Rhs-Object),width=10em]
{\ell = \regval{r_o} \\ o = \heap(\ell)}
{\regval{r_o.f} = o.f}

\inferrule*[lab=(Rhs-Static)]
{ }
{\regval{c.f} = \sheap(c.f)}

\inferrule*[lab=(Rhs-Prim)]
{ }
{\regval{\prim} = \prim}
\end{mathpar}
\textbf{Convention:} in all the rules, let $\Sigma = \tmethconf{\callstack_c}{\pi}{\threadstack}{\heap}{\sheap}{\ell_r}$ with $\callstack_c = \locstate{\pp}{\stm^*}{R}{\_} :: \callstack'$ or $\callstack_c = \abnormal{\locstate{\pp}{\stm^*}{R}{\_} :: \callstack'}$.
\caption{\label{tab:rhs-eval} Evaluation of Right-hand Sides ($\regval{\rhs} = v$)}
\end{table}

\subsubsection{Instruction Fetching} We recall that the definition of the local reduction relation uses an auxiliary relation $\Sigma,\stm \Downarrow \Sigma'$, which means that the execution of the statement $\stm$ in $\Sigma$ produces $\Sigma'$. The simplest rule defining a local reduction $\Sigma \rightsquigarrow \Sigma'$ just fetches the next statement $\stm$ to run and performs a look-up on the auxiliary relation $\Sigma,\stm \Downarrow \Sigma'$. Formally:
\[
\inferrule[(R-NextStm)]
{\Sigma, \getst{\Sigma} \Downarrow \Sigma'}
{\Sigma \rightsquigarrow \Sigma'}\]

 We are finally ready to give the semantics of the Dalvik bytecode relation: the standard operation are in Table~\ref{tab:semantics}, while the new operations are given in Table~\ref{tab:exception-semantics}

\begin{table}[p]
\fontsize{8pt}{10pt}
\begin{mathpar}
\inferrule*[lab=(R-Goto)]
{ }
{\Sigma, \goto{\pc'} \Downarrow \Sigma[\pc \mapsto \pc']} 

\inferrule*[lab=(R-True)]
{\regval{r_1} \comp \regval{r_2}}
{\Sigma, \ifbr{r_1}{r_2}{\pc'} \Downarrow \Sigma[\pc \mapsto \pc']}

\inferrule*[lab=(R-False)]
{\neg(\regval{r_1} \comp \regval{r_2})}
{\Sigma, \ifbr{r_1}{r_2}{\pc'} \Downarrow \Sigma^+}

\inferrule*[width=25em,lab=(R-MoveReg)]
{v = \regval{\rhs} \\ R' = R[r \mapsto v]}
{\Sigma, \move{r}{\rhs} \Downarrow \Sigma^+[R \mapsto R']} 

\inferrule*[width=32em,lab=(R-MoveFld)]
{v = \regval{\rhs} \\ 
\ell = \regval{r_o} \\
o = \heap(\ell) \\ 
\heap' = \heap[\ell \mapsto o[f \mapsto v]]}
{\Sigma, \move{r_o.f}{\rhs} \Downarrow \Sigma^+[\heap \mapsto \heap']} 

\inferrule*[width=32em,lab=(R-MoveArr)]
{v = \regval{\rhs} \\ \ell = \regval{r_a} \\
\gettype{\heap}{\ell} = \arrtype{\tau} \\
\gettype{\heap}{v} \subtype \tau \\ 
a =  \heap(\ell) \\ j = \regval{r_\ind} \\
\heap' = \heap[\ell \mapsto a[j \mapsto v]] }
{\Sigma, \move{r_a[r_\ind]}{\rhs} \Downarrow \Sigma^+[\heap \mapsto \heap']}

\inferrule*[width=25em,lab=(R-MoveSFld)]
{v = \regval{\rhs} \\ \sheap' = \sheap[c'.f \mapsto v]}
{\Sigma, \move{c'.f}{\rhs} \Downarrow \Sigma^+[\sheap \mapsto \sheap']} 

\inferrule*[width=25em,lab=(R-UnOp)]
{v = \odot \regval{r_s} \\ R' = [r_d \mapsto v]}
{\Sigma, \unop{r_d}{r_s} \Downarrow \Sigma^+[R \mapsto R']}

\inferrule*[width=25em,lab=(R-BinOp)]
{v = \regval{r_1} \oplus \regval{r_2} \\ R' = R[r_d \mapsto v] }
{\Sigma, \binop{r_d}{r_1}{r_2} \Downarrow \Sigma^+[R \mapsto R']} 

\inferrule*[width=25em,lab=(R-NewObj)]
{o =  \obj{c'}{(f_{\tau} \mapsto \defvalue_{\tau})^*} \\ 
\ell = \pointer{p}{c,m,\pc} \notin \dom(\heap) \\
\heap' = \heap[\ell \mapsto o] \\ R' = R[r_d \mapsto \ell] }
{\Sigma, \new{r_d}{c'} \Downarrow \Sigma^+[\heap \mapsto \heap', R \mapsto R']}

\inferrule*[width=30em,lab=(R-NewArr)]
{\size = \regval{r_l} \\ 
a = \arr{\tau}{(\defvalue_{\tau})^{j \leq \size}} \\
\ell = \pointer{p}{c,m,\pc} \notin \dom(\heap) \\ 
\heap' = \heap[\ell \mapsto a] \\ 
R' = R[r_d \mapsto \ell]}
{\Sigma, \newarray{r_d}{r_l}{\tau}  \Downarrow \Sigma^+[\heap \mapsto \heap', R \mapsto R']}

\inferrule*[width=25em,lab=(R-Cast)]
{\ell = \regval{r_s} \\ \gettype{\heap}{\ell} \subtype \tau}
{\Sigma, \checkcast{r_s}{\tau} \Downarrow \Sigma^+}

\inferrule*[width=25em,lab=(R-InstOfTrue)]
{\ell = \regval{r_s} \\ \gettype{\heap}{\ell} \subtype \tau \\
R' = R[r_d \mapsto \true]}
{\Sigma, \instanceof{r_d}{r_s}{\tau} \Downarrow \Sigma^+[R \mapsto R']}

\inferrule*[width=25em,lab=(R-InstOfFalse)]
{\ell = \regval{r_s} \\ \gettype{\heap}{\ell} \not\subtype \tau \\
R' = R[r_d \mapsto \false]}
{\Sigma, \instanceof{r_d}{r_s}{\tau} \Downarrow \Sigma^+[R \mapsto R']}

\inferrule*[width=25em,lab=(R-Return)]
{\callstack = \locstate{c,m,\pc}{\_}{R}{\_} :: \locstate{c',m',\pc'}{\stm^*}{R'}{v^*} :: \callstack_0 \\
\callstack'' = \locstate{c',m',\pc'+1}{\stm^*}{R'[r_{\res} \mapsto \regval{r_{\res}}]}{v^*} :: \callstack_0}
{\Sigma, \return \Downarrow \Sigma[\callstack \mapsto \callstack'']}

\inferrule*[width=35em,lab=(R-SCall)] 
{\lookup(c',m') = (c',\stm^*) \\ 
\sign(c',m') = \methsign{\tau_1,\ldots,\tau_n}{\tau}{\loc} \\
R' = ((r_j \mapsto \defvalue)^{j \leq \loc}, (r_{\loc+k} \mapsto \regval{r_k'})^{k \leq n}) \\ 
\callstack'' = \locstate{c',m',0}{\stm^*}{R'}{(\regval{r_k'})^{k \leq n}} :: \callstack}
{\Sigma, \sinvoke{c'}{m'}{r_1',\ldots,r_n'} \Downarrow \Sigma[\callstack \mapsto \callstack'']}

\inferrule*[width=55em,lab=(R-Call)]
{\ell = \regval{r_o} \\ \lookup(\gettype{\heap}{\ell},m') = (c',\stm^*) \\
\sign(c',m') = \methsign{\tau_1,\ldots,\tau_n}{\tau}{\loc} \\
R' = ((r_j \mapsto \defvalue)^{j \leq \loc}, r_{\loc + 1} \mapsto \ell, (r_{\loc+1+k} \mapsto 
\regval{r_k'})^{k \leq n}) \\ 
\callstack'' = \locstate{c',m',0}{\stm^*}{R'}{(\regval{r_k'})^{k \leq n}} :: \callstack}
{\Sigma, \invoke{r_o}{m'}{r_1',\ldots,r_n'} \Downarrow \Sigma[\callstack \mapsto \callstack'']}

\inferrule*[width=25em,lab=(R-NewIntent)]
{i =  \intent{c'}{\cdot} \\ 
\ell = \pointer{p}{c,m,\pc} \notin \dom(\heap) \\
\heap' = \heap[\ell \mapsto i] \\ R' = R[r_d \mapsto \ell]}
{\Sigma, \newintent{r_d}{c'} \Downarrow \Sigma^+[\heap \mapsto \heap', R \mapsto R']}

\inferrule*[width=30em,lab=(R-PutExtra)]
{\ell = \regval{r_i} \\ i = \heap(\ell) \\
k = \regval{r_k} \\ v = \regval{r_v} \\
\heap' = \heap[\ell \mapsto i[k \mapsto v]] }
{\Sigma, \putextra{r_i}{r_k}{r_v} \Downarrow \Sigma^+[\heap \mapsto \heap']}

\inferrule*[width=30em,lab=(R-GetExtra)]
{\ell = \regval{r_i} \\ k = \regval{r_k} \\
\heap(\ell) = i \\
\gettype{\heap}{i.k} \subtype \tau \\
v = i.k \\
R' = R[r_{\res} \mapsto v]}
{\Sigma, \getextra{r_i}{r_k}{\tau} \Downarrow \Sigma^+[R \mapsto R']}

\inferrule*[width=25em,lab=(R-StartAct)]
{\ell = \regval{r_i} \\ H(\ell) = i \\
\pi' = i :: \pi }
{\Sigma, \startact{r_i} \Downarrow \Sigma^+[\pi \mapsto \pi']}
\end{mathpar}

\textbf{Convention:} let $\pp = c,m,\pc$ and let $\Sigma = \tmethconf{\callstack}{\pi}{\threadstack}{\heap}{\sheap}{\_}$ with $\callstack = \locstate{c,m,\pc}{\_}{R}{\_} :: \callstack'$. We recall that $\Sigma^+$ stands for $\Sigma$ where $\pc$ is replaced by $\pc + 1$.
\caption{\label{tab:semantics} Small step semantics of \sem{} - Standard Statements}
\end{table}

\begin{table}[p]
\fontsize{8pt}{10pt}
\textbf{Exception Rules}
\begin{mathpar}
\inferrule*[width=30em,lab=(R-Throw)]
{\ell = \regval{r_i} \\ 
\heap(\ell) = \obj{c'}{(f \mapsto v)^*}
}
{\Sigma, \throw{r_i} \Downarrow \Sigma[\callstack \mapsto \abnormal{\callstack}][r_\excpt \mapsto \ell]}

\inferrule*[width=30em,lab=(R-MoveException)]
{\ell = \regval{r_\excpt} \\
}
{\Sigma, \moveexcpt{r_d} \Downarrow \Sigma^+[r_d \mapsto \ell]}

\inferrule*[width=30em,lab=(R-Caught)]
{\ell = \aregval{r_\excpt}\\ 
\heap(\ell) = \obj{c'}{(f \mapsto v)^*}\\
\excpttable{c,m,\pc}{c'} = \pc'\\
\callstack_c = \locstate{c,m,\pc'}{\_}{R}{\_} :: \callstack'} 
{\Sigma_A \rightsquigarrow \Sigma_A[\callstack_A \mapsto \callstack_c]}

\inferrule*[width=30em,lab=(R-UnCaught)]
{\ell = \aregval{r_\excpt}\\
\heap(\ell) = \obj{c'}{(f \mapsto v)^*}\\
\excpttable{c,m,\pc}{c'} = \bot}
{\Sigma_A \rightsquigarrow \Sigma_A[\callstack_A \mapsto \abnormal{\callstack'}][r_\excpt \mapsto \ell]}
\end{mathpar}
\textbf{Thread Rules}
\begin{mathpar}
\inferrule*[width=32em,lab=(R-StartThread)]
{\ell = \regval{r_i} \\ \heap(\ell) = \threadobj{c'}{(f \mapsto v)^*} \\
\threadstack' = \ell :: \threadstack }
{\Sigma, \startthread{r_i} \Downarrow \Sigma^+[\threadstack \mapsto \threadstack']}

\inferrule*[width=30em,lab=(R-InterruptThread)]
{\ell = \regval{r_i} \\ \heap(\ell) = \threadobj{c'}{(f \mapsto v)^*,\interrupted \mapsto \_} \\
\heap' = \heap[\ell \mapsto \threadobj{c'}{(f \mapsto v)^*,\interrupted \mapsto \true}]
}
{\Sigma, \interruptthread{r_i} \Downarrow \Sigma^+[\heap \mapsto \heap']}

\inferrule*[width=30em,lab=(R-InterruptedThread)]
{\ell = \regval{r_i} \\ \heap(\ell) = \threadobj{c'}{(f \mapsto v)^*,\interrupted \mapsto u} \\
\heap' = \heap[\ell \mapsto \threadobj{c'}{(f \mapsto v)^*,\interrupted \mapsto \false}]}
{\Sigma, \interruptedthread{r_i} \Downarrow \Sigma^+[r_\res \mapsto u, \heap \mapsto \heap']}

\inferrule*[width=30em,lab=(R-IsInterruptedThread)]
{\ell = \regval{r_i} \\
 \heap(\ell) = \threadobj{c'}{(f \mapsto v)^*,\interrupted \mapsto u}
}
{\Sigma, \isinterruptedthread{r_i} \Downarrow \Sigma^+[r_\res \mapsto u]}

\inferrule*[width=25em,lab=(R-JoinThread)]
{\heap(\ell_r) = \obj{c_r}{(f_r \mapsto v_r)^*,\interrupted \mapsto \false}\\
\ell = \regval{r_i} \\ 
\heap(\ell) = \threadobj{c'}{(f \mapsto v)^*,\finished \mapsto \true}
}
{\Sigma, \jointhread{r_i} \Downarrow \Sigma^+}

\inferrule*[width=35em,lab=(R-InterruptJoin)]
{\heap(\ell_r) = \obj{c_r}{(f_r \mapsto v_r)^*,\interrupted \mapsto \true}\\
o = \obj{c_r}{(f_r \mapsto v_r)^*,\interrupted \mapsto \false}\\
\pointer{p}{c,m,\pc} \not\in \dom(H) \\
\heap' = \heap, \pointer{p}{c,m,\pc} \mapsto \obj{\interruptedexception}{} \\
\callstack_c = \abnormal{\callstack[r_\excpt \mapsto \pointer{p}{c,m,\pc}]}}
{\Sigma, \jointhread{r_i} \Downarrow \Sigma[\callstack \mapsto \callstack_c, \heap \mapsto \heap'[\ell_r \mapsto o]]}
\end{mathpar}
\textbf{Monitor and Wait Rules}
\begin{mathpar}
\inferrule*[width=30em,lab=(R-MonitorEnter1)]
{\ell = \regval{r_i} \\ 
\heap(\ell) = \obj{c'}{(f \mapsto v)^*,\monitor \mapsto \_,\monitorcounter \mapsto 0}\\
o' =  \obj{c'}{(f \mapsto v)^*,\monitor \mapsto \ell_{r},\monitorcounter \mapsto 1}}
{\Sigma, \monitorenter{r_i} \Downarrow \Sigma^+[\heap \mapsto \heap[\ell \mapsto o']]}

\inferrule*[width=30em,lab=(R-MonitorEnter2)]
{\ell = \regval{r_i} \\
\heap(\ell) = \obj{c'}{(f \mapsto v)^*,\monitor \mapsto \ell_{r},\monitorcounter \mapsto j}\\
o' =  \obj{c'}{(f \mapsto v)^*,\monitor \mapsto \ell_{r},\monitorcounter \mapsto j + 1}\\j > 0}
{\Sigma, \monitorenter{r_i} \Downarrow \Sigma^+[\heap \mapsto \heap[\ell \mapsto o']]}

\inferrule*[width=30em,lab=(R-MonitorExit)]
{\ell = \regval{r_i} \\
\heap(\ell) = \obj{c'}{(f \mapsto v)^*,\monitor \mapsto \ell_{r},\monitorcounter \mapsto j + 1}\\
o' = \obj{c'}{(f \mapsto v)^*,\monitor \mapsto \ell_{r},\monitorcounter \mapsto j}\\j \ge 0}
{\Sigma, \monitorexit{r_i} \Downarrow \Sigma^+[\heap \mapsto \heap[\ell \mapsto o']]}

\inferrule*[width=30em,lab=(R-StartWait)]
{\ell = \regval{r_i} \\
\heap(\ell) = \obj{c'}{(f \mapsto v)^*,\monitor \mapsto \ell_{r},\monitorcounter \mapsto j}\\
o' =  \obj{c'}{(f \mapsto v)^*,\monitor \mapsto \ell_{r},\monitorcounter \mapsto 0} \\j > 0}
{\Sigma, \startwait{r_i} \Downarrow \Sigma[\callstack \mapsto \waiting{\ell}{j}::\callstack,\heap \mapsto \heap[\ell \mapsto o']]}

\inferrule*[width=30em,lab=(R-StopWait)]
{\heap(\ell_r) = \obj{c_r}{(f_r \mapsto v_r)^*,\interrupted \mapsto \false}\\
\callstack = \waiting{\ell_o}{j}:: \callstack_0 \\
\heap(\ell_o) = \obj{c'}{(f \mapsto v)^*,\monitor \mapsto \_,\monitorcounter \mapsto 0}\\
o' =  \obj{c'}{(f \mapsto v)^*,\monitor \mapsto \ell_r,\monitorcounter \mapsto j}}
{\Sigma \rightsquigarrow \Sigma^+[\callstack \mapsto \callstack_0, \heap \mapsto \heap[\ell_o \mapsto o']]}

\inferrule*[width=32em,lab=(R-InterruptWait)]
{\heap(\ell_r) = \obj{c_r}{(f_r \mapsto v_r)^*,\interrupted \mapsto \true}\\
\callstack = \waiting{\_}{\_}:: \callstack_0 \\
\pointer{p}{c,m,\pc} \not\in \dom(H) \\
o = \obj{c_r}{(f_r \mapsto v_r)^*,\interrupted \mapsto \false}\\
o_e = \obj{\interruptedexception}{}}
{\Sigma \rightsquigarrow \Sigma[\callstack \mapsto \abnormal{\callstack_0[r_\excpt \mapsto \ell_e]}, \heap \mapsto \heap[\pointer{p}{c,m,\pc} \mapsto o_e,\ell_r \mapsto o]]}
\end{mathpar}
\textbf{Convention:} let $\Sigma = \tmethconf{\callstack}{\pi}{\threadstack}{\heap}{\sheap}{\ell_{r}}$ with $\callstack = \locstate{c,m,\pc}{\_}{R}{\_} :: \callstack'$ (apart when specified otherwise), and $\Sigma_A = \tmethconf{\callstack_A}{\pi}{\threadstack}{\heap}{\sheap}{\ell_{r}}$ with $\callstack_A = \abnormal{\callstack}$. We recall that $\Sigma^+$ stands for $\Sigma$ where $\pc$ is replaced by $\pc + 1$.
\caption{\label{tab:exception-semantics} Small step semantics of \sem{} - New Statements}
\end{table}


\FloatBarrier

\subsection{Global Rules Descriptions}
\subsubsection{Serialization}

All the activities running on some Android device are \emph{sand-boxed}, in order to provide some security guarantees. Inter-component communications are still allowed through the intent mechanism: activities can exchanged objects using intents, which are a special kind of object storing data in a dictionary-like structure. When an activity sends an intent to some activity, a copy of this intent is given to the receiver activity. This copying is performed by a recursive \emph{serialization} procedure, and there is therefore no object-sharing between different activities.

We model serialization using a set of derivation rules for fact of the form $\Gamma \vdash \serval{\heap}(v) = (v',\heap',\Gamma')$  and $\Gamma \vdash \serblock{\heap}(b) = (b',\heap',\Gamma')$, where $\Gamma$ and $\Gamma'$ are serialization context consisting a of list of key-value bindings of locations of the form $(\pointer{p}{\absloc} \mapsto \pointer{p'}{\absloc})$ (notice that both location have the same annotation). Serialization contexts store, for each already serialized location $\ell$, the fresh location $\ell'$ that was used to replace $\ell$. This way if the same location is encountered twice (or more) during the serialization process, it will be serialized by the same location each time. Intuitively, if $\serval{\heap}(v) = (v',\heap',\Gamma')$ (resp.  $\Gamma \vdash \serblock{\heap}(b) = (b',\heap',\Gamma')$) is derivable then $v'$ (resp. $b'$) is the serialized version of the value $v$ (resp. block $b$), $\heap'$ is the heap containing all the serialized version of the objects encountered, and $\Gamma'$ is the history of all serialized locations. We refer to Table~\ref{tab:activity-serialization} for the formal statement of the serialization rules.

\begin{table}[H]
\begin{mathpar}
\inferrule
{ }
{\serialized \vdash \serval{\heap}({\prim}) = (\prim, \cdot, \serialized)}

\inferrule
{(\pointer{p}{\ann} \mapsto \pointer{p'}{\ann}) \in \serialized}
{\serialized \vdash \serval{\heap}(\pointer{p}{\ann}) = (\pointer{p'}{\ann}, \cdot, \serialized)}

\inferrule
{\pointer{p}{\ann} \notin dom(\serialized) \\
\pointer{p'}{\ann} \text{ fresh location} \\
\serialized ,\pointer{p}{\ann} \mapsto \pointer{p'}{\ann} \vdash \serblock{\heap}(\heap(\pointer{p}{\ann})) = (b, \heap'',\serialized') \\
\heap' = \heap'',\pointer{p'}{\ann} \mapsto b}
{\serialized \vdash \serval{\heap}(\pointer{p}{\ann}) = (\pointer{p'}{\ann}, \heap',\serialized')}

\inferrule
{\serialized_0 = \serialized \\
\forall i \in [1,n]: \serialized_{i-1} \vdash \serval{\heap}(v_i) = {(u_i,\heap_i,\serialized_i)} \\
\heap' = \heap_1,\dots,\heap_n }
{\serialized \vdash \serblock{\heap}(\obj{c'}{(f_i \mapsto v_i)^{i \leq n}}) = (\obj{c'}{(f_i \mapsto u_i)^{i \leq n}}, \heap',\serialized_n)}

\inferrule
{\serialized_0 = \serialized \\
\forall i \in [1,n]: \serialized_{i-1} \vdash \serval{\heap}(v_i) = {(u_i,\heap_i,\serialized_i)} \\
\heap' = \heap_1, \ldots, \heap_n}
{\serialized \vdash \serblock{\heap}(\arr{\tau}{(v_i)^{i \leq n}}) = (\arr{\tau}{(u_i)^{i \leq n}}, \heap',\serialized_n)}

\inferrule
{\serialized_0 = \serialized \\
\forall i \in [1,n]: \serialized_{i-1} \vdash \serval{\heap}(v_i) = {(u_i,\heap_i,\serialized_i)} \\
\heap' = \heap_1, \ldots, \heap_n }
{\serialized \vdash \serblock{\heap}(\intent{c'}{(k_i \mapsto v_i)^{i \leq n}}) = (\intent{c'}{(k_i \mapsto u_i)^{i \leq n}}, \heap',\serialized_n)}
\end{mathpar}
\textbf{Conventions:} environments (denoted by $\Gamma, \Gamma' \dots$) are partial mappings from the set of all locations to itself.
\caption{\label{tab:activity-serialization} Serialization rules}
\end{table}

\subsubsection{Threads and Activities}
Before giving the global reduction relation, we need some definitions. We start by formally define what is a thread class and an activity class.
\begin{definition}
A class $\class$ is a \emph{thread class} if and only if $\class = \cls{c}{c'}{c^*}{\field^*}{\method^*}$ for some $c' \subtype \thread$. A \emph{thread} is an instance of a thread class. We stipulate that each thread implements the method $\threadrun$, has a boolean field $\interrupted$ stating whether the thread was interrupted and a boolean field $\finished$ stating whether the thread has finished or not.
\end{definition}

\begin{definition}
A class $\class$ is an \emph{activity class} if and only if $\class = \cls{c}{c'}{c^*}{\field^*}{\method^*}$ for some $c' \subtype \activity$. An \emph{activity} is an instance of an activity class. We stipulate that each activity has the following fields: (1) $\finished$: a boolean flag stating whether the activity has finished or not; (2) $\fintent$: a location to the intent which started the activity; (3) $\result$: a location to an intent storing the result of the activity computation; and (4) $\parent$: a location to the parent activity, i.e., the activity which started the present one.
\end{definition}



Each activity provides a set of \emph{event handlers} which are callbacks methods used to respond to user inputs: for all activity class $c$, let  $\handlers(c) = \{m_1,\ldots,m_n\}$ be the set of callback method names of $c$. We model the activity life-cycle (see \cite{PayetS14}) by a set of activity states $\actstates$ and a transition relation $\lifecycle \subseteq \actstates \times \actstates$. For each activity state $s$, we let $\cb(c,s)$ be the set of callbacks for the activity $c$ in the state $s$. Moreover we assume that for the $\actstate{running}$ state, $\cb(c,\actstate{running}) = \handlers(c)$.

We also need the notion of \emph{callback stack}: a callback stack is the initial call stack of an new activity frame, created upon a callback method invocation:
\begin{definition}
Given a location $\ell$ pointing to an activity of class $c$, we let $\getcb{\ell}{s}$ stand for an arbitrary \emph{callback stack} for state $s$, i.e., any call stack $\locstate{c',m,0}{\stm^*}{R}{\cdot} :: \varepsilon$, where $(c',\stm^*) = \lookup(c,m)$ for some $m \in \cb(c,s)$, $\sign(c',m) = \methsign{\tau_1,\ldots,\tau_n}{\tau}{\loc}$ and:
\[ 
R = ((r_i \mapsto \defvalue)^{i \leq \loc}, r_{\loc+1} \mapsto \ell, (r_{\loc+1+j} \mapsto v_j)^{j \leq n}),
\]
for some values $v_1,\ldots,v_n$ of the correct type $\tau_1,\ldots,\tau_n$.
\end{definition}

\subsubsection{Global Reduction Relation}

We are now ready to give the global reduction relation. First we will describe two new rules which were not given in the body and can be found in Table~\ref{tab:activity-tsemantics-appendix}: rule \irule{T-Intent} allows a thread to transfer an intent to the activity that spawned it, and rule \irule{T-Thread} allows a thread to transfer a location in its pending thread stack to the activity that spawned it. 

\begin{table*}[h]
\begin{mathpar}
\inferrule[(T-Reduce)]
{\tmethconf{\callstack}{\pi}{\threadstack}{\heap}{\sheap}{\ell_t} \rightsquigarrow \tmethconf{\callstack'}{\pi'}{\threadstack'}{\heap'}{\sheap'}{\ell_t}}
{\tactconf{\actstack}{\threadpool :: \threadframe{\ell}{\ell_t}{\pi}{\threadstack}{\callstack} :: \threadpool'}{\heap}{\sheap} \Rightarrow \tactconf{\actstack}{\threadpool :: \threadframe{\ell}{\ell_t}{\pi'}{\threadstack'}{\callstack'} :: \threadpool'}{\heap'}{\sheap'}}

\inferrule[(T-Kill)]
{ \heap(\ell') = \obj{c}{(f\mapsto v)^*,\finished \mapsto \_}\\
\heap'= \ \heap[\ell' \mapsto \obj{c}{(f\mapsto v)^*,\finished \mapsto \true}]
}
{\tactconf{\actstack}{\threadpool :: \threadframe{\ell}{\ell'}{\varepsilon}{\varepsilon}{\ocallstack} :: \threadpool'}{\heap}{\sheap} \Rightarrow \tactconf{\actstack}{\threadpool :: \threadpool'}{\heap'}{\sheap} }

\inferrule[(T-Intent)]
{(\varphi,\varphi') \in \{(\tactframe{\ell}{s}{\pi}{\threadstack}{\callstack},\tactframe{\ell}{s}{i :: \pi}{\threadstack}{\callstack}),(\tuactframe{\ell}{s}{\pi}{\threadstack}{\callstack},\tuactframe{\ell}{s}{i :: \pi}{\threadstack}{\callstack})\}}
{\tactconf{\actstack :: \varphi :: \actstack'}{\threadpool :: \threadframe{\ell}{\ell'}{i :: \pi'}{\threadstack'}{\callstack'} :: \threadpool'}{\heap}{\sheap} \Rightarrow \tactconf{\actstack :: \varphi' :: \actstack'}{\threadpool :: \threadframe{\ell}{\ell'}{\pi'}{\threadstack'}{\callstack'} :: \threadpool'}{\heap}{\sheap}}

\inferrule[(T-Thread)]
{(\varphi,\varphi') \in \{(\tactframe{\ell}{s}{\pi}{\threadstack}{\callstack},\tactframe{\ell}{s}{\pi}{\ell_t :: \threadstack}{\callstack}),(\tuactframe{\ell}{s}{\pi}{\threadstack}{\callstack},\tuactframe{\ell}{s}{\pi}{\ell_t :: \threadstack}{\callstack})\}}
{\tactconf{\actstack :: \varphi :: \actstack'}{\threadpool :: \threadframe{\ell}{\ell'}{\pi'}{\threadstack' :: \ell_t :: \threadstack''}{\callstack'} :: \threadpool'}{\heap}{\sheap} \Rightarrow \tactconf{\actstack :: \varphi' :: \actstack'}{\threadpool :: \threadframe{\ell}{\ell'}{\pi'}{\threadstack':: \threadstack''}{\callstack'} :: \threadpool'}{\heap}{\sheap}}

\inferrule[(A-ThreadStart)]
{\varphi = \tuactframe{\ell}{s}{\pi}{\threadstack :: \ell' :: \threadstack'}{\callstack}\\
\varphi' =\tuactframe{\ell}{s}{\pi}{\threadstack :: \threadstack'}{\callstack}\\
\psi =\threadframe{\ell}{\ell'}{\varepsilon}{\varepsilon}{\callstack'}\\
\heap(\ell') = \threadobj{c'}{(f\mapsto v)^*}\\
\lookup(c',\threadrun) = (c'',\stm^*)\\
\sign(c'',\threadrun) = \methsign{\thread}{\void}{\loc}\\
\callstack' = \locstate{c'',\threadrun,0}{\stm^*}{(r_k \mapsto \defvalue)^{k \leq loc},r_{loc + 1} \mapsto \ell'}{\ell'}}
{\tactconf{\actstack :: \varphi:: \actstack'}{\threadpool}{\heap}{\sheap} \Rightarrow \tactconf{\actstack :: \varphi' :: \actstack'}{\psi :: \threadpool}{\heap}{\sheap}}
\end{mathpar}
\caption{\label{tab:activity-tsemantics-appendix}New Global Reduction Rules}
\end{table*}

Table~\ref{tab:activity-semantics} recalls the rules introduced by~\cite{CalzavaraGM16} to model the activity life-cycle mechanism, with only minor modifications to include the thread pool. Rule \irule{A-Active} executes the statements of the active frame in the activity stack, using the reduction relation for local configurations. Rule \irule{A-Deactivate} stops an activity frame from being active when it has completed its computations. Rule \irule{A-Step} models the transition of the top-most activity frame from one activity state to one of its successor in the activity life-cycle, and executes a callback method from this new activity state, provided some side conditions related to the pending activity stack and the $\finished$ field of the activity object are met. Rule \irule{A-Destroy} models the removal of a finished activity from the activity stack. Rule \irule{A-Back} is used by the system to finished the top-most activity when the user hits the back button. Rule \irule{A-Replace} models the screen orientation changing, by destroying and restarting the top-most activity. Rule \irule{A-Hidden} allows an activity in the background to take precedence over the foreground activity, stopping or destroying it. Rule \irule{A-Start} allows to start a new activity: the top-most activity must be paused or stopped, and must have an intent $i$ sent to some activity $c$ in its pending activity stack: a new activity of class $c$ is added to the top of the activity stack, its $\fintent$ field is set to a serialized copy of $i$ and its $\parent$ field is set to the starting activity. Rule \irule{A-Swap} allows a parent activity to come back to the foreground, assuming the foreground activity is finished and is one of its child activity. Finally, rule \irule{A-Result} allows the top-most activity  to return the result of its computation to the parent activity, provided that the top-most activity is finished: a serialized copy of the result is sent to the parent activity, which becomes active and executes the $\actstate{onActivityResult}$ callback.

\begin{table}[htp]
\begin{mathpar}
\inferrule[(A-Active)]
{\tmethconf{\callstack}{\pi}{\threadstack}{\heap}{\sheap}{\ell} \rightsquigarrow \tmethconf{\callstack'}{\pi'}{\threadstack'}{\heap'}{\sheap'}{{\ell}}}
{\tactconf{\actstack :: \tuactframe{\ell}{s}{\pi}{\threadstack}{\callstack} :: \actstack'}{\threadpool}{\heap}{\sheap} \Rightarrow \tactconf{\actstack :: \tuactframe{\ell}{s}{\pi'}{\threadstack'}{\callstack'} :: \actstack'}{\threadpool}{\heap'}{\sheap'}}

\inferrule[(A-Deactivate)]
{ }
{\tactconf{\actstack :: \tuactframe{\ell}{s}{\pi}{\threadstack}{\ocallstack} :: \actstack'}{\threadpool}{\heap}{\sheap} \Rightarrow \tactconf{\actstack :: \tactframe{\ell}{s}{\pi}{\threadstack}{\ocallstack} :: \actstack'}{\threadpool}{\heap}{\sheap}}

\inferrule[(A-Step)]
{(s,s') \in \lifecycle \\
\pi \neq \varepsilon \Rightarrow (s,s') = (\actstate{running},\actstate{onPause}) \\
\heap(\ell).\finished = \true \Rightarrow (s,s') \in \{(\actstate{running},\actstate{onPause}),(\actstate{onPause},\actstate{onStop}),(\actstate{onStop},\actstate{onDestroy})\}}
{\tactconf{\tactframe{\ell}{s}{\pi}{\threadstack}{\ocallstack} :: \actstack}{\threadpool}{\heap}{\sheap} \Rightarrow \tactconf{\tuactframe{\ell}{s'}{\pi}{\threadstack}{\getcb{\ell}{s'}} :: \actstack}{\threadpool}{\heap}{\sheap}}

\inferrule[(A-Destroy)]
{\heap(\ell).\finished = \true}
{\tactconf{\actstack :: \tactframe{\ell}{\actstate{onDestroy}}{\pi}{\threadstack}{\ocallstack} :: \actstack'}{\threadpool}{\heap}{\sheap} \Rightarrow \tactconf{\actstack :: \actstack'}{\threadpool}{\heap}{\sheap}}

\inferrule[(A-Back)]
{\heap' = \heap[\ell \mapsto \heap(\ell)[\finished \mapsto \true]]}
{\tactconf{\tactframe{\ell}{\actstate{running}}{\varepsilon}{\threadstack}{\ocallstack} :: \actstack}{\threadpool}{\heap}{\sheap} \Rightarrow \tactconf{\tactframe{\ell}{\actstate{running}}{\varepsilon}{\threadstack}{\ocallstack} :: \actstack}{\threadpool}{\heap'}{\sheap}}

\inferrule[(A-Replace)]
{\heap(\ell) = \obj{c}{(f_{\tau} \mapsto v)^*,\finished \mapsto u} \\ 
 \pointer{p}{c} \not \in dom(\heap)\\
o = \obj{c}{(f_{\tau} \mapsto \defvalue_{\tau})^*,\finished \mapsto \false} \\ 
\heap' = \heap, \pointer{p}{c} \mapsto o}
{\tactconf{\tactframe{\ell}{\actstate{onDestroy}}{\pi}{\threadstack}{\ocallstack} :: \actstack}{\threadpool}{\heap}{\sheap} \Rightarrow \tactconf{\tuactframe{\pointer{p}{c}}{\actstate{constructor}}{\pi}{\threadstack}{\getcb{\pointer{p}{c}}{\actstate{constructor}}} :: \actstack}{\threadpool}{\heap'}{\sheap}}

\inferrule[(A-Hidden)]
{\varphi = \tactframe{\ell}{s}{\pi}{\threadstack}{\ocallstack} \\
s \in \{\actstate{onResume},\actstate{onPause}\} \\
(s',s'') \in \{(\actstate{onPause},\actstate{onStop}),(\actstate{onStop},\actstate{onDestroy})\} }
{\tactconf{\varphi :: \actstack :: \tactframe{\ell'}{s'}{\pi'}{\threadstack'}{\ocallstack'} :: \actstack'}{\threadpool}{\heap}{\sheap} \Rightarrow \tactconf{\varphi :: \actstack :: \tuactframe{\ell'}{s''}{\pi'}{\threadstack'}{\getcb{\ell'}{s''}} :: \actstack'}{\threadpool}{\heap}{\sheap}}

\inferrule[(A-Start)]
{s \in \{\actstate{onPause},\actstate{onStop}\} \\
i = \intent{c}{(k \mapsto v)^*} \\
\emptyset \vdash \serblock{\heap}(i) = (i',\heap') \\
\pointer{p}{c},\pointer{p'}{\astart{c}} \not\in \dom(\heap,\heap') \\
o = \obj{c}{(f_{\tau} \mapsto \defvalue_{\tau})^*,\finished \mapsto \false, \fintent \mapsto \pointer{p'}{\astart{c}}, \parent \mapsto \ell} \\
\heap'' = \heap,\heap',\pointer{p}{c} \mapsto o, \pointer{p'}{\astart{c}} \mapsto i'}
{\tactconf{\tactframe{\ell}{s}{i :: \pi}{\threadstack}{\ocallstack} :: \actstack}{\threadpool}{\heap}{\sheap} \Rightarrow \tactconf{\tuactframe{\pointer{p}{c}}{\actstate{constructor}}{\varepsilon}{\varepsilon}{\getcb{\pointer{p}{c}}{\actstate{constructor}}} :: \tactframe{\ell}{s}{\pi}{\threadstack}{\ocallstack} :: \actstack}{\threadpool}{\heap''}{\sheap}}

\inferrule*[width=30em,lab=(A-Swap)]
{\varphi' = \tactframe{\ell'}{\actstate{onPause}}{\varepsilon}{\threadstack'}{\ocallstack'} \\
\heap(\ell').\finished = \true \\
\varphi = \tactframe{\ell}{s}{i :: \pi}{\threadstack}{\ocallstack} \\
s \in \{\actstate{onPause},\actstate{onStop}\} \\
\heap(\ell').\parent = \ell}
{\tactconf{\varphi' :: \varphi :: \actstack}{\threadpool}{\heap}{\sheap} \Rightarrow \tactconf{\varphi :: \varphi' :: \actstack}{\threadpool}{\heap}{\sheap}}

\inferrule*[width=50em,lab=(A-Result)]
{\varphi' = \tactframe{\ell'}{\actstate{onPause}}{\varepsilon}{\threadstack'}{\ocallstack'} \\
\heap(\ell').\finished = \true \\
\varphi = \tactframe{\ell}{s}{\varepsilon}{\threadstack}{\ocallstack} \\
s \in \{\actstate{onPause},\actstate{onStop}\} \\
\heap(\ell').\parent = \ell \\
\emptyset \vdash \serval{\heap}(\heap(\ell').\result) = (w',\heap') \\
\heap'' = (\heap,\heap')[\ell \mapsto \heap(\ell)[\result \mapsto w']]}
{\tactconf{\varphi' :: \varphi :: \actstack}{\threadpool}{\heap}{\sheap} \Rightarrow \tactconf{\tuactframe{\ell}{s}{\varepsilon}{\threadstack}{\getcb{\ell}{\actstate{onActivityResult}}} :: \varphi' :: \actstack}{\threadpool}{\heap''}{\sheap}}
\end{mathpar}
\textbf{Conventions:} the activity stack on the left-hand side does
not contain underlined frames, with the exception of \irule{A-Deactivate} and \irule{A-Activate}
\caption{\label{tab:activity-semantics} Reduction Rules for Configurations ($\tactconf{\actstack}{\threadpool}{\heap}{\sheap} \Rightarrow \tactconf{\actstack'}{\threadpool'}{\heap'}{\sheap'}$)}
\end{table}


\FloatBarrier
\clearpage

\section{Abstract semantics}
\label{sec:abs-sem}
\subsubsection{Lifting functions} We first give the formal definition of the $\lhlift{}{}$ and $\afunion$ functions, that we informally described in the body of the paper.
\begin{mathpar}
\absfi \afunion \absfi' = \left(\pp \mapsto \mmax{\absfi(\pp)}{\absfi'(\pp)} \right)^*

\lhlift{\abslh}{\absfi} = \left(\pp \mapsto 
\begin{cases}
\absobj{c}{(f \mapsto \lift{\absual}{\absfi})^*} &\text{ if }\absfi(\pp) = 0 \wedge \abslh(\pp) = \absobj{c}{(f \mapsto \absual)^*}\\
\absintent{c}{\lift{\absual}{\absfi}} &\text{ if }\absfi(\pp) = 0 \wedge \abslh(\pp) = \absintent{c}{\absual}\\
\absarray{\tau}{\lift{\absual}{\absfi}} &\text{ if }\absfi(\pp) = 0 \wedge \abslh(\pp) = \absarray{\tau}{\absual}\\
\bot &\text{ otherwise}
\end{cases}\right)^*
\end{mathpar}

\subsubsection{Right-Hand Side} We can now present the rules for the abstract evaluation  of right-hand sides (a formal description is given in Table~\ref{tab:abs-rhs}): to abstract a primitive value $\prim$ at a program point $\spp$, we take the corresponding element $\absprim$ from the underlying abstract domain. To abstract the content of a register $r_i$ at program point $\spp$, we take the abstract local state fact $\absreg{\spp}{ \_ }{\absval^*}{\_}{\_}$ and we return the $i$-th abstract value $\absval_i$. To abstract, at program point $\spp$, the content of the field $\mathsf{f}$ of an object whose location is stored in register $r_i$, we retrieve the $i$-th abstract value $\absval_i$ from the abstract fact $\absreg{\spp}{ \_ }{\absval^*}{\abslh}{\_}$: if $\absval_i$ contains any location abstraction $\abslab$, we look whether it is an abstract \flowsensitive location $\absl{\absloc}$ or an abstract \flowinsensitive location $\absg{\absloc}$ : in the former case, we get the entry $(\absloc \mapsto \hat{o})$  from the abstract \flowsensitive heap $\abslh$, and we return the abstract value stored in the field $\mathsf{f}$ of the abstract object $\hat{o}$; in the latter case, we try to find a matching \flowinsensitive heap fact  $\absheap(\absloc,\hat{o})$ and we return the \emph{lifted} value of the field $\mathsf{f}$ of the abstract object $\hat{o}$ contained therein. We similarly abstract the content of array cells, but in a field-insensitive fashion. To abstract the content of a static field $\mathsf{c.f}$ at program point $\spp$, we take any fact $\abssheap_{\mathsf{c},\mathsf{f}}(\absval)$ and we return the \emph{lifted} abstract value $\absval$.

\begin{remark} When getting an abstract value from a \flowinsensitive heap fact, a static field fact or an array we lift it, by returning $\lift{\absval}{1^*}$~\footnote{We abuse the notation here: $1^*$ should be interpreted as $(\_ \mapsto 1)^*$.}. This is due to the fact that, by definition, a \flowinsensitive memory block cannot contain a location to a \flowsensitive memory block. Therefore we chose that instead of lifting abstract locations before putting them in abstract \flowinsensitive facts, arrays or static fields, we lift abstract locations when performing look-ups. We believe this to (slightly) simplify the abstract semantics and the soundness proof.
\end{remark}

\begin{table}[htb]
\begin{mathpar}
\arhs{\prim} = \{\prhs{\absprim}\}

\arhs{r_i} = \{\absreg{\spp}{ \_ }{\absval^*}{\_}{\_} \implies \prhs{\absval_i }\}

\arhs{c.f} = \{\abssheap_{\mathsf{c},\mathsf{f}}(\absval) \implies \prhs{\lift{\absval}{1^*}}\}

\arhs{r_i.f} = \{\absreg{\spp}{ \_ }{\absval^*}{\abslh}{\_} \wedge \rlookup{i}{\absval^*}{\abslh}{\absg{\absloc}}{\absobj{c}{(f' \mapsto \absval')^*, f \mapsto \absual}}  \implies \prhs{\lift{\absual}{1^*}}\} 

\cup \{\absreg{\spp}{ \_ }{\absval^*}{\abslh}{\_} \wedge   \rlookup{i}{\absval^*}{\abslh}{\absl{\absloc}}{\absobj{c}{(f' \mapsto \absval')^*, f \mapsto \absual}} \implies \prhs{\absual}\}

\arhs{r_i[r_j]} = \{\absreg{\spp}{ \_ }{\absval^*}{\abslh}{\_} \wedge  \rlookup{i}{\absval^*}{\abslh}{\absg{\absloc}}{ \absarray{\tau}{\absual}} \implies \prhs{\lift{\absual}{1^*}}\}

\cup \{\absreg{\spp}{ \_ }{\absval^*}{\abslh}{\_} \wedge  \rlookup{i}{\absval^*}{\abslh}{\absl{\absloc}}{ \absarray{\tau}{\absual}} \implies \prhs{\absual}\}
\end{mathpar}
\caption{\label{tab:abs-rhs} Abstract Evaluation of Right-hand Sides}
\end{table}

\subsubsection{Activity Abstraction} We will now describe the rules abstracting the activity life-cycle and thread management mechanisms, which are given in Table~\ref{tab:abs-activity}. The rule \irule{Tstart} over-approximates the spawning of a new thread $\abstdispatch(\absloc,\absobj{c}{(f \mapsto \_)^*})$ by generating an abstract local state running the method $\threadrun$ of the corresponding thread object. The rule \irule{Cbk} abstracts the callback invocation by generating an abstract local heap fact for all the callbacks of a started activity. Observe that the initial arguments supplied are over-approximated by $\top$, since they depend on user-inputs and are not statistically known. The rule \irule{Fin} roughly over-approximates  whether an activity is finished or not: it always replaces the $\finished$ field of an activity object by $\top_{\type{bool}}$. The rule \irule{Rep} restarts abstract activity objects at any time, by re-setting their fields to their default initial abstract value $\adefvalue_{\tau}$ (this over-approximates the restarting of an activity when the screen orientation changes). The rule \irule{Act} handles the starting of new activities: if an intent $\absdispatch_{\mathsf{c}'}(\absintent{\astart{c}}{\absval^*})$ has been sent to an activity $c$ by an activity $c'$, the rule creates a new abstract activity object of class $c$ with properly bound and initialized fields. It also creates a new special abstract heap fact $\absheap(\astart{c},\absintent{c}{\absval^*})$ that contains a copy of the sent intent: this over-approximates the serialization mechanism, and is sound because the intent contains only abstract \flowinsensitive locations, that are updated with weak updates. The rule \irule{Res} over-approximates the mechanism by which an child activity returns a result to its parent activity. Finally rule \irule{Sub} contains  subtyping judgments for classes, and rule \irule{Po} contain partial ordering rules for abstract values.

\begin{table}[htb]
\begin{mathpar}
\begin{array}{lcl}
\rulename{Tstart} &=& \{\abstdispatch(\absloc,\absobj{c}{(f \mapsto \_)^*}) \wedge c \subtype \mathsf{c'} \wedge c \le \thread\\
&& \implies \absreg{\mathsf{c'},\threadrun,\mathsf{0}}{(\absg{\absloc},\absg{\absloc})}{(\adefvalue_k)^{k \leq \loc},\absg{\absloc}}{(\bot)^*}{0^*} ~|~  \mathsf{c'} \in \abslookup(\threadrun) \wedge \sign(\mathsf{c'},\threadrun) = \methsign{\thread}{\void}{\loc}\}\\

\rulename{Cbk} & = & \{\absheap(c,\absobj{c}{(f \mapsto \_)^*}) \wedge c \subtype \mathsf{c'} \implies \absreg{\mathsf{c'},\mathsf{m},\mathsf{0}}{(\absg{c},(\top_{\tau_j})^{j \leq n})}{(\adefvalue_k)^{k \leq \loc},\absg{c},(\top_{\tau_j})^{j \leq n}}{(\bot)^*}{0^*} ~|~ \\
& & \mathsf{c'} \text{ is an activity class} \wedge \exists s: m \in \cb(\mathsf{c'},s) \wedge \sign(\mathsf{c'},m) = \methsign{\tau_1,\ldots,\tau_n}{\tau}{\loc}\} \\

\rulename{Fin} & = & \{\absheap(c,\absobj{c}{(f \mapsto \_)^*,\finished \mapsto \_}) \implies \absheap(c,\absobj{c}{(f \mapsto \_)^*,\finished \mapsto \top_{\type{bool}}})\} \\

\rulename{Rep} & = & \{\absheap(c,\absobj{c}{(f_{\tau} \mapsto \_)^*}) \implies  \absheap(c,\absobj{c}{(f_{\tau} \mapsto \adefvalue_{\tau})^*})\} \\

\rulename{Act} & = & \{\absdispatch_{\mathsf{c'}}(\absintent{c}{\absval})) \implies \absheap(\astart{c},\absintent{c}{\absval})\}\, \cup \\
& & \{\absdispatch_{\mathsf{c}'}(\absintent{c}{\absval})) \implies \absheap(c, \absobj{c}{(f_{\tau} \mapsto \adefvalue_{\tau})^*, \finished \mapsto \widehat{\false}, \parent \mapsto c', \fintent \mapsto \astart{c}})\} \\

\rulename{Res} & = & \{\absheap(c',\absobj{c'}{(f' \mapsto \_)^*,\parent \mapsto c,\result \mapsto \abswal} \wedge \absheap(c,\absobj{c}{(f \mapsto \_)^*,\result \mapsto \_} \\&&\implies  \absheap(c,\absobj{c}{(f \mapsto \_)^*,\result \mapsto \abswal}\} \\

\rulename{Sub} & = & \{\tau \subtype \tau' ~|~ \tau \subtype \tau' \text{ is a valid subtyping judgment} \}\\

\rulename{Po} & = & \{\absval \abspo \absval' ~|~ \absval \abspo \absval' \text{ is a valid partial ordering} \}
\end{array}
\end{mathpar}
\caption{\label{tab:abs-activity} Abstract Semantics of \sem{} - Activity Rules}
\end{table}

\subsubsection{Statement  Abstraction} Before giving the abstract rule for Dalvik statements, we need to define the abstract counter-part of the $\gettype{H}{b}$ function:
\begin{definition} Given an abstract memory block $\hat{b}$, we define a function $\absgettype(\hat{b})$ as follows:
\[
\absgettype(\hat{b}) =
\begin{cases}
c & \text{if } \hat{b} = \absobj{c}{(f \mapsto \absval)^*} \\
\arrtype{\tau} & \text{if } \hat{b} = \absarray{\tau}{\absval} \\
\texttt{Intent} & \text{if } \hat{b} = \absintent{c}{\absval}
\end{cases}
\]
\end{definition}

For all standard Dalvik statement $\stm$ and program point $\pp$, the rule $\ainst{\stm}$ abstracts the action of $\stm$ at program point $\pp$. The most important rules have already been described in the main body of this paper, and the full set of rules is given in  Table~\ref{tab:abs-statements}, Table~\ref{tab:abs-statements-inv} and Table~\ref{tab:abs-statements-thread}. A few points are worth mentioning:
\begin{itemize}
\item $\ainst{\startwait{r_i}}$: We just check whether the $\interrupted$ field of the abstract object over-approximating the running thread or activity is over-approximating $\widehat{\true}$, in which case an abstract abnormal local state throwing an $\interruptedexception$ is generated, or $\widehat{\false}$, in which case the abstract local state is propagated to the next program point;
\item $\ainst{\monitorenter{r_i}}$ and $\ainst{\monitorexit{r_i}}$: Given that monitors are synchronization constructs, it is sound to ignore them when checking reachability properties, which is the target of the present work. There are of course more precise ways of abstracting monitors, but they would make the analysis more complicated and their practical benefits are unclear.
\item  $\ainst{\startact{r_i}}$: When an abstract intent $\absintent{c'}{\absual}$ stored in the \flowsensitive heap at program point $\abslab$ is used to start a new (abstract) activity, every abstract \flowsensitive location reachable from $\abslab$ in $\abslh$ (represented by the abstract filter $\absfi'$ computed by $\cfilter{\absl{\absloc}}{\abslh}{\absfi'}$) is being lifted, to make sure that these heap entries are abstract in a \flowinsensitive fashion, since they are being shared between the parent and the started child activity. 
\end{itemize}

\begin{table}[p]
\[\begin{array}{lcl}
\ainst{\goto \pc'}  &=&  \{\absreg{\spp}{ \_ }{\absval^*}{\_}{\_} \implies \absreg{\apcp}{ \_ }{\absval^*}{\_}{\_}\}\\

\ainst{\ifbr {r_i}{r_j} {\pc'}}  &=&
 \{\absreg{\spp}{ \_ }{\absval^*}{\_}{\_} \wedge \absval_i\ \acomp\ \absval_j \implies \absreg{\apcp}{ \_ }{\absval^*}{\_}{\_}\} \cup\\
&&  \{\absreg{\spp}{ \_ }{\absval^*}{\_}{\_} \wedge \absval_i\ \ancomp\ \absval_j \implies \absreg{\apcn}{ \_ }{\absval^*}{\_}{\_}\}\\

\ainst{\binop{r_d}{r_i}{r_j}}  &=&  \{\absreg{\spp}{ \_ }{\absval^*}{\_}{\_} \implies \absreg{\apcn}{ \_ }{\absval^*[d \mapsto \absval_i\ \abinop\ \absval_j]}{\_}{\_}\} \\

\ainst{\unop{r_d}{r_i}}  &=&  \{\absreg{\spp}{ \_ }{\absval^*}{\_}{\_} \implies \absreg{\apcn}{ \_ }{\absval^*[d \mapsto \aunop\, \absval_i]}{\_}{\_}\}\\

\ainst{\move{r_d}{\rhs}}  &=&  \arhs{\rhs} \cup  \{\prhs{\absval'} \wedge \absreg{\spp}{\_}{\absval^*}{\_}{\_} \implies \absreg{\apcn}{ \_} {\absval^*[d \mapsto \absval']}{\_}{\_}\} \\

\ainst{\move{r_a[r_\ind]}{\rhs}}  &=& \arhs{\rhs} \cup
   \{\prhs{\absval''} \wedge \absreg{\spp}{ \_ }{\absval^*}{\abslh}{\absfi} \wedge \rlookup{a}{\absval^*}{\abslh}{ \absg{\absloc}}{\absarray{\tau}{\absval'}} \wedge \cfilter{\absval''}{\abslh}{\absfi'} \\
&&\implies \absheap(\absloc,\absarray{\tau}{\absval' \absjoin \absval''}) \wedge \liftlh{\abslh}{\absfi'} \wedge \absreg{\apcn}{ \_ }{\lift{\absval^*}{\absfi'}}{\lhlift{\abslh}{\absfi'}}{\absfi \afunion \absfi'}\}\,\cup\\
&& \{\prhs{\absval''}  \wedge \absreg{\spp}{ \_} {\absval^*}{\abslh}{\absfi} \wedge \rlookup{a}{\absval^*}{\abslh}{ \absl{\absloc}}{\absarray{\tau}{\absval'}} \\
&&\implies\absreg{\apcn}{ \_ }{\absval^*}{\abslh[\absloc \mapsto \absarray{\tau}{\absval' \absjoin \absval''}}{\absfi}\} \\

\ainst{\move{r_o.f}{\rhs}}  &=& \arhs{\rhs} \cup
 \{ \prhs{\absval''} \wedge \absreg{\spp}{ \_} {\absval^*}{\abslh}{\absfi} \\
&& \wedge \rlookup{o}{\absval^*}{\abslh}{\absg{\absloc}}{\absobj{c'}{(f' \mapsto \absual')^*, f \mapsto \absval'}} \wedge \cfilter{\absval''}{\abslh}{\absfi'} \implies\\
&&  \absheap(\absloc,\absobj{\mathsf{c'}}{(f' \mapsto \absual')^*, f \mapsto {\absval''})})\wedge \liftlh{\abslh}{\absfi'}\wedge \absreg{\apcn}{ \_ }{\lift{\absval^*}{\absfi'}}{\lhlift{\abslh}{\absfi'}}{\absfi \afunion \absfi'}\} \cup\\
&&  \{\prhs{\absval''}  \wedge \absreg{\spp}{ \_} {\absval^*}{\abslh}{\absfi} \wedge \rlookup{o}{\absval^*}{\abslh}{\absl{\absloc}}{\absobj{c'}{(f' \mapsto \absual')^*, f \mapsto \absval'}}  \\
&& \implies\absreg{\apcn}{ \_ }{\absval^*}{\abslh[\absloc \mapsto \absobj{c'}{(f' \mapsto \absual')^*, f \mapsto \absval''}}{\absfi}\} \\

\ainst{\move{c'.f}{\rhs}}  &=& \arhs{\rhs} \cup
\{\prhs{\absval'} \wedge \absreg{\spp}{ \_ }{\absval^*}{\abslh}{\absfi} \wedge  \cfilter{\absval'}{\abslh}{\absfi'} \\
&&\implies \abssheap_{\mathsf{c'},\mathsf{f}}({\absval'}) \wedge \liftlh{\abslh}{\absfi'}\wedge\absreg{\apcn}{ \_ }{\lift{\absval^*}{\absfi'}}{\lhlift{\abslh}{\absfi'}}{\absfi \afunion \absfi'} \}\\

\ainst{\instanceof{r_d}{r_s}{\tau}}  &=& 
 \{\absreg{\spp}{ \_}{\absval^*}{\abslh}{\absfi} \wedge \rlookup{s}{\absval^*}{\abslh}{\_}{\hat{b}} \wedge \absgettype(\hat{b}) \subtype \tau\\
&& \implies \absreg{\apcn}{ \_ }{\absval^*[d \mapsto \widehat{\true}]}{\abslh}{\absfi}\}\, \cup \\
&& \{\absreg{\spp}{ \_ }{\absval^*}{\abslh}{\absfi} \wedge \rlookup{s}{\absval^*}{\abslh}{\_}{\hat{b}} \wedge \absgettype(\hat{b}) \not\subtype \tau \\
&&\implies \absreg{\apcn}{ \_ }{\absval^*[d \mapsto \widehat{\false}]}{\abslh}{\absfi}\} \\

\ainst{\checkcast{r_s}{\tau}}  &=& 
 \{\absreg{\spp}{ \_ }{\absval^*}{\abslh}{\absfi} \wedge \rlookup{s}{\absval^*}{\abslh}{\_}{\hat{b}} \wedge \absgettype(\hat{b}) \subtype \tau \implies \absreg{\apcn}{ \_}{\absval^*}{\abslh}{\absfi}\} \\

\ainst{\new{r_d}{\mathsf{c'}}}  &=&  \{ \absreg{\spp}{\_}{\absval^*}{\abslh}{\absfi} \wedge \cfilter{\absl{\spp}}{\abslh}{\absfi'} \implies \\
&&\liftlh{\abslh}{\absfi'}\wedge\absreg{\apcn}{\_}{\lift{\absval^*}{\absfi'}[d \mapsto \absl{\spp}]}{\lhlift{\abslh}{\absfi'}[\spp \mapsto \absobj{\mathsf{c'}}{(f \mapsto \adefvalue_{\tau})^*}]}{\absfi \afunion \absfi'}\} \\

\ainst{\newintent{r_d}{\mathsf{c'}}}  &=& 
\{ \absreg{\spp}{\_}{\absval^*}{\abslh}{\absfi} \wedge \cfilter{\absl{\spp}}{\abslh}{\absfi'} \\
&&\implies \liftlh{\abslh}{\absfi'} \wedge\absreg{\apcn}{\_}{\lift{\absval^*}{\absfi'}[d \mapsto \absl{\spp}]}{\lhlift{\abslh}{\absfi'}[\spp \mapsto \absintent{\mathsf{c'}}{\bot})]}{\absfi \afunion \absfi'}\} \\

\ainst{\newarray{r_d}{r_l}{\tau}}  &=&
\{ \absreg{\spp}{\_}{\absval^*}{\abslh}{\absfi} \wedge \cfilter{\absl{\spp}}{\abslh}{\absfi'} \\
&&\implies \liftlh{\abslh}{\absfi'}\wedge\absreg{\apcn}{\_}{\lift{\absval^*}{\absfi'}[d \mapsto \absl{\spp}]}{\lhlift{\abslh}{\absfi'}[\spp \mapsto \absarray{\tau}{\adefvalue_{\tau}})]}{\absfi \afunion \absfi'} \}  \\

\ainst{\startact{r_i}}  &=& 
  \{\absreg{\spp}{\_}{\absval^*}{\abslh}{\absfi} \wedge  \rlookup{i}{\absval^*}{\abslh}{\absg{\absloc}}{\absintent{c'}{\absual}} \\
&&\implies \absdispatch_{\const{c}}(\absintent{c'}{\absual}) \wedge \absreg{\apcn}{\_}{\absval^*}{\abslh}{\absfi}\}\,\cup\\
&&\{\absreg{\spp}{\_}{\absval^*}{\abslh}{\absfi} \wedge \rlookup{i}{\absval^*}{\abslh}{\absl{\absloc}}{\absintent{c'}{\absual}} \wedge \cfilter{\absl{\absloc}}{\abslh}{\absfi'} \\
&&\implies \absdispatch_{\const{c}}(\absintent{c'}{\absual}) \wedge \liftlh{\abslh}{\absfi'} \wedge \absreg{\apcn}{\_}{\lift{\absval^*}{\absfi'}}{\lhlift{\abslh}{\absfi'}}{\absfi \afunion \absfi'}\}\\

\ainst{\putextra{r_i}{r_k}{r_j}} &=& 
\{\absreg{\spp}{\_}{\absval^*}{\abslh}{\absfi} \wedge  \rlookup{i}{\absval^*}{\abslh}{\absg{\absloc}}{\absintent{c'}{\absval'}} \wedge  \cfilter{\absval_j}{\abslh}{\absfi'} \implies\\
&& \absheap(\absloc, \absintent{c'}{\absval'\absjoin\absval_j}) \wedge  \liftlh{\abslh}{\absfi'} \wedge  \absreg{\apcn}{\_}{\lift{\absval^*}{\absfi'}}{\lhlift{\abslh}{\absfi'}}{\absfi \afunion \absfi'}\} \cup  \\
&&\{\absreg{\spp}{\_}{\absval^*}{\abslh}{\absfi} \wedge  \rlookup{i}{\absval^*}{\abslh}{\absl{\absloc}}{\absintent{c'}{\absval'}} \\
&&\implies \absreg{\apcn}{\_}{\absval^*}{\abslh[\absloc \mapsto  \absintent{c'}{\absval'\absjoin \absval_j} ]}{\absfi}\}\\

\ainst{\getextra{r_i}{r_k}{\tau}}  &=& \{\absreg{\spp}{\_}{\absval^*}{\abslh}{\absfi} \wedge\rlookup{i}{\absval^*}{\abslh}{\_}{\absintent{c'}{\absval'}}\implies  \absreg{\apcn}{\_}{\absval^*[\res \mapsto \absval']}{\abslh}{\absfi}\}\\

\ainst{\return}  &=& \{\absreg{\spp}{(\absthread,\absval^*_{call})}{\absval^*}{\abslh}{\absfi} \implies \absresult{\mathsf{c},\mathsf{m}}{(\absthread,\absval^*_{call})}{\absval_{\res}}{\abslh}{\absfi}\}
\end{array}\]
\textbf{Conventions:} $\spp = \mathsf{c,m,\spc}$
\caption{\label{tab:abs-statements} Abstract Semantics of \sem{} - Standard Statements}
\end{table}

\begin{table}
\begin{itemize}
\item $\ainst{\invoke{r_o}{m'}{(r_{i_j})^{j \leq n}}}  = $\\
\( \{\absreg{\spp}{(\absthread, \_ )}{\absval^*}{ \abslh }{\absfi} \wedge \rlookup{o}{\absval^*}{\abslh}{\_}{\absobj{c'}{(f \mapsto \absual)^*}} \wedge c' \subtype \const{c''} \)\\
\(\null\hfill\implies \absreg{\const{c''},\mathsf{m'},\mathsf{0}}{(\absthread,(\absval_{i_j})^{j \leq n})}{(\adefvalue_k)^{k \leq \loc}, (\absval_{i_j})^{j \leq n}}{ \abslh }{0^*}  ~|~ \const{c''} \in \abslookup(m') \wedge \sign(\const{c''},m') = \methsign{(\tau_j)^{j \leq n}}{\tau}{\loc}\}\, \cup \)\\
\(  \{ \absreg{\spp}{ (\absthread, \_ )}{\absval^*}{ \abslh }{\absfi} \wedge \rlookup{o}{\absval^*}{\abslh}{\_}{\absobj{c'}{(f \mapsto \absual)^*}}  \wedge c' \subtype \const{c''} \wedge \absresult{\const{c''},\mathsf{m'}}{(\absthread',\abswal^*)}{\absval'_{\res}}{\abslh_{{\res}}}{\absfi_\res}  \)\\
\(\null\hfill\wedge \absthread = \absthread' \wedge \left(\bigwedge_{j \le n} \absval_{i_j} \absmeet \abswal_j \absnpo\bot\right) \implies \absreg{\apcn}{ (\absthread, \_ )} {\lift{\absval^*}{\absfi_\res}[\res \mapsto \absval'_{\res}]}{ \abslh_{{\res}} }{\absfi \afunion \absfi_\res} ~|~ \const{c''} \in \abslookup(\mathsf{m'})\} \)\\
\(  \{ \absreg{\spp}{ (\absthread,\_) }{\absval^*}{ \abslh }{\absfi} \wedge \rlookup{o}{\absval^*}{\abslh}{\_}{\absobj{c'}{(f \mapsto \absual)^*}}  \wedge c' \subtype \const{c''} \wedge \absuncaught{\const{c''},\mathsf{m'}}{(\absthread',\abswal^*))}{\absval'_\excpt}{\abslh_{{\res}}}{\absfi_\res}\)\\
\(\null\hfill\wedge \absthread = \absthread' \wedge \left(\bigwedge_{j \le n} \absval_{i_j} \absmeet \abswal_j \absnpo \bot\right) \implies \absabnormal{\apc}{ (\absthread, \_ )} {\lift{\absval^*}{\absfi_\res}[\excpt \mapsto \absval'_{\excpt}]}{ \abslh_{{\res}} }{\absfi \afunion \absfi_\res} ~|~ \const{c''} \in \abslookup(\mathsf{m'})\}\)

\item $\ainst{\sinvoke{c'}{m'}{(r_{i_j})^{j \leq n}}} =$\\
$ \{\absreg{\spp}{(\absthread, \_ )}{\absval^*}{\abslh}{\absfi} \implies \absreg{\mathsf{c'},\mathsf{m'},\mathsf{0}}{(\absthread,(\absval_{i_j})^{j \leq n})}{(\adefvalue_k)^{k \leq \loc},(\absval_{i_j})^{j \leq n}}{\abslh}{0^*} ~|~ \sign(\mathsf{c'}, m') = \methsign{(\tau_j)^{j \leq n}}{\tau}{\loc} \}\,  \cup$\\
$\{\absreg{\spp}{ (\absthread, \_ ) }{\absval^*}{\abslh}{\absfi} \wedge \absresult{\mathsf{c'},\mathsf{m'}}{(\absthread',\abswal^*)}{\absval'_{\res}}{\abslh_{{\res}}}{\absfi_\res}\wedge \absthread = \absthread' \wedge \left(\bigwedge_{j \le n} \absval_{i_j} \absmeet \abswal_j \absnpo \bot\right)$\\
$\null\hfill\implies \absreg{\apcn}{ (\absthread, \_ )} {\lift{\absval^*}{\absfi_\res}[\res \mapsto \absval'_{\res}]}{\abslh_{{\res}}}{\absfi \afunion \absfi_\res}\} $\\
$\{\absreg{\spp}{ (\absthread, \_ ) }{\absval^*}{\abslh}{\absfi} \wedge \absuncaught{\mathsf{c'},\mathsf{m'}}{(\absthread',\abswal^*)}{\absval'_{\excpt}}{\abslh_{{\res}}}{\absfi_\res}\wedge \absthread = \absthread' \wedge \left(\bigwedge_{j \le n} \absval_{i_j} \absmeet \abswal_j \absnpo \bot\right)$\\
$\null\hfill\implies \absabnormal{\apc}{ (\absthread, \_ )} {\lift{\absval^*}{\absfi_\res}[\excpt \mapsto \absval'_{\excpt}]}{\abslh_{{\res}}}{\absfi \afunion \absfi_\res}\} $
\end{itemize}
\textbf{Conventions:} $\spp = \mathsf{c,m,\spc}$
\caption{\label{tab:abs-statements-inv} Abstract Semantics of \sem{} - Invoke Statements}
\end{table}

\begin{table}[htb]
\textbf{Statement Abstractions:}

\(\begin{array}[t]{lcl}

\ainst{\startthread{r_i}}  &= &
 \{\absreg{\spp}{\_}{\absval^*}{\abslh}{\absfi} \wedge \rlookup{i}{\absval^*}{\abslh}{\absg{\absloc}}{\absthreadobj{c'}{(f \mapsto \absual)^*}} \wedge c' \le \thread\\
&&\implies \abstdispatch(\absloc,\absthreadobj{c'}{(f \mapsto \absual)^*}) \wedge  \absreg{\apcn}{\_}{\absval^*}{\abslh}{\absfi}\}\, \cup\\ 
&&\{\absreg{\spp}{\_}{\absval^*}{\abslh}{\absfi} \wedge\rlookup{i}{\absval^*}{\abslh}{\absl{\absloc}}{\absthreadobj{c'}{(f \mapsto \absual)^*}} \wedge c' \le \thread\wedge \cfilter{\absl{\absloc}}{\abslh}{\absfi'} \\
&&\implies \abstdispatch(\absloc,\absthreadobj{c'}{(f \mapsto \absual)^*})\wedge \liftlh{\abslh}{\absfi'} \wedge \absreg{\apcn}{\_}{\lift{\absval^*}{\absfi'}}{\lhlift{\abslh}{\absfi'}}{\absfi \afunion \absfi'}\}\\

\ainst{\interruptthread{r_i}}  &=&
 \{\absreg{\spp}{\_}{\absval^*}{\abslh}{\absfi}  \wedge  \rlookup{i}{\absval^*}{\abslh}{\absg{\absloc}}{\absthreadobj{c'}{(f \mapsto \absual)^*,\interrupted \mapsto \_}}  \\
&&\implies \absheap(\absloc,\absthreadobj{c'}{(f \mapsto \absual)^*,\interrupted \mapsto \widehat{\true}} \wedge  \absreg{\apcn}{\_}{\absval^*}{\abslh}{\absfi}\} \, \cup \\ 
&& \{\absreg{\spp}{\_}{\absval^*}{\abslh}{\absfi}  \wedge   \rlookup{i}{\absval^*}{\abslh}{\absl{\absloc}}{\absthreadobj{c'}{(f \mapsto \absual)^*,\interrupted \mapsto \_}} \\
&&\implies  \absreg{\apcn}{\_}{\absval^*}{\abslh[\absloc \mapsto \absthreadobj{c'}{(f \mapsto \absual)^*,\interrupted \mapsto \widehat{\true}}]}{\absfi}\}\\

\ainst{\interruptedthread{r_i}}  &=&
 \{\absreg{\spp}{\_}{\absval^*}{\abslh}{\absfi}  \wedge  \rlookup{i}{\absval^*}{\abslh}{\absg{\absloc}}{\absthreadobj{c'}{(f \mapsto \absual)^*,\interrupted \mapsto \absval'}}  \\
&&\implies \absheap(\absloc,\absthreadobj{c'}{(f \mapsto \absual)^*,\interrupted \mapsto \widehat{\false}} \wedge  \absreg{\apcn}{\_}{\absval^*[\res \mapsto \absval']}{\abslh}{\absfi}\}\cup \\ 
&&  \{ \absreg{\spp}{\_ }{\absval^*}{\abslh}{\absfi} \wedge \rlookup{i}{\absval^*}{\abslh}{\absl{\absloc}}{\absthreadobj{c'}{(f \mapsto \absual)^*,\interrupted \mapsto \absval'}}\\
&&\implies \absreg{\apcn}{\_}{\absval^*[\res \mapsto \absval']}{\abslh[\absloc \mapsto \absthreadobj{c'}{(f \mapsto \absual)^*,\interrupted \mapsto \widehat{\false}}]}{\absfi}\}\\

\ainst{\isinterruptedthread{r_i}}  &=&
  \{ \absreg{\spp}{\_ }{\absval^*}{\abslh}{\absfi} \wedge \rlookup{i}{\absval^*}{\abslh}{\_}{\absthreadobj{c'}{(f \mapsto \absual)^*,\interrupted \mapsto \absval'}}\\
&&\implies \absreg{\apcn}{\_}{\absval^*[\res \mapsto \absval']}{\abslh}{\absfi}\}\\
\end{array}$

$\begin{array}[t]{lcl}
\ainst{\jointhread{r_i}}  &=&
 \{ \absreg{\spp}{(\absg{\absloc_t},\_)}{\absval^*}{\abslh}{\absfi}   \wedge \absheap(\absloc_t,\absthreadobj{c'}{(f \mapsto \absual)^*,\interrupted \mapsto \absval'})\wedge \widehat{\false} \abspo \absval' \\
&&  \implies \absreg{\apcn}{(\absg{\absloc_t},\_)}{\absval^*}{\abslh}{\absfi}\}\cup\\
&& \{ \absreg{\spp}{(\absg{\absloc_t},\_)}{\absval^*}{\abslh}{\absfi} \wedge\absheap(\absloc_t,\absthreadobj{c'}{(f \mapsto \absual)^*,\interrupted \mapsto \absval'})  \wedge \widehat{\true} \abspo \absval'  \implies\\
&& \absheap(\spp;\absobj{\interruptedexception}{})\wedge \absabnormal{\spp}{(\absg{\absloc_t},\_)}{\absval^*[\excpt \mapsto \absg{\spp}]}{\abslh}{\absfi} \wedge \absheap(\absloc_t,\absthreadobj{c'}{(f \mapsto \absual)^*,\interrupted \mapsto \widehat{\false}})\}\\

\ainst{\startwait{r_i}}  &=&
\{ \absreg{\spp}{(\absg{\absloc_t},\_)}{\absval^*}{\abslh}{\absfi}  \wedge\absheap(\absloc_t,\absthreadobj{c'}{(f \mapsto \absual)^*,\interrupted \mapsto \absval'}) \wedge \widehat{\false} \abspo \absval'\\
&&  \implies \absreg{\apcn}{(\absg{\absloc_t},\_)}{\absval^*}{\abslh}{\absfi}\}\cup\\
&&\{ \absreg{\spp}{(\absg{\absloc_t},\_)}{\absval^*}{\abslh}{\absfi}\wedge  \absheap(\absloc_t,\absthreadobj{c'}{(f \mapsto \absual)^*,\interrupted \mapsto \absval'}) \wedge \widehat{\true} \abspo \absval'  \implies\\
&& \absheap(\spp;\absobj{\interruptedexception}{})\wedge\absabnormal{\spp}{(\absg{\absloc_t},\_)}{\absval^*[\excpt \mapsto \absg{\spp}]}{\abslh}{\absfi} \wedge \absheap(\absloc_t,\absthreadobj{c'}{(f \mapsto \absual)^*,\interrupted \mapsto \widehat{\false}})\}\\
\end{array}$

$\begin{array}[t]{lcl}
\ainst{\monitorenter{r_i}}  &=&
  \{ \absreg{\spp}{\_ }{\absval^*}{\abslh}{\absfi} \implies \absreg{\apcn}{\_}{\absval^*}{\abslh}{\absfi}\}\\

\ainst{\monitorexit{r_i}}  &=& 
  \{ \absreg{\spp}{\_ }{\absval^*}{\abslh}{\absfi} \implies \absreg{\apcn}{\_}{\absval^*}{\abslh}{\absfi}\}\\

\ainst{\throw{r_i}}  &=&
\{ \absreg{\apc}{\_ }{\absval^*}{\abslh}{\absfi} \implies \absabnormal{\apc'}{\_}{\absval^*[\excpt \mapsto \absval_i]}{\abslh}{\absfi}\}\\

\ainst{\moveexcpt{r_d}}  &=& 
 \{ \absreg{\apc}{\_ }{\absval^*}{\abslh}{\absfi} \implies \absreg{\apc+1}{\_}{\absval^*[d \mapsto  \absval_\excpt]}{\abslh}{\absfi}\}\\
\end{array}\)\\[1em]

\textbf{Global Abstractions:}

\(\begin{array}{lcl}
\rulename{AbState} &=&
\{ \absabnormal{\apc}{\_ }{\absval^*}{\abslh}{\absfi}  \wedge \rlookup{\excpt}{\absval^*}{\abslh}{\_}{\absobj{\mathsf{c'}}{\_}} \wedge \mathsf{c'} \le \throwable\\
&&\implies \absreg{\apc'}{\_}{\absval^*}{\abslh}{\absfi}~|~ \excpttable{\apc}{\mathsf{c'}} = \spc'\}\\
&& \{ \absabnormal{\apc}{\_ }{\absval^*}{\abslh}{\absfi}  \wedge \rlookup{\excpt}{\absval^*}{\abslh}{\_}{\absobj{\mathsf{c'}}{\_}}\wedge \mathsf{c'} \le \throwable\\
&&\implies \absuncaught{\mathsf{c},\mathsf{m}}{\_}{\absval_\excpt}{\abslh}{\absfi}~|~ \excpttable{\apc}{\mathsf{c'}} = \bot\}
\end{array}\)

\textbf{Conventions:} $\spp = \mathsf{c,m,\spc}$
\caption{\label{tab:abs-statements-thread} Abstract Semantics of \sem{} - Rules for New Statements}
\end{table}


\FloatBarrier
\clearpage
\section{Proofs}
\label{sec:proof}
Before entering in the formalism, we are going to give an informal description of the difficulties. The main problem is that knowing which locations are going to be abstracted as abstract \flowsensitive locations and which locations are going to be abstracted as abstract \flowinsensitive locations is \emph{dynamically} determined by the analysis: this is not a property of the concrete semantics that is abstracted. That is, given a snapshot of an execution (a configuration $\Psi$), there is no \emph{unique} correct way of choosing which locations should be handled in a \flowsensitive fashion, since the information about who are the most-recently allocated locations is not stored in $\Psi$. Therefore  there are several ways of abstracting a configuration: there is one possible abstraction of a configuration for each decomposition of the set of locations into locations that are handled in a \flowsensitive fashion and location that are handled in a \flowinsensitive fashion, and for each \emph{history} of the heap. An \emph{history} is a record of which locations used to be abstracted as abstract \flowsensitive locations, and when they were lifted. To see why it is necessary to take into account the history, consider the following example.

\begin{example} 
Consider the following call-stack: $\callstack = \locstate{c,m,\pc}{\stm^*}{u}{R} :: \locstate{c',m',\pc'}{\stm'^*}{\_}{R'}$ with $R = (r_1 \mapsto \pointer{p}{\spp},r_2 \mapsto \pointer{p'}{\spp})$, $u = \pointer{p}{\spp}$ and $R' = (r \mapsto \pointer{p}{\spp})$.

Here there are several possible abstractions of this call-stack: for example, $\pointer{p}{\spp}$ could have been lifted before $c',m'$ invoked $c,m$, and $c,m$ could have just allocated a new object at location $\pointer{p'}{\spp}$, in which case $\pointer{p}{\spp}$ is abstracted in a \flowinsensitive fashion in both $c,m$ and $c',m'$.

But another possibility is that, when $c',m'$ invoked $c,m$, the location $\pointer{p}{\spp}$ was abstracted in a \flowsensitive fashion. Then later on $c,m$ allocated a new object with location $\pointer{p'}{\spp}$ at program point $\spp$, and $\pointer{p}{\spp}$ was lifted. In that case, $\pointer{p}{\spp}$ would abstracted in a \flowsensitive fashion in $c',m'$ and in a \flowinsensitive fashion in $c,m$. Therefore we need to record that $\pointer{p}{\spp}$ used to be abstract in a \flowsensitive fashion, and that lifting occurred somewhere between $c',m'$ and $c,m$: this will be done using \emph{filters} (which are the concrete counter-part of abstract filters).
\end{example}

\subsection{Heap decompositions}
We are now going to define formally what is the decomposition of a heap between a sub-heap (that will be handled in a \flowinsensitive fashion) and  local heaps (that will be handled in a \flowsensitive fashion). To do so we first need several definitions.

\paragraph{Heap} Formally we defined heaps as finite sequences of key-value bindings between a location and a memory block. We can then state that some location $\ell$ maps to $b$ by $(\ell \mapsto b) \in \heap$. The active domain of a heap $\heap$, denoted by $dom(\heap)$, is the finite set of locations having a mapping in $\heap$.

For convenience reasons, we would like to see a heap $\heap$ as a function from the set of locations to memory block: to do so we use the special symbol $\bot$ that we introduced for abstract \flowsensitive heap entries. We will see the heap as a function that maps any location to a memory block or $\bot$. Since the heap is a finite sequence of key-value bindings between a location and a memory block, this function has a finite support. To summarize, if one reads $(\ell \mapsto b) \in \heap$ then we know that $\ell$ is in the active domain of $\heap$ and that it points to the memory block $b$, whereas $\heap(\ell)$ may be either a memory block, or the empty block $\bot$.

\paragraph{Local heap} Intuitively a local heap $\lheap$ is a heap such that for all $\spp$, there is at most one memory block $b$ such that $(\spp \mapsto b) \in \lheap$. For technical reasons we will consider a slightly different definition: a local heap is a finite sequence of key-value bindings from locations to memory block or $\bot$ such that there is \emph{exactly} one key-value binding for all $\spp$. Formally we have:
\begin{definition} A heap $\lheap$ is a local heap if and only if it satisfies the following equations:
\begin{itemize}
\item $\forall \spp, p,p'.\; \pointer{p}{\spp} \in dom(\lheap) \wedge \pointer{p'}{\spp} \in dom(\lheap) \Rightarrow p = p'$
\item $\forall \spp. \exists p. (\pointer{p}{\spp} \mapsto \_) \in \lheap$
\end{itemize}
\end{definition}
\begin{remark}
Observe that if a heap $\heap$ and some local heaps $(\lheap_i)_{i \le n}$ have disjoint domains then we can easily define their union.
\end{remark}

We define the relation $\heap \heapto \gheap$ between two heaps (local or not), to holds if the heap $\heap$ contains an memory block storing a location to an element of $\gheap$.

\begin{definition}
$\heap \heapto \gheap$ if and only if there exists $(\_ \mapsto b) \in \heap$ such that one of the following cases holds:
\begin{itemize}
\item  $b = \obj{c}{(f_i \mapsto v_i)^*} \in \heap$ and there exists $j$ such that $v_j \in dom(\gheap)$.
\item  $b = \intent{c}{(f_i \mapsto v_i)^*} \in \heap$ and there exists $j$ such that $v_j \in dom(\gheap)$.
\item  $b = \arr{\tau}{v^*} \in \heap$ and there exists $j$ such that $v_j \in dom(\gheap)$.
\end{itemize}
\end{definition}

\noindent
\begin{minipage}{0.5\linewidth}
Now we can define what the heap decomposition of a heap together with a static heap is. Intuitively it is a partitioning of the heap $\heap$ into a heap $\gheap$ and a finite set of local heaps $(\lheap_i)_{i \le n}$ such we have no locations going from $\gheap$ to any $\lheap_i$, or from $\lheap_i$ to $\lheap_j$ for any $i \ne j$ (we allow locations from $\lheap_i$ to $\lheap_i$ or to $\gheap$, and locations from $\gheap$ to itself). Formally:

\begin{definition}
$(\gheap,(\lheap_i)_{i \le n})$ is a heap decomposition of $\heap\cdot\sheap$ if and only if:
\begin{itemize}
\item $\heap = \gheap \cup \bigcup_{i \le n} \lheap_i$
\item $\forall i. dom(\gheap) \cap dom(\lheap_i) = \emptyset$
\item $\forall i\ne j. dom(\lheap_i) \cap dom(\lheap_j) = \emptyset$
\item $\forall i. \gheap \cup \sheap \not \heapto \lheap_i$ and $\forall j\ne i. \lheap_i \not \heapto \lheap_j$
\end{itemize}
\end{definition}
\end{minipage}
\begin{minipage}{0.45\linewidth}
\begin{center}
\begin{tikzpicture}[>=stealth,,transform shape,scale=0.6]
\tikzset{state/.style={rounded corners,draw,minimum width=1cm,minimum height=1cm}};

\node[state] (k1) at (0,0){$\lheap_1$};
\node[state] (k2) at (3,6){$\lheap_2$};
\node[state] (k3) at (6,0){$ \lheap_3$};

\node[state] (g) at (3,2.25){$\gheap$};

\draw[<->,red] (k1.north) to[bend left] node[midway,rotate=60]{$|$} (k2.west);

\draw[<->,red] (k2.east) to[bend left] node[midway,rotate=120]{$|$} (k3.north);

\draw[<-,red] (k1.east) to[bend right] node[midway]{$|$} (k3.west);

\draw[<-,red] (k1.35) to[bend right=20] node[rotate=30,midway]{$|$} (g.235);
\draw[->,blue] (k1.55) to[bend left=20] node[midway]{$ $} (g.215);

\draw[<-,red] (k2.260) to[bend right=25] node[rotate=90,midway]{$|$} (g.100);
\draw[->,blue] (k2.280) to[bend left=20] node[midway]{$ $} (g.80);

\draw[<-,red] (k3.125) to[bend right=20] node[rotate=-30,midway]{$|$} (g.325);
\draw[->,blue] (k3.155) to[bend left=20] node[midway]{$ $} (g.305);

\draw[rounded corners] (-1,-1.5) rectangle (7,7);
\node () at (6,6) {$\heap$};


\end{tikzpicture}

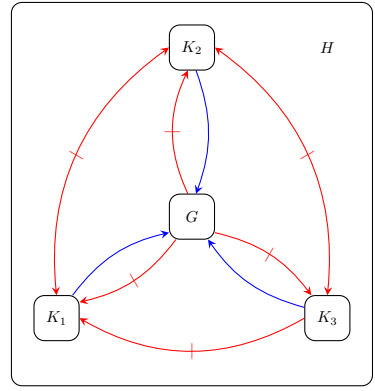
\textbf{Example:} a local heap decomposition with three local heaps.
\end{center}


\end{minipage}

\subsection{Filter history}

We are now going to define formally what the history of a configuration is. As we mentioned earlier, this is used to determine which locations were lifted, and when (in a given call-stack). It turns out that this definition is quite technical, because we need to make sure that the history of a configuration respected some properties: no locations should have been lifted twice, and a location to an object cannot appear in a local state that is situated in the call-stack \emph{before} the local state that  allocated this object.

First, we are going to define what a filter is. Filters are going to be used to represent one \emph{layer} of the history, that is which locations were lifted between two local states.
\begin{definition}
A filter $\lfilter$ is a mapping from locations to $\{0,1\}$ such that for all $\spp$, there exists at most one $p$ such that $\lfilter(\pointer{p}{\spp}) = 1$. Besides we define the following function:
\[\lfilter \lfunion \lfilter' = \left(\pointer{p}{\spp} \mapsto
\begin{cases} 
1 \text{ if }\lfilter'(\pointer{p}{\spp}) = 1\\
1 \text{ if }\lfilter(\pointer{p}{\spp}) = 1 \text{ and }\forall \pointer{p'}{\spp}, \lfilter'(\pointer{p'}{\spp}) = 0\\
0 \text{ otherwise}
\end{cases}
\right)^*
\]
\end{definition}

\begin{proposition}
The binary operation $\lfunion$ admits $(\spp \mapsto 0)^*$ as left and right neuter and is associative.
\end{proposition}

\begin{remark}
$\lfunion$ is \textbf{not} commutative.
\end{remark}

The history of a call-stack $\callstack = L_1 :: \dots :: L_n$ is going to be recorded using a list of filters $\flist$, such that for all $i$, $\lfilter_i$ records which locations were lifted between $L_i$ and $L_{i+1}$. We then define, for all $i$, the function $\fhistget{\fhist}{i}$ that, given a local heap and an history, give us which for all program point $\spp$ the location which is handled in a \flowsensitive fashion in the local state $L_{i}$.
\begin{definition}
For all $i \in \mathbb{N}\cup\{ +\infty\}$, $\fhistget{\fhist}{i}$ is the function defined as follows: let $\lfilter = \lfilter^1 \lfunion \dots \lfunion \lfilter^{i-1}$, then
\[
\fhistget{\fhist}{i} = \left(\spp \mapsto
\begin{cases}
\pointer{p}{\spp} \text{ if } \lfilter(\pointer{p}{\spp}) = 1\\
\pointer{p}{\spp} \text{ if } \pointer{p}{\spp} \in dom(\lheap_a) \wedge \forall \pointer{p'}{\spp}, \lfilter(\pointer{p'}{\spp}) = 0\\
\end{cases}
\right)^*
\]
\end{definition}
A graphical representation of $\Gamma$ on an example can be found in Figure~\ref{fig:filter-history}.
\begin{table}[h]
\begin{center}
\begin{tikzpicture}
\tikzset{n/.style={color=blue}}
\tikzset{rn/.style={draw,thick,color=red, rounded corners=4, minimum height=1.6em, minimum width=1.6em}}
\tikzset{gn/.style={draw,thick,color=goodgreen, rounded corners=4, minimum height=2em, minimum width=2em}}

\draw[step=1] (0,0) grid (7,5);
\draw[step=1] (0,-2) grid (7,-1);

\foreach \i in {1,...,5}{
  \pgfmathsetmacro{\label}{\i - 0.5};
  \node (lf\i) at (-0.5,\label) {$\lfilter_{\i}$};
}
\node (ka) at (-0.5,-1.5){$\lheap_a$};

\foreach \i in {1,...,7}{
  \pgfmathsetmacro{\label}{\i - 0.5};
  \node (pp\i) at (\label,5.5) {$\spp_{\i}$};
  \node[n] (lh\i) at (\label,-1.5) {$\ell_{\i}$};

}

\node[rn] (ra1) at (0.5,-1.5) {};
\node[rn] (ra4) at (3.5,-1.5) {};
\node[rn] (ra5) at (4.5,-1.5) {};
\node[rn] (ra7) at (6.5,-1.5) {};
\node[gn] (ga7) at (6.5,-1.5) {};

\node[n] (a8) at (1.5,0.5) {$\ell_{8}$};
\node[rn] (ra8) at (1.5,0.5) {};
\node[gn] (ga8) at (1.5,0.5) {};
\node[n] (a9) at (2.5,0.5) {$\ell_{9}$};
\node[rn] (ra9) at (2.5,0.5) {};
\node[gn] (ga9) at (2.5,0.5) {};
\node[n] (a10) at (5.5,0.5) {$\ell_{10}$};
\node[rn] (ra10) at (5.5,0.5) {};

\node[n] (a11) at (0.5,1.5) {$\ell_{11}$};
\node[gn] (ga11) at (0.5,1.5) {};

\node[n] (a12) at (3.5,2.5) {$\ell_{12}$};
\node[gn] (ga12) at (3.5,2.5) {};
\node[n] (a13) at (4.5,2.5) {$\ell_{13}$};
\node[gn] (ga13) at (4.5,2.5) {};
\node[n] (a14) at (5.5,2.5) {$\ell_{14}$};
\node[gn] (ga14) at (5.5,2.5) {};

\node[n] (a15) at (1.5,3.5) {$\ell_{15}$};
\node[n] (a16) at (4.5,3.5) {$\ell_{16}$};

\node[n] (a17) at (0.5,4.5) {$\ell_{17}$};
\node[n] (a18) at (2.5,4.5) {$\ell_{18}$};

\draw[very thick,red] (-0.5,1) -- ++(8,0) node[right] {$\fhistget{(\lheap_a,(\lfilter_i)_{i \le 5})}{2}$};
\draw[very thick,goodgreen] (-0.5,3) -- ++(8,0) node[right] {$\fhistget{(\lheap_a,(\lfilter_i)_{i \le 5})}{4}$};
\end{tikzpicture}
\end{center}
\textbf{Convention:} Each line of the table represents one local filter, by having a pointer $\ell$ in position $(\lfilter_i,\spp_j)$ if and only if there exists $p$ such that $\ell = \pointer{p}{\spp}$ and $\lfilter_i(\ell) = 1$. The last line represent the domain of the local heap $\lheap_a$.

The pointer framed by red (resp. green) in column $\spp_i$ is the image of $\spp_i$ by $\fhistget{(\lheap_a(\lfilter_i)_{i \le 5})}{2}$ (resp. $\fhistget{(\lheap_a,(\lfilter_i)_{i \le 5})}{4}$).
\caption{Graphical representation of the $\fhistget{(\lheap_a,(\lfilter_i)_{i \le n})}{j}$ functions}
\label{fig:filter-history}
\end{table}

\begin{proposition}[Properties of $\fhistget{}{}$]
\label{prop:fhsitget}
For all  $(\lheap_a,(\lfilter_i)_{1 \le i \le n})$  we have :
\begin{enumerate}
\item For all $i \in \{n + 1, n + 2,\dots\} \cup \{\infty\}$, $\fhistget{(\lheap_a,(\lfilter_j)_{1 \le j \le n})}{i} = \fhistget{(\lheap_a,(\lfilter_j)_{1 \le j \le n})}{n + 1}$ 
\item If $n \ge 2$, then for all $i > 1$, $\fhistget{(\lheap_a,(\lfilter_j)_{1 \le j \le n})}{i + 1} = \fhistget{(\lheap_a,(\lfilter_1 \lfunion \lfilter_2) :: (\lfilter_j)_{3 \le j \le n})}{i}$
\item For all $i \ge 0$, $\fhistget{(\lheap_a,(\lfilter_j)_{1 \le j \le n})}{i} = \fhistget{(\lheap_a,(\spp \mapsto 0)^*  :: (\lfilter_j)_{1 \le j \le n})}{i + 1}$
\item Let $\lheap_a'$ be a local heap such that $dom(\lheap_a) = dom(\lheap_a')$. Then for all $j$ we have:
\[ \fhistget{(\lheap_a,(\lfilter_j)_{1 \le j \le n})}{i} = \fhistget{(\lheap'_a,(\lfilter_j)_{1 \le j \le n})}{i}\]
\item Let $\lfilter_a$ be a filter such that $\forall \ell,\lfilter_a(\ell) = 1 \implies \ell \in dom(\lheap_a)$. Let $\lheap_a'$ be a local heap such that :
\[dom(\lheap'_a)\backslash \left\{\pointer{p}{\spp}\in dom(\lheap'_a) ~|~ \exists p' , \lfilter_a(\pointer{p'}{\spp}) = 1\right\} \subseteq dom(\lheap_a)\]
Then for all $i \ge 2$ we have:
\[ \fhistget{(\lheap_a,(\lfilter_j)_{1 \le j \le n})}{i} = \fhistget{(\lheap'_a,(\lfilter_a \lfunion \lfilter_1) :: (\lfilter_j)_{2 \le j \le n})}{i}\]
\end{enumerate}
\end{proposition}

We can now define when  $(\lheap,(\lfilter^j)_j)$ is a filter history of a call-stack $\callstack$. Equation~\eqref{eq:fhsitdef1} expresses that a location never appears before it was allocated: this is done by stating that if, for a given $\spp$, the location $\pointer{p}{\spp}$ being handled in a \flowsensitive fashion in the local state $L_i$ is not the same one than in local state $L_j$ (where $L_j$ appears before $L_i$ in the call-stack), then no object was stored at location $\pointer{p}{\spp}$ when $L_j$ was the top-most element of the call-stack. Therefore $\pointer{p}{\spp}$ cannot appear in any of the local state $L_j :: \dots L_n$. Equation~\eqref{eq:fhsitdef2} expresses the fact that no location was lifted twice, and that if a location is in the local heap then it was never lifted.

\begin{definition}
 $(\lheap,(\lfilter^j)_j)$ is a filter history of $\callstack = L_1 :: \dots :: L_n$ if and only if for all $1 \le i < l \le n$ and for all $\spp$ we have:
\begin{gather}
\fhistget{(\lheap,(\lfilter^j)_j)}{i}(\spp) \ne \fhistget{(\lheap,(\lfilter^j)_j)}{l}(\spp) \implies \fhistget{(\lheap,(\lfilter^j)_j)}{i}(\spp) \not \in dom(L_{l} :: \ldots :: L_n)\label{eq:fhsitdef1}\\
\forall i, \forall \pointer{p}{\spp}, \left (\left(i = 0 \wedge \pointer{p}{\spp} \in dom(\lheap) \right) \vee \lfilter^i(\pointer{p}{\spp}) = 1 \right) \implies \forall j \ne i, \lfilter^j(\pointer{p}{\spp}) = 0\label{eq:fhsitdef2}
\end{gather}
\end{definition}

The following (rather technical) lemma gives sufficient conditions to show that $\fhistp$ is a filter history, knowing that $\fhist$ is a filter history and that $\fhist$ and $\fhistp$ coincide everywhere except on the top-most filter and on the local heap.

\begin{lemma}
\label{lem:fhist-char}
Let $(\lheap,(\lfilter^j)_j)$ be a filter history of $\callstack = L_1 :: \callstack_t$. Let $\callstack' = L_1' :: \callstack_t$, and $\fhistp$ be such that $(\lfilter'^j)_{j} = \lfilter'^1 :: (\lfilter^j)_{j > 1}$, and let  $n$ be the length of $\callstack'$. If the four following conditions holds:
\begin{alignat}{2}
\forall i > 1, \forall \spp, \fhistget{(\lheap,(\lfilter^j)_j)}{i}(\spp) \quad&=&& \fhistget{(\lheap',(\lfilter'^j)_j)}{i}(\spp)\label{eq:fhist1}\\
\left( dom(\lheap') \backslash  dom(\lheap)\right) \cap dom(\callstack_t) \quad&=\quad&& \emptyset\label{eq:fhist2}\\
\left( dom(\lheap') \backslash  dom(\lheap)\right) \cap \{ \ell ~|~ \exists j, \lfilter^j(\ell) = 1\} \quad&=\quad&& \emptyset\label{eq:fhistp1}\\
\{ \ell ~|~ \lfilter'^1(\ell) = 1 \wedge \lfilter'^1(\ell) \ne \lfilter^1(\ell) \} \quad&\subseteq\quad&& dom(\lheap) \backslash dom(\lheap')\label{eq:fhistp2}
\end{alignat}
then $(\lheap',(\lfilter'^j)_j)$ is a filter history of $\callstack'$.
\end{lemma}

\begin{IEEEproof}This proof is done in two steps:
\begin{itemize}
\item First we are going to show that for all $1 \le i < j < n$ we have:
\begin{equation}
\fhistget{(\lheap',(\lfilter'^j)_j)}{i}(\spp) \ne \fhistget{(\lheap',(\lfilter'^j)_j)}{l}(\spp) \implies \fhistget{(\lheap',(\lfilter'^j)_j)}{i}(\spp) \not \in dom(\callstack'_{l} :: \ldots :: \callstack'_{n})\label{eq:fhist3}
\end{equation}
\begin{itemize}
\item For $1 < i < l \le n$, using Equation~\eqref{eq:fhist1} we have that $\fhistget{\fhistp}{i}(\spp) \ne \fhistget{\fhistp}{l}(\spp)$ implies that $\fhistget{\fhist}{i}(\spp) \ne \fhistget{\fhist}{l}(\spp)$. Since $\fhist$ is a filter history of $L_1 :: \callstack_t$, this implies that $\fhistget{\fhist}{i}(\spp) \not \in dom(\callstack_{l} :: \ldots :: \callstack_{n})$. Since $l > 1$, $dom(\callstack_{l} :: \ldots :: \callstack_{n}) = dom(\callstack'_{l} :: \ldots :: \callstack'_{n})$. Moreover using Equation~\eqref{eq:fhist1} again we know that $\fhistget{\fhist}{i}(\spp) = \fhistget{\fhistp}{i}(\spp)$, therefore Equation~\eqref{eq:fhist3} holds.

\item For $i = 1$, and $1 < l \le n$. If $\fhistget{(\lheap',(\lfilter'^j)_j)}{1}(\spp) = \fhistget{(\lheap,(\lfilter^j)_j)}{1}(\spp)$ then the same argument works. If $\fhistget{(\lheap',(\lfilter'^j)_j)}{1}(\spp) \ne \fhistget{(\lheap,(\lfilter^j)_j)}{1}(\spp)$, then since locations are annotated by their allocation point, and each local heap domain contains at most one location for each allocation point, we have $\fhistget{(\lheap',(\lfilter'^j)_j)}{1}(\spp) \in \left( dom(\lheap') \backslash  dom(\lheap)\right)$. Therefore by applying Equation~\eqref{eq:fhist2} we get that $\fhistget{(\lheap',(\lfilter'^j)_j)}{1}(\spp) \not \in dom(\callstack_t)$, which shows that Equation~\eqref{eq:fhist3} holds.
\end{itemize}

\item Now we are going to show that:
\[\forall i, \forall \pointer{p}{\spp}, \left (\left(i = 0 \wedge \pointer{p}{\spp} \in dom(\lheap') \right) \vee \lfilter'^i(\pointer{p}{\spp}) = 1 \right) \implies \forall j \ne i, \lfilter'^j(\pointer{p}{\spp}) = 0\]
Since we know that $(\lheap,(\lfilter^j)_j)$ is a filter history, we just need to show it for $i = 0$ and $i = 1$.
\begin{itemize}
\item $i = 0$. Let $\ell = \pointer{p}{\spp} \in dom(\lheap')$. In a first time assume that $\ell \in dom(\lheap)$. Since $(\lheap,(\lfilter)^j)_j$ is a filter history we know that for all $j > 2, \lfilter'^j(\ell) = \lfilter^j(\ell) = 0$. It remains to show that $\lfilter'^1(\ell) = \lfilter^1(\ell) = 0$: if $\lfilter'^1(\ell) = 0$ then we have nothing to prove, and if $\lfilter'^1(\ell) \ne 0$ then since $\ell \in dom(\lheap')$, Equation~\eqref{eq:fhistp2} gives us  that $\lfilter^1(\ell) = \lfilter'^1(\ell) \ne 0$, which contradicts the fact that $(\lheap,(\lfilter)^j)_j$ is a filter history.

Now assume that $\ell \not \in dom(\lheap)$. Then by Equation~\eqref{eq:fhistp1} we know that  $\forall j > 2, \lfilter'^j(\ell) = \lfilter^j(\ell)$. Besides by Equation~\eqref{eq:fhistp2} we know that either $\lfilter'^1(\ell) = 0$, in which case we have nothing to prove, or that $\lfilter'^1(\ell) = \lfilter^1(\ell) = 1$, which contradict Equation~\eqref{eq:fhistp1}.

\item $i = 1$. Let $\ell = \pointer{p}{\spp}$ be such that $\lfilter'^1(\ell) = 1$. If $\lfilter'^1(\ell) = \lfilter^1(\ell)$ then since $(\lheap,(\lfilter)^j)_j$ is a filter history we know that for all $j > 2, \lfilter'^j(\ell) = \lfilter^j(\ell) = 0$. If $\lfilter'^1(\ell) \ne \lfilter^1(\ell)$ then by Equation~\eqref{eq:fhistp2} we know that $\ell \in dom(\lheap)$ and we conclude again by using the fact that  $(\lheap,(\lfilter)^j)_j$ is a filter history.
\end{itemize}
\end{itemize}
\end{IEEEproof}

\subsection{Configuration Decomposition}
The heap decomposition notion is relative to a \emph{heap}, and the filter history notion is relative to a \emph{call-stack}. We then link these two notions into the local configuration decomposition notion, that is relative to a \emph{local configuration}.

\begin{definition}
$\lheapdh$ is a local configuration decomposition of $\Sigma = \tmethconf{\callstack}{\pi}{\threadstack}{\heap}{\sheap}{\ell}$ if and only if:
\begin{itemize}
\item $\lheapd$ is a heap decomposition of $\heap\cdot\sheap$ and $\lheap \in (\lheap_i)_i$
\item $dom(\callstack) \subseteq dom(\gheap) \cup dom(\lheap)$
\item $(\lheap,\flist)$ is a filter history of $\callstack$
\item $\forall i \in \pi, \exists \pointer{p}{\ann}, (\pointer{p}{\ann} \mapsto i) \in \gheap$
\item $\forall \ell \in \threadstack, \ell \in dom(\gheap)$
\item $\ell \in dom(\gheap)$
\end{itemize}
\end{definition}

Finally we use the local configuration decomposition notion to define what is a \emph{configuration decomposition}.
\begin{definition}
Let $\actstack = \phi_1 :: \dots :: \phi_n $ and $\threadpool = \psi_1 :: \dots :: \psi_m$. Then $(\gheap,(\lheap_i,(\lfilter^{i,j})_j)_{i \le n+m})$ is a configuration decomposition of $\tactconf{\actstack}{\threadpool}{\heap}{\sheap}$ if and only if:
\begin{itemize}
\item $\lheapd$ is a heap decomposition of $\heap\cdot\sheap$.
\item for all $i \le n$, if $\phi_i \in \{  \tactframe{\ell}{s}{\pi}{\threadstack}{\callstack},\tuactframe{\ell}{s}{\pi}{\threadstack}{\callstack}\}$ then $(\gheap,(\lheap_j)_j,\lheap_i,(\lfilter^{i,j})_j)$ is a heap decomposition history of $\tmethconf{\callstack}{\pi}{\threadstack}{\heap}{\sheap}{\ell}$ with local heap $\lheap_i$.
\item for all $n+1 \le i \le m+n$, if $\psi_i =  \threadframe{\ell}{\ell'}{\pi}{\threadstack}{\callstack}$ then $(\gheap,(\lheap_j)_j,\lheap_i,(\lfilter^{i,j})_j)$ is a heap decomposition history of $\tmethconf{\callstack}{\pi}{\threadstack}{\heap}{\sheap}{\ell}$ with local heap $\lheap_i$.
\end{itemize}
\end{definition}

\subsection{Well-Formedness}
First we are going to make some assumptions on the program $P$, which are guaranteed by the Java type system: we assume that the exception table built by the compiler only contain entries for exception class, and that the compiler guarantee type soundness for the thread and exception rules.
\begin{assumption}[Exception Table Correction]
\label{asm:excpt-table}
If $\excpttable{c,m,\pc}{c'}$ is defined (i.e is equal to some $\pc'$ or to $\bot$) then $c' \le \throwable$.
\end{assumption}

\begin{assumption}[Type Soundness Guarrantee]\null\hfill
\label{asm:thread-excpt-sound}
\begin{itemize}
\item If $\Sigma, \throw{r_e} \Downarrow \Sigma'$ and $\heap(\regval{r_e}) = \obj{c'}{(f \mapsto v)^*}$ then $c' \le \throwable$.
\item If $\Sigma, \stm \Downarrow \Sigma'$ where $\stm \in \{\startthread{r_t},\interruptthread{r_t},\jointhread{r_t}\}$ and $\heap(\regval{r_t}) = \obj{c'}{(f \mapsto v)^*}$ then $c' \le \thread$.
\end{itemize}
\end{assumption}

We are going to need some \emph{well-formedness} properties in the proof, that are preserved by the local configuration and configuration reductions.

\begin{definition}
A local configuration $\Sigma = \tmethconf{\callstack}{\pi}{\threadstack}{\heap}{\sheap}{\ell}$ is \emph{well-formed}  if and only if, whenever $\callstack = L_1 :: \ldots :: L_n$ or $\callstack = \abnormal{L_1 :: \ldots :: L_n}$, we have:
\begin{itemize}
\item For all $i$, $L_i = \waiting{\_}{\_}$ implies that $i = 1$ and $\callstack = \abnormal{L_1 :: \ldots :: L_n}$.
\item If $L_1 = \waiting{\ell_o}{\_}$ then $L_2 = \locstate{c,m,\pc}{\stm^*}{\_}{\_}$ with $\stm_\pc = \startwait{r_i}$ and $\ell_o = \regval{r_i}$.
\item For all $i \le n$,  if $L_i = \locstate{c,m,\pc}{\stm^*}{R}{v^*}$ and $R(r) = \ell$ then $\ell \in dom(\heap)$.
\item For all $\ell \in \gamma$, if $\heap(\ell) = \obj{c'}{\_}$ then $c' \le \thread$.
\item Either $n \in \{0,1\}$, or $n \geq 2$ and for each $i \in [2,n]$, either of the following conditions hold true:
\begin{itemize}
\item $L_i = \locstate{c',m',\pc'}{\stm'^*}{R'}{v'^*}$ and $L_{i-1} = \locstate{c,m,\pc}{\stm^*}{R}{\_}$ with $\stm_{\pc} = \invoke{r_o}{m'}{r_1,\ldots,r_n}$, \\ $\lookup(\gettype{\heap}{R({r_o})},m') = (c',\stm'^*)$, $\sign(c',m') = \methsign{\tau_1,\ldots,\tau_n}{\tau}{\loc}$ and $v'^* = (R({r_k}))^{k \leq n}$
\item $L_i = \locstate{c',m',\pc'}{\stm'^*}{R'}{v'^*}$ and $L_{i-1} = \locstate{c,m,\pc}{\stm^*}{R}{\_}$ with $\stm_{\pc} = \sinvoke{c'}{m'}{r_1,\ldots,r_n}$, \\ $\lookup(c',m') = (c',\stm'^*)$, $\sign(c',m') = \methsign{\tau_1,\ldots,\tau_n}{\tau}{\loc}$ and $v'^* = (R({r_k}))^{k \leq n}$.
\end{itemize}
\end{itemize}
\end{definition}

\begin{lemma}[Preserving Local Well-formation]
\label{lem:preserve-local}
If $\Sigma$ is well-formed and $\Sigma \rightsquigarrow^* \Sigma'$, then $\Sigma'$ is well-formed.
\end{lemma}
\begin{IEEEproof}
By induction on the length of the reduction sequence and a case analysis on the last rule applied.
\end{IEEEproof}

\begin{definition}
A heap $\heap$ is \emph{well-typed} if and only if, whenever $\heap(\ell) = \obj{c}{(f_i \mapsto v_i)^{i \leq n}}$, for all $i \in [1,n]$ we have $\gettype{\heap}{v_i} \subtype \tau_i$, where $\tau_i$ is the declared type of field $f_i$ for an object of type $c$ according to the underlying program.
\end{definition}

\begin{assumption}[Java Type Soundness]
\label{asm:java-sound}
\item If $\tmethconf{\callstack}{\pi}{\threadstack}{\heap}{\sheap}{\ell} \rightsquigarrow \tmethconf{\callstack'}{\pi'}{\threadstack'}{\heap'}{\sheap'}{\ell'}$, then for any value $v$ we have $\gettype{\heap'}{v} \subtype \gettype{\heap}{v}$. Moreover, if $\heap$ is well-typed, then also $\heap'$ is well-typed.
\end{assumption}

\begin{definition}
A configuration $\Psi = \tactconf{\actstack}{\threadpool}{\heap}{\sheap}$ is \emph{well-formed}  if and only if:
\begin{itemize}
\item whenever $\actstack = \actstack_0 :: \varphi :: \actstack_1$ with $\varphi \in \{\tactframe{\ell}{s}{\pi}{\threadstack}{\callstack},\tuactframe{\ell}{s}{\pi}{\threadstack}{\callstack}\}$, we have
\begin{itemize}
\item $\heap(\ell) = \obj{c}{(f \mapsto v)^*}$ for some activity class $c$ and $\ell = \pointer{p}{c}$ for some pointer $p$
\item $\Sigma = \tmethconf{\callstack}{\pi}{\threadstack}{\heap}{\sheap}{\ell}$ is a well-formed local configuration
\end{itemize}

\item whenever $\threadframe{\ell}{\ell'}{\pi}{\threadstack}{\callstack} \in \threadpool$ , we have
\begin{itemize}
\item $\heap(\ell) = \obj{c}{(f \mapsto v)^*}$ for some activity class $c$ and $\ell = \pointer{p}{c}$ for some pointer $p$
\item $\heap(\ell') = \obj{c'}{(f' \mapsto v')^*}$ for some  thread class $c'$
\item $\Sigma = \tmethconf{\callstack}{\pi}{\threadstack}{\heap}{\sheap}{\ell}$ is a well-formed local configuration
\end{itemize}

\item $\heap$ is a well-typed heap.
\end{itemize}
\end{definition}

\begin{lemma}[Preserving Well-formation]
\label{lem:preserve-well}
If $\Psi$ is well-formed and $\Psi \Rightarrow^* \Psi'$, then $\Psi'$ is well-formed.
\end{lemma}
\begin{IEEEproof}
By induction on the length of the reduction sequence and a case analysis on the last rule applied, using Lemma~\ref{lem:preserve-local} and Assumption~\ref{asm:java-sound} to deal with case \irule{A-Active}.
\end{IEEEproof}
From now on, we tacitly focus only on well-formed configurations. All the formal results only apply to them: notice that well-formed configurations always reduce to well-formed configurations by Lemma~\ref{lem:preserve-well}.

\subsection{Representation Functions}

From now on, we will consider only ground abstract values, and we will identify these values with their evaluation in the abstract domain $\absdom$.

We are now ready to define the \emph{representation functions} that we will use in the proof. A representation function is a (possibly parametrized) function that takes as input a concrete value and returns an abstraction of this value. The final goal of this section is to define the representation function $\rfconf(\Psi)$ that takes as input a configuration $\Psi$ and returns a set of sets of abstract facts, where each set of abstract facts $X$ in $\rfconf(\Psi)$ is an abstraction of $\Psi$ for a given configuration decomposition.

\subsubsection{Basic Representation Functions}
First we presuppose the existence of a representation function $\rfprim$ which associates to each primitive value $\prim$ a corresponding abstract value $\{\absprim\}$. We then define the following representation function, that abstracts a filter $\lfilter$ into an abstract filter $\absfi$, where the $\absfi$ is the abstract filters that maps a program point $\spp$ to $1$ iff there exists a locations $\ell$ annotated with $\spp$ (i.e. $\ell = \pointer{p}{\spp}$) such that $\lfilter(\ell) = 1$.  
\[\rffilter(\lfilter) =\left( \spp \mapsto\begin{cases}  1 &\text{ if } \exists \pointer{p}{\spp}, \lfilter(\pointer{p}{\spp}) = 1\\0 &\text{ otherwise}\end{cases}\right)^*\]

We then define the \flowsensitive and \flowinsensitive location and value representation functions. The \flowsensitive representation functions are going to be used when the analysis is flow-sensitive (for example one registers), and the \flowinsensitive representation functions are going to be used when the analysis is \emph{not} flow-sensitive (for example on the static heap).

\noindent\resizebox{\textwidth}{!}{
$\begin{array}{|l|lcl|lcl|}
\cline{2-7}
\multicolumn{1}{c|}{\rule{0pt}{1.1em}}&\multicolumn{3}{c|}{\textsf{\flowsensitive abstraction}}&\multicolumn{3}{c|}{\textsf{\flowinsensitive abstraction}}\\
\hline
\rotatebox[origin=c]{90}{\;\textsf{location}\;\null}& \rfloc{}(\pointer{p}{\absloc},\lheap_a,\flist) &=&
\begin{cases}
\absl{\absloc} &\text{ if }  \absloc = \spp \wedge \pointer{p}{\spp} = \fhistget{\fhist}{\infty}(\spp)\\
\absg{\absloc} &\text{ otherwise}\\
\end{cases} \null &
\rflab(\pointer{p}{\absloc}) &=& \absloc \\
\hline
\rule{0pt}{2.3em}\rotatebox[origin=c]{90}{\;\textsf{value}\;\null}&\rflval{}(v,\lheap_a,(\lfilter^j)_j) &=&
\begin{cases}
\rfprim(v) & \text{if } v = \prim \\
\rfloc{}(v,\lheap_a,(\lfilter^j)_j) & \text{if } v = \ell
\end{cases}&
\rfval{}(v) &=&
\begin{cases}
\rfprim(v) & \text{if } v = \prim \\
\absg{\rflab(v)} & \text{if } v = \ell
\end{cases}\\
\hline
\end{array}$
}

We typically omit brackets around singleton abstract values, and we will write $\rflval{}(v,\lheap_a)$ instead of the more verbose $\rflval{}(v,\lheap_a,\varepsilon)$ when the filter list is empty.
\begin{remark}
Recall that by definition, only locations annotated with \emph{program points} can be abstracted as \flowsensitive abstract location. In particular activity object and their intents are always \flowinsensitive.
\end{remark}

With these representation functions, we can define the \flowsensitive  representation function $\rflblock{}$  for local blocks, and the \flowinsensitive representation function $\rfblock{}$  for blocks.
\begin{alignat*}{3}
&\rflblock{}(l,\lheap_a) &\quad=&\quad
\begin{cases}
\absobj{c}{(f \mapsto \absval)^*} & \text{if } l =  \obj{c}{(f \mapsto v)^*} \text{ and } \forall i: \rflval{}(v_i,\lheap_a) = \absval_i \\
\absintent{c}{\absval} & \text{if } l = \intent{c}{(f \mapsto v)^*} \text{ and } \absval = \absjoin_i\, \rflval{}(v_i,\lheap_a) \\
\absarray{\tau}{\absval} & \text{if } l = \arr{\tau}{v^*} \text{ and } \absval = \absjoin_i\, \rflval{}(v_i,\lheap_a)\\
\bot & \text{if } l = \bot
\end{cases}\\
&\rfblock{}(b) &\quad=&\quad
\begin{cases}
\absobj{c}{(f \mapsto \absval)^*} & \text{if } b =  \obj{c}{(f \mapsto v)^*} \text{ and } \forall i: \rfval{}(v_i) = \absval_i \\
\absintent{c}{\absval} & \text{if } b = \intent{c}{(f \mapsto v)^*} \text{ and } \absval = \absjoin_i\, \rfval{}(v_i) \\
\absarray{\tau}{\absval} & \text{if } b = \arr{\tau}{v^*} \text{ and } \absval = \absjoin_i\, \rfval{}(v_i)\\
\end{cases}
\end{alignat*}

\subsubsection{Advanced Representation Functions}

We define the representation function $\rflheap{}(\lheap_a)$ abstracting a local heap into an abstract \flowsensitive heap as follows:
\begin{equation*}
 \rflheap{}(\lheap_a) = \left\{\left(\spp \mapsto \rflblock{}\left(\lheap_a(\pointer{p}{\spp}),\lheap_a\right)\right) \;|\; \pointer{p}{\spp} \in dom(\lheap_a)\right\}
\end{equation*}

We have three representation functions used to abstract a local state $L$ taken from the call-stack $\callstack$ of a local configuration $\Sigma$, where $\ell$  is the pointer to the activity or thread object and $\lheap_a,(\lfilter^n)_n$ is a filter history of $\Sigma$:
\begin{itemize}
\item If a local state $L$ is not the top-most local state in its call-stack then we use $\rfinvoke{\ell}(L,n_0, c',\lheap_a,(\lfilter^n)_n)$ where $n_0$ is the position is the call-stack and $c'$ is the class of the object that $L$ invoked a method upon.
\begin{multline*}
\rfinvoke{\ell}(\locstate{\spp}{\stm^*}{R}{u^*},n_0, c',\lheap_a,(\lfilter^n)_n)  = \Big\{\absinvoke{\spp}{c'}{(\absthread,\absual^*)}{\absval^*}{\absfi}  ~|~\absfi = \rffilter(\lfilter^{n_0}) \\\wedge \forall j: \absual_j = \rflval{}(u_j,\lheap_a,(\lfilter^n)_{\mathbf{n \le {n_0}}}) \wedge \absthread = \rfval(\ell)  \wedge \forall k: \absval_k = \rflval{}(R(r_k),\lheap_a,(\lfilter^n)_{\mathbf{n < {n_0}}})\Big\} 
\end{multline*}
\item If $L$ is the top-most local state, and $\callstack$ is \textbf{not} abnormal, then we use $\rflocstate{\ell}(L,\lheap_a,(\lfilter^n)_n)$.
\begin{multline*}
\rflocstate{\ell}(\locstate{\spp}{\stm^*}{R}{u^*},\lheap_a,(\lfilter^n)_n)  =  \Big\{\absreg{\spp}{(\absthread,\absual^*)}{\absval^*}{\abslh}{\absfi} ~|~\absfi = \rffilter(\lfilter^{1})\\
\wedge \forall j: \absual_j = \rflval{}(u_j,\lheap_a,(\lfilter^n)_{\mathbf{n \le {1}}}) \wedge \absthread = \rfval(\ell)\wedge \forall k: \absval_k = \rflval{}(R(r_k),\lheap_a,(\lfilter^n)_{\mathbf{n < {1}}}) \wedge \abslh = \rflheap{}(\lheap_a)\Big\} 
\end{multline*}
\item If $L$ is the top-most local state, and $\callstack$ is  abnormal, then we use $\rfalocstate{\ell}(\locstate{\spp}{\stm^*}{R}{u^*},\lheap_a,(\lfilter^n)_n)$.
\begin{multline*}
\rfalocstate{\ell}(\locstate{\spp}{\stm^*}{R}{u^*},\lheap_a,(\lfilter^n)_n)  =  \Big\{\absabnormal{\spp}{(\absthread,\absual^*)}{\absval^*}{\abslh}{\absfi} ~|~\absfi = \rffilter(\lfilter^{1})\\
\wedge \forall j: \absual_j = \rflval{}(u_j,\lheap_a,(\lfilter^n)_{\mathbf{n \le {1}}}) \wedge \absthread = \rfval(\ell)\wedge \forall k: \absval_k = \rflval{}(R(r_k),\lheap_a,(\lfilter^n)_{\mathbf{n < {1}}}) \wedge \abslh = \rflheap{}(\lheap_a)\Big\} 
\end{multline*}
\end{itemize}

Using these, we can define how the call-stack $\callstack$ is abstracted. For all $i \le n$, let $L_i = \locstate{c_i,m_i,\pc_i}{\_}{\_}{\_}$. If $\callstack = L_1 :: \cdots :: L_n$ and $n \ge 1$ then:
\begin{alignat*}{3}
& \rfcall{\ell}(\waiting{\_}{\_} :: \callstack,\lheap_a,(\lfilter^n)_n) &\qquad&=\qquad&& \rfcall{\ell}(\callstack,\lheap_a,(\lfilter^n)_n)\\
&&   &=&&   \rflocstate{\ell}(L_1, \lheap_a,(\lfilter^n)_n) \cup \bigcup_{i \in [2,n]} \rfinvoke{\ell}(L_i, i, c_{i-1},\lheap_a,(\lfilter^n)_n)\\
& \rfcall{\ell}(\abnormal{\callstack},\lheap_a,(\lfilter^n)_n)   &\qquad&=&&   \rfalocstate{\ell}(L_1, \lheap_a,(\lfilter^n)_n) \cup \bigcup_{i \in [2,n]} \rfinvoke{\ell}(L_i, i, c_{i-1},\lheap_a,(\lfilter^n)_n)\\
&\rfcall{\ell}(\varepsilon,\lheap_a,(\lfilter^n)_n) &\qquad&=&&\rfcall{\ell}(\abnormal{\varepsilon},\lheap_a,(\lfilter^n)_n) \qquad=\qquad  \emptyset
\end{alignat*}

We can now define the following representation functions:
\begin{alignat*}{3}
&\begin{array}{lcl}
\rfheap{\gheap}(\heap) & = & \Big\{\absheap(\absloc,\absblock) ~|~ \heap(\ell') = b \wedge \absloc = \rflab(\ell')\wedge \absblock = \rfblock{}(b) \wedge \ell' \in dom(\gheap)\Big\} \\
\rfstat{}(\sheap) & = & \left\{\abssheap(c,f,\absval) ~|~ \sheap = \sheap',c.f \mapsto v \wedge \absval = \rfval{}(v)\right\} \\
\rfdispatch{\ell}(\pi) & = & \left\{\absdispatch_{\mathsf{c}}(\absblock) ~|~ c = \rflab(\ell) \wedge \pi = \pi_0 :: i :: \pi_1 \wedge \absblock = \rfblock{}(i)\right\} \\
\rftdispatch{\gheap}(\threadstack) & = & \left\{\abstdispatch(\absloc,\absblock) ~|~ \threadstack = \threadstack_0 :: \ell :: \threadstack_1 \wedge \absloc = \rflab(\ell) \wedge (\ell \mapsto b)\in \gheap \wedge \absblock = \rfblock(b)\right\}
\end{array}\\
&\begin{array}{lcl}
\rfframe{\gheap}(\tactframe{\ell}{s}{\pi}{\threadstack}{\callstack},\lheap_a,(\lfilter^j)_j)  & = & \rfframe{}(\tuactframe{\ell}{s}{\pi}{\threadstack}{\callstack},\lheap_a,(\lfilter^j)_j) \\
& = & \rfframe{}(\threadframe{\ell}{\ell'}{\pi}{\threadstack}{\callstack},\lheap_a,(\lfilter^j)_j) \\
&=& \rfcall{\ell}(\callstack,\lheap_a,(\lfilter^j)_j) \cup \rfdispatch{\ell}(\pi) \cup \rftdispatch{\gheap}(\threadstack)
\end{array}
\end{alignat*}

Let $\actstack = \varphi_1 :: \ldots :: \varphi_n$ and $\threadpool = \psi_1 :: \ldots :: \psi_m$. We then define the representation function $\rfastk{\gheap}$ abstracting the activity stack and the thread pool as follows:
\[
\rfastk{\gheap}(\actstack,\threadpool,(\lheap_i,(\lfilter^{i,j})_j)_i)  =  \left( \bigcup_{i \in [1,n]} \rfframe{\gheap}(\varphi_i,\lheap_i,(\lfilter^{i,j})_j) \right) \cup  \left( \bigcup_{l \in [1,m]} \rfframe{\gheap}(\psi_l,\lheap_{n+l},(\lfilter^{{n+l},j})_j) \right)
\]

The representation function $\rflconf$ abstracts a local configuration $\Sigma$ into a set of sets of abstract facts, one for each local configuration decomposition of $\Sigma$:
\begin{multline*}
\rflconf(\tmethconf{\callstack}{\pi}{\threadstack}{\heap}{\sheap}{\ell})  =  \Big\{\rfcall{\ell}(\callstack,\lheap_a,(\lfilter^{j})_j) \cup \rfdispatch{\ell}(\pi) \cup \rftdispatch{\gheap}(\threadstack) \cup \rfheap{\gheap}(\heap) \cup \rfstat{}(\sheap) \\
|\,(\gheap,(\lheap_i)_i,\lheap_a,\flist) \text{ is a local configuration decomposition} \text{ of } \tmethconf{\callstack}{\pi}{\threadstack}{\heap}{\sheap}{\ell}\Big\}
\end{multline*}

The representation function $\rfconf$ abstracts a configuration $\Psi$ into a set of sets of abstract facts, one for each configuration decomposition of $\Psi$:
\begin{multline*}
\rfconf(\tactconf{\actstack}{\threadpool}{\heap}{\sheap})  =  \Big\{\rfastk{\gheap}(\actstack,(\lheap_i,(\lfilter^{i,j})_{j})_i) \cup \rfheap{\gheap}(\heap) \cup \rfstat{}(\sheap)\\
 \,|\,(\gheap,(\lheap_i,(\lfilter^{i,j})_j)_i) \text{ is a configuration decomposition of }\tactconf{\actstack}{\threadpool}{\heap}{\sheap}\Big\}
\end{multline*}

\begin{remark}
The predicates $\absinvoke{\spp}{c'}{(\absthread,\absual^*)}{\absval^*}{\absfi}$ are used to abstract local states of function which have invoked some other method and are waiting for it to return. There are two differences with $\absreg{\spp}{(\absthread,\absual^*)}{\absval^*}{\abslh}{\absfi}$: the first one is that we drop the local heap, which is no longer needed since it will be replaced by the callee's local heap when it will return. The second difference is that we have extra information about the class $c'$ implementing the invoked method.

Also observe that this invoke predicate does not appear in any rules, and that it is \emph{only} used in the proof. Therefore it can be ignored in an implementation.
\end{remark}

\subsection{Pre-Orders}
We will now define several pre-orders and relations used to compare abstract elements. Some abstract syntactic domains, such as abstract values and abstract memory blocks, have two different pre-orders used to compare them, that we distinguish by decorating one with a $\mathsf{nfs}$ superscript. The pre-order with the $\mathsf{nfs}$ superscript is a \emph{flow-insensitive} pre-order.
\subsubsection{Abstract Values Pre-Orders}

We define the pre-order $\polab$ on abstract location by:
\[
 \abslab \polab \abslab' \;\text{ iff }\;
\begin{cases}
\abslab = \absg{\spp} \wedge \abslab' = \absl{\spp}\\
\abslab = \absl{\spp} \wedge \abslab' = \absg{\spp}\\
\abslab = \abslab'
\end{cases}
\]

Based on this, we define the pre-order $\poval$ on abstract values to the reflexive and transitive closure of $\abspo \cup\polab $. We then build the pre-orders $\poseq$ and $\poseqp$  on sequences of abstract values by having $\absual^* \poseq \absval^*$ (resp. $\absual^* \poseqp \absval^*$)  iff $\absual^*$ and $\absval^*$ have the same length and \(\forall i: \absual_i \poval \absval_i\) (resp. \(\forall i: \absual_i \povalp \absval_i\)). We then define a pre-order $\poblk$ on abstract memory blocks as follows:
\begin{itemize}
\item if $\absblock = \absobj{c}{(f \mapsto \absual)^*}$ and $\absblock' = \absobj{c}{(f \mapsto \absval)^*}$ and $\absual^* \poseq \absval^*$, then $\absblock \poblk \absblock'$
\item if $\absblock = \absintent{c}{\absual}$ and $\absblock' = \absintent{c}{\absval}$ and $\absual \poval \absval$, then $\absblock \poblk \absblock'$
\item if $\absblock = \absarray{\tau}{\absual}$ and $\absblock' = \absarray{\tau} {\absval}$ and $\absual \poval \absval$, then $\absblock \poblk \absblock'$
\end{itemize}

We also define the  pre-order $\polblk$ on abstract memory blocks, which is the the flow-sensitive counterpart of $\poblk$.
\begin{itemize}
\item if $\absblock = \absobj{c}{(f \mapsto \absual)^*}$ and $\absblock' = \absobj{c}{(f \mapsto \absval)^*}$ and $\absual^* \poseqp \absval^*$, then $\absblock \polblk \absblock'$
\item if $\absblock = \absintent{c}{\absual}$ and $\absblock' = \absintent{c}{\absval}$ and $\absual \povalp \absval$, then $\absblock \polblk \absblock'$
\item if $\absblock = \absarray{\tau}{\absual}$ and $\absblock' = \absarray{\tau} {\absval}$ and $\absual \povalp \absval$, then $\absblock \polblk \absblock'$
\end{itemize}

Finally we define the relation $\pofilter$ on abstract filters to be the equality order. Next, we state some simple properties satisfied by these pre-orders.

\begin{proposition}
\label{prop:coarseblk}
$\poblk$ is coarser than $\polblk$, and $\poval$ is coarser than $\povalp$.
\end{proposition}

\begin{proposition}
\label{prop:nonempty}
If $\absual \ne \bot$ and $\absual \povalp \absval$ and $\absual \povalp \abswal$ then $\absval \absmeet \abswal \ne \bot$
\end{proposition}

\begin{IEEEproof}
Since $(\absdom,\abspo,\absjoin,\absmeet,\top,\bot)$ is a lattice we know that $\absual \abspo \absval \absmeet \abswal$. Moreover $\absual \ne \bot$, therefore $\absval \absmeet \absual \ne \bot$.
\end{IEEEproof}

\begin{proposition}
\label{prop:fieldr}
For any abstract memory blocks $\absblock, \absblock'$, for any abstract values $\absual,\absval$ and for any field $f$ we have
\begin{gather*}
\absblock \poblk \absblock' \wedge \absual \poval \absval \implies \absblock[f \mapsto \absual] \poblk \absblock'[f \mapsto \absval] \\
\absblock \polblk \absblock' \wedge \absual \povalp \absval \implies \absblock[f \mapsto \absual] \polblk \absblock'[f \mapsto \absval] 
\end{gather*}
\end{proposition}

\subsubsection{Facts Pre-Orders}
For all register $r_o$, class $c''$, abstract heap $\abslh$ and sequence of abstract values $\absval^*$ we define the formula:
\begin{multline*}
\callinv{r_o,c'',m'}{\Delta}{\absval^*}{\abslh} =\exists \spp', c',\left(\left(\absg{\spp'} \abspo \absval_{o} \wedge \absheap(\spp',\absobj{c'}{\_}) \in \Delta\right) \vee \left(\absl{\spp'} \abspo \absval_{o} \wedge \abslh(\spp') = \absobj{c'}{\_}\right)\right) \\
\wedge c' \le c'' \wedge c'' \in \abslookup(m')
\end{multline*}
Intuitively this states that element $o$ of the abstract registers $\absval^*$ over-approximates an abstract location to an abstract object $\absobj{c'}{\_}$ in $\abslh$ or $\absprog$, such abstract virtual dispatch resolution on $c',m'$ return $c''$. We are now ready to define more complex relation between abstract facts, using the pre-orders defined in the previous subsection. Let $\Delta,\Delta'$ be two finite sets of facts. We define the relations $\poreg$, $\poabnormal$ and $\poinvoke{\absprog'}$ as follows:
\begin{itemize}
\item $\absreg{c,m,\pc}{(\absthread^1,\absual_{call}^*)}{\absual^*}{\abslh}{\absfi} \poreg \absreg{c,m,\pc}{(\absthread^2,\absval_{call}^*)}{\absval^*}{\abslh'}{\absfi'}$ iff
\begin{itemize}
\item $\absthread^1 = \absthread^2$ and $\absual_{call}^* \poseqp \absval_{call}^*$
\item $\absual^* \poseqp \absval^*$
\item $\absfi \pofilter \absfi'$ 
\item $\forall \spp,\abslh(\spp) \ne \bot \implies \abslh(\spp) \polblk \abslh'(\spp)$
\end{itemize}

\item $\absabnormal{c,m,\pc}{(\absthread^1,\absual_{call}^*)}{\absual^*}{\abslh}{\absfi} \poabnormal \absabnormal{c,m,\pc}{(\absthread^2,\absval_{call}^*)}{\absval^*}{\abslh'}{\absfi'}$ iff :
\[\absreg{c,m,\pc}{(\absthread^1,\absual_{call}^*)}{\absual^*}{\abslh}{\absfi} \poreg \absreg{c,m,\pc}{(\absthread^2,\absval_{call}^*)}{\absval^*}{\abslh'}{\absfi'}\]

\item $\absinvoke{c,m,\pc}{c''}{(\absthread^1,\absual_{call}^*)}{\absual^*}{\absfi} \poinvoke{\Delta} \absreg{c,m,\pc}{(\absthread^2,\absval_{call}^*)}{\absval^*}{\abslh'}{\absfi'}$ iff:
\begin{itemize}
\item $\absthread^1 = \absthread^2$ and $\absual_{call}^* \poseqp \absval_{call}^*$
\item $\absual^* \poseqp \absval^*$
\item $\absfi \pofilter \absfi'$ 
\item $\lookup(c,m) = (\_,\stm^*)$, $\stm_\pc = \invoke{r_o}{m'}{\_}$ and $\callinv{r_o,c'',m'}{\Delta}{\absval'^*}{\abslh'}$
\end{itemize}

\end{itemize}

Finally, we define the pre-order $<:$ by having $\Delta <: \Delta'$ (where $\Delta,\Delta'$ are two finite sets of facts) if and only if:
\begin{itemize}
\item $\forall\absreg{c,m,\pc}{(\absthread^1,\absual_{call}^*)}{\absual^*}{\abslh}{\absfi} \in \Delta$, $\exists \absreg{c,m,\pc}{(\absthread^2,\absval_{call}^*)}{\absval^*}{\abslh'}{\absfi'} \in \Delta'$ s.t.
\[\absreg{c,m,\pc}{(\absthread^1,\absual_{call}^*)}{\absual^*}{\abslh}{\absfi} \poreg\absreg{c,m,\pc}{(\absthread^2,\absval_{call}^*)}{\absval^*}{\abslh'}{\absfi'}\]

\item $\forall\absabnormal{c,m,\pc}{(\absthread^1,\absual_{call}^*)}{\absual^*}{\abslh}{\absfi} \in \Delta$, $\exists \absabnormal{c,m,\pc}{(\absthread^2,\absval_{call}^*)}{\absval^*}{\abslh'}{\absfi'} \in \Delta'$ s.t.
\[\absabnormal{c,m,\pc}{(\absthread^1,\absual_{call}^*)}{\absual^*}{\abslh}{\absfi} \poabnormal\absabnormal{c,m,\pc}{(\absthread^2,\absval_{call}^*)}{\absval^*}{\abslh'}{\absfi'}\]

\item $\forall\absinvoke{c,m,\pc}{c''}{(\absthread^1,\absual_{call}^*)}{\absual^*}{\absfi} \in \Delta$, $\exists \absreg{c,m,\pc}{(\absthread^2,\absval_{call}^*)}{\absval^*}{\abslh'}{\absfi'} \in \Delta'$ s.t.
\[\absinvoke{c,m,\pc}{c''}{(\absthread^1,\absual_{call}^*)}{\absual^*}{\absfi} \poinvoke{\Delta'} \absreg{c,m,\pc}{(\absthread^2,\absval_{call}^*)}{\absval^*}{\abslh'}{\absfi'}\]

\item $\forall\absheap(\absloc,\absblock) \in \Delta$, $\exists \absheap(\absloc,\absblock') \in\Delta'$ such that $\absblock \poblk \absblock'$

\item $\forall\abssheap(c,f,\absual) \in \Delta$, $\exists\abssheap(c,f,\absval)\in\Delta'$ such that $\absual \poval \absval$

\item $\forall\absdispatch_{\mathsf{c}}(\absblock) \in \Delta$, $\exists\absdispatch_{\mathsf{c}}(\absblock')\in\Delta'$ such that $\absblock \poblk \absblock'$

\item $\forall\abstdispatch(\absloc,\absblock) \in \Delta$, $\exists\abstdispatch(\absloc,\absblock')\in\Delta'$ such that $\absblock \poblk \absblock'$
\end{itemize}

\subsection{Preliminary Lemmas}

\subsubsection{Pre-orders}

\begin{lemma}
\label{lem:call}
For all set of facts $\Delta$ and $\Delta'$, if $\Delta \subseteq \Delta'$ then 
\[
\callinv{r_o,c'',m'}{\Delta}{\absval^*}{\abslh} \implies \callinv{r_o,c'',m'}{\Delta'}{\absval^*}{\abslh}
\]
As a direct corollary, $\poinvoke{\Delta'}$ is coarser than $\poinvoke{\Delta}$.
\end{lemma}

\begin{lemma}
\label{lem:sub-order}
If $\absprog \subseteq \absprog'$, and $\absprog' <: \absprog''$ then $\absprog <: \absprog''$.
\end{lemma}

\begin{lemma}
\label{lem:join-order}
If $\absprog_1 <: \absprog_2$ and $\absprog_3 <: \absprog_4$, then $\absprog_1 \cup \absprog_3 <: \absprog_2 \cup \absprog_4$.
\end{lemma}

\begin{lemma}
\label{lem:trans-order}
If $\absprog <: \absprog'$ and $\absprog' <: \absprog''$, then $\absprog <: \absprog''$.
\end{lemma}

\begin{IEEEproof}
All cases are very easy, except for the following one: 

Let $\absinvoke{c,m,\pc}{c''}{(\absthread,\absual_{call}^*)}{\absval^*}{\absfi}  \in \Delta$, $\absreg{c,m,\pc}{(\absthread',\absual_{call}'^*)}{\absval'^*}{\abslh'}{\absfi'}  \in \Delta'$,  $\absreg{c,m,\pc}{(\absthread'',\absual_{call}''^*)}{\absval''^*}{\abslh''}{\absfi''} \in \Delta''$. Assume that:
\begin{align*}
\absinvoke{c,m,\pc}{c''}{(\absthread,\absual_{call}^*)}{\absval^*}{\absfi} 
\poinvoke{\Delta'} \absreg{c,m,\pc}{(\absthread',\absual_{call}'^*)}{\absval'^*}{\abslh'}{\absfi'} 
\poreg \absreg{c,m,\pc}{(\absthread'',\absual''^*_{call})}{\absval''^*}{\abslh''}{\absfi''}
\end{align*}
We want to prove that:
\begin{align*}
\absinvoke{c,m,\pc}{c''}{(\absthread,\absual_{call}^*)}{\absval^*}{\absfi} 
\poinvoke{\Delta''} \absreg{c,m,\pc}{(\absthread'',\absual_{call}''^*)}{\absval''^*}{\abslh''}{\absfi''}
\end{align*}
To this end we need to prove that the following four conditions holds:
\begin{itemize}
\item $\absthread,\absual_{call}^* \poseqp \absthread'',\absual_{call}''^*$: follows directly from transitivity of $\poseqp$
\item $\absval^* \poseqp \absval''^*$: follows directly from transitivity of $\poseqp$
\item $\absfi \pofilter \absfi''$ : follows directly from transitivity of $\pofilter$
\item $\lookup(c,m) = (\_,\stm^*)$, $\stm_\pc = \invoke{r_o}{m'}{\_}$ and $\callinv{r_o,c'',m'}{\Delta''}{\absval''^*}{\abslh''}$: 

The fact that $\lookup(c,m) = (\_,\stm^*)$, $\stm_\pc = \invoke{r_o}{m'}{\_}$ is easy. It remains to check that  $\callinv{r_o,c'',m'}{\Delta''}{\absval''^*}{\abslh''}$. First we know that $\callinv{r_o,c'',m'}{\Delta'}{\absval'^*}{\abslh'}$ holds, therefore there exist $\spp'$ and $c'$ such that:
\begin{gather*}
\hspace{-0.5cm} \Big(\overbrace{\left(\absg{\spp'} \abspo \absval'_{r_o} \wedge \absheap(\spp',\absobj{c'}{\_}) \in \Delta'\right)}^{A} \vee \overbrace{\left(\absl{\spp'} \abspo \absval'_{r_o} \wedge \abslh'(\spp') = \absobj{c'}{\_}\right)}^{B}\Big) 
\wedge c' \le c'' \wedge c'' \in \abslookup(m')
\end{gather*}
\begin{itemize}
\item Assume that $A$ holds: we have $\absheap(\spp',\absobj{c'}{\_}) \in \Delta'$ and $\absg{\spp'} \abspo \absval'_{r_o}$. Then since $\Delta' <: \Delta''$ we know that there exists $\absheap(\spp',\absobj{c'}{\_}) \in \Delta''$. Moreover since $\absval'^* \poseqp \absval''^*$ and $\absg{\spp'} \abspo \absval'_{r_o}$ we know that $\absg{\spp'} \abspo \absval''_{r_o}$. Therefore $\callinv{r_o,c'',m'}{\Delta''}{\absval''^*}{\abslh''}$ holds.
\item Assume that $B$ holds:  we have $\absl{\spp'} \abspo \absval'_{r_o} $ and $\abslh'(\spp') = \absobj{c'}{\_}$. First, since $\absval'^* \poseqp \absval''^*$ and $\absl{\spp'} \abspo \absval'_{r_o}$ we know that $\absl{\spp'} \abspo \absval''_{r_o}$. Moreover $\abslh'(\spp') = \absobj{c'}{\_}$ and $\abslh'(\spp') \ne \bot \implies \abslh'(\spp') \polblk \abslh''(\spp')$, hence $\abslh''({\spp'}) = \absobj{c'}{\_}$. Therefore $\callinv{r_o,c'',m'}{\Delta''}{\absval''^*}{\abslh''}$ holds.
\end{itemize}
\end{itemize}
\end{IEEEproof}

\subsubsection{Representation Function}

\begin{proposition}
\label{prop:brf}
For all filter history $\lheap,\flist$ we have:
\begin{itemize}
\item For any block $b$, $\rflblock{}(b,\lheap) \poblk \rfblock{}(b)$ and  $\rfblock{}(b) \poblk \rflblock{}(b,\lheap)$.
\item For any value $v$, $\rflval{}(v,\lheap,\flist) \poval \rfval{}(v)$ and $\rfval{}(v) \poval \rflval{}(v,\lheap,\flist)$.
\end{itemize}
\end{proposition}

\begin{IEEEproof}
This is following from the fact that the pre-orders $\poblk$ and $\poval$ ignore the \flowsensitive and \flowinsensitive  annotations of the abstract labels.
\end{IEEEproof}

\begin{assumption}[Soundness of the Abstract Operations] $\acomp,\aunop$ and $\abinop$ are monotonous operators, and soundly over-approximate the concrete operators $\comp,\odot$ and $\oplus$: for all local heap $\lheap$, we have:
\label{asm:sound-op}
\begin{itemize}
\item $u \comp v$ implies that $\rflval{}(u,\lheap)\acomp \rflval{}(v,\lheap)$
\item $ \rflval{}(\odot v,\lheap) \abspo \aunop \rflval{}(v,\lheap)$
\item $ \rflval{}(u \oplus v,\lheap) \abspo \rflval{}(u,\lheap)\abinop \rflval{}(v,\lheap)$
\end{itemize}
\end{assumption}

This carry over to all the representation functions $\rflval{}(\cdot,\lheap,(\lfilter^i)_i)$ (with order $\abspo$) and $\rfval{}(\cdot)$ (with order $\poval$):
\begin{proposition}For all concrete values $u$ and $v$, and for all filter history $\lheap,(\lfilter^i)_i$ we have:
\label{prop:sound:op2}
\begin{itemize}
\item $u \comp v$ implies that $\rflval{}(u,\lheap,(\lfilter^i)_i)\acomp \rflval{}(v,\lheap,(\lfilter^i)_i)$ and that $\rfval{}(u)\acomp \rfval{}(v)$
\item $ \rflval{}(\odot v,\lheap,(\lfilter^i)_i) \abspo \aunop \rflval{}(v,\lheap,(\lfilter^i)_i)$ and $ \rfval{}(\odot v) \poval \aunop \rfval{}(v)$
\item $ \rflval{}(u \oplus v,\lheap,(\lfilter^i)_i) \abspo \rflval{}(u,\lheap,(\lfilter^i)_i)\abinop \rflval{}(v,\lheap,(\lfilter^i)_i)$ and  $ \rfval{}(u \oplus v) \poval \rfval{}(u)\abinop \rfval{}(v)$
\end{itemize}
\end{proposition}

\begin{IEEEproof}
Observe that for all filter history  $\lheap,(\lfilter^i)_i$, we have that for all concrete value $u$:
\[ \rflval{}(u,\lheap,(\lfilter^i)_i) = \rflval{}\left(u,\left( \spp \mapsto \fhistget{\fhist}{\infty}(\spp) \right)^*\right)\]
This together with Assumption~\ref{asm:sound-op} shows the first point of each item bullet.

The second point of each item bullet follows from the fact that if $\poval$ is coarser than $\abspo$, and the monotonicity of the abstract operators. We are going to detail the proof of the second item bullet (the other cases work exactly in the same way). Let $\lheap$ be an arbitrary local heap:
\[\begin{array}{lr}
 \rflval{}(\odot v,\lheap) \abspo \aunop \rflval{}(v,\lheap) & \text{by Assumption~\ref{asm:sound-op}}\\
 \rflval{}(\odot v,\lheap) \poval \aunop \rflval{}(v,\lheap) & \text{by Proposition~\ref{prop:coarseblk}}\\
 \rfval{}(\odot v) \poval \rflval{}(\odot v,\lheap) \poval \aunop \rflval{}(v,\lheap) & \text{by Proposition~\ref{prop:brf}}
\end{array}
\]
By Proposition~\ref{prop:brf} we know that $\rflval{}(v,\lheap) \poval \rfval{}(v)$, therefore by monotonicity of $\aunop$ we get that $\aunop\rflval{}(v,\lheap) \poval \aunop\rfval{}(v)$. This concludes the $\aunop$ case by showing that:
\[  \rfval{}(\odot v) \poval \rflval{}(\odot v,\lheap) \poval \aunop \rflval{}(v,\lheap) \poval \aunop \rfval{}(v)\]
\end{IEEEproof}

\begin{assumption}[Overriding]
\label{asm:overriding}
If $\lookup(c,m) = (c',\stm^*)$, then $c \subtype c'$.
\end{assumption}

In the next results, let $\absprog \vdash \absprog'$ whenever $\absprog \vdash \fact$ for each $\fact \in \absprog'$.

\begin{proposition}
\label{prop:exactfilter}
$\afunion$ is an exact abstraction of $\lfunion$: for all filters $\lfilter^1$ and $\lfilter^2$ we have $\rffilter(\lfilter^1 \lfunion \lfilter^2) = \rffilter(\lfilter^1) \afunion \rffilter(\lfilter^2)$. 
\end{proposition}

\begin{proposition}
\label{prop:vallift}For all abstract filter $\absfi$, for all abstract values $\absual$ and $\absval$ we have:
\begin{itemize}
\item if $\absual \povalp \absval$ then  $\lift{\absual}{\absfi} \povalp \lift{\absval}{\absfi}$.
\item if $\absual \polab \absval$ then  $\lift{\absual}{\absfi} \polab \lift{\absval}{\absfi}$.
\item if $\absual \poval \absval$ then  $\lift{\absual}{\absfi} \poval \lift{\absval}{\absfi}$.
\item for all abstract heap $\abslh$ and $\abslh'$,  if $\forall \spp,\abslh(\spp) \polblk \abslh'(\spp)$ then:
  \[\forall \spp,\lhlift{\abslh}{\absfi}(\spp) \polblk \lhlift{\abslh'}{\absfi}(\spp)\]
\end{itemize}
\end{proposition}

\begin{IEEEproof}
The first point is an assumption made on the $\lift{\cdot}{\cdot}$ function, and the second point is trivial. Observe that for all $\absual,\absval$, if $\absual \polab \absval$ then  $\lift{\absual}{\absfi} \polab \lift{\absval}{\absfi}$. Since $\poval$ is the transitive and reflexive closure of $\abspo$ and $\polab$, this third point is a direct consequence of the first and second points. The fourth point is an easy consequence of $\lhlift{\cdot}{\cdot}$ definition and of the first point.
\end{IEEEproof}

\begin{proposition}
\label{prop:valtovalp}
$\absual \poval \absval$ implies that $\lift{\absual}{1^*} \abspo \lift{\absval}{1^*}$.
\end{proposition}

\begin{IEEEproof}
By definition of $\poval$, we know that there exists $(\absval_i)_{i \le n},(\absval'_i)_{i \le n}$ such that:
\[ \absual = \absval_1 \polab \absval_1' \abspo \absval_2 \polab \absval_2' \dots \absval_{n-1}' \abspo \absval_n \polab \absval_n'= \absval\]
By Proposition~\ref{prop:vallift}.2, we know that for all $i \le n$,  $\absval_i \polab \absval_i'$ implies that $\lift{\absval_i}{1^*} \polab \lift{\absval_i'}{1^*}$. Moreover $\lift{\absval_i}{1^*} \polab \lift{\absval_i'}{1^*}$ implies that there exists $\absloc$ such that $\lift{\absval_i}{1^*} = \absg{\absloc}$ and $\lift{\absval_i'}{1^*} = \absg{\absloc}$. Therefore $\lift{\absval_i}{1^*} \abspo \lift{\absval_i'}{1^*}$. By Proposition~\ref{prop:vallift}.1, for all $i < n$,  $\absval_i' \abspo \absval_{i+1}$ implies that $\lift{\absval_i'}{1^*} \abspo \lift{\absval_{i+1}}{1^*}$, hence we have:
\[   \lift{\absual}{1^*} = \lift{\absval_1}{1^*} \abspo \lift{\absval_1'}{1^*} \abspo \lift{\absval_2}{1^*}  \dots  \lift{\absval_n}{1^*} \abspo \lift{\absval_n'}{1^*} = \lift{\absval}{1^*}\]
Which concludes this proof.
\end{IEEEproof}

\begin{proposition}
\label{prop:samefhistget}
If for some $i$ we have :
\[\fhistget{(\flist,\lheap_a)}{i} = \fhistget{(\flistp,\lheap'_a)}{i + k} \text{ and }\fhistget{(\flist,\lheap_a)}{i + 1} = \fhistget{(\flistp,\lheap'_a)}{i + k + 1}\]
 then for all local state $L$ and class $c'$ we have:
\[\rfinvoke{\ell}(L,i,c',\lheap_a,(\lfilter^n)_n) = \rfinvoke{\ell}(L,i + k,c',\lheap_a',(\lfilter'^n)_n)\]
\end{proposition}

\begin{proposition}
\label{prop:lslh}
Let $\Sigma = \tmethconf{\callstack}{\pi}{\threadstack}{\heap}{\sheap}{\ell}$ and let $\regval{\rhs} = \ell$, then for any $X \in \rflconf(\Sigma)$ with local configuration decomposition \lheapdh, $v \in  dom(\heap)$ implies that $v \in dom(\lheap)$.
\end{proposition}
\begin{IEEEproof}
By a case analysis on the structure of $\rhs$, and using the fact that we have a local configuration decomposition.
\end{IEEEproof}

\begin{proposition}
\label{prop:rfconfiguration}
Let \lheapdh and \lheapdhp be two local configuration decomposition of $\actstack_i$ such that $\lheap = \lheap'$ and $\forall j, \lfilter^j = \lfilter'^j$. Then we have:
\begin{gather*}
 \rfframe{}(\actstack_i,\lheap,(\lfilter'^{j})_{j}) = \rfframe{}(\actstack_i,\lheap,(\lfilter^{j})_{j})
\end{gather*}
\end{proposition}

\subsubsection{Technical lemmas}

\begin{lemma}[Right-hand Sides]
\label{lem:rhs}
Let $\Sigma = \tmethconf{\callstack}{\pi}{\threadstack}{\heap}{\sheap}{\ell}$ with $\callstack = \locstate{\pp}{\stm^*}{R}{u^*} :: \callstack_0$, let $\regval{\rhs} = v$, $X \in \rflconf(\Sigma)$ with local configuration decomposition \lheapdh, let $\absprog :> X$.

Let $\absreg{{c,m,\pc}}{(\absthread',\absual'^*)}{\absval'^*}{\abslh'}{\absfi'} \in \absprog$ be such that :
\[\rflocstate{\ell}(\locstate{c,m,\pc}{\stm^*}{R}{u^*},\lheap,(\lfilter^j)_j) \poreg \absreg{{c,m,\pc}}{(\absthread',\absual'^*)}{\absval'^*}{\abslh'}{\absfi'}\]
 Then there exists $\absval$ such that $\rflval{}(v,\lheap) \povalp\absval$ and $\absprog \cup \arhs{\rhs} \vdash \prhs{\absval}$.

Moreover if $\rhs$ is a register $r_i$ then we can take $\absval = \absval'_i$.
\end{lemma}
\begin{IEEEproof}
By a case analysis on the structure of $\rhs$. We are going to detail the object field look-up case, which is the more complicated one. Let $\absreg{{c,m,\pc}}{(\absthread,\absual^*)}{\absval^*}{\abslh}{\absfi}$ be such that:
\begin{equation}
\rflocstate{\ell}(\locstate{c,m,\pc}{\stm^*}{R}{u^*},\lheap,(\lfilter^j)_j) = \absreg{{c,m,\pc}}{(\absthread,\absual^*)}{\absval^*}{\abslh}{\absfi}\label{eq:rhs1}
\end{equation}

Let $\regval{r_i} = \ell = \pointer{p}{\absloc}$. Since $\gheap,(\lheap_i)_i$ is a heap decomposition of $\heap$ we know that $\ell \in dom(\gheap)$ or $\ell \in \bigcup_i dom(\lheap_i)$. Moreover by Proposition~\ref{prop:lslh}, $\ell \in \bigcup_i dom(\lheap_i)$ implies that $\ell \in dom(\lheap)$. Therefore we are in one of the two following cases:
\begin{itemize}
\item $\ell \in dom(\gheap)$: from Equation~\ref{eq:rhs1} we get that $\absval_i = \rflval{}(\ell,\lheap) = \absg{\absloc}$. Moreover since:
\[\rflocstate{\ell}(\locstate{c,m,\pc}{\stm^*}{R}{u^*},\lheap,(\lfilter^j)_j) \poreg \absreg{{c,m,\pc}}{(\absthread',\absual'^*)}{\absval'^*}{\abslh'}{\absfi'}\]
we know that $\absg{\absloc} = \absval_i  \povalp \absval'_i$. We know that there exists $o$ such that $o = \heap(\ell) = \obj{c}{(f_j \mapsto u_j)^*, f \mapsto v}$. Since $\absprog :> X$, there exists $\absheap(\absloc,\absobj{c}{(f_i \mapsto \absual_i)^*, f \mapsto \absval_f}) \in \absprog$ such that $\rfval{}(v) \poval \absval_f$. Let $\absval = \lift{\absval_f}{1^*}$, then we have $\absprog \cup \arhs{\rhs} \vdash \prhs{\absval}$ by applying the rule:
\[ \absreg{{c,m,\pc}}{(\absthread',\absual'^*)}{\absval'^*}{\abslh'}{\absfi'} \wedge \absg{\absloc} \abspo \absval'_i \wedge \absheap(\absloc,\absobj{c}{(f_i \mapsto \absual_i)^*, f \mapsto \absval_f}) \implies \prhs{\lift{\absval_f}{1^*}}\]
which is in $\arhs{r_i.f}$. It remains to check that  $\rflval{}(v,\lheap) \povalp\absval$: if $v$ is a primitive value then this is trivial. The value $v$ is stored in a field of an object referenced to by $\ell$, which is a \flowinsensitive location and cannot contain \flowsensitive locations. Therefore $v$ cannot be a \flowsensitive location. If $v$ is a \flowinsensitive location $\pointer{p'}{\absloc'}$ then $\rflval{}(v,\lheap) = \absg{\absloc'}$, and $\rfval{}(v) = \absg{\absloc'}$. Moreover by Proposition~\ref{prop:valtovalp} we know that  $ \rfval{}(v) \poval \absval_f$ implies that  $ \lift{\rfval{}(v)}{1^*} \poval \lift{\absval_f}{1^*}$. Since $\lift{\rfval{}(v)}{1^*} = \absg{\absloc'} = \rflval{}(v,\lheap)$, we proved that $\rflval{}(v,\lheap) \povalp \absval$.

\item $\ell \in dom(\lheap)$: from Equation~\ref{eq:rhs1} we get that $\absval_i = \rflval{}(\ell,\lheap) = \absl{\absloc}$. Moreover since:
\begin{equation}
\absreg{{c,m,\pc}}{(\absthread,\absual^*)}{\absval^*}{\abslh}{\absfi} \poreg \absreg{{c,m,\pc}}{(\absthread',\absual'^*)}{\absval'^*}{\abslh'}{\absfi'}\label{eq:rhs2}
\end{equation}
we know that $\absl{\absloc} = \absval_i \povalp \absval'_i$. We know that there exists $o$ such that $o = \heap(\ell) = \obj{c}{(f_j \mapsto u_j)^*, f \mapsto v}$, hence by definition of $\rflheap{}$ we get that $\abslh(\absloc) = \absobj{c}{(f_i \mapsto \absual_i)^*, f \mapsto \absval_f}$ where $\rflval{}(v,\lheap) \povalp \absval_f$. Moreover from Equation~\ref{eq:rhs2} and the fact that $\abslh(\absloc) \ne \bot$ we get that $\abslh(\absloc) \polblk \abslh'(\absloc)$, which in turns implies that $\abslh'(\absloc) = \absobj{c}{(f_i \mapsto \absual''_i)^*, f \mapsto \absval'_f}$ where $\absval_f \povalp \absval'_f$. By transitivity of $\povalp$ we have $\rflval{}(v,\lheap) \povalp \absval'_f$.

It just remains to show that $\absprog \cup \arhs{\rhs} \vdash \prhs{\absval_f'}$ by applying the following rule, which is in  $\arhs{r_i.f}$:
\[ \absreg{{c,m,\pc}}{(\absthread',\absual'^*)}{\absval'^*}{\abslh'}{\absfi'} \wedge \absl{\absloc} \abspo \absval'_i  \wedge \abslh'(\absloc) = \absobj{c}{(f_i \mapsto \absual''_i)^*, f \mapsto \absval'_f} \implies \prhs{\absval'_f}\]
\end{itemize}
\end{IEEEproof}

\begin{lemma}[Reachability]
\label{lem:reach}
For any abstract value $\absual$ and abstract heap $\abslh$, there exists an abstract filter $\absfi_a$ such that $\vdash \cfilter{\absual}{\abslh}{\absfi_a}$ and $\absfi_a$ is the indicator function of the set of reachable elements starting from $\absual$ in the points-to graph of $\abslh$.
\end{lemma}

\begin{IEEEproof}

We define $Reach^n_{\absloc}$ and $Reach^n_{\absval}$ as follows:
\begin{itemize}
\item $Reach^n_{\absval} =  \bigcup_{\absl{\absloc'} \abspo \absval} Reach^{n}_{\absloc'}$
\item $Reach^0_{\absloc} = \{ \absloc\}$
\item $ Reach^{n+1}_{\absloc} = Reach^{n}_{\absloc} \cup \bigcup_i  Reach^{n}_{\absval_i}$ if $\abslh(\absloc) = \absobj{c}{(f_i \mapsto \absval_i)_i}$
\item $ Reach^{n+1}_{\absloc} = Reach^{n}_{\absloc} \cup Reach^{n}_{\absval}$ if $\abslh(\absloc) = \absarray{\tau}{\absval}$
\item $ Reach^{n+1}_{\absloc} = Reach^{n}_{\absloc} \cup Reach^{n}_{\absval}$ if $\abslh(\absloc) = \absintent{\tau}{\absval}$
\end{itemize}
For all $\absloc$ (resp. $\absval$), $(Reach^{n}_{\absloc})_{n \ge 0}$ (resp. $(Reach^{n}_{\absval})_{n \ge 0}$) is an non-decreasing sequence, and the set $Reach_\absloc$ (resp. $Reach_{\absval}$)  of reachable elements starting from $\absloc$ (resp. $\absval$) in the points-to graph of $\abslh$ is $Reach_\absloc = \bigcup_{n \ge 0} Reach^{n}_{\absloc}$ (resp. $Reach_{\absval} = \bigcup_{n \ge 0} Reach^{n}_{\absval}$). Moreover since $\abslh$ is finite, this limit is reached in a finite number of steps. Therefore there exists $N$ such that $ Reach_{\absloc} = \bigcup_{n\le N} Reach^{n}_{\absloc}$ and $ Reach_{\absval} = \bigcup_{n\le N} Reach^{n}_{\absval}$.

We define $I_n^\absloc$ to be the indicator function of $Reach^{n}_{\absloc}$, and  $I_n^{\absval}$ to be the indicator function of $Reach^{n}_{\absval}$. We will see $I_n^\absloc$ and $I_n^{\absval}$ as abstract filters. It is easy to show by induction over $n$ that for all $n \ge 0$, for all $\absloc$ and for all $\absval$ we have $\vdash \cfilter{\absl{\absloc}}{\abslh}{I_n^\absloc}$ and $\vdash \cfilter{\absval}{\abslh}{I_n^\absloc}$ (observe that the second point uses the fact that there is a finite number of $\absloc$). Therefore we have $\vdash \cfilter{\absual}{\abslh}{I_N^{\absual}}$, where $I_N^{\absual}$ is the indicator function of $Reach^{N}_{\absual}  = Reach_{\absual}$.
\end{IEEEproof}

\begin{lemma}[Abstract Value Lifting]
\label{lem:tedious}
Let $\lheap$ and $\lheap'$ be two local heaps, $u$ be a concrete value and $S$ be a set of locations such that $dom(\lheap') \backslash dom(\lheap) = S$ and $u \not \in S$.

Let $\absval = \rflval{}(u,\lheap)$,  $\lfilter_a = \{(\pointer{p}{\absloc} \mapsto 1)  ~|~  \pointer{p}{\absloc} \in dom(\lheap) \wedge \exists \pointer{p'}{\absloc} \in S\}$ and $\absfi_a = \rffilter(\lfilter_a)$. Then we have:
\begin{equation*}
\rflval{}(u,\lheap') = \lift{\absval}{\absfi_a}
\end{equation*}
\end{lemma}

\begin{IEEEproof}
If $u$ is a primitive value then this is trivial. Assume $u = \ell = \pointer{p}{\absloc}$, then one of the following cases holds:
\begin{itemize}
\item  $\ell \in dom(\lheap') \cap dom(\lheap)$. Then we have:
\[ \rfloc{}(\pointer{p}{\absloc},\lheap') = \absl{\absloc} = \rfloc{}(\pointer{p}{\absloc},\lheap)\]
Moreover since $S \subseteq dom(\lheap')$, we know that $\ell \not \in S$. Assume that there exists a location $\pointer{p'}{\absloc} \in S$, then since $dom(\lheap') \backslash dom(\lheap) = S$ we know that $\pointer{p'}{\absloc} \in dom(\lheap')$. Since $\pointer{p'}{\absloc} \in dom(\lheap')$ and $p \ne p'$, this implies that $dom(\lheap')$ contains two locations with the same allocation point, which contradicts the fact that $\lheap'$ is a local heap. Therefore there exists no $p'$ such that $\pointer{p'}{\absloc} \in dom(\lheap')$, which in turn implies that implies that $\absfi_a(\absloc) = 0$. Hence $\lift{\absval}{\absfi_a} = \lift{\absl{\absloc}}{\absfi_a} = \absl{\absloc}$, which concludes this case.

\item  $\ell \in dom(\lheap') \backslash dom(\lheap)$. Then since $dom(\lheap') \backslash dom(\lheap) = S$ we have $\ell \in S$. Besides by hypothesis $\ell \not \in S$. Absurd.

\item $\ell \in dom(\lheap) \backslash dom(\lheap')$. Therefore  $\pointer{p}{\absloc} \not \in dom(\lheap')$, and since $\lheap'$ is a local heap there exists $p' \ne p$ such that $\pointer{p'}{\absloc} \in dom(\lheap')$. Moreover since  $\lheap$ is a local heap we have $\pointer{p'}{\absloc} \not \in dom(\lheap)$. Therefore $\pointer{p'}{\absloc} \in S$, which implies that $\absfi_a(\absloc) = 1$. By consequence we have:
\[ \rfloc{}(\pointer{p}{\absloc},\lheap') = \absg{\absloc} = \lift{\absl{\absloc}}{\absfi_a} = \lift{\rfloc{}(\pointer{p}{\absloc},\lheap')}{\absfi_a} = \lift{\absval}{\absfi_a}\]

\item $\ell \not \in dom(\lheap') \cup dom(\lheap)$. Then we trivially have:
\[ \rfloc{}(\pointer{p}{\absloc},\lheap') = \absg{\absloc} = \lift{\absg{\absloc}}{\absfi_a} = \lift{\rfloc{}(\pointer{p}{\absloc},\lheap)}{\absfi_a} = \lift{\absval}{\absfi_a}\]
\end{itemize}
\end{IEEEproof}

\begin{lemma}[Abstract Local State Lifting]
\label{lem:tedious2}
Let $\Sigma = \tmethconf{\callstack}{\pi}{\threadstack}{\heap}{\sheap}{\ell}$ with $\callstack = \locstate{\pp}{\stm^*}{R}{u^*} :: \callstack_0$. Let \lheapdh be a local configuration decomposition of $\Sigma$, and assume that: 
\[\rflocstate{\ell_r}(\locstate{c,m,\pc}{\stm^*}{R}{u^*},\lheap,(\lfilter^n)_n) = \absreg{c,m,\pc}{(\absthread,\absual^*)}{\absval^*}{\abslh}{\absfi}\]
Let $\lheap'$ be a local heap, and $S$ a set of locations such that:
\begin{itemize}
\item $dom(\lheap') \backslash dom(\lheap) = S$
\item $\forall \pointer{p}{\absloc} \in S, \lheap'(\pointer{p}{\absloc}) = \bot$ and  $\forall \pointer{p}{\absloc} \not \in S, \lheap'(\pointer{p}{\absloc}) = \lheap(\pointer{p}{\absloc})$
\item $S$ is fresh in $\Sigma$
\end{itemize}
Let  $\lfilter_a = \{(\pointer{p}{\absloc} \mapsto 1)  ~|~  \pointer{p}{\absloc} \in dom(\lheap) \wedge \exists \pointer{p'}{\absloc} \in S\}$ and $\absfi_a = \rffilter(\lfilter_a)$. Then we have:
\begin{enumerate}
\item \(\rflocstate{\ell_r}(\locstate{c,m,\pc+1}{\stm^*}{R}{u^*},\lheap',(\lfilter_a \funion \lfilter^1) :: (\lfilter^n)_{n > 1}))\; = \; \absreg{c,m,\pc+1}{(\absthread,\absual^*)}{\lift{\absval^*}{\absfi_a}}{\lhlift{\abslh}{\absfi_a}}{\absfi_a \afunion \absfi}\)

\item for all register $r_d$, concrete value $w$, locations $\pointer{p}{\absloc'}$ and memory block $b$ we have:
\begin{alignat*}{2}
& \rflocstate{\ell_r}(\locstate{c,m,\pc+1}{\stm^*}{R[r_d \mapsto w]}{u^*},\lheap'[\pointer{p}{\absloc'} \mapsto b],(\lfilter_a \funion \lfilter^1) :: (\lfilter^n)_{n > 1}))\\
=\quad& \; \absreg{c,m,\pc+1}{(\absthread,\absual^*)}{\lift{\absval^*}{\absfi_a}[d \mapsto \rflval{}(w,\lheap')]}{\lhlift{\abslh}{\absfi_a}[\absloc' \mapsto \rflblock{}(b,\lheap')]}{\absfi_a \afunion \absfi}
\end{alignat*}
\end{enumerate}
\end{lemma}

\begin{IEEEproof}
We are only going to prove  1), as  2) is a rather simple extension of 1). We want to show the four following points:
\begin{itemize}
\item  We know that $dom(\lheap') \backslash S \subseteq dom(\lheap)$. Moreover by definition of $\lfilter_a$ we know that $S = \{\pointer{p}{\absloc} ~|~ \exists \pointer{p'}{\absloc}, \lfilter_a(\pointer{p'}{\absloc}) = 1 \}$. Moreover for all $\ell$, $\lfilter(\ell) = 1$ implies that $\ell \in dom(\lheap)$. Hence by Proposition~\ref{prop:fhsitget}.5 we have:
\[ \fhistget{(\lheap,(\lfilter_j)_{j \ge 1})}{2} = \fhistget{(\lheap',(\lfilter_a \lfunion \lfilter^1) :: (\lfilter^j)_{j \ge 2})}{2}\]
It is then easy to check that for all $l \le |u^*|$, we have $\rflval{}(u_l,\lheap',(\lfilter_a \funion \lfilter^1)) = \rflval{}(u_l,\lheap, \lfilter^1) =\absual_l$.

\item Let $r_k$ be a register of $R$. Since $S$ is fresh in $\Sigma$, we know that $R(r_k) \not \in S$, therefore by Lemma~\ref{lem:tedious} we get that $\rflval{}(R(r_k),\lheap') = \lift{\absval_k}{\absfi_a}$. 
\item Let $\spp$ be an allocation point. We want to show that there exists $\pointer{p}{\spp} \in dom(\lheap')$ such that \(\lhlift{\abslh}{\absfi_a}(\spp) =  \rflblock{}\left(\lheap'(\pointer{p}{\spp}),\lheap'\right))\). Since $\lheap'$ is a local heap, we know that there exists $\ell = \pointer{p}{\spp} \in dom(\lheap')$. One of the two following cases holds:
\begin{itemize}
\item $\ell \in S$. By hypothesis, we know that $\lheap'(\ell) = \bot$. Moreover by definition of $\absfi_a$ we know that $\absfi_a(\spp) = 1$, therefore we have:
\[ \rflblock{}\left(\lheap'(\ell),\lheap'\right) = \rflblock{}\left(\bot,\lheap'\right) = \bot = \lhlift{\abslh}{\absfi_a}(\spp) \]
\item $\ell \not \in S$. Then by hypothesis we know that $\lheap'(\ell) = \lheap(\ell)$. Assume that $\lheap(\ell) = \obj{c}{(f_i \mapsto u_i)_{i \le n}}$ (the array and intent cases are similar). Then we have:
\[ \rflblock{}\left(\lheap'(\ell),\lheap'\right) = \absobj{c}{(f_i \mapsto \rflval{}(u_i,\lheap'))_{i \le n}}\]
Since $S$ is fresh in $\Sigma$ we know that for all $i \le n$, $u_i \not \in S$. Therefore by Lemma~\ref{lem:tedious}, for all $i \le n$, we have $\rflval{}(u_i,\lheap'))_{i \le n} = \lift{\rflval{}(u_i,\lheap)}{\absfi_a}$. Moreover since $\ell \in dom(\lheap') \backslash S$, we know that $\absfi_a(\absloc) = 0$. Therefore:
\[\absobj{c}{(f_i \mapsto \rflval{}(u_i,\lheap'))_{i \le n}} = \absobj{c}{(f_i \mapsto \lift{\rflval{}(u_i,\lheap)}{\absfi_a})_{i \le n}} = \lhlift{\abslh}{\absfi_a}(\absloc)\]
\end{itemize}
\item $\absfi_a \afunion \absfi = \rffilter(\lfilter_a \funion \lfilter^1)$: this is trivial.
\end{itemize}
\end{IEEEproof}

We can now state the local preservation lemma, which shows that our abstraction soundly over-approximates the concrete reduction $\rightsquigarrow^*$ between local reduction.
\begin{lemma}[Local Preservation]
\label{thm:local}
If $\Sigma \rightsquigarrow^* \Sigma'$ under a given program $P$, then for any $X \in \rflconf(\Sigma)$ with local configuration decomposition $(\gheap,(\lheap_i)_{i \le n}, \lheap,(\lfilter^j)_j)$, for any $\absprog :> X$ there exists $\absprog'$ and $X' \in \rflconf(\Sigma')$  with local configuration decomposition  $(\gheap',(\lheap'_i)_{i \le n}, \lheap',(\lfilter'^j)_j)$ such that  $\forall i, \lheap_i \ne \lheap \implies \lheap_i = \lheap_i'$,  $\absprog' :> X'$ and $\translate{P} \cup \absprog \vdash \absprog'$.
\end{lemma}

The proof is postponed in Section~\ref{subsec:local}.

\subsection{Serialization}
To state and prove the global soundness theorem, we are going to need some lemmas to handle heap serialization. Basically these lemmas state that if you serialize only memory blocks that are abstracted in a flow-insensitive fashion, then the serialized versions are still properly over-approximated. The serialization lemmas will be applicable in the global soundness theorem proof because the concrete semantics use serialization for inter-components communications and because our analysis always abstract shared memory blocks in a flow-insensitive fashion.

\begin{lemma}
\label{lem:serialization}
 The following statements hold:
\begin{itemize}
\item if $\serialized \vdash \serval{\heap}(v) = (v',\heap',\serialized')$
then $\rfval{}(v) = \rfval{}(v')$
\item if $\serialized \vdash \serblock{\heap}(b) = (b',\heap',\serialized')$
then $\rfblock{}(b) = \rfblock{}(b')$
\end{itemize}
\end{lemma}

\begin{IEEEproof}
If $v = \prim$, then $v' = \prim$ and $\rfval{}(v) = \rfval{}(v') = \rfprim(\prim)$. If $v = \pointer{p}{\ann}$ then $v' = \pointer{p'}{\ann}$ for some pointer $p'$ and $\rfval{}(v) = \absg{\ann} = \rfval{}(v')$. The second point is a direct consequence of the first one.
\end{IEEEproof}

Let $image(\serialized) = \{\ell' ~|~ \exists \ell. (\ell \mapsto \ell') \in \serialized\}$.
\begin{lemma}
\label{lem:serialization-prop}
If $image(\serialized) \cap dom(\heap) = \emptyset$ then :
\begin{itemize}
\item if $\serialized \vdash \serval{\heap}(v) = (v',\heap',\serialized')$ then $image(\serialized') \cap dom(\heap) = \emptyset$.
\item if $\serialized \vdash \serblock{\heap}(b) = (b',\heap',\serialized')$ then  $image(\serialized') \cap dom(\heap) = \emptyset$.
\end{itemize}
\end{lemma}

\begin{IEEEproof}
We prove the first two points by mutual induction on the proof derivation:
\begin{itemize}
\item $\inferrule { } {\serialized \vdash \serval{\heap}({\prim}) = (\prim, \cdot, \serialized)}:$ by lemma's hypothesis.

\item $\inferrule {(\pointer{p}{\ann} \mapsto \pointer{p'}{\ann}) \in \serialized} {\serialized, \vdash \serval{\heap}(\pointer{p}{\ann}) = (\pointer{p'}{\ann}, \cdot, \serialized)}: $ idem.

\item $\inferrule {\pointer{p}{\ann} \notin dom(\serialized) \\ \pointer{p'}{\ann} \text{ fresh pointer} \\\serialized ,\pointer{p}{\ann} \mapsto \pointer{p'}{\ann} \vdash \serblock{\heap}(\heap(\pointer{p}{\ann})) = (b, \heap'',\serialized') \\ \heap' = \heap'',\pointer{p'}{\ann} \mapsto b } {\serialized \vdash \serval{\heap}(\pointer{p}{\ann}) = (\pointer{p'}{\ann}, \heap',\serialized')}: $ 

$\pointer{p'}{\ann}$ is fresh and  $image(\serialized) \cap dom(\heap) = \emptyset$, therefore  $image(\serialized,\pointer{p}{\ann} \mapsto \pointer{p'}{\ann}) \cap dom(\heap) = \emptyset$. Hence by induction we know that  $image(\serialized') \cap dom(\heap) = \emptyset$.

\item $\inferrule {\serialized_0 = \serialized \\ \forall i \in [1,n]: \serialized_{i-1} \vdash \serval{\heap}(v_i) = {(u_i,\heap_i,\serialized_i)} \\ \heap' = \heap_1,\dots,\heap_n } {\serialized \vdash \serblock{\heap}(\obj{c'}{(f_i \mapsto v_i)^{i \leq n}}) = (\obj{c'}{(f_i \mapsto u_i)^{i \leq n}}, \heap',\serialized_n)}: $

We do an induction over $i \in [0,n]$ to prove that  $image(\serialized_i) \cap dom(\heap) = \emptyset$: $\serialized_0 = \serialized$ hence by lemma's hypothesis  $image(\serialized_0) \cap dom(\heap) = \emptyset$. Now assume that $image(\serialized_{i - 1}) \cap dom(\heap) = \emptyset$, then by outer induction hypothesis we have  $image(\serialized_i) \cap dom(\heap) = \emptyset$. 

\item Block serialization of arrays and intents works exactly like the object case.
\end{itemize}
\end{IEEEproof}

\begin{lemma}
\label{lem:serialization-prop2}
If $image(\serialized) \cap dom(\heap) = \emptyset$ then
\begin{itemize}
\item if $\serialized \vdash \serval{\heap}(u) = (u',\heap',\serialized')$ then $u \not \in dom(\heap)$.
\item if $\serialized \vdash \serblock{\heap}(b) = (b',\heap',\serialized')$ then $(\_ \mapsto b') \not \heapto \heap$.
\end{itemize}
\end{lemma}

\begin{IEEEproof}
Simple proof by case analysis on the last (or two last) derivation rule(s) applied.
\end{IEEEproof}

\begin{lemma}
\label{lem:heap-serialization}
Let \lheapd be a heap decomposition of $\heap$. If $\absprog :> \rfheap{\gheap}(\heap)$ and $image(\serialized) \cap dom(\heap) = \emptyset$ then:
\begin{itemize}
\item if $\serialized \vdash \serval{\heap}(v) = (v',\heap',\serialized')$ and $v \in dom(\gheap)$ or $v$ is a primitive value then $\absprog :> \rfheap{\gheap \cup \heap'}( \heap')$
\item if $\serialized \vdash \serblock{\heap}(b) = (b',\heap',\serialized')$  and there exists $\ell$ such that $(\ell \mapsto b) \in \gheap$ then $\absprog :> \rfheap{\gheap \cup \heap'}( \heap')$
\end{itemize}
Moreover $\gheap \cup \heap' \cdot (\lheap_i)_i$ is a heap decomposition of $\heap \cup \heap'$.
\end{lemma}
\begin{IEEEproof}
We prove this by mutual induction on the serialization proof derivation.
\begin{itemize}
\item $\inferrule { } {\serialized \vdash \serval{\heap}({\prim}) = (\prim, \cdot, \serialized)}:$  in that case $\rfheap{\gheap \cup \heap'}(\heap') = \emptyset$

\item $\inferrule {(\pointer{p}{\ann} \mapsto \pointer{p'}{\ann}) \in \serialized} {\serialized, \vdash \serval{\heap}(\pointer{p}{\ann}) = (\pointer{p'}{\ann}, \cdot, \serialized)}: $ idem here we have $\rfheap{\gheap\cup\heap'}(\heap') = \emptyset$

\item $\inferrule {\pointer{p}{\ann} \notin dom(\serialized) \\ \pointer{p'}{\ann} \text{ fresh pointer} \\\serialized ,\pointer{p}{\ann} \mapsto \pointer{p'}{\ann} \vdash \serblock{\heap}(\heap(\pointer{p}{\ann})) = (b, \heap'',\serialized') \\ \heap' = \heap'',\pointer{p'}{\ann} \mapsto b } {\serialized \vdash \serval{\heap}(\pointer{p}{\ann}) = (\pointer{p'}{\ann}, \heap',\serialized')}: $ 

Since $\pointer{p}{\ann} \in dom(\gheap)$ we know that $(\pointer{p}{\ann} \mapsto \heap(\pointer{p}{\ann})) \in \gheap$. Therefore by induction  we know that $\absprog >: \rfheap{\gheap\cup\heap''}(\heap'')$. Observe the following:
\[ \rfheap{\gheap\cup\heap'}(\heap') = \rfheap{\gheap\cup\heap''}(\heap'') \cup \rfheap{\gheap\cup\heap'}(\newpointer{\pointer{p}{\ann}} \mapsto b)\]
Therefore to show that $\absprog :> \rfheap{\gheap\cup\heap'}(\heap')$ we just need to show that:
\[\begin{array}{lcll}
\absprog & :> & \rfheap{\gheap\cup\heap'}(\pointer{p'}{\ann} \mapsto b) \\
 & = & \{\absheap(\ann,\rfblock{}(b))\} & \\
 & = & \{\absheap(\ann,\rfblock{}(\heap(\pointer{p}{\ann})))\} & \text{by Lemma~\ref{lem:serialization}} \\
 & = & \rfheap{\gheap}(\pointer{p}{\ann} \mapsto \heap(\pointer{p}{\ann})) &\text{since }\pointer{p}{\ann} \in dom(\gheap) \\
\end{array}\]

The last point is implied by the fact that $\absprog :> \rfheap{\gheap}(\heap)$.

Moreover by induction we know that $\gheap \cup \heap'' \cdot (\lheap_i)_i$ is a heap decomposition of $\heap \cup \heap''$. By Lemma~\ref{lem:serialization-prop2} we know that $(\_ \mapsto b) \not \heapto H$. Moreover $\pointer{p'}{\ann}$ is a fresh location, therefore it is easy to check that $\gheap \cup \heap' \cdot (\lheap_i)_i$ is a heap decomposition of $\heap \cup \heap'$.

\item $\inferrule {\serialized_0 = \serialized \\ \forall i \in [1,n]: \serialized_{i-1} \vdash \serval{\heap}(v_i) = {(u_i,\heap_i,\serialized_i)} \\ \heap' = \heap_1,\dots,\heap_n } {\serialized \vdash \serblock{\heap}(\obj{c'}{(f_i \mapsto v_i)^{i \leq n}}) = (\obj{c'}{(f_i \mapsto u_i)^{i \leq n}}, \heap',\serialized_n)}: $

By applying repeatedly Lemma~\ref{lem:serialization-prop} we get that for all $i \in [1,n]$, $image(\serialized_i) \cap dom(\heap) = \emptyset$.

We know that there exists $\pointer{p}{\ann}$ such that $(\pointer{p}{\ann} \mapsto \obj{c'}{(f_i \mapsto v_i)^{i \leq n}})) \in \gheap$. Since \lheapd is a heap decomposition, we know that for all $i \in [1,n]$, $u_i \in dom(\gheap)$ or $u_i$ is a primitive value. Therefore by induction we know that for all $i \in [1,n]$ $\absprog :> \rfheap{\gheap \cup \heap_i}(\heap_i)$, which implies that :
\[\absprog :> \bigcup_{1 \le i \le n} \rfheap{\gheap \cup \heap_i}(\heap_i) =  \rfheap{\gheap \cup (\bigcup_{1 \le i \le n} \heap_i)}\left( \bigcup_{1 \le i \le n} \heap_i\right)\]

Moreover the induction hypothesis gives us the fact that  for all $i \in [1,n], \gheap \cup \heap_i \cdot (\lheap_i)_i$ is a heap decomposition of $\heap \cup \heap_i$. It is rather simple to check that this implies that  $\gheap \cup \left( \bigcup_{1 \le i \le n} \heap_i\right) \cdot (\lheap_i)_i$ is a heap decomposition of $\heap  \left( \bigcup_{1 \le i \le n} \heap_i\right)$.

\item Block serialization of arrays and intents works exactly like the object case.
\end{itemize}
\end{IEEEproof}

\subsection{Proof of Theorem~\ref{thm:preservation}}

The global preservation theorem states that our analysis is soundly over-approximating the configuration reduction relation. To prove it, we need an extra assumption on the values that can be given by the Android system to a callback:
\begin{assumption}
\label{asm:bundle}
For all configuration decomposition \cheapd, for all location $\ell$ pointing to an activity object, for all life-cycle state $s$, for any arbitrary callback state $\getcb{\ell}{s} = \locstate{\_}{\_}{R}{\_} :: \varepsilon$,  the callback register $R$ contains only locations in $\gheap$.
\end{assumption}

This is because callback arguments are supplied by the system, and are either primitive values, locations pointing to running $\activity$ objects (which are always global), or locations to $\mathsf{Bundle}$. $\mathsf{Bundle}$ are special objects (that we did not model), which are used to save an activity state in order to be able to restore it after it has been destroyed (for example by a screen orientation change). To properly handle callbacks, we would need to model these $\mathsf{Bundle}$ objects, and to always abstract them in a flow-insensitive fashion.

\begin{theorem*}[Global Preservation]
\label{thm:global}
If $\Psi \Rightarrow^* \Psi'$ under a given program $P$, then for any $X \in \rfconf(\Psi)$, for any $\Delta :> X$ there exists $\Delta'$ and $X' \in \rfconf(\Psi')$  such that  $\Delta' :> X'$ and $\translate{P} \cup \absprog \vdash \absprog'$.
\end{theorem*}

The proof can be found in Section~\ref{subsec:global}.

\subsection{Application to Taint Tracking}
\label{subsec:taint-tracking} 

\begin{lemma}[Taint Abstraction Soundness] 
\label{lem:taint-sound}
For all configuration $\Psi = \tactconf{\actstack}{\threadpool}{\heap}{\sheap}$, for all $\phi = \tactframe{\ell}{s}{\pi}{\threadstack}{\callstack} \in \actstack$ or $\phi = \threadframe{\ell}{\ell'}{\pi}{\threadstack}{\callstack} \in \threadpool$, if $\callstack = \locstate{c,m,\pc}{\stm^*}{R}{u^*} :: \_$ then for all register $r_k$ we have that all $\absprog  \in \rfconf(\Psi)$ with configuration decomposition $\cheapd$ such that $\lheap_n$ is $\phi$'s local heap, for all $\absprog' :> \absprog$, there exist two abstract local state facts $\absreg{{c,m,\pc}}{(\absthread,\absual^*)}{\absval^*}{\abslh}{\absfi}$ and  $\absreg{{c,m,\pc}}{(\absthread',\absual'^*)}{\absval'^*}{\abslh'}{\absfi'}$ such that:
\begin{align*}
&\quad\rflocstate{\ell_r}(\locstate{c,m,\pc}{\stm^*}{R}{u^*},\lheap_n,(\lfilter^{n,j})_j) \\
=&\quad \absreg{c,m,\pc}{(\absthread,\absual^*)}{\absval^*}{\abslh}{\absfi} &\in \absprog\\
\poreg&\quad \absreg{{c,m,\pc}}{(\absthread',\absual'^*)}{\absval'^*}{\abslh'}{\absfi'}&\in \absprog'
\end{align*}
and there exists  $\ataint$ such that $\taintf{\Psi}(R(r_k)) \taintpo \ataint$ and :
\[
  \translate{P} \cup \absprog' \vdash \ataintf{\absval'_i}{\abslh'}{\ataint}
\]
\end{lemma}

\begin{IEEEproof}
The first part is easy, the only difficulty lies in proving that there exists  $\ataint$ such that $\taintf{\Psi}(R(r_k)) \taintpo \ataint$ and :
\[
  \translate{P} \cup \absprog' \vdash \ataintf{\absval'_i}{\abslh'}{\ataint}
\]
We let:
\[
  \taintf{\Psi}^0(u) = 
  \begin{cases} 
    \taint & \text{if } u = \prim^\taint\\
    \public & \text{otherwise}
  \end{cases}
\]
For all $n$ we define the following functions:
\begin{equation*}
  \taintf{\Psi}^{n+1}(u) = 
  \begin{cases}
    \taintcup_{i}\; \taintf{\Psi}^n(v_i) & \text{if } u = \ell \wedge H(\ell) = \obj{c}{(f_i \mapsto v_i)^*} \\
    \taintcup_{i}\; \taintf{\Psi}^n(v_i) & \text{if } u = \ell \wedge H(\ell) = \arr{\tau}{v^*} \\
    \taintcup_{i}\; \taintf{\Psi}^n(v_i) & \text{if } u = \ell \wedge H(\ell) = \intent{c}{(k_i \mapsto v_i)^*} \\
    \taint & \text{if } u = \prim^\taint
  \end{cases}
\end{equation*}
We know that $\taintf{\Psi}(v) = \lim_{n\in \mathbb{N}} \taintf{\Psi}^n(v)$ and that this limit is reached in a finite number of step (since the lattice and the heap are finite). We then show by induction on $n$ that for all $u$, for all $u \povalp \absual$, there exists $\ataint$ such that $\taintf{\Psi}^n(u) \taintpo \ataint$ and:
\[
  \translate{P} \cup \absprog' \vdash \ataintf{\absual}{\abslh'}{\ataint}
\]
Applying the previous result to $\taintf{\Psi}(R(r_k))$ conclude this proof.
\end{IEEEproof}

\begin{lemma}
If for all sinks $(c, m) \in \sinks{}$, $\absprog \in \rfconf(\Psi)$:
\[
 \translate{P} \cup \absprog \vdash  \absreg{{c,m,\pc}}{\_}{\absval^*}{\abslh}{\absfi} \wedge \ataintf{\absval_i}{\abslh}{\secret}
\] 
is unsatisfiable for each $i$, then $P$ does not leak from $\Psi$.
\end{lemma}

\begin{IEEEproof}
We prove the contraposition. Assume that a program $P$ satisfies Definition~\ref{def:leak}, then there exists a configuration $\Psi'$ starting from $\Psi$ where one of the registers $r_k$ in a sink $(c,m)$ contains a $\secret$ value. By Theorem~\ref{thm:preservation}, for all $\absprog \in \rfconf(\Psi)$ there exists $\absprog' \in \rfconf(\Psi')$ and $\absprog'' :> \absprog'$ such that $\translate{P} \cup \absprog \vdash \absprog''$. 

Let $\cheapd$ be the configuration decomposition of $\absprog'$ and $\lheap_n$ be the local heap of $\phi$. By Lemma~\ref{lem:taint-sound} there exist two abstract local state facts $\absreg{{c,m,\pc}}{(\absthread,\absual^*)}{\absval^*}{\abslh}{\absfi}$ and  $\absreg{{c,m,\pc}}{(\absthread',\absual'^*)}{\absval'^*}{\abslh'}{\absfi'}$ such that:
\begin{align*}
&\quad\rflocstate{\ell_r}(\locstate{c,m,\pc}{\stm^*}{R}{u^*},\lheap_n,(\lfilter^{n,j})_j) \\
=&\quad \absreg{c,m,\pc}{(\absthread,\absual^*)}{\absval^*}{\abslh}{\absfi} &\in \absprog'\\
\poreg&\quad \absreg{{c,m,\pc}}{(\absthread',\absual'^*)}{\absval'^*}{\abslh'}{\absfi'}&\in \absprog''
\end{align*}
and there exists  $\ataint$ such that $\taintf{\Psi'}(R(r_k)) \taintpo \ataint$ and :
\[
  \translate{P} \cup \absprog'' \vdash \ataintf{\absval'_i}{\abslh'}{\ataint}
\]
Since $\taintf{\Psi'}(R(r_k)) = \secret$ we know that $\ataint = \secret$. This implies that the following formula is derivable:
\begin{equation*}
 \translate{P} \cup \absprog \vdash
 \absreg{{c,m,\pc}}{(\absthread',\absual'^*)}{\absval'^*}{\abslh'}{\absfi'} \wedge \ataintf{\absval'_i}{\abslh}{\secret}
\end{equation*}
\end{IEEEproof}

\subsection{Proof of Lemma~\ref{thm:local}}

\label{subsec:local}
\begin{IEEEproof}

If $\Sigma = \Sigma'$ then it suffices to take $\Delta' = \Delta$.

We are just going to prove that this is true if $\Sigma$ reduces to $\Sigma'$ in one step. The lemma proof is then obtained by a straightforward induction on the reduction length.

Let $X \in \rflconf(\Sigma)$ with local configuration decomposition $(\gheap,(\lheap_i)_{i \le n}, \lheap,(\lfilter^j)_j)$. Let $\absprog$ be such that $\absprog :> X$.
 
\paragraph{Notation Conventions:} When not explicitly mentioned otherwise, we let $\Sigma = \tmethconf{\callstack}{\pi}{\threadstack}{\heap}{\sheap}{\ell_r}$ with $\callstack = L_1 :: \callstack_0$ 
, and let $\Sigma' = \tmethconf{\callstack'}{\pi'}{\threadstack'}{\heap'}{\sheap'}{\ell_r}$ with $\callstack' = L_1' :: \callstack'_0$.
We also let $L_1 = \locstate{c,m,\pc}{\stm^*}{R}{u^*}$, 
and $L_1' = \locstate{c',m',\pc'}{\stm'^*}{R'}{u'^*}$. 

\paragraph{Proof Structure} First we are going to describe each case structure:
\begin{enumerate}
\item Define $(\gheap',(\lheap'_i)_{i \le n}, \lheap',(\lfilter'^j)_j)$ and show that it is a local configuration decomposition  of $\Sigma'$, and that $\forall i, \lheap_i \ne \lheap \implies \lheap_i = \lheap_i'$
\item Define $\dcall,\dheap, \dstat, \ddispatch$ and $ \dtdispatch$ such that:
\begin{itemize}
\item $ \rfcall{\ell_r}(\callstack',\lheap',(\lfilter'^j)_j) \backslash \rfcall{\ell_r}(\callstack,\lheap,(\lfilter^j)_j)\subseteq \dcall $
\item $ \rfheap{\gheap'}(\heap') \backslash \rfheap{\gheap}(\heap) \subseteq \dheap $
\item $ \rfstat{}(\sheap') \backslash \rfstat{}(\sheap) \subseteq \dstat$
\item $ \rfdispatch{\ell_r}(\pi') \backslash \rfdispatch{\ell_r}(\pi) \subseteq \ddispatch $
\item $ \rftdispatch{\gheap}(\threadstack') \backslash \rftdispatch{\gheap}(\threadstack) \subseteq \dtdispatch$
\end{itemize}
\item Define $\delcall, \delheap, \delstat$, $\deldispatch$ and $\deltdispatch$.
\item Show that:
\begin{itemize}
\item $ \dcall <: \Delta \cup \delcall$
\item $\dheap <: \delheap$
\item $\dstat <: \delstat$
\item $\ddispatch <: \deldispatch$
\item $\dtdispatch <: \deltdispatch$
\end{itemize}
\item Show that:
\begin{itemize}
\item $\translate{P} \cup \absprog \vdash \delcall$
\item $\translate{P} \cup \absprog \vdash \delheap$
\item $\translate{P} \cup \absprog \vdash \delstat$
\item $\translate{P} \cup \absprog \vdash \deldispatch$
\item $\translate{P} \cup \absprog \vdash \deltdispatch$
\end{itemize}
\end{enumerate}

This is enough to prove the lemma. Indeed by point $1)$ we know that $X' =  \rfcall{\ell_r}(\callstack',\lheap',(\lfilter'^j)_j) \cup \rfheap{\gheap'}(\heap') \cup  \rfstat{}(\sheap') \cup  \rfdispatch{\ell_r}(\pi') \cup \rfdispatch{\gheap'}(\threadstack')$ is in $\rflconf(\Sigma')$. Let $\absprog' = \absprog \cup \delcall \cup \delheap \cup \delstat \cup \deldispatch \cup \deltdispatch$. 

 Using the fact that $\absprog :> X$ and point $4)$ we get by applying Lemma~\ref{lem:join-order} that $X \cup \dcall \cup \dheap \cup \dstat \cup \ddispatch <: \absprog'$. We know that $X' \subseteq X \cup \dcall \cup \dheap \cup \dstat \cup \ddispatch \cup \dtdispatch$ by the definitions in point $2)$. Then by applying Lemma~\ref{lem:sub-order} we have $X' <: X \cup \dcall \cup \dheap \cup \dstat \cup \ddispatch \cup \dtdispatch$, and by applying Lemma~\ref{lem:trans-order} we have $X' <: \absprog'$.

The fact that  $\translate{P} \cup \absprog \vdash \absprog$ and point $5)$ implies that  $\translate{P} \cup \absprog \vdash \absprog'$, which concludes the proof.\\

We apply this method to each case,  and detail the most important cases in the next following items.

\begin{itemize}
\item \textbf{\irule{R-Goto}:} 
The rule applied is $\goto{\pc'}$.
\begin{enumerate}
\item Let $\lheapdp = \lheapd$ and $(\lfilter'^j)_j = (\lfilter^j)_j$. It is trivial to check that $\lheapdhp$ is a local configuration decomposition of $\Sigma'$.

\item Since $\lheapdp = \lheapd$ and $(\lfilter'^j)_j = (\lfilter^j)_j$ we know that for all $i \ge 2$ we have $\fhistget{(\lheap,\flist)}{i} = \fhistget{(\lheap',\flistp)}{i}$. Therefore using Proposition~\ref{prop:samefhistget} we know that for all $i \ge 2$ we have:
\[\rfinvoke{\ell_r}(\callstack_i,i,\_,\lheap,(\lfilter^n)_n) = \rfinvoke{\ell_r}(\callstack_i,i,\_,\lheap',(\lfilter'^n)_n)\]
Hence $\dcall = \rflocstate{\ell_r}(\locstate{c,m,\pc'}{\stm^*}{R}{u^*},\lheap',(\lfilter'^n)_n)$ satisfies the wanted properties.

\item We know that $\rflocstate{\ell_r}(\locstate{c,m,\pc}{\stm^*}{R}{u^*},\lheap,(\lfilter^n)_n) = \absreg{c,m,\pc}{(\absthread,\absual^*)}{\absval^*}{\abslh}{\absfi}$ is in $X$ and $X <: \absprog$. Therefore there exists $\absreg{{c,m,\pc}}{(\absthread',\absual'^*)}{\absval'^*}{\abslh'}{\absfi'}$ in $\absprog$ such that :
\[\absreg{{c,m,\pc}}{(\absthread,\absual^*)}{\absval^*}{\abslh}{\absfi} \poreg \absreg{{c,m,\pc}}{(\absthread',\absual'^*)}{\absval'^*}{\abslh'}{\absfi'}\]
Then we define $\delcall = \absreg{{c,m,\pc'}}{(\absthread',\absual'^*)}{\absval'^*}{\abslh'}{\absfi'}$.

\item We are going to show that $ \dcall <: \Delta \cup \delcall$. First one can check that:
\[\rflocstate{\ell_r}(\locstate{c,m,\pc'}{\stm^*}{R}{u^*},\lheap',(\lfilter'^n)_n) = \absreg{c,m,\pc'}{(\absthread,\absual^*)}{\absval^*}{\abslh}{\absfi}\]
The fact that $\absreg{c,m,\pc'}{(\absthread,\absual^*)}{\absval^*}{\abslh}{\absfi} \poreg \absreg{c,m,\pc'}{(\absthread',\absual'^*)}{\absval'^*}{\abslh'}{\absfi'}$ is then trivial.

\item We are going to show that $\translate{P} \cup \absprog \vdash \delcall$. We know that $\translate{\goto{\pc'}}_\spp$ is included in $\translate{P}$, therefore we have the following rule:
\[\absreg{c,m,\pc}{(\absthread',\absual'^*)}{\absval'^*}{\abslh'}{\absfi'} \implies \absreg{c,m,\pc'}{(\absthread',\absual'^*)}{\absval'^*}{\abslh'}{\absfi'}\} \in \translate{P}\]
Moreover $\absreg{{c,m,\pc}}{(\absthread',\absual'^*)}{\absval'^*}{\abslh'}{\absfi'}$ is in $\absprog$, therefore by resolution we get:
\[\translate{P} \cup \absprog \vdash\absreg{{c,m,\pc'}}{(\absthread',\absual'^*)}{\absval'^*}{\abslh'}{\absfi'}\]
This concludes this proof.
\end{enumerate}

\item \textbf{\irule{R-MoveFld}}
The rule applied is $\move{r_o.f}{\rhs}$. We know that there exist two abstract local state facts $\absreg{{c,m,\pc}}{(\absthread,\absual^*)}{\absval^*}{\abslh}{\absfi}$ and  $\absreg{{c,m,\pc}}{(\absthread',\absual'^*)}{\absval'^*}{\abslh'}{\absfi'}$ such that:
\begin{equation*}
\rflocstate{\ell_r}(\locstate{c,m,\pc}{\stm^*}{R}{u^*},\lheap,(\lfilter^n)_n) = \absreg{c,m,\pc}{(\absthread,\absual^*)}{\absval^*}{\abslh}{\absfi}
\poreg \absreg{{c,m,\pc}}{(\absthread',\absual'^*)}{\absval'^*}{\abslh'}{\absfi'}\in \Delta \numberthis\label{eq:move1}
\end{equation*}

Let $\regval{r_o} = \ell''$, we know by Proposition~\ref{prop:lslh} we know that either $\ell'' \in \gheap$ or $\ell'' \in \lheap$.
\begin{itemize}
\item[Case 1:] $\ell'' \in \gheap$

By Lemma~\ref{lem:rhs} we know that  $\rflval{}(\regval{r_o},\lheap) \povalp \absval'_{r_o}$. 
Moreover by applying Lemma~\ref{lem:rhs} to $\rhs$ we know that there exists $\absval''$ such that  $\rflval{}(\regval{\rhs},\lheap) \povalp \absval''$ and that $\absprog \cup \arhs{\rhs} \vdash \prhs{\absval''}$. By Lemma~\ref{lem:reach} there exists $\absfi_a$ such that $\vdash \cfilter{\absval''}{\abslh'}{\absfi_a}$ and $\absfi_a$ is the indicator function of the set of reachable elements starting from $\absval''$ in the points-to graph of $\abslh'$.

\begin{enumerate}
\item For all $j \ne a$, let $\lheap_j' = \lheap_j$. Let $Reach_a$ be the subset of $\lheap$ defined as follows:
\begin{gather*}
Reach_a = \{ (\pointer{p}{\absloc} \mapsto b) \in \lheap~|~\absfi_a(\absloc) = 1\}
\end{gather*}
Let $M$ be the partial mapping containing, for all $\absloc$,  exactly one entry $(\pointer{p}{\absloc} \mapsto \bot)$ if there exists a pointer $\pointer{p'}{\absloc} $ in the domain of $Reach_a$. Moreover we assume that the location  $\pointer{p}{\absloc}$ is a fresh location. 
Let $\lheap' = (\lheap)_{|dom(\lheap) \backslash dom(Reach_a)} \cup M$, and $\gheap' = \left(\gheap[\ell'' \mapsto \gheap(\ell'')[f \mapsto \regval{\rhs}]]\right)\cup Reach_a$.

 We define $\lfilter_a$ to be the indicator function of $Reach_a$, $\lfilter'^1 = \lfilter_a \lfunion \lfilter^1$ and $(\lfilter'^j)_{j>1} = (\lfilter^j)_{j > 1}$. One can check that $\lheapdp$ is a heap decomposition of $\heap'\cdot\sheap'$. We know that:
\begin{alignat*}{2}
&\quad dom(\lheap') \backslash \left\{\pointer{p}{\spp}\in dom(\lheap') ~|~ \exists p' , \lfilter_a(\pointer{p'}{\spp}) = 1\right\} \\
=&\quad dom(\lheap') \backslash \left\{\pointer{p}{\spp}\in dom(\lheap') ~|~ \exists p' , \pointer{p'}{\spp} \in dom(Reach_a)\right\} \\
=&\quad dom(\lheap') \backslash dom(M) \\
\subseteq&\quad dom(\lheap)
\end{alignat*}

Therefore by  Proposition~\ref{prop:fhsitget}.5 we get that for all $i \ge 2$, $\fhistget{(\lheap,\flist)}{i} = \fhistget{(\lheap',\flistp)}{i}$. Moreover $dom(\lheap') \backslash dom(\lheap) = dom(M)$, hence by  Lemma~\ref{lem:fhist-char}  we know that  $(\lheap',\flistp)$ is a filter history of $\callstack'$.

The fact that $\lheapdhp$ is a local configuration decomposition of $\Sigma'$ follows easily.

\item Let $L_2,\dots,L_n$ be such that $\callstack = \locstate{c,m,\pc}{\stm^*}{R}{u^*} :: L_2 :: \dots :: L_n$. By Proposition~\ref{prop:samefhistget} we know that for all $j \ge 2$:
\[\rfinvoke{\ell_r}(L_j,j,\_,\lheap,(\lfilter^i)_i) = \rfinvoke{\ell_r}(L_j,j,\_,\lheap',(\lfilter'^i)_i)\]
One can then show that the following definitions of $\dcall$ and $\dheap$ satisfy the wanted properties:
\begin{itemize}
\item $\dcall = \rflocstate{\ell_r}(\locstate{c,m,\pc+1}{\stm^*}{R}{u^*},\lheap',(\lfilter'^i)_i)$
\item $\dheap = \{\absheap(\absloc,\absblock) ~|~ \heap(\ell')= b \wedge \absloc = \rflab(\ell') \wedge \absblock = \rfblock{}(b) \wedge \ell' \in dom(Reach_a)\}$\\
$\cup \{\absheap(\absloc,\absblock) ~|~  \absloc = \rflab(\ell'') \wedge \absblock = \rfblock{}(H(\ell'')[f \mapsto \rfval{}(\regval{\rhs})])\}$
\end{itemize}

\item
\begin{itemize}
\item $\delcall = \absreg{{c,m,\pc+1}}{(\absthread',\absual'^*)}{\lift{\absval'^*}{\absfi_a}}{\lhlift{\abslh'}{\absfi_a}}{\absfi_a \afunion \absfi'}$.
\item We define $\delheap$ as follows: for all $\spp$, if $ \absfi_a(\spp) = 1$ and $\abslh'(\spp) \ne \bot$  then $\absheap(\spp,\abslh'(\spp)) \in \delheap$.

Moreover we add to $\delheap$ the following formula: since $ \rfheap{\gheap}(H) <: \absprog$ and $H(\ell'') \ne \bot$ we know that there exists $\absheap(\absloc_o,\absblock_o) \in \absprog $ such that $\rfblock{}(H(\ell'')) \poblk \absblock_o$ and $\absloc_o = \rflab(\ell'')$. Then we add $H(\absloc_o,\absblock_o[f \mapsto\absval'' ]) $ to $\delheap$.
\end{itemize}

\item We are going to show that:
\begin{itemize}
\item $ \dcall <: \Delta \cup \delcall :$ by applying Lemma~\ref{lem:tedious2}.1 we know that:
\begin{equation*}
 \rflocstate{\ell_r}(\locstate{c,m,\pc+1}{\stm^*}{R}{u^*},\lheap',(\lfilter'^n)_n))
= \; \absreg{c,m,\pc+1}{(\absthread,\absual^*)}{\lift{\absval^*}{\absfi_a}}{\lhlift{\abslh}{\absfi_a}}{\absfi_a \afunion \absfi}
\end{equation*}

Therefore we just have to prove that:
\begin{multline*}
\numberthis\label{eq:heapsim}
 \absreg{c,m,\pc+1}{(\absthread,\absual^*)}{\lift{\absval^*}{\absfi_a}}{\lhlift{\abslh}{\absfi_a}}{\absfi_a \afunion \absfi}\\
\poreg\;\absreg{{c,m,\pc+1}}{(\absthread',\absual'^*)}{\lift{\absval'^*}{\absfi_a}}{\lhlift{\abslh'}{\absfi_a}}{\absfi_a \afunion \absfi'}
\end{multline*}

From Equation~\eqref{eq:move1} we know that $\absthread = \absthread'$, $\absual^* \poseqp \absual'^*$, $\absval^* \poseqp \absval'^*$, $\absfi \pofilter \absfi'$ and that  $\forall \spp,\abslh(\spp) \ne \bot \implies \abslh(\spp) \polblk \abslh'(\spp)$.

To show that Equation~\eqref{eq:heapsim} holds we have four conditions to check:
\begin{itemize}
\item We already know that $\absthread = \absthread'$ and $\absual^* \poseqp \absual'^*$.
\item Since $\absval^* \poseqp \absval'^*$, we know by applying Proposition~\ref{prop:vallift} that $\lift{\absval^*}{\absfi_a} \poseqp \lift{\absval'^*}{\absfi_a}$.
\item Since $\absfi \pofilter \absfi'$, it is straightforward to check that $\absfi_a \afunion \absfi \pofilter \absfi_a \afunion \absfi'$.
\item By applying Proposition~\ref{prop:vallift} we know that $\forall \spp,\lhlift{\abslh}{\absfi_a}(\spp) \polblk \lhlift{\abslh'}{\absfi_a}(\spp)$.
\end{itemize}

\item $\delheap :> \dheap$:
\begin{itemize}
\item In a first time we are going to show that:
\begin{equation*}
\delheap >: \{\absheap(\absloc,\absblock) ~|~ \heap = \heap',\ell' \mapsto b \wedge \absloc = \rflab(\ell') \wedge \absblock = \rfblock{}(b) 
\wedge \ell' \in dom(Reach_a)\}
\end{equation*}

Let $\absheap(\absloc,\absblock)$ be an element of the right set of the above relation. We know that there exists $b,\ell'$ such that $\heap(\ell')= b$, $ \absloc = \rflab(\ell')$, $ \absblock = \rfblock{}(b)$ and $\ell' \in dom(Reach_a)$. Besides $\ell' \in Reach_{a}$ implies that $\absfi_a(\absloc) = 1$. We have:
\[\rflocstate{\ell_r}(\locstate{c,m,\pc}{\stm^*}{R}{u^*},\lheap,(\lfilter^n)_n) = \absreg{c,m,\pc}{(\absthread,\absual^*)}{\absval^*}{\abslh}{\absfi}\]
Therefore by definitions of $\rflocstate{\ell_r}$ and of $\rflheap{}$ we know that :
\[\abslh = \left\{\left(\spp \mapsto \rflblock{}\left(\lheap(\pointer{p}{\spp}),\lheap\right)\right) \;|\; \pointer{p}{\spp} \in dom(\lheap)\right\}\]

Since $(\ell' \mapsto b) \in \lheap$ we have $\abslh(\absloc) = \rflblock{}(b,\lheap)$. Besides by applying Proposition~\ref{prop:brf} we know that $\rfblock{}(b) \poblk  \rflblock{}(b,\lheap)$. In summary:
\begin{equation}
\absblock = \rfblock{}(b) \poblk \rflblock{}(b,\lheap) = \abslh(\absloc)\label{eq:blk}
\end{equation}

By Equation~\eqref{eq:move1} we know that $\forall \spp, \abslh(\spp) \ne \bot \implies \abslh(\spp) \polblk \abslh'(\spp)$. Since $(\ell' \mapsto b) \in \heap$, we know that $\abslh(\absloc) \ne \bot$, which implies that $\abslh(\spp) \polblk \abslh'(\spp)$, and by Proposition~\ref{prop:coarseblk} we get that$\abslh(\spp) \poblk \abslh'(\spp)$. Putting Equation~\eqref{eq:blk} together with this we get that:
\[ \absblock \poblk \abslh(\absloc) \poblk \abslh'(\absloc)\]

We know that $\absfi_a(\absloc) = 1$. Besides  $\abslh(\absloc) \poblk \abslh'(\absloc)$ and $\abslh(\absloc) \ne \bot$ implies that $\abslh'(\absloc) \ne \bot$. Therefore $\absheap(\absloc,\abslh'(\absloc)) \in \delheap$, which concludes this case by showing that $\absheap(\absloc,\absblock) <: \absheap(\abslh'(\absloc)) \in \delheap$.

\item It remains to show that:
\[ \{\absheap(\absloc,\absblock) ~|~  \absloc = \rflab(\ell'') \wedge \absblock = \rfblock{}(H(\ell'')[f \mapsto \regval{\rhs}])\} <: \delheap\]

Recall that $\rflval{}(\regval{\rhs},\lheap) \povalp \absval''$, $\absheap(\absloc_o,\absblock_o) \in \absprog $, $\rfblock{}(H(\ell'')) \poblk \absblock_o$, $\absloc_o = \rflab(\ell'')$ and $H(\absloc_o,\absblock_o[f \mapsto\absval'' ]) \in \delheap$.

By Proposition~\ref{prop:coarseblk} we have $\rflval{}(\regval{\rhs},\lheap) \poval \absval''$, and by Proposition~\ref{prop:brf} we have $\rfval{}(\regval{\rhs}) \poval \rflval{}(\regval{\rhs},\lheap)$. Therefore by transitivity of $\poval$ we have $\rfval{}(\regval{\rhs}) \poval \absval''$.  Finally by definition of $\rfblock{}$ we have that:
\[\rfblock{}(H(\ell'')[f \mapsto \regval{\rhs}]) = \rfblock{}(H(\ell''))[f \mapsto \rfval{}(\regval{\rhs})]\]
Applying Proposition~\ref{prop:fieldr} to $\rfblock{}(H(\ell'')) \poblk \absblock_o$ and $\rfval{}(\regval{\rhs}) \poval \absval''$ we get that :
\[\rfblock{}(H(\ell''))[f \mapsto \rfval{}(\regval{\rhs})] \poblk \absblock_o[f \mapsto \absval'']\]
Which proves that :
\[\absheap(\absloc_o,\rfblock{}(H(\ell'')[f \mapsto \regval{\rhs}])) <: \absheap(\absloc_o,\absblock_o[f \mapsto \absval'']) <: \delheap\]
This concludes the proof of $\dheap <: \delheap$.
\end{itemize}
\end{itemize}

\item
\begin{itemize}
\item $\translate{P} \cup \absprog \vdash \delcall$:  Recall that $\absreg{{c,m,\pc}}{(\absthread,\absual^*)}{\absval^*}{\abslh}{\absfi} \poreg \absreg{{c,m,\pc}}{(\absthread',\absual'^*)}{\absval'^*}{\abslh'}{\absfi'} \in \absprog$ and that:
\[\delcall = \absreg{{c,m,\pc+1}}{(\absthread',\absual'^*)}{\lift{\absval'^*}{\absfi_a}}{\lhlift{\abslh'}{\absfi_a}}{\absfi_a \afunion \absfi'}\]

We proved at the beginning of this case that $\absprog \cup \arhs{\rhs} \vdash \prhs{\absval''}$ and  $\vdash \cfilter{\absval''}{\abslh'}{\absfi_a}$. 

Recall that $\absloc_o = \rflab(\ell'')$ and that $\ell'' \in dom(\gheap)$. Lemma~\ref{lem:rhs} applied to $\ell''$ and $\absreg{\spp}{ (\absthread',\absual'^*)} {\absval'^*}{\abslh'}{\absfi'}$ gives us that $\absg{\absloc_o} = \rflval{}(\ell'',\lheap) \povalp \absval_o'$. Moreover we know that $\absheap(\absloc_o,\absblock_o) \in \Delta$, hence we can apply the following rule:
\[ \absg{\absloc_o} \povalp \absval_o' \wedge \absheap(\absloc_o,\absblock_o) \implies \rlookup{o}{\absval'^*}{\abslh'}{\absg{\absloc_o}}{\absblock_o}\]
Finally we apply the following rule:
\begin{align*}
& \prhs{\absval''} \wedge \absreg{\spp}{ (\absthread',\absual'^*)} {\absval'^*}{\abslh'}{\absfi'} \wedge \rlookup{o}{\absval'^*}{\abslh'}{\absg{\absloc_o}}{\absblock_o}  \wedge \cfilter{\absval''}{\abslh'}{\absfi_a} \\
&\implies \absreg{\apcn}{ (\absthread',\absual'^*) }{\lift{\absval'^*}{\absfi'}}{\lhlift{\abslh}{\absfi'}}{\absfi_a \afunion \absfi'}
\end{align*}
This concludes this case.

\item $\translate{P} \cup \absprog \vdash \delheap$: $\translate{P}$ contains the two following rules:
\begin{align*}
& \prhs{\absval''} \wedge\absreg{\spp}{ (\absthread',\absual'^*)} {\absval'^*}{\abslh'}{\absfi'} \wedge \rlookup{o}{\absval'^*}{\abslh'}{\absg{\absloc_o}}{\absblock_o}  \wedge \cfilter{\absval''}{\abslh'}{\absfi_a}
\\& \wedge  \absheap(\absloc_o,\absobj{c'}{(f' \mapsto \absual'')^*, f \mapsto \_}) \implies \absheap(\absloc_o,\absobj{c'}{(f' \mapsto \absual'')^*, f \mapsto {\absval''})})\numberthis\label{eq:delheap1}
\\& \prhs{\absval''} \wedge \absreg{\spp}{ (\absthread',\absual'^*)} {\absval'^*}{\abslh'}{\absfi'} \wedge \rlookup{o}{\absval'^*}{\abslh'}{\absg{\absloc_o}}{\absblock_o}  \\
&\wedge \cfilter{\absval''}{\abslh'}{\absfi_a} \wedge \cfilter{\absval''}{\abslh'}{\absfi_a}\implies  \liftlh{\abslh'}{\absfi_a}\numberthis\label{eq:delheap2}
\end{align*}

$\delheap$ is the set defined by:
\begin{itemize}
\item for all $\spp$, if $ \absfi_a(\spp) = 1 \wedge \abslh'(\spp) \ne \bot$ then $\absheap(\spp,\abslh'(\spp)) \in \delheap$:

Let $\spp$ satisfying the above conditions. The following rules is in $\translate{P}$:
\[\liftlh{\abslh'}{\absfi_a} \wedge \abslh'(\spp) = \absblock \wedge \absfi_a(\spp) = 1 \implies \absheap(\spp,{\absblock})\]
Rule \eqref{eq:delheap2} plus the above rule yield $\translate{P} \cup \absprog \vdash \absheap(\spp,\abslh'(\spp)) $.
\item $\absheap(\absloc_o,\absblock_o[f \mapsto\absval'' ]) $ is in $\delheap$: directly entailed by the rule \eqref{eq:delheap1}.
\end{itemize}

\end{itemize}
\end{enumerate}

\item[Case 2:] $\ell''  \in \lheap$.

Let $\absloc_o = \rflab(\ell'')$, since $\ell''\in dom(\lheap)$ we have that $\absval_o = \absl{\absloc_o}$. We know from Equation~\eqref{eq:move1} that $\absval_o \povalp \absval_o'$, therefore $\absl{\absloc_o} \abspo \absual_o'$.

Let $b$ be such that  $(\ell'' \mapsto b) \in \heap$. This implies that $\abslh({\absloc_o}) \ne \bot$, hence from Equation~\eqref{eq:move1} we get that $\abslh({\absloc_o}) \polblk \abslh'({\absloc_o})$, which in turn implies that there exists $\absblock_o = \absobj{c'}{(f \mapsto \absual'')}$ such that $\absblock_o = \abslh'({\absloc_o})$.
\begin{enumerate}
\item Let $\lheap' = \lheap[\ell'' \mapsto \lheap(\ell'')[f \mapsto \regval{\rhs}]]$ , $\gheap' = \gheap$ and for all $i \ne a$, $\lheap_i' = \lheap_i$. Let $(\lfilter'^j)_j = (\lfilter^j)_j$. Observe that  $dom(\lheap) = dom(\lheap')$, and that $\flist = \flistp$, therefore by Proposition~\ref{prop:fhsitget}.4 we know that for all $j \ge 2$, $\fhistget{(\lheap,\flist)}{j} = \fhistget{(\lheap',\flistp)}{j}$. By applying Lemma~\ref{lem:fhist-char} we get that $\fhistp$ is a filter history of $\callstack'$. It is then rather easy to check that $\lheapdhp$ is a local configuration decomposition of $\Sigma'$.

\item By Proposition~\ref{prop:samefhistget} we get that for all $j \ge 2$:
\[\rfinvoke{\ell_r}(\callstack_j,j,\_,\lheap,(\lfilter^i)_i) = \rfinvoke{\ell_r}(\callstack_j,j,\_,\lheap',(\lfilter'^i)_i)\]
It is then easy to check that $\dcall = \rflocstate{\ell_r}(\locstate{c,m,\pc+1}{\stm^*}{R}{u^*},\lheap',(\lfilter'^n)_n)$ satisfies the wanted property.

\item By Lemma~\ref{lem:rhs} we know that there exists $\absval''$ such that $\rflval{}(\regval{\rhs},\lheap) \povalp \absval''$ and $\absprog \cup \arhs{\rhs} \vdash \prhs{\absval''}$. Then we define  $\delcall$ to be the set containing the predicate:
\[ \absreg{{c,m,\pc + 1}}{(\absthread',\absual'^*)}{\absval'^*}{\underbrace{\abslh'[\absloc_o \mapsto \absblock_o[f \mapsto \absval'']]}_{\abslh_1'}}{\absfi'}\]

\item We are going to show that $ \dcall <: \delcall \cup \Delta$: first one can check that:
\begin{multline*}
\rflocstate{\ell_r}(\locstate{c,m,\pc+1}{\stm^*}{R}{u^*},\lheap,(\lfilter^n)_n)\\
\;=\;\absreg{c,m,\pc + 1}{(\absthread,\absual^*)}{\absval^*}{\underbrace{\abslh[\absloc_o \mapsto \abslh({\absloc_o})[f \mapsto \rflval{}(\regval{\rhs},\lheap)]]}_{\abslh_1}}{\absfi}
\end{multline*}
We are trying to prove that:
\begin{equation*}
\absreg{c,m,\pc + 1}{(\absthread,\absual^*)}{\absval^*}{{\abslh_1}}{\absfi} \poreg  \absreg{{c,m,\pc + 1}}{(\absthread',\absual'^*)}{\absval'^*}{{\abslh_1'}}{\absfi'}
\end{equation*}
Since we already know that:
\begin{equation}
\absreg{c,m,\pc}{(\absthread,\absual^*)}{\absval^*}{{\abslh}}{\absfi} \poreg  \absreg{{c,m,\pc}}{(\absthread',\absual'^*)}{\absval'^*}{{\abslh'}}{\absfi'}\label{eq:moveE}
\end{equation}
We just need to prove that  $\forall \spp,\abslh_1(\spp) \ne \bot \implies \abslh_1(\spp) \polblk \abslh_1'(\spp)$:
\begin{itemize}
\item Equation~\ref{eq:moveE} gives us that  $\forall \spp,\abslh(\spp) \ne \bot \implies \abslh(\spp) \polblk \abslh'(\spp)$, and we know that for all $\spp \ne \absloc_o$ we have $\abslh(\spp) = \abslh_1(\spp)$ and $\abslh'(\spp) = \abslh_1'(\spp)$. Hence $\forall \spp \ne \absloc_o,(\abslh_1)(\spp) \ne \bot \implies \abslh_1(\spp) \polblk \abslh_1'(\spp)$.
\item $ \abslh_1({\absloc_o}) = \abslh({\absloc_o})[f \mapsto \rflval{}(\regval{\rhs},\lheap)]$ and $\abslh_1'({\absloc_o}) = \absblock_o[f \mapsto \absval'']$. Moreover $\abslh({\absloc_o}) \ne \bot$, so $\abslh({\absloc_o}) \polblk \abslh'({\absloc_o}) = \absblock_o$. Therefore by Proposition~\ref{prop:fieldr} we have $\abslh_1({\absloc_o}) \polblk \abslh_1'({\absloc_o})$.

\end{itemize}
\item We are going to show that $\translate{P} \cup \absprog \vdash \delcall$: Recall that  $\absprog \cup \arhs{\rhs} \vdash \prhs{\absval''}$.

We know that  $\vdash \absl{\absloc_o} \abspo \absval'_o$. Moreover recall that  $\absblock_o = \absobj{c'}{(f \mapsto \absual'')} = \abslh'({\absloc_o})$. Therefore we can apply the following two rules:
\begin{align*} & \absl{\absloc_o} \abspo \absval'_o \wedge \absblock_o = \abslh'({\absloc_o}) \implies \rlookup{o}{\absval'^*}{\abslh}{\absl{\absloc_o}}{\absblock_o}\\
&\prhs{\absval''}  \wedge \absreg{\spp}{ (\absthread',\absual'^*)} {\absval'^*}{\abslh'}{\absfi'} \wedge \rlookup{o}{\absval'^*}{\abslh}{\absl{\absloc_o}}{\absblock_o} \\
& \implies\absreg{\apcn}{ (\absthread',\absual'^*) }{\absval'^*}{\abslh[\absloc \mapsto \absblock_o[ f \mapsto \absval'']}{\absfi'}
\end{align*}
Which conclude this case.
\end{enumerate}
\end{itemize}

\item \textbf{\irule{R-Call}} 

Since $\Sigma$ reduces to $\Sigma'$ by applying the rule $\invoke{r_o}{m'}{(r_{i_k})^{k \leq n}}$ we know that $\regval{r_o} = \ell$ and that
\begin{mathpar}
\lookup(\gettype{\heap}{\ell},m') = (c',\stm'^*)

\sign(c',m') = \methsign{\tau_1,\ldots,\tau_n}{\tau}{\loc}

R' = ((r_j \mapsto \defvalue)^{j \leq \loc}, r_{\loc + 1} \mapsto \ell, (r_{\loc+1+k} \mapsto \regval{r_{i_k}})^{k \leq n})

\callstack' = \locstate{c',m',0}{\stm'^*}{R'}{(\regval{r_{i_k}})^{k \leq n}} :: \callstack
\end{mathpar}
\begin{enumerate}
\item Let $\lheapdp = \lheapd$ and $(\lfilter'^j)_j = (\spp \mapsto 0)^* :: (\lfilter^l)_l$ (we have one more filter in the list).

It is easy to check that $\lheapdp$ is a heap decomposition of $\heap'\cdot\sheap'$. By Proposition~\ref{prop:fhsitget}.3 we know that for all $j \ge 1$, $\fhistget{(\lheap,\flist)}{j} = \fhistget{(\lheap',\flistp)}{j + 1}$. Moreover $\fhistget{(\lheap,\flist)}{1} = \fhistget{(\lheap',\flistp)}{1}$.

Let us show that $\fhistp$ is a filter history $\callstack'$. The fact that:
\[\forall i, \forall \pointer{p}{\spp}, \left (\left(i = 0 \wedge \pointer{p}{\spp} \in dom(\lheap') \right) \vee \lfilter'^i(\pointer{p}{\spp}) = 1 \right) \implies \forall j \ne i, \lfilter'^j(\pointer{p}{\spp}) = 0\]
is rather obvious here, so we are going to focus on showing that:
\begin{equation*}
\fhistget{(\lheap',(\lfilter'^j)_j)}{i}(\spp) \ne \fhistget{(\lheap',(\lfilter'^j)_j)}{l}(\spp) \implies \fhistget{(\lheap',(\lfilter'^j)_j)}{i}(\spp) \not \in dom(\callstack'_{|\ge l})
\end{equation*}
\begin{itemize}
\item If $1 < i < l \le n$. For all  $\spp$ we have:
\[\fhistget{\fhistp}{i}(\spp) \ne \fhistget{\fhistp}{l}(\spp) \text{ iff } \fhistget{\fhist}{i-1}(\spp) \ne \fhistget{\fhist}{l-1}(\spp)\]
Moreover since $\fhist$ is a filter history of $\callstack$ we know that:
\[\fhistget{\fhist}{i-1}(\spp) \ne \fhistget{\fhist}{l-1}(\spp) \text{ implies } \fhistget{\fhist}{i-1}(\spp) \not \in dom(\callstack_{|\ge l - 1})\]
Since $l > 2$, $\callstack_{|\ge l - 1} = \callstack'_{| \ge l}$. Moreover $\fhistget{\fhist}{i-1}(\spp) = \fhistget{\fhistp}{i}(\spp)$, so:
\[ \fhistget{\fhist}{i-1}(\spp) \not \in dom(\callstack_{|\ge l - 1}) \implies \fhistget{\fhistp}{i}(\spp) \not \in dom(\callstack'_{| \ge l})\]
Hence we have:
\[\fhistget{\fhistp}{i}(\spp) \ne \fhistget{\fhistp}{l}(\spp) \implies \fhistget{\fhistp}{i}(\spp) \not \in dom(\callstack'_{| \ge l})\]

\item If $i = 1$ and $1 < l \le n$. For all $\spp$ we have:
\[\fhistget{\fhistp}{1}(\spp) \ne \fhistget{\fhistp}{l}(\spp) \text{ iff } \fhistget{\fhist}{1}(\spp) \ne \fhistget{\fhist}{l-1}(\spp)\]
If $l = 2$ then $\fhistget{\fhist}{1}(\spp) \ne \fhistget{\fhist}{l-1}(\spp)$ is never true, so the result holds. If $l > 2$ then the same reasoning that we did in the previous case works.
\end{itemize}

The fact that $\lheapdhp$ is a local configuration decomposition of $\Sigma'$ follows easily.

\item By Proposition~\ref{prop:samefhistget} we get that for all $j > 2$:
\[\rfinvoke{\ell_r}(\callstack_j,j,\_,\lheap,(\lfilter^i)_i) = \rfinvoke{\ell_r}(\callstack_j,j + 1,\_,\lheap',(\lfilter'^i)_i)\]

One can then show that the following set $\dcall$ satisfies the wanted property:
\begin{equation*}
\dcall = \{\rflocstate{\ell_r}(\locstate{c',m',0}{\stm'^*}{R'}{(\regval{r_{i_k}})^{k \leq n}},\lheap',(\lfilter'^j)_j)
\}\cup\{\rfinvoke{\ell_r}(\locstate{c,m,\pc}{\stm^*}{R}{u^*},2,c',\lheap',(\lfilter'^j)_j)\}
\end{equation*}

\item We know that there exist  $\absreg{c,m,\pc}{(\absthread',\absual'^*)}{\absval'^*}{{\abslh'}}{\absfi'} \in \Delta$ and $\absreg{c,m,\pc}{(\absthread,\absual^*)}{\absval^*}{{\abslh}}{\absfi}$ such that
\begin{equation*}
\rflocstate{\ell_r}(\locstate{c,m,\pc}{\stm^*}{R}{u^*},\lheap,(\lfilter^n)_n)= \absreg{c,m,\pc}{(\absthread,\absual^*)}{\absval^*}{{\abslh}}{\absfi}
\poreg  \absreg{c,m,\pc}{(\absthread',\absual'^*)}{\absval'^*}{{\abslh'}}{\absfi'}\numberthis\label{eq:call0}
\end{equation*}

Let $\absloc_o = \rflab(\ell)$. Let $\absual_{call}^* = (\absual_{i_k})^{k \le n}$ and  $\absual_{call}'^* = (\absual'_{i_k})^{k \le n}$. One can  check that:
\begin{alignat*}{2}
&\rflocstate{\ell_r}(\locstate{c',m',0}{\stm'^*}{R'}{(\defvalue_k)^{k \leq \loc},(\regval{r_{i_k}})^{k \leq n}},\lheap',(\lfilter'^j)_j) &\;=&\;\absreg{c',m',0}{(\absthread,\absual_{call}^*)}{(\adefvalue_k)^{k \leq \loc},\absual_{call}^*}{\abslh}{0^*}\numberthis\label{eq:call1}\\
&\rfinvoke{\ell_r}(\locstate{c,m,\pc}{\stm^*}{R}{u^*},2,c',\lheap',(\lfilter'^j)_j) &=&\;\absinvoke{c,m,\pc}{c'}{(\absthread,\absual^*)}{\absval^*}{\absfi}\numberthis\label{eq:call11}
\end{alignat*}

We define $\delcall =\{\absreg{c',m',0}{(\absthread',\absual_{call}'^*)}{(\adefvalue_k)^{k \leq \loc},\absual_{call}'^*}{\abslh'}{0^*}\}\cup\{ \absreg{c,m,\pc}{(\absthread',\absual'^*)}{\absval'^*}{{\abslh'}}{\absfi'}\}$

\item We are going to show that $ \dcall <: \Delta \cup \delcall $, or more specifically that:
\begin{alignat}{3}
&\absinvoke{c,m,\pc}{c'}{(\absthread,\absual^*)}{\absval^*}{\absfi}&&\poinvoke{\Delta}\quad&&  \absreg{c,m,\pc}{(\absthread',\absual'^*)}{\absval'^*}{{\abslh'}}{\absfi'}\label{eq:calld2}\\
& \absreg{c',m',0}{(\absthread,\absual_{call}^*)}{(\adefvalue_k)^{k \leq \loc},\absual_{call}^*}{{\abslh}}{0^*} \quad&&\poreg\quad&& \absreg{c',m',0}{(\absthread',\absual_{call}'^*)}{(\adefvalue_k)^{k \leq \loc},\absual_{call}'^*}{\abslh'}{0^*}\label{eq:calld1}
\end{alignat}

\begin{itemize}
\item[\textbf{Eq. \eqref{eq:calld2}:}] All conditions are trivial consequences of  Equation~\eqref{eq:call0}, except for $\callinv{r_o,c',m'}{\Delta \cup \delcall}{\absval'^*}{\abslh'}$, that we are going to show.

 We know by Lemma~\ref{lem:rhs} that  $\rflval{}(\regval{r_o},\lheap) \povalp \absval'_o$. The fact that $\lookup(\gettype{\heap}{\ell},m') = (c',\stm'^*)$  implies that $\heap(\ell) = \obj{c''}{\_}$ for some class $c''$ such that $c'' \subtype c'$, and that $c' \in \abslookup(m')$. By definition of $\rflconf(\Sigma)$  we know that if $\ell \in dom(\gheap)$ then there exists $\absheap(\absloc_o,\absobj{c''}{\_}) \in X$, and if $\ell \in dom(\lheap)$ then $\abslh({\absloc_o}) = \absobj{c''}{\_}$.

\begin{itemize}
\item If $\ell \in dom(\lheap)$ and  $\abslh({\absloc_o}) = \absobj{c'}{\_}$: then by definition of $\rflval{}$ we have $ \rflval{}(\regval{r_o},\lheap) = \absl{\absloc_o}$, hence $\absl{\absloc_o} \abspo \absval_o'$. Besides since $\abslh({\absloc_o})  = \absobj{c''}{\_} \polblk \abslh'({\absloc_o})$ we know that there exists some $\absblock$ such that $\abslh'({\absloc_o})  = \absobj{c''}{\absblock}$.

\item If $\ell \in dom(\gheap)$ and $\absheap(\absloc_o,\absobj{c''}{\_}) \in X$, then there exists $\hat b$ such that $\absheap(\absloc_o,\absobj{c''}{\hat b}) \in \Delta$. Besides by definition of $\rflval{}$ we have $ \rflval{}(\regval{r_o},\lheap) = \absg{\absloc_o}$, which implies that $ \absval_o' \abspo \absg{\absloc_o}$.
\end{itemize}
This concludes the proof that $\callinv{r_o,c',m'}{\Delta \cup \delcall}{\absval'^*}{\abslh'}$ holds.

\item[\textbf{Eq. \eqref{eq:calld1}:}] The fact that $0^* \pofilter 0^*$ is trivial. From Equation~\eqref{eq:call0} we know that $\forall \spp,\abslh(\spp) \ne \bot \implies \abslh(\spp) \polblk \abslh'(\spp)$ and that $\absual^* \poseqp \absval^*$. The latter implies that $\absual_{call}^* = (\absual_{i_k})^{k \le n} \poseqp  (\absual'_{i_k})^{k \le n} =\absval_{call}^*$. This concludes this case.
\end{itemize}

\item We are going to show that $\translate{P} \cup \absprog \vdash \delcall$. Since $\absreg{c,m,\pc}{(\absthread',\absual'^*)}{\absval'^*}{{\abslh'}}{\absfi'} \in \Delta$ we just need to check that $\translate{P} \cup \absprog \vdash\absreg{c',m',0}{(\absthread',\absual_{call}'^*)}{(\adefvalue_k)^{k \leq \loc},\absual_{call}'^*}{\abslh'}{0^*}$

As in case $4.$ we know that one of the  following  holds:
\begin{itemize}
\item  if $\vdash \absl{{\absloc_o}} \abspo \absval'_o$ and $\abslh'({\absloc_o})  = \absobj{c''}{\absblock}$ then we can apply the following rule:
\[\absl{{\absloc_o}} \abspo \absval'_o \wedge \abslh'({\absloc_o}) = \absobj{c''}{\hat b} \implies \rlookup{o}{\absval'^*}{\abslh'}{\absl{\absloc_o}}{\absobj{c''}{\hat b}}\]

\item if  $\vdash \absg{\absloc_o} \abspo \absval'_o$ and $\absheap(\absloc_o,\absobj{c''}{\hat b}) \in \Delta$ then we can apply the rule:
\[\absg{{\absloc_o}} \abspo \absval'_o \wedge \absheap({\absloc_o},\absobj{c''}{\hat b}) \implies \rlookup{o}{\absval'^*}{\abslh'}{\absg{\absloc_o}}{\absobj{c''}{\hat b}}\]
\end{itemize}
Hence $\Delta \vdash \rlookup{o}{\absval'^*}{\abslh'}{\_}{\absobj{c''}{\hat b}}$. Moreover we already knew that $c'' \subtype c'$ and that $c' \in \abslookup(m')$, therefore we can apply the following rule, which is included in $\translate{P}$:
\begin{multline*}
\absreg{\spp}{ (\absthread',\absual'^*) }{\absval'^*}{\abslh'}{\absfi'} \wedge \rlookup{o}{\absval'^*}{\abslh'}{\_}{\absobj{c''}{\hat b}} \wedge c'' \subtype c' \implies\\
\absreg{c',\mathsf{m'},\mathsf{0}}{(\absthread',\absual_{call}')}{(\adefvalue_k)^{k \leq \loc}, \absual_{call}'}{\abslh'}{0^*}
\end{multline*}
This concludes the proof that  $\translate{P} \cup \absprog \vdash \delcall$.
\end{enumerate}

\item \textbf{\irule{R-Return}}
\begin{enumerate}
\item Let $\lheapdp = \lheapd$ and $(\lfilter'^j)_j = (\lfilter_1 \lfunion \lfilter_2 ) :: (\lfilter_i)_{i > 2}$.

The fact that $\lheapdp$ is a heap decomposition of $\Sigma'$ is easy to prove. 

Since $\Sigma \rightsquigarrow \Sigma'$ we know that $\callstack = \locstate{c,m,\pc}{\stm^*}{R}{v^*} :: \locstate{c',m',\pc'}{\stm'^*}{R'}{u'^*} :: \callstack_1$ and that $\callstack' = \locstate{c',m',\pc'+1}{\stm'^*}{R'[r_{\res} \mapsto \regval{r_{\res}}]}{u'^*} :: \callstack_1$. By Proposition~\ref{prop:fhsitget}.2 we know that for all $j > 1$, $\fhistget{(\lheap,\flist)}{j + 1} = \fhistget{(\lheap',\flistp)}{j}$. Moreover $\fhistget{(\lheap,\flist)}{1} = \fhistget{(\lheap',\flistp)}{1}$.

Let us show that $\fhistp$ is a filter history $\callstack'$. Let us show that $\fhistp$ is a filter history $\callstack'$. The fact that:
\[\forall i, \forall \pointer{p}{\spp}, \left (\left(i = 0 \wedge \pointer{p}{\spp} \in dom(\lheap') \right) \vee \lfilter'^i(\pointer{p}{\spp}) = 1 \right) \implies \forall j \ne i, \lfilter'^j(\pointer{p}{\spp}) = 0\]
is easy to prove, so we are going to focus on showing that:
\begin{equation*}
\fhistget{(\lheap',(\lfilter'^j)_j)}{i}(\spp) \ne \fhistget{(\lheap',(\lfilter'^j)_j)}{l}(\spp) \implies \fhistget{(\lheap',(\lfilter'^j)_j)}{i}(\spp) \not \in dom(\callstack'_{|\ge l})
\end{equation*}
\begin{itemize}
\item If $1 < i < l \le n$, then for all  $\spp$ we have:
\[\fhistget{\fhistp}{i}(\spp) \ne \fhistget{\fhistp}{l}(\spp) \text{ iff } \fhistget{\fhist}{i+1}(\spp) \ne \fhistget{\fhist}{l+1}(\spp)\]
Moreover since $\fhist$ is a filter history of $\callstack$ we know that:
\[\fhistget{\fhist}{i+1}(\spp) \ne \fhistget{\fhist}{l+1}(\spp) \text{ implies } \fhistget{\fhist}{i+1}(\spp) \not \in dom(\callstack_{|\ge l + 1})\]
$\callstack_{| \ge l + 1} = \callstack'_{| \ge l}$, and $\fhistget{\fhist}{i+1}(\spp) = \fhistget{\fhistp}{i}(\spp)$, hence:
\[ \fhistget{\fhist}{i+1}(\spp) \not \in dom(\callstack_{| > l + 1}) \implies \fhistget{\fhistp}{i}(\spp) \not \in dom(\callstack'_{| \ge l})\]
Therefore we have:
\[\fhistget{\fhistp}{i}(\spp) \ne \fhistget{\fhistp}{l}(\spp) \implies \fhistget{\fhistp}{i}(\spp) \not \in dom(\callstack'_{| \ge l})\]

\item If $i = 1$ and $1 < l \le n$. For all $\spp$ we have:
\[\fhistget{\fhistp}{1}(\spp) \ne \fhistget{\fhistp}{l}(\spp) \text{ iff } \fhistget{\fhist}{1}(\spp) \ne \fhistget{\fhist}{l+1}(\spp)\]
The same reasoning that we did in the previous case works.
\end{itemize}

The fact that $\lheapdhp$ is a local configuration decomposition of $\Sigma'$ follows easily.

\item By Proposition~\ref{prop:samefhistget} we get for all $j \ge 1$:
\[\rfinvoke{\ell_r}(\callstack_j,j + 1,\_,\lheap,(\lfilter^i)_i) = \rfinvoke{\ell_r}(\callstack_j,j,\_,\lheap',(\lfilter'^i)_i)\]

One can then check that the following definition of $\dcall$ satisfies the wanted property:
\[\dcall = \{\rflocstate{\ell_r}(\locstate{c',m',\pc' + 1}{\stm'^*}{R'[r_{\res} \mapsto \regval{r_{\res}}]}{u'^*},\lheap',(\lfilter'^j)_j)\}\]

\item We know that:
\begin{align*}
 \rflocstate{\ell_r}(\locstate{c,m,\pc}{\stm^*}{R}{u^*},\lheap,(\lfilter^j)_j) \quad=\quad& \absreg{c,m,pc}{(\absthread,\absual_1^*)}{\absval_1^*}{\abslh_1}{\absfi_1}\numberthis\label{eq:ret1}\\
\quad\poreg\quad& \absreg{c,m,pc}{(\abswal'_1,\absual_1'^*)}{\absval_1'^*}{\abslh_1'}{\absfi_1'} \in \Delta\\
\phantom{m}\\
 \rfinvoke{\ell_r}(\locstate{c',m',\pc'}{\stm'^*}{R'}{u'^*},2,c,\lheap,(\lfilter^j)_j) \quad=\quad& \absinvoke{c',m',pc'}{c}{(\absthread,\absual_2^*)}{\absval_2^*}{\absfi_2}\numberthis\label{eq:ret2}\\
\quad\poinvoke{\Delta}\quad& \absreg{c',m',pc'}{(\abswal'_2,\absual_2'^*)}{\absval_2'^*}{\abslh_2'}{\absfi_2'} \in \Delta
\end{align*}

Let $\delcall = \{\absreg{c',m',pc'+1}{(\abswal'_2,\absual_2'^*)}{\lift{\absval_2'^*}{\absfi_1'}[\res \mapsto ({\absval'^*_1})_{\res}]}{\abslh_1'}{\absfi_1' \afunion \absfi_2'}\}$.

\item By Proposition~\ref{prop:fhsitget}.1 and Proposition~\ref{prop:fhsitget}.2 we have $\fhistget{(\lheap,\lfilter^1 :: \lfilter^2)}{3} = \fhistget{(\lheap,\lfilter^1 \lfunion \lfilter^2)}{2}$, therefore for all $k \le |u_2^*|$ we have 
\begin{equation}
\label{eq:pain1}
\rflval{}((u_2^*)_k,\lheap,\lfilter^1 :: \lfilter^2) = \rflval{}((u_2^*)_k,\lheap,\lfilter^1 \lfunion \lfilter^2)
\end{equation}

Let $r_d$ be a register different from $r_\res$, we want to show that:
\begin{equation}
\label{eq:pain2}
\rflval{}(R'(r_d),\lheap) = \lift{\rflval{}(R'(r_d),\lheap,\lfilter^1)}{\absfi_1}
\end{equation}

 If $R'(r_d)$ is a primitive value then this is trivial, so assume $R'(r_d) = \ell = \pointer{p}{\absloc}$. Let $\ell' = \pointer{p'}{\absloc} \in dom(\lheap)$ (it exists because $\lheap$ is a local heap). Then we have several cases:
\begin{itemize}
\item Case 1: for all $\pointer{p''}{\absloc}$, we have, $\lfilter^1(\pointer{p''}{\absloc}) = 0$.  Then $\fhistget{(\lheap,\lfilter^1)}{\infty}(\absloc) = \fhistget{(\lheap,\varepsilon)}{\infty}(\absloc) = \ell'$, therefore :
\[\rflval{}(\ell,\lheap,\lfilter^1) = \rfloc{}(\ell,\lheap,\lfilter^1) = \rfloc{}(\ell,\lheap) = \rflval{}(\ell,\lheap)\]
Moreover $\forall \pointer{p''}{\absloc}, \lfilter^1(\pointer{p''}{\absloc}) = 0$ also implies that $\absfi_1(\absloc) = 0$, hence :
\[\lift{ \rflval{}(\ell,\lheap,\lfilter^1)}{\absfi_1} = \rflval{}(\ell,\lheap,\lfilter^1)\]
This concludes this case.
\item Case 2: there exists $\ell'' = \pointer{p''}{\absloc}$ such that $\lfilter^1(\pointer{p''}{\absloc}) = 1$. Then $\fhistget{(\lheap,\lfilter^1)}{\infty}(\absloc) = \ell''$ and $ \fhistget{(\lheap,\varepsilon)}{\infty}(\absloc) = \ell'$. We know that $\lfilter^1(\ell'') = 1$ and that $\ell' \in dom(\lheap)$, therefore since $(\lheap,(\lfilter^j)_j)$ is a filter history we have $\ell' \ne \ell''$.

This implies that  $\fhistget{(\lheap,\lfilter^1)}{2}(\absloc) \ne \fhistget{(\lheap,\varepsilon)}{1}(\absloc)$, therefore since $(\lfilter^i)_i$ is a filter history of $\Sigma$ we know that $\ell' = \fhistget{(\lheap,\varepsilon)}{\infty}(\absloc) \ne R'(r_d) = \ell$. Hence one of the two following cases holds:
\begin{itemize}
\item $\ell \ne \ell''$. Then $\rflval{}(\ell,\lheap) =\rflval{}(\ell,\lheap,\lfilter^1)  =  \absg{\absloc} = \lift{\rflval{}(\ell,\lheap,\lfilter^1)}{\absfi_1}$.
\item $\ell = \ell''$. Then we have:
 \[\rflval{}(\ell,\lheap,\lfilter^1) = \absl{\absloc} \text{ and }  \rfloc{}(\ell,\lheap) = \absg{\absloc}\]
Moreover $\lfilter^1(\ell'') = 1$ implies that $\absfi_1(\absloc) = 1$, therefore :
\[\lift{\rflval{}(\ell,\lheap,\lfilter^1)}{\absfi_1} = \lift{\absl{\absloc}}{\absfi_1} = \absg{\absloc} = \rfloc{}(\ell,\lheap)\]
\end{itemize}
\end{itemize}

Using Equation~\ref{eq:pain1} and Equation~\ref{eq:pain2} one can easily show that:
\begin{equation*}
\dcall = \absreg{c',m',pc'+1}{(\absthread,\absual_2^*)}{\lift{\absval_2^*}{\absfi_1}[\res \mapsto ({\absval_1^*})_{\res}]}{\abslh_1}{\rffilter(\lfilter^1 \lfunion \lfilter^2)}
\end{equation*}

We want to show that $ \dcall <: \Delta \cup \delcall$: by definition of $\poreg$ we need to check the four following conditions:
\begin{itemize}
\item $\absthread =  \abswal'_2$ and $\absual_2^* \poseqp \absual_2'^*$: this is trivially implied by Equation~\eqref{eq:ret2}. 

\item $\forall i,\lift{\absval_2^*}{\absfi_1}[\res \mapsto ({\absval_1^*})_{\res}]\povalp \lift{\absval_2'^*}{\absfi_1'}[\res \mapsto ({\absval_1'^*})_{\res}]$: the case where $i = r_{\res}$ is a trivial consequence of Equation~\eqref{eq:ret2}. 

Assume $i \ne r_{\res}$: from Equation~\eqref{eq:ret1} we get that $\absfi_1\pofilter \absfi_1'$, which implies that $\absfi_1 = \absfi_1'$. Let $\abswal = \lift{(\absval_2^*)_i}{\absfi_1})$ and $\abswal' = \lift{(\absval_2'^*)_i}{\absfi_1'} = \lift{(\absval_2'^*)_i}{\absfi_1}$. We also know from Equation~\eqref{eq:ret2} that $\absval_2 \poseqp \absval_2'^*$, therefore by applying Proposition~\ref{prop:vallift} we get that $\abswal \povalp \abswal'$.

\item $\rffilter(\lfilter^1 \lfunion \lfilter^2) \pofilter \absfi_1' \afunion \absfi_2'$: from Equation~\eqref{eq:ret1}, Equation~\eqref{eq:ret2} and $\rflocstate{\ell_r}$ definition we know that $\absfi_1 = \rffilter(\lfilter^1) \pofilter \absfi_1'$ and that $\absfi_2 = \rffilter(\lfilter^2) \pofilter \absfi_2'$. By Proposition~\ref{prop:exactfilter} we know that $\rffilter(\lfilter^1 \lfunion \lfilter^2) = \rffilter(\lfilter^1) \afunion \rffilter(\lfilter^2)$. Therefore $\rffilter(\lfilter^1 \lfunion \lfilter^2) = \absfi_1 \afunion \absfi_2$. It directly follows that $\absfi_1 \afunion \absfi_2 \pofilter \absfi_1' \afunion \absfi_2'$.

\item  $\forall \spp,\abslh_1(\spp) \ne \bot \implies \abslh_1(\spp) \polblk \abslh'_1(\spp)$: this is trivially implied by Equation~\eqref{eq:ret2}.
\end{itemize}

\item We are going to show that $\translate{P} \cup \absprog \vdash \delcall$. First observe that  the following rule is included in $\translate{P}$:
\[\absreg{c,m,pc}{(\abswal'_1,\absual_1'^*)}{\absval_1'^*}{\abslh_1'}{\absfi_1'} \implies \absresult{c,m}{(\abswal'_1,\absual_1'^*)}{(\absval_1'^*)_{\res}}{\abslh_1'}{\absfi_1'}\]
Therefore $\Delta \vdash \absresult{c,m}{(\abswal'_1,\absual_1'^*)}{(\absval_1'^*)_{\res}}{\abslh_1'}{\absfi_1'}$.

By well-formedness of $\Sigma$ we know that  $\sign(c',m') = \methsign{(\tau_{i})_{i \le n}}{\tau}{\loc}$, $\stm'_{\pc'} = \invoke{r_o}{m}{(r_{j_i})_{i \le n}}$ and $u^* = (R'(r_{j_i})))_{i \le n}$. Moreover from Equation~\eqref{eq:ret1} we get that $\forall i \le n, (\absual_1^*)_i = \rflval{}((u^*)_i,\lheap,\lfilter^1) \povalp (\absual_1'^*)_i$, and from Equation~\eqref{eq:ret2} we get that $\forall k, (\absval_1^*)_k = \rflval{}((R'(r_k)),\lheap,\lfilter^1) \povalp (\absval_2'^*)_k$. Therefore for all $i \le n$ we have $(\absual_1^*)_i = \rflval{}((u^*)_i,\lheap,\lfilter^1) = \rflval{}((R'(r_{j_i})),\lheap,\lfilter^1) = (\absval_1^*)_{j_i}$, which implies that  $(\absual_1^*)_i \povalp (\absual_1'^*)_{i}$ and $(\absual_1^*)_i \povalp  (\absval_2'^*)_{j_i}$. By Proposition~\ref{prop:nonempty} we get that $(\absval_2'^*)_{j_i} \absmeet (\absual_1'^*)_{i} \ne \bot$.

Similarly from Equation~\eqref{eq:ret1} we get that $\absthread = \rfval{}(\ell_r) = \abswal_1'$, and from  Equation~\eqref{eq:ret2} we get that $\absthread = \rfval{}(\ell_r) = \abswal_2'$, hence we have $\abswal_1' = \abswal'_2$.

From Equation~\eqref{eq:ret2} we get that $\callinv{r_o,c',m'}{\Delta}{\absval_2'^*}{\abslh_2'}$ holds. Therefore there exist $\absloc_o$ and $c''$ such that:
\begin{equation*}
\Big(\overbrace{\left(\absg{\absloc_o} \abspo (\absval_2'^*)_{o} \wedge \absheap(\absloc_o,\absobj{c''}{\_}) \in \Delta\right)}^A \vee \overbrace{\left(\absl{\absloc_o} \abspo (\absval_2'^*)_{o} \wedge \abslh_2'({\absloc_o}) = \absobj{c''}{\_}\right)}^B\Big) \wedge c'' \le c' \wedge c' \in \abslookup(m')
\end{equation*}

Hence one of the following cases holds:
\begin{itemize}
\item If $\absl{\absloc_o} \abspo (\absval_2'^*)_{o} \wedge \abslh_2'({\absloc_o}) = \absobj{c''}{\_}$ then we can apply the following rule:
\[ \absl{\absloc_o} \abspo (\absval_2'^*)_{o} \wedge \abslh_2'({\absloc_o}) = \absobj{c''}{\_} \implies \rlookup{o}{\absval_2'^*}{\abslh_2'}{\absl{\absloc_o}}{\absobj{c''}{\_}}\]

\item If $\absg{\absloc_o} \in (\absval_2'^*)_{o} \wedge \absheap(\absloc_o,\absobj{c''}{\_}) \in \Delta$ then we can apply the rule:
\[ \absg{\absloc_o} \abspo (\absval_2'^*)_{o} \wedge \absheap(\absloc_o,\absobj{c''}{\_}) \implies \rlookup{o}{\absval_2'^*}{\abslh_2'}{\absg{\absloc_o}}{\absobj{c''}{\_}}\]
\end{itemize}

Therefore we can apply the following rule, which is included in $\translate{P}$:
\begin{align*}
&\absreg{c',m',\pc'}{ (\abswal'_2,\absual_2'^*) }{\absval_2'^*}{\abslh_2'}{\absfi_2'} \wedge \rlookup{o}{\absval_2'^*}{\abslh_2'}{\_}{\absobj{c''}{\_}} \wedge c'' \subtype c' \\
&\wedge\absresult{c,m}{(\abswal'_1,\absual_1'^*)}{(\absval_1'^*)_{\res}}{\abslh_1'}{\absfi_1'} \wedge \abswal_1' = \abswal'_2  \wedge \left(\bigwedge\nolimits_{j \le n} (\absval_2'^*)_{i_j} \absmeet (\absual_1'^*)_j \ne \bot\right)\\
\implies\;&\absreg{c',m',pc'+1}{(\abswal'_2,\absual_2')}{\lift{\absval_2'^*}{\absfi_1'}[\res \mapsto ({\absval_1'^*})_{\res}]}{\abslh_1'}{\absfi_1' \afunion \absfi_2'}
\end{align*}
This shows that $\translate{P} \cup \absprog \vdash \delcall$.
\end{enumerate}

\item \textbf{\irule{R-NewObj}}
\[
\inferrule*[width=20em,lab=(R-NewObj)]
{o =  \obj{c'}{(f_{\tau} \mapsto \defvalue_{\tau})^*} \\ 
\ell = \pointer{p}{c,m,\pc} \notin \dom(\heap) \\
\heap' = \heap[\ell \mapsto o] \\ R' = R[r_d \mapsto \ell] }
{\Sigma, \new{r_d}{c'} \Downarrow \Sigma^+[\heap \mapsto \heap', R \mapsto R']}
\]
We know that there exist $\absreg{{c,m,\pc}}{(\absthread,\absual^*)}{\absval^*}{\abslh}{\absfi}$ and  $\absreg{{c,m,\pc}}{(\absthread',\absual'^*)}{\absval'^*}{\abslh'}{\absfi'}$ such that:
\begin{align*}
\rflocstate{\ell_r}(\locstate{c,m,\pc}{\stm^*}{R}{u^*},\lheap,(\lfilter^n)_n) &= \absreg{c,m,\pc}{(\absthread,\absual^*)}{\absval^*}{\abslh}{\absfi}\\
 &\poreg \absreg{{c,m,\pc}}{(\absthread',\absual'^*)}{\absval'^*}{\abslh'}{\absfi'}\in \Delta \numberthis\label{eq:new1}
\end{align*}

By Lemma~\ref{lem:reach} there exists $\absfi_a$ such that $\vdash \cfilter{\absl{\spp}}{\abslh'}{\absfi_a}$ and $\absfi_a$ is the indicator function of the set of reachable elements starting from $\absl{\spp}$ in the points-to graph of $\abslh'$.

\begin{enumerate}
\item  For all $j \ne a$, let $\lheap_j' = \lheap_j$. Let $Reach_a$  the subset of $\lheap$ defined as follows:
\begin{gather*}
Reach_a = \{ (\pointer{p}{\absloc} \mapsto b) \in \lheap\;|\;\absfi_a(\absloc) = 1\}
\end{gather*}

Let $M$ be the partial mapping containing, for all $\absloc$,  exactly one entry $(\pointer{p}{\absloc} \mapsto \bot)$ if there exists a location $\pointer{p'}{\absloc} $ in the domain of $Reach_a$. Besides we assume that the location $\pointer{p}{\absloc}$ is a fresh location.
Let  $\gheap' = \gheap \cup Reach_a$, and  $\lheap'$ be the local heap defined by:
\[\lheap' = \left((\lheap)_{|dom(\lheap) \backslash dom(Reach_a)} \cup M\right)[\ell \mapsto o]\]
Let $\lfilter_a$  be the indicator function of $Reach_a$, $\lfilter'^1 = \lfilter_a \lfunion \lfilter^1$ and $(\lfilter'^j)_{j>1} = (\lfilter^j)_{j > 1}$.

One can check that $\lheapdp$ is a heap decomposition of $\heap'\cdot\sheap'$. Besides we have:
\begin{alignat*}{2}
&\quad dom(\lheap') \backslash \left\{\pointer{p}{\spp}\in dom(\lheap') ~|~ \exists p' , \lfilter_a(\pointer{p'}{\spp}) = 1\right\} \\
=&\quad dom(\lheap') \backslash \left\{\pointer{p}{\spp}\in dom(\lheap') ~|~ \exists p' , \pointer{p'}{\spp} \in dom(Reach_a)\right\} \\
=&\quad dom(\lheap') \backslash \left( dom(M) \cup \{\ell\}\right)\\
\subseteq&\quad dom(\lheap)
\end{alignat*}

Hence by  Proposition~\ref{prop:fhsitget}.5 we know that for all $i \ge 2$, $\fhistget{(\lheap,\flist)}{i} = \fhistget{(\lheap',\flistp)}{i}$. For all $\ell_x \in dom(\callstack)$, we have by well-formedness of $\Sigma$ that $\ell_x \in dom(\heap)$. Therefore since $\ell \not \in dom(\heap)$ we know that $\ell \not \in dom(\callstack)$. Moreover $dom(M)$ is a set of fresh locations, therefore $\left(dom(\lheap') \backslash dom(\lheap)\right) \cap \dom(\callstack_{|>1})  = \emptyset$. 

We know that $dom(\lheap') \backslash dom(\lheap) \subseteq dom(M) \cup \{\ell\}$, and $dom(M)$ is a set of fresh locations so it is easy to check that $dom(M) \cap \{\ell' ~|~\exists j, \lfilter^j(\ell') = 1 \} = \emptyset$. Besides we are going to assume that $\ell$ is not only not appearing in $\Sigma$, but that it is also not appearing in any of the filters, i.e. $\ell \not \in \{\ell' ~|~\exists j, \lfilter^j(\ell') = 1 \}$. Basically this means that $\ell$ is not only a location that was never used yet in the heap $\heap$, but also a location that was never introduced as a ``dummy'' location for proof purposes. We could modify the \irule{R-NewObj} rule, and the configuration decomposition definition, so as to avoid this, but that would make the definitions even lengthier than they are.

Hence we can apply Lemma~\ref{lem:fhist-char}, which shows us that $\fhistp$ is a filter history of $\callstack'$. The fact that $\lheapdhp$ is a local configuration decomposition of $\Sigma'$ follows easily.

\item Let $L_2,\dots,L_n$ be such that $\callstack = \locstate{c,m,\pc}{\stm^*}{R}{u^*} :: L_2 :: \dots :: L_n$. By Proposition~\ref{prop:samefhistget} we know that for all $j \ge 2$,
\[\rfinvoke{\ell_r}(L_j,j,\_,\lheap,(\lfilter^i)_i) = \rfinvoke{\ell_r}(L_j,j,\_,\lheap',(\lfilter'^i)_i)\]
One can then show that the following definitions satisfy the wanted property:
\begin{itemize}
\item $\dcall = \rflocstate{\ell_r}(\locstate{c,m,\pc+1}{\stm^*}{R[r_d \mapsto \ell]}{u^*},\lheap',(\lfilter'^i)_i)$
\item $\dheap = \{\absheap(\absloc,\absblock) ~|~ \heap(\ell')= b \wedge \absloc = \rflab(\ell') \wedge \absblock = \rfblock{}(b) \wedge \ell' \in dom(Reach_a)\}$
\end{itemize}

\item
\begin{itemize}
\item $\delcall = \absreg{{c,m,\pc+1}}{(\absthread',\absual'^*)}{\lift{\absval'^*}{\absfi_a}[d \mapsto \absl{\spp}]}{\lhlift{\abslh'}{\absfi_a}[\spp \mapsto \absobj{c'}{(f \mapsto \adefvalue_{\tau})^*}]}{\absfi_a \afunion \absfi'}$
\item We define $\delheap$ as follows: for all $\spp$, if $ \absfi_a(\spp) = 1 \wedge \abslh'(\spp) \ne \bot$  then $\absheap(\spp,\abslh'(\spp)) \in \delheap$.
\end{itemize}

\item We are going to show that:
\begin{itemize}
\item $ \dcall <:  \delcall :$ by applying Lemma~\ref{lem:tedious2}.2 we get that:
\begin{align*}
&  \rflocstate{\ell_r}(\locstate{c,m,\pc+1}{\stm^*}{R[r_d \mapsto \ell]}{u^*},\lheap',(\lfilter'^n)_n))\\
= &\; \absreg{c,m,\pc+1}{(\absthread,\absual^*)}{\lift{\absval^*}{\absfi_a}[d \mapsto \absl{\spp}]}{\lhlift{\abslh}{\absfi_a}[\spp \mapsto \absobj{c'}{(f \mapsto \adefvalue_{\tau})^*}]}{\absfi_a \afunion \absfi}
\end{align*}

Therefore we just have to prove that:
\begin{align*}
& \absreg{c,m,\pc+1}{(\absthread,\absual^*)}{\lift{\absval^*}{\absfi_a}[d \mapsto \absl{\spp}]}{\overbrace{\lhlift{\abslh}{\absfi_a}[\spp \mapsto \absobj{c'}{(f \mapsto \adefvalue_{\tau})^*}]}^{\abslh_1}}{\absfi_a \afunion \absfi}\numberthis\label{eq:newreg}\\
\poreg&\;\absreg{{c,m,\pc+1}}{(\absthread',\absual'^*)}{\lift{\absval'^*}{\absfi_a}[d \mapsto \absl{\spp}]}{\underbrace{\lhlift{\abslh'}{\absfi_a}[\spp \mapsto \absobj{c'}{(f \mapsto \adefvalue_{\tau})^*}]}_{\abslh_1'}}{\absfi_a \afunion \absfi'}
\end{align*}

From Equation~\eqref{eq:new1} we know that $\absthread = \absthread'$, $\absual^* \poseqp \absual'^*$, $\absval^* \poseqp \absval'^*$, $\absfi \pofilter \absfi'$ and that  $\forall \spp,\abslh(\spp) \ne \bot \implies \abslh(\spp) \polblk \abslh'(\spp)$. To show that Equation~\eqref{eq:newreg} holds we have four conditions to check:
\begin{itemize}
\item We already know that $\absthread = \absthread'$ and $\absual^* \poseqp \absual'^*$.
\item Since $\absval^* \poseqp \absval'^*$, we know by applying Proposition~\ref{prop:vallift} that $\lift{\absval^*}{\absfi_a} \poseqp \lift{\absval'^*}{\absfi_a}$.
\item Since $\absfi \pofilter \absfi'$, it is straightforward to check that $\absfi_a \afunion \absfi \pofilter \absfi_a \afunion \absfi'$.
\item For all $\spp' \ne \spp$,  $\abslh_1({\spp'}) = \lhlift{\abslh}{\absfi_a}({\spp'})$ and $\abslh_1'({\spp'}) = \lhlift{\abslh'}{\absfi_a}(\spp')$. Therefore by applying Proposition~\ref{prop:vallift} we know that $\abslh_1({\spp'}) \polblk \abslh_1'({\spp'})$. Moreover $\abslh_1({\spp}) = \abslh_1'({\spp}) = \absobj{c'}{(f \mapsto \adefvalue_{\tau})^*}$, hence we have $\abslh_1({\spp}) \polblk \abslh_1'({\spp})$.
\end{itemize}

\item $\delheap :> \dheap$: we want to show that:
\begin{equation*}
\delheap >: \{\absheap(\absloc,\absblock) ~|~ \heap(\ell') = b \wedge \absloc = \rflab(\ell') \wedge \absblock = \rfblock{}(b) 
\wedge \ell' \in dom(Reach_a)\}
\end{equation*}

Let $\absheap(\absloc,\absblock)$ be an element of the right set of the above relation. We know that there exists $b,\ell'$ such that $\heap(\ell')= b$,$ \absloc = \rflab(\ell')$,$ \absblock = \rfblock{}(b)$ and $\ell' \in dom(Reach_a)$. Observe that $\ell' \in Reach_{a}$ implies that $\absfi_a(\absloc) = 1$. We have:
\[\rflocstate{\ell_r}(\locstate{c,m,\pc}{\stm^*}{R}{u^*},\lheap,(\lfilter^n)_n) = \absreg{c,m,\pc}{(\absthread,\absual^*)}{\absval^*}{\abslh}{\absfi}\]
Therefore by definitions of $\rflocstate{\ell_r}$ and of $\rflheap{}$ we know that :
\[\abslh = \left\{\left(\spp \mapsto \rflblock{}\left(\lheap(\pointer{p}{\spp}),\lheap\right)\right) \;|\; \pointer{p}{\spp} \in dom(\lheap)\right\}\]

Since $(\ell' \mapsto b) \in \lheap$ we have $\abslh(\absloc) = \rflblock{}(b,\lheap)$. Besides by applying Proposition~\ref{prop:brf} we know that $\rfblock{}(b) \poblk  \rflblock{}(b,\lheap)$. In summary:
\begin{equation}
\absblock = \rfblock{}(b) \poblk \rflblock{}(b,\lheap) = \abslh(\absloc)\label{eq:newblk}
\end{equation}

By Equation~\eqref{eq:new1} we know that $\forall \spp, \abslh(\spp) \ne \bot \implies \abslh(\spp) \polblk \abslh'(\spp)$. Since $(\ell' \mapsto b) \in dom(\heap)$, we know that $\abslh(\absloc) \ne \bot$, which implies that $\abslh(\absloc) \polblk \abslh'(\absloc)$. Putting Equation~\eqref{eq:newblk} together with this we get that $\absblock \poblk \abslh(\absloc) \poblk \abslh'(\absloc)$.

We know that $\absfi_a(\absloc) = 1$. Besides  $\abslh(\absloc) \poblk \abslh'(\absloc)$ and $\abslh(\absloc) \ne \bot$ implies that $\abslh'(\absloc) \ne \bot$. Therefore $\absheap(\absloc,\abslh'(\absloc)) \in \delheap$, which concludes this case.
\end{itemize}

\item
\begin{itemize}
\item $\translate{P} \cup \absprog \vdash \delcall$:  recall that $\absreg{{c,m,\pc}}{(\absthread,\absual^*)}{\absval^*}{\abslh}{\absfi} \poreg \absreg{{c,m,\pc}}{(\absthread',\absual'^*)}{\absval'^*}{\abslh'}{\absfi'} \in \absprog$ and that
\[\delcall = \absreg{{c,m,\pc+1}}{(\absthread',\absual'^*)}{\lift{\absval'^*}{\absfi_a}[d \mapsto \absl{\spp}]}{\lhlift{\abslh'}{\absfi_a}[\spp \mapsto \absobj{c'}{(f \mapsto \adefvalue_{\tau})^*}]}{\absfi_a \afunion \absfi'}\]

We already know that  $\vdash \cfilter{\absl{\spp}}{\abslh'}{\absfi_a}$, hence we can apply the following rule which is included in $\translate{P}$:
\begin{align*}
& \absreg{\spp}{ (\absthread',\absual'^*)} {\absval'^*}{\abslh'}{\absfi'} \wedge \cfilter{\absl{\spp}}{\abslh'}{\absfi_a} \\
&\implies\absreg{{c,m,\pc+1}}{(\absthread',\absual'^*)}{\lift{\absval'^*}{\absfi_a}[d \mapsto \absl{\spp}]}{\lhlift{\abslh'}{\absfi_a}[\spp \mapsto \absobj{c'}{(f \mapsto \adefvalue_{\tau})^*}]}{\absfi_a \afunion \absfi'}
\end{align*}
This concludes this case.

\item $\translate{P} \cup \absprog \vdash \delheap$: we can apply the following rule, which is included in $\translate{P}$:
\begin{equation}
\absreg{\spp}{ (\absthread',\absual'^*)} {\absval'^*}{\abslh'}{\absfi'}  \wedge \cfilter{\absl{\spp}}{\abslh'}{\absfi_a} \implies  \liftlh{\abslh'}{\absfi_a}\label{eq:newrule1}
\end{equation}

$\delheap$ is the set defined by: for all $\spp$, if $ \absfi_a(\spp) = 1 \wedge \abslh'(\spp) \ne \bot$ then $\absheap(\spp,\abslh'(\spp)) \in \delheap$. Let $\spp$ be a program point satisfying those conditions. The following rules is in included in $\translate{P}$:
\[\liftlh{\abslh'}{\absfi_a'^*} \wedge \abslh'(\spp) = \absblock \wedge \absfi_a(\spp) = 1 \implies \absheap(\spp,\absblock)\]
Equation~\eqref{eq:newrule1} plus the above rule yield $\translate{P} \cup \absprog \vdash \absheap(\spp,\abslh'(\spp)) $.
\end{itemize}
\end{enumerate}

\item \textbf{\irule{R-StartThread}}
\[
\inferrule*[lab=(R-StartThread)]
{\ell = \regval{r_i} \\ \heap(\ell) = \threadobj{c'}{(f \mapsto v)^*} \\
\threadstack' = \ell :: \threadstack }
{\Sigma, \startthread{r_i} \Downarrow \Sigma^+[\threadstack \mapsto \threadstack']}
\]

We know that there exist $\absreg{{c,m,\pc}}{(\absthread,\absual^*)}{\absval^*}{\abslh}{\absfi}$ and  $\absreg{{c,m,\pc}}{(\absthread',\absual'^*)}{\absval'^*}{\abslh'}{\absfi'}$ such that:
\begin{equation*}
\rflocstate{\ell_r}(\locstate{c,m,\pc}{\stm^*}{R}{u^*},\lheap,(\lfilter^n)_n) = \absreg{c,m,\pc}{(\absthread,\absual^*)}{\absval^*}{\abslh}{\absfi}
\poreg \absreg{{c,m,\pc}}{(\absthread',\absual'^*)}{\absval'^*}{\abslh'}{\absfi'}\in \Delta \numberthis\label{eq:st1}
\end{equation*}

Let $\ell = \regval{r_i}$, $\heap(\ell) = b = \threadobj{c'}{(f \mapsto w)^*}$. By Assumption~\ref{asm:thread-excpt-sound} we know that with $c' \le \thread$. Let $\lheap$ be the local heap of $\Sigma$. Also let $\absloc = \rflab(\ell)$ and $\absblock = \rfblock(b)$.
\begin{itemize}
\item[Case 1:] $(\ell \mapsto b)\in \gheap$.
\begin{enumerate}
\item Let $\lheapdhp = \lheapdh$. This is trivially a local configuration decomposition of $\Sigma'$.
\item We take:
\begin{itemize}
\item $\dcall = \rflocstate{\ell_r}(\locstate{c,m,\pc+1}{\stm^*}{R}{u^*},\lheap,(\lfilter^n)_n)$
\item $\dtdispatch = \abstdispatch(\absloc,\absblock)$
\end{itemize}
\item We define:
\begin{itemize}
\item $\delcall = \absreg{{c,m,\pc+1}}{(\absthread',\absual'^*)}{\absval'^*}{\abslh'}{\absfi'}$
\item $(\ell \mapsto b)\in \gheap$, therefore $\absheap(\absloc,\absblock) \in X$. Since $X <: \absprog$ we have $\absblock'$  such that $\absheap(\absloc,\absblock') \in \Delta$ and $\absblock \poblk \absblock'$. We then define  $\deltdispatch = \abstdispatch(\absloc,\absblock')$.
\end{itemize}
\item We are going to show that:
\begin{itemize}
\item $ \dcall <: \delcall$. We first check that $\dcall = \absreg{c,m,\pc+1}{(\absthread,\absual^*)}{\absval^*}{\abslh}{\absfi}$. This case then follows directly from Equation~\eqref{eq:st1}.
\item $\dtdispatch <: \deltdispatch$: this case is trivial since $\absblock \poblk \absblock'$.
\end{itemize}
\item  We know by Lemma~\ref{lem:rhs} that  $\rflval{}(\regval{r_i},\lheap) \povalp \absval'_i$. Moreover since $\regval{r_i} = \ell \in dom(\gheap)$ we have $\rflval{}(\regval{r_i},\lheap) = \absg{\absloc}$. We already knew that $\absheap(\absloc,\absblock') \in \Delta$, therefore we have $\absprog \vdash \absg{\absloc} \abspo \absval'_i \wedge \absheap(\absloc,\absblock')$, which implies that $\absprog \vdash \rlookup{i}{\absval'^*}{\abslh'}{\absg{\absloc}}{\absblock'}$. Since $\rfblock(b) = \rfblock(\threadobj{c'}{(f \mapsto w)^*}) \poblk \absblock'$ we know that $\absblock' = \absobj{c'}{(f\mapsto \abswal)}$. Moreover we know that $\translate{P}$ contains the two following rules:
\begin{multline*}
\absreg{\spp}{(\absthread',\absual'^*)}{\absval'^*}{\abslh'}{\absfi'} \wedge \rlookup{i}{\absval'^*}{\abslh'}{\absg{\absloc}}{\absobj{c'}{(f\mapsto \abswal)}} \wedge c' \le \thread\\
\implies \abstdispatch(\absloc,\absthreadobj{c'}{(f \mapsto \abswal)^*})
\end{multline*}
\vspace{-3em}
\begin{multline*}
 \absreg{\spp}{(\absthread',\absual'^*)}{\absval'^*}{\abslh'}{\absfi'} \wedge\rlookup{i}{\absval'^*}{\abslh'}{\absg{\absloc}}{\absobj{c'}{(f\mapsto \abswal)}} \wedge c' \le \thread\\
\implies \absreg{\apcn}{(\absthread',\absual'^*)}{\absval'^*}{\abslh'}{\absfi'}
\end{multline*}
By applying them we get that $\translate{P} \cup \absprog \vdash \delcall$ and  $\translate{P} \cup \absprog \vdash \deltdispatch$, which concludes this case.
\end{enumerate}

\item[Case 2:] $\ell \in dom(\lheap)$
\begin{enumerate}
\item By Lemma~\ref{lem:reach} there exists $\absfi_a$ such that $\vdash \cfilter{\absl{\absloc}}{\abslh'}{\absfi_a}$ and $\absfi_a$ is the indicator function of the set of reachable elements starting from $\absl{\absloc}$ in the points-to graph of $\abslh'$. For all $j \ne a$, let $\lheap_j' = \lheap_j$, and let $Reach_a$  be the subset of $\lheap$ defined as follows:
\begin{gather*}
Reach_a = \{ (\pointer{p}{\absloc} \mapsto b) \in \lheap\;|\;\absfi_a(\absloc) = 1\}
\end{gather*}
Let $M$ be the partial mapping containing, for all $\absloc'$,  exactly one entry $(\pointer{p}{\absloc'} \mapsto \bot)$ if there exists a location $\pointer{p'}{\absloc'} $ in the domain of $Reach_a$. Besides we assume that the location  $\pointer{p}{\absloc'}$ is a fresh location.

Let $\lheap' = \left((\lheap)_{|dom(\lheap) \backslash dom(Reach_a)} \cup M\right)$ and $\gheap' = \gheap \cup Reach_a$, and we define $\lfilter_a$ to be the indicator function of $Reach_a$, $\lfilter'^1 = \lfilter_a \lfunion \lfilter^1$ and $(\lfilter'^j)_{j>1} = (\lfilter^j)_{j > 1}$ .

One can check that $\lheapdp$ is a heap decomposition of $\heap\cdot\sheap$. As we did in \irule{R-MoveFld}, we can apply By  Proposition~\ref{prop:fhsitget}.5 to get that for all $i \ge 2$, $\fhistget{(\lheap,\flist)}{i} = \fhistget{(\lheap',\flistp)}{i}$.  $dom(M)$ is a set of fresh locations, therefore we can apply Lemma~\ref{lem:fhist-char}, which shows us that $\fhistp$ is a filter history of $\callstack'$. The fact that $\lheapdhp$ is a local configuration decomposition of $\Sigma'$ follows easily.

\item  Let $L_2,\dots,L_n$ be such that $\callstack = \locstate{c,m,\pc}{\stm^*}{R]}{u^*} :: L_2 :: \dots :: L_n$. By Proposition~\ref{prop:samefhistget} we know that for all $j \ge 2$:
\[\rfinvoke{\ell_r}(L_j,j,\_,\lheap,(\lfilter^i)_i) = \rfinvoke{\ell_r}(L_j,j,\_,\lheap',(\lfilter'^i)_i)\]
One can then show that the following sets satisfy the wanted property:
\begin{itemize}
\item $\dcall = \rflocstate{\ell_r}(\locstate{c,m,\pc+1}{\stm^*}{R}{u^*},\lheap',(\lfilter'^n)_n))$
\item $\dheap = \{\absheap(\absloc'',\absblock'') ~|~ \heap(\ell'') = b'' \wedge \absloc'' = \rflab(\ell'') \wedge \absblock'' = \rfblock{}(b'') \wedge \ell'' \in dom(Reach_a)\}$
\item $\dtdispatch = \abstdispatch(\absloc,\absblock)$
\end{itemize}

\item We define:
\begin{itemize}
\item $\delcall = \absreg{{c,m,\pc+1}}{(\absthread',\absual'^*)}{\lift{\absval'^*}{\absfi_a}}{\lhlift{\abslh'}{\absfi_a}}{\absfi_a \afunion \absfi'}$
\item We define $\delheap$ as follows: for all $\spp$, if $ \absfi_a(\spp) = 1 \wedge \abslh'(\spp) \ne \bot$  then $\absheap(\spp,\abslh'(\spp)) \in \delheap$.
\item $\ell \in dom(\lheap)$, therefore we know that $\abslh(\absloc) = \rflblock{}(b,\lheap) \ne \bot$. From \eqref{eq:st1} and the definition of $\poreg$  we get that $\abslh(\absloc) \polblk \abslh'(\absloc)$. We define $\deltdispatch = \abstdispatch(\absloc,\abslh'(\absloc))$.
\end{itemize}

\item We are going to show that:
\begin{itemize}
\item $ \dcall <:  \delcall$. By applying Lemma~\ref{lem:tedious2}.1 we get that:
\begin{equation*}
\rflocstate{\ell_r}(\locstate{c,m,\pc+1}{\stm^*}{R}{u^*},\lheap',(\lfilter'^n)_n))
=\absreg{{c,m,\pc+1}}{(\absthread,\absual^*)}{\lift{\absval^*}{\absfi_a}}{\lhlift{\abslh}{\absfi_a}}{\absfi_a \afunion \absfi}
\end{equation*}

Therefore we just have to prove that:
\begin{align*}
\numberthis\label{eq:st4}
& \absreg{c,m,\pc+1}{(\absthread,\absual^*)}{\lift{\absval^*}{\absfi_a}}{\lhlift{\abslh}{\absfi_a}}{\absfi_a \afunion \absfi}\\
\poreg&\;\absreg{{c,m,\pc+1}}{(\absthread',\absual'^*)}{\lift{\absval'^*}{\absfi_a}}{\lhlift{\abslh'}{\absfi_a}}{\absfi_a \afunion \absfi'}
\end{align*}

From Equation~\eqref{eq:st1} we know that $\absthread = \absthread'$, $\absual^* \poseqp \absual'^*$, $\absval^* \poseqp \absval'^*$, $\absfi \pofilter \absfi'$ and that  $\forall \spp,\abslh(\spp) \ne \bot \implies \abslh(\spp) \polblk \abslh'(\spp)$. To show that Equation~\eqref{eq:st4} holds we have four conditions to check:
\begin{itemize}
\item We already know that $\absthread = \absthread'$ and $\absual^* \poseqp \absual'^*$.
\item Since $\absval^* \poseqp \absval'^*$, we know by applying Proposition~\ref{prop:vallift} that $\lift{\absval^*}{\absfi_a} \poseqp \lift{\absval'^*}{\absfi_a}$.
\item Since $\absfi \pofilter \absfi'$, it is straightforward to check that $\absfi_a \afunion \absfi \pofilter \absfi_a \afunion \absfi'$.
\item For all $\spp$, by applying Proposition~\ref{prop:vallift} we know that $\lhlift{\abslh}{\absfi_a}({\spp}) \polblk \lhlift{\abslh'}{\absfi_a}({\spp})$.
\end{itemize}

\item $\dheap <: \delheap$: we want to show that
\begin{equation*}
\delheap >: \{\absheap(\absloc'',\absblock'') ~|~ \heap(\ell'')= b'' \wedge \absloc'' = \rflab(\ell'') \wedge \absblock'' = \rfblock{}(b'') \wedge \ell'' \in dom(Reach_a)\}
\end{equation*}

Let $\absheap(\absloc,\absblock)$ be an element of the right set of the above relation. We know that there exists $b'',\ell''$ such that $\heap(\ell'')= b''$,$ \absloc'' = \rflab(\ell'')$,$ \absblock'' = \rfblock(b'')$ and $\ell'' \in dom(Reach_a)$. Besides $\ell'' \in Reach_{a}$ implies that $\absfi_a({\absloc''}) = 1$. We have:
\[\rflocstate{\ell_r}(\locstate{c,m,\pc}{\stm^*}{R}{u^*},\lheap,(\lfilter^n)_n) = \absreg{c,m,\pc}{(\absthread,\absual^*)}{\absval^*}{\abslh}{\absfi}\]
Therefore by definitions of $\rflocstate{\ell_r}$ and of $\rflheap{}$ we know that :
\[\abslh = \left\{\left(\spp \mapsto \rflblock{}\left(\lheap(\pointer{p}{\spp}),\lheap\right)\right) \;|\; \pointer{p}{\spp} \in dom(\lheap)\right\}\]

Since $(\ell'' \mapsto b'') \in \lheap$ we have $\abslh_{\absloc''} = \rflblock{}(b'',\lheap)$. Besides by applying Proposition~\ref{prop:brf} we know that $\rfblock(b'') \poblk  \rflblock{}(b'',\lheap)$. In summary:
\begin{equation}
\absblock'' = \rfblock(b'') \poblk \rflblock{}(b'',\lheap) = \abslh({\absloc''})\label{eq:stblk}
\end{equation}

From Equation~\eqref{eq:st1} we get that $\forall \spp, \abslh(\spp) \ne \bot \implies \abslh(\spp) \polblk \abslh'(\spp)$. Since $(\ell'' \mapsto b'') \in \heap$, we know that $\abslh({\absloc''}) \ne \bot$, which implies that $\abslh({\absloc''}) \polblk \abslh'({\absloc''})$. Putting  Equation~\eqref{eq:stblk} together with this we get that $\absblock'' \poblk \abslh({\absloc''}) \poblk \abslh'({\absloc''})$.

We know that $\absfi_a(\absloc'') = 1$. Besides  $\abslh({\absloc''}) \poblk \abslh'({\absloc''})$ and $\abslh({\absloc''}) \ne \bot$ implies that $\abslh'({\absloc''}) \ne \bot$. Therefore $\absheap(\absloc'',\abslh'({\absloc''})) \in \delheap$, which concludes this case.

\item $\ell \in dom(\lheap)$, therefore $\abslh(\absloc) = \rflblock{}(b,\lheap) \ne \bot$. Hence by Equation~\eqref{eq:st1} we know that $\abslh(\absloc) \polblk \abslh'(\absloc)$. By Proposition~\ref{prop:brf} we know that $\absblock = \rfblock(b) \poblk \rflblock{}(b,\lheap) = \abslh(\absloc)$, and by Proposition~\ref{prop:coarseblk} we get that  $\abslh(\absloc) \poblk \abslh'(\absloc)$. Therefore $\absblock \poblk \abslh'(\absloc)$, which shows that $\dtdispatch <: \deltdispatch$. 
\end{itemize}

\item We are going to show that:
\begin{itemize}

\item $\translate{P} \cup \absprog \vdash \delcall$:  recall that $\absreg{{c,m,\pc}}{(\absthread',\absual'^*)}{\absval'^*}{\abslh'}{\absfi'} \in \absprog$ and that:
\[\delcall = \absreg{{c,m,\pc+1}}{(\absthread',\absual'^*)}{\lift{\absval'^*}{\absfi_a}}{\lhlift{\abslh'}{\absfi_a}}{\absfi_a \afunion \absfi'}\]

We know by Lemma~\ref{lem:rhs} that  $\rflval{}(\regval{r_i},\lheap) \povalp \absval'_i$. Moreover since $\regval{r_i} = \ell \in dom(\lheap)$ we have $\absl{\absloc} = \rflval{}(\regval{r_i},\lheap)$. We saw previously that $\rfblock(b) \poblk \abslh'(\absloc)$, and since $b = \threadobj{c'}{(f \mapsto w)^*}$, we have $\abslh'(\absloc) = \absobj{c'}{(f \mapsto \abswal)^*}$. Hence we have the following abstract heap look-up fact:
\[\vdash \rlookup{i}{\absval'^*}{\abslh'}{\absl{\absloc}}{\absobj{c'}{(f \mapsto \abswal)^*}}\]

Finally $c' \le \thread$ and  $\vdash \cfilter{\absl{\absloc}}{\abslh'}{\absfi_a}$, which allows us to apply the following rule, which is included in $\translate{P}$:
\begin{multline*}
\absreg{c,m,\pc}{(\absthread',\absual'^*)}{\absval'^*}{\abslh'}{\absfi'} \wedge \rlookup{i}{\absval'^*}{\abslh'}{\absl{\absloc}}{\absobj{c'}{(f \mapsto \abswal)^*}} \wedge \cfilter{\absl{\absloc}}{\abslh'}{\absfi_a}\\
 \wedge c' \le \thread \implies \absreg{c,m,\pc + 1}{(\absthread',\absual'^*)}{\lift{\absval'^*}{\absfi_a}}{\lhlift{\abslh'}{\absfi_a}}{\absfi' \afunion \absfi_a}
\end{multline*}
This concludes this case.

\item $\translate{P} \cup \absprog \vdash \delheap$: We can apply the following rule, which is in $\translate{P}$:
\begin{multline*}
\absreg{c,m,\pc}{(\absthread',\absual'^*)}{\absval'^*}{\abslh'}{\absfi'} \wedge\rlookup{i}{\absval'^*}{\abslh'}{\absl{\absloc}}{\absobj{c'}{(f \mapsto \abswal)^*}} \wedge \cfilter{\absl{\absloc}}{\abslh'}{\absfi_a}\\
 \wedge c' \le \thread \implies \liftlh{\abslh'}{\absfi_a} \numberthis\label{eq:strule1}
\end{multline*}

$\delheap$ is the set defined by: for all $\spp$, if $ \absfi_a(\spp) = 1 \wedge \abslh'(\spp) \ne \bot$ then $\absheap(\spp,\abslh'(\spp)) \in \delheap$. Let $\spp$ satisfying those conditions. $\translate{P}$ contains the following rule:
\[\liftlh{\abslh'}{\absfi_a} \wedge \abslh'(\spp) = \absblock'' \wedge \absfi_a(\spp) = 1 \implies \absheap(\spp,\absblock'')\]
Rule Equation~\eqref{eq:strule1} plus the above rule yield $\translate{P} \cup \absprog \vdash \absheap(\spp,\abslh'(\spp)) $.

\item $\translate{P} \cup \absprog \vdash \deltdispatch$: directly obtained by applying:
\begin{multline*}
\absreg{c,m,\pc}{(\absthread',\absual'^*)}{\absval'^*}{\abslh'}{\absfi'} \wedge\rlookup{i}{\absval'^*}{\abslh'}{\absl{\absloc}}{\absobj{c'}{(f \mapsto \abswal)^*}} \wedge c' \le \thread \\
\implies \abstdispatch(\absloc,\absobj{c'}{(f \mapsto \abswal)^*})
\end{multline*}
\end{itemize}
\end{enumerate}
\end{itemize}

\item \textbf{\irule{R-InterruptWait}}
\[
\inferrule*[width=35em,lab=(R-InterruptWait)]
{\heap(\ell_r) = \obj{\absloc_r}{(f_r \mapsto u_r)^*,\interrupted \mapsto \true}\\
\pointer{p}{c,m,\pc} \not\in \dom(\heap)\\
o = \obj{c_r}{(f_r \mapsto u_r)^*,\interrupted \mapsto \false}\\
\callstack = \waiting{\_}{\_}:: \callstack_0\\
o_e = \obj{\interruptedexception}{}}
{\Sigma \Downarrow \Sigma[\callstack \mapsto \abnormal{\callstack_0[r_\excpt \mapsto \ell_e]}, \heap \mapsto \heap[\pointer{p}{c,m,\pc} \mapsto o_e,\ell_r \mapsto o]]}
\]

\begin{enumerate}
\item Let $\spp = c,m,\pc$. Let $\gheap' = \gheap[\ell_r \mapsto o] \cup \{(\pointer{p}{c,m,\pc} \mapsto o_e)\}$ and $((\lheap'_i)_{i \le n}, \lheap',(\lfilter'^j)_j) = ((\lheap_i)_{i \le n}, \lheap,(\lfilter^j)_j)$. Since $\lheapdh$ is a local configuration decomposition of $\Sigma$, we know that $\ell_r \in dom(\gheap)$. Besides $\pointer{p}{c,m,\pc}$ is a fresh location, hence it is quite easy to check that $\lheapdhp$ is a  local configuration decomposition  of $\Sigma'$, and that $\forall i, \lheap_i \ne \lheap \implies \lheap_i = \lheap_i'$.

\item Let $\callstack = L_1 ::\ldots::L_n$. By  Proposition~\ref{prop:fhsitget}.4 we know that for all $i \ge 2$, $\fhistget{(\lheap,\flist)}{i} = \fhistget{(\lheap',\flistp)}{i}$. Therefore by Proposition~\ref{prop:samefhistget} we know that for all $j \ge 2$:
\[\rfinvoke{\ell_r}(L_j,j,\_,\lheap,(\lfilter^i)_i) = \rfinvoke{\ell_r}(L_j,j,\_,\lheap',(\lfilter'^i)_i)\]
One can then show that the following definitions satisfy the wanted property:
\begin{itemize}
\item $\dcall = \rfalocstate{\ell_r}(\locstate{c,m,\pc}{\stm^*}{R[r_\excpt \mapsto \pointer{p}{c,m,\pc}]}{u^*},\lheap',(\lfilter'^n)_n))$
\item $\dheap = \{\heap(\rflab(\ell_r),\rfblock{}(o))\}\cup\{\heap(\rflab(\pointer{p}{c,m,\pc}),\rfblock{}(o_e))\}$
\end{itemize}

\item We know that there exist $\absreg{{c,m,\pc}}{(\absthread,\absual^*)}{\absval^*}{\abslh}{\absfi}$ and  $\absreg{{c,m,\pc}}{(\absthread',\absual'^*)}{\absval'^*}{\abslh'}{\absfi'}$ such that:
\begin{equation*}
\rflocstate{\ell_r}(\locstate{c,m,\pc}{\stm^*}{R}{u^*},\lheap,(\lfilter^n)_n) = \absreg{c,m,\pc}{(\absthread,\absual^*)}{\absval^*}{\abslh}{\absfi}
\poreg \absreg{{c,m,\pc}}{(\absthread',\absual'^*)}{\absval'^*}{\abslh'}{\absfi'}\in \Delta \numberthis\label{eq:iw1}
\end{equation*}
We define:
\begin{itemize}
\item $\delcall = \absabnormal{c,m,\pc}{(\absthread',\absual'^*)}{\absval'^*[\excpt \mapsto \spp]}{\abslh'}{\absfi'}$ 
\item Since $X <: \absprog$ and $\ell_r \in dom(\gheap)$ we know that there exists $\absheap({\absloc_r},\absblock) \in \absprog$ such that $\heap(\ell_r) \poblk \absblock$. This implies that $\absblock = \absobj{c_r}{(f_r \mapsto \absual_r)^*,\interrupted \mapsto \absval_i}$ and that $(\rfval{}(u_r))^* \poseq \absval^*_r$ and $\rfval{}(\true) \poval \absval_i$. We define :
\[\delheap = \{\absheap(\absloc_r,\absobj{c_r}{(f_r \mapsto \absual_r)^*,\interrupted \mapsto \widehat{\false}})\}\cup\{ \absheap(\spp;\absobj{\interruptedexception}{})\}\]
\end{itemize}
\item Show that:
\begin{itemize}
\item $ \dcall <: \delcall$: one can check that:
\begin{equation}
\rfalocstate{\ell_r}(\locstate{c,m,\pc}{\stm^*}{R[r_\excpt \mapsto \pointer{p}{c,m,\pc}]}{u^*},\lheap',(\lfilter'^n)_n)) = \absabnormal{c,m,\pc}{(\absthread,\absual^*)}{\absval^*[\excpt \mapsto \spp]}{\abslh}{\absfi}\label{eq:iw2}
\end{equation}
From Equation~\eqref{eq:iw1} we know that:
\[\absreg{c,m,\pc}{(\absthread,\absual^*)}{\absval^*}{\abslh}{\absfi} \poreg \absreg{{c,m,\pc}}{(\absthread',\absual'^*)}{\absval'^*}{\abslh'}{\absfi'}\]
This implies that:
\[\absreg{c,m,\pc}{(\absthread,\absual^*)}{\absval^*[\excpt \mapsto \spp]}{\abslh}{\absfi} \poreg \absreg{{c,m,\pc}}{(\absthread',\absual'^*)}{\absval'^*[\excpt \mapsto \spp]}{\abslh'}{\absfi'}\]
Hence by definition of $\poabnormal$ we have:
\[\absabnormal{c,m,\pc}{(\absthread,\absual^*)}{\absval^*[\excpt \mapsto \spp]}{\abslh}{\absfi} \poabnormal \absabnormal{{c,m,\pc}}{(\absthread',\absual'^*)}{\absval'^*[\excpt \mapsto \spp]}{\abslh'}{\absfi'}\]
Equation~\eqref{eq:iw2} and the above relation shows that $\dcall <: \delcall$.

\item $\dheap <: \delheap$: we know that $(\rfval{}(u_r))^* \poseq \absual^*_r$. Besides $\rfval{}(\false) \poval \widehat{\false}$, therefore we have $\rfblock{}(o) \poblk \absobj{c_r}{(f_r \mapsto \absual_r)^*,\interrupted \mapsto \widehat{\false}})$, which in turn implies that :
\[\{\heap(\rflab(\ell_r),\rfblock{}(o))\} <: \{\absheap(\absloc_r,\absobj{c_r}{(f_r \mapsto \absual_r)^*}\} \subseteq \delheap\]
The fact that $\{\heap(\rflab(\ell_r),\rfblock{}(o_e))\} <: \{ \absheap(\spp;\absobj{\interruptedexception}{}) \} \subseteq\delheap$ is trivial.

\end{itemize}
\item By definition of $\rflocstate{}$, we get from Equation~\eqref{eq:iw1} that $\absthread = \rfval{}(\ell_r) = \absg{\absloc_r}$, and that $\absthread = \absthread'$. Besides we know that $\absheap({\absloc_r},\absblock) \in \absprog$, where $\absblock = \absobj{c_r}{(f_r \mapsto \absual_r)^*,\interrupted \mapsto \absval_i}$ and $\rfval{}(\true) = \widehat{\true} \poval \absval_i$, which implies that $\widehat{\true} \abspo \absval_i$. Moreover Equation~\eqref{eq:iw1} gives us that $\absreg{{c,m,\pc}}{(\absthread',\absual'^*)}{\absval'^*}{\abslh'}{\absfi'}\in \Delta$, therefore we have :
\begin{equation}
\absprog \vdash \absreg{{c,m,\pc}}{(\absg{\absloc_r},\absual'^*)}{\absval'^*}{\abslh'}{\absfi'}  \wedge \absheap({\absloc_r},\absobj{c_r}{(f_r \mapsto \absual_r)^*,\interrupted \mapsto \absval_i}) \wedge \widehat{\true} \abspo \absval_i \label{eq:iwp}
\end{equation}

Since $\Sigma$ is well-formed, and since $L_1 = \waiting{\_}{\_}$ we know that $\stm_\pc = \startwait{\_}\,$. Therefore $\translate{P}$ contains the following rules:
\begin{align*}
\absreg{\spp}{(\absg{\absloc_r},\absual'^*)}{\absval'^*}{\abslh'}{\absfi'}\wedge \absheap(\absloc_r,\absobj{c_r}{(f_r \mapsto \absual_r)^*,\interrupted \mapsto \absval_i}) \wedge \widehat{\true} \abspo \absval_i \\
\implies \absabnormal{\spp}{(\absg{\absloc_r},\absual'^*)}{\absval'^*[\excpt \mapsto \spp]}{\abslh'}{\absfi'}\numberthis\label{eq:iw3}\\
 \absreg{\spp}{(\absg{\absloc_r},\absual'^*)}{\absval'^*}{\abslh'}{\absfi'} \wedge \absheap(\absloc_r,\absobj{c_r}{(f_r \mapsto \absual_r)^*,\interrupted \mapsto \absval_i}) \wedge \widehat{\true} \abspo \absval_i \\
\implies \absheap({\absloc_r},\absthreadobj{c'}{(f \mapsto \absual)^*,\interrupted \mapsto \widehat{\false}})\numberthis\label{eq:iw4}\\
 \absreg{\spp}{(\absg{\absloc_r},\absual'^*)}{\absval'^*}{\abslh'}{\absfi'} \wedge \absheap(\absloc_r,\absobj{c_r}{(f_r \mapsto \absual_r)^*,\interrupted \mapsto \absval_i}) \wedge \widehat{\true} \abspo \absval_i \\
\implies \absheap(\spp;\absobj{\interruptedexception}{})\numberthis\label{eq:iw5}
\end{align*}

\begin{itemize}
\item $\translate{P} \cup \absprog \vdash \delcall$: this is trivially implied by Equation~\eqref{eq:iwp} and Equation~\eqref{eq:iw3}.
\item $\translate{P} \cup \absprog \vdash \delheap$: Equation~\eqref{eq:iwp} and Equation~\eqref{eq:iw4} gives us that $\translate{P} \cup \absprog \vdash  \absheap({\absloc_r},\absthreadobj{c'}{(f \mapsto \absual)^*,\interrupted \mapsto \widehat{\false}})$, and abstract fact $\absheap(\spp;\absobj{\interruptedexception}{})$ is obtained by Equation~\eqref{eq:iw5}.
\end{itemize}
\end{enumerate}

\item \textbf{\irule{R-Caught}}
\[
\inferrule*[width=35em,lab=(R-Caught)]
{\ell = \regval{r_\excpt} \\ 
\heap(\ell) = \obj{c'}{(f \mapsto v)^*}\\
\excpttable{c,m,\pc}{c'} = \pc'\\
\callstack' = \locstate{c,m,\pc'}{\_}{R}{\_} :: \callstack_0}
{\Sigma \Downarrow \Sigma[\callstack \mapsto \callstack']}
\]
Here call-stack is abnormal and of the form  $\callstack = \abnormal{\locstate{c,m,\pc}{\stm^*}{R}{u^*} :: \callstack_0}$.

\begin{enumerate}
\item We take $\lheapdhp = \lheapdh$. It is trivially a local configuration decomposition  of $\Sigma'$, and $\forall i, \lheap_i \ne \lheap \implies \lheap_i = \lheap_i'$

\item Let $L_1 ::\ldots::L_n = \abnormal{\locstate{c,m,\pc}{\stm^*}{R}{u^*} :: \callstack_0}$. By  Proposition~\ref{prop:fhsitget}.4 we know that for all $i \ge 2$, $\fhistget{(\lheap,\flist)}{i} = \fhistget{(\lheap',\flistp)}{i}$. Therefore by Proposition~\ref{prop:samefhistget} we know that for all $j \ge 2$:
\[\rfinvoke{\ell_r}(L_j,j,\_,\lheap,(\lfilter^i)_i) = \rfinvoke{\ell_r}(L_j,j,\_,\lheap',(\lfilter'^i)_i)\]
One can then show that $\dcall = \rfalocstate{\ell_r}(\locstate{c,m,\pc'}{\stm^*}{R}{u^*},\lheap',(\lfilter'^n)_n))$ satisfies the wanted property.

\item We know that there exist $\absabnormal{{c,m,\pc}}{(\absthread,\absual^*)}{\absval^*}{\abslh}{\absfi}$ and  $\absabnormal{{c,m,\pc}}{(\absthread',\absual'^*)}{\absval'^*}{\abslh'}{\absfi'}$ such that:
\begin{equation*}
\rfalocstate{\ell_r}(\locstate{c,m,\pc}{\stm^*}{R}{u^*},\lheap,(\lfilter^n)_n) = \absabnormal{c,m,\pc}{(\absthread,\absual^*)}{\absval^*}{\abslh}{\absfi}
\poabnormal \absabnormal{{c,m,\pc}}{(\absthread',\absual'^*)}{\absval'^*}{\abslh'}{\absfi'}\in \Delta \numberthis\label{eq:c1}
\end{equation*}
We take $\delcall =  \absreg{{c,m,\pc'}}{(\absthread',\absual'^*)}{\absval'^*}{\abslh'}{\absfi'}$.

\item $ \dcall <:  \delcall$: this is a trivial consequence of Equation~\eqref{eq:c1}.

\item We want to show that $\translate{P} \cup \absprog \vdash \delcall$. First recall that $\excpttable{c,m,\pc}{c'} = \pc'$, hence $c' \le \throwable$ by Assumption~\ref{asm:excpt-table}. We know by Lemma~\ref{lem:rhs} that  $\rflval{}(\ell,\lheap) \povalp \absval'_\excpt$. Let $\absloc = \rflab(\ell)$.
\begin{itemize}
\item If $\ell \in dom(\gheap)$ then we have $\rflval{}(\ell,\lheap) = \absg{\absloc}$. Moreover since $X <: \absprog$ we know that there exists $\absheap(\absloc,\absobj{c'}{(f\mapsto \abswal)^*}) \in \absprog$. Therefore we have:
\[\absprog\vdash  \rlookup{\excpt}{\absval'^*}{\abslh'}{\absg{\absloc}}{\absobj{c'}{(f\mapsto \abswal)^*}} \wedge c' \le \throwable\]

\item If $\ell \in dom(\lheap)$ then we have $\rflval{}(\regval{r_\excpt},\lheap) = \absl{\absloc}$. Since $\ell \in dom(\lheap)$, we know that $\abslh(\absloc) = \rflblock{}(\heap(\ell),\lheap) \ne \bot$. Therefore from Equation~\eqref{eq:c1} we get that $\abslh(\absloc) \polblk \abslh'(\absloc)$, which in turns implies that $\abslh'(\absloc) = \absobj{c'}{(f\mapsto \abswal)^*}$. Hence we have:
\[\absprog\vdash \rlookup{\excpt}{\absval'^*}{\abslh'}{\absl{\absloc}}{\absobj{c'}{(f\mapsto \abswal)^*}}\wedge c' \le \throwable\]
\end{itemize}
In both case we can apply the rule below, which is included in $\translate{P}$:
\begin{multline*}
\absabnormal{c,m,\pc}{\absual'^*}{\absval'^*}{\abslh'}{\absfi'} \wedge \rlookup{\excpt}{\absval'^*}{\abslh'}{\_}{\absobj{c'}{(f\mapsto \abswal)^*}} \wedge c' \le \throwable\\
\implies \absreg{c,m,\pc'}{\absual'^*}{\absval'^*}{\abslh'}{\absfi'}
\end{multline*}
This concludes this case.
\end{enumerate}

\item \textbf{\irule{R-UnCaught}}
\[
\inferrule*[width=25em,lab=(R-UnCaught)]
{\ell = \regval{r_\excpt} \\
\heap(\ell) = \obj{c_e}{(f \mapsto v)^*}\\
\excpttable{c,m,\pc}{c_e} = \bot}
{\Sigma \Downarrow \Sigma[\callstack \mapsto \abnormal{\callstack_0[r_\excpt \mapsto \ell]}]}
\]

Here the call-stack is abnormal  $\callstack = \abnormal{\locstate{c,m,\pc}{\stm^*}{R}{u^*} :: \callstack_0}$. If $\callstack_0$ is the empty list, then this case is easy. Hence we assume that :
\begin{alignat*}{2}
\callstack \quad&=\quad&& \abnormal{\locstate{c,m,\pc}{\stm^*}{R}{v^*} :: \locstate{c',m',\pc'}{\stm'^*}{R'}{u'^*} :: \callstack_1}\\
\callstack' \quad&=\quad&& \abnormal{\locstate{c',m',\pc'}{\stm'^*}{R'[r_{\excpt} \mapsto \ell]}{u'^*} :: \callstack_1}
\end{alignat*}

\begin{enumerate}
\item Let $\lheapdp = \lheapd$ and $(\lfilter'^j)_j = (\lfilter_1 \lfunion \lfilter_2 ) :: (\lfilter_i)_{i > 2}$.

The proof that $\lheapdhp$ is a local configuration decomposition of $\Sigma'$ is the same than in the \irule{R-Return} case.

\item By Proposition~\ref{prop:samefhistget} we get for all $j \ge 1$:
\[\rfinvoke{\ell_r}(\callstack_j,j,\_,\lheap,(\lfilter^i)_i) = \rfinvoke{\ell_r}(\callstack_j,j,\_,\lheap',(\lfilter'^i)_i)\]

One can then check that:
\[\dcall = \rfalocstate{\ell_r}(\locstate{c',m',\pc'}{\stm'^*}{R'[r_{\excpt} \mapsto \ell]}{u'^*},\lheap',(\lfilter'^j)_j)\]

\item We know that:
\begin{align*}
 \rfalocstate{\ell_r}(\locstate{c,m,\pc}{\stm^*}{R}{u^*},\lheap,(\lfilter^j)_j) \quad=\quad& \absabnormal{c,m,pc}{(\absthread,\absual_1^*)}{\absval_1^*}{\abslh_1}{\absfi_1}\numberthis\label{eq:ue1}\\
\quad\poreg\quad& \absabnormal{c,m,pc}{(\abswal'_1,\absual_1'^*)}{\absval_1'^*}{\abslh_1'}{\absfi_1'} \in \Delta\\
\phantom{m}\\
 \rfinvoke{\ell_r}(\locstate{c',m',\pc'}{\stm'^*}{R'}{u'^*},2,c,\lheap,(\lfilter^j)_j) \quad=\quad& \absinvoke{c',m',pc'}{c}{(\absthread,\absual_2^*)}{\absval_2^*}{\absfi_2}\numberthis\label{eq:ue2}\\
\quad\poinvoke{\Delta}\quad& \absreg{c',m',pc'}{(\abswal'_2,\absual_2'^*)}{\absval_2'^*}{\abslh_2'}{\absfi_2'} \in \Delta
\end{align*}

Let $\delcall = \absabnormal{c',m',pc'}{(\abswal'_2,\absual_2'^*)}{\lift{\absval_2'^*}{\absfi_1'}[\excpt \mapsto ({\absval'^*_1})_{\excpt}]}{\abslh_1'}{\absfi_1' \afunion \absfi_2'}$.

\item The proof that  $ \dcall <: \Delta \cup \delcall$ is exactly the same than in the \irule{R-Return} case.

\item We are going to show that $\translate{P} \cup \absprog \vdash \delcall$. Since $\excpttable{c,m,\pc}{c_e} = \bot$ we know that $c_e \le \throwable$ by Assumption~\ref{asm:excpt-table}. Therefore we have the following rule in $\translate{P}$:
\begin{multline*}
\absabnormal{c,m,\pc}{(\abswal_1',\absual'^*_1)}{\absval'^*_1}{\abslh'_1}{\absfi'_1}  \wedge \rlookup{\excpt}{\absval'^*_1}{\abslh'_1}{\_}{\absobj{c_e}{\_}}\wedge c_e \le \throwable\\
\implies \absuncaught{c,m}{(\abswal_1',\absual'^*_1)}{(\absval'^*_1)_\excpt}{\abslh'_1}{\absfi'_1}
\end{multline*}

As it was done in \irule{R-Caught}, one can show that:
\[\absprog \vdash \rlookup{\excpt}{\absval'^*_1}{\abslh'_1}{\_}{\absobj{c_e}{\_}}\wedge c_e \le \throwable\]

Therefore $\Delta\vdash \absuncaught{c,m}{(\abswal_1',\absual'^*_1)}{\abslab}{\abslh'_1}{\absfi'_1}$.

By well-formedness of $\Sigma$ we know that  $\sign(c',m') = \methsign{(\tau_{i})_{i \le n}}{\tau}{\loc}$, $st'_{\pc'} = \invoke{r_o}{m}{(r_{j_i})_{i \le n}}$ and $u^* = (R'(r_{j_i})))_{i \le n}$. By using the same reasoning that we did in \irule{R-Return} we can show that:
\begin{alignat*}{2}
\absprog \quad\vdash\quad& \rlookup{o}{\absval_2'^*}{\abslh_2'}{\_}{\absobj{c''}{\_}} \wedge c'' \subtype c' \wedge \abswal_1'= \abswal'_2\wedge \left(\bigwedge\nolimits_{j \le n} (\absval_2'^*)_{i_j} \absmeet (\absual_1'^*)_j \ne \bot\right)
\end{alignat*}

Hence we can apply the following rule, which is included in $\translate{P}$:
\begin{align*}
&\absreg{c',m',\pc'}{ (\abswal'_2,\absual_2'^*) }{\absval_2'^*}{\abslh_2'}{\absfi_2'} \wedge \rlookup{o}{\absval_2'^*}{\abslh_2'}{\_}{\absobj{c''}{\_}} \wedge c'' \subtype c' \\
&\wedge\absuncaught{c,m}{(\abswal'_1,\absual_1'^*)}{(\absval_1'^*)_{\excpt}}{\abslh_1'}{\absfi_1'} \wedge \abswal_1' = \abswal'_2\wedge \left(\bigwedge\nolimits_{j \le n} (\absval_2'^*)_{i_j} \absmeet (\absual_1'^*)_j \ne \bot\right)\\
\implies\;&\absreg{c',m',pc'}{(\abswal'_2,\absual_2')}{\lift{\absval_2'^*}{\absfi_1'}[\excpt \mapsto ({\absval_1'^*})_{\excpt}]}{\abslh_1'}{\absfi_1' \afunion \absfi_2'}
\end{align*}
This shows that $\translate{P} \cup \absprog \vdash \delcall$.
\end{enumerate}

\item \textbf{Remaining cases}
The remaining cases are straightforward or very similar to cases we already analyzed. For example: 
\begin{itemize}
\item \irule{R-SCall}: Similar to the \irule{R-Call} case
\item \irule{R-NewIntent}: Similar to the \irule{R-NewObj} case
\item \irule{R-NewArr}: Similar to the \irule{R-NewObj} case
\item \irule{R-MoveSFld}: Similar to the \irule{R-MoveFld} case
\item \irule{R-MoveArr}: Similar to the \irule{R-MoveFld} case
\item \irule{R-PutExtra}: Similar to the \irule{R-MoveFld} case
\item \irule{R-MoveException} Similar to the \irule{R-MoveFld} case
\item \irule{R-InterruptJoin}: Similar to the \irule{R-InterruptWait} case
\end{itemize}
\end{itemize}

\end{IEEEproof}


\subsection{Proof of Lemma~\ref{thm:global}}

\label{subsec:global}
\begin{IEEEproof}
If $\Psi = \Psi'$ then it suffices the take $\Delta = \Delta'$.

We are just going to prove that this is true if $\Psi$ reduces to $\Psi'$ in one step. The lemma's proof is then obtained by a straightforward induction on the reduction length.

Let $X \in \rfconf(\Psi)$ with  \cheapd its  configuration decomposition.
\begin{itemize}
\item Rule applied is \irule{A-Active}:
\[ \inferrule[(A-Active)] {\tmethconf{\callstack}{\pi}{\threadstack}{\heap}{\sheap}{\ell} \rightsquigarrow \tmethconf{\callstack'}{\pi'}{\threadstack'}{\heap'}{\sheap'}{\ell}} {\tactconf{\actstack :: \tuactframe{\ell}{s}{\pi}{\threadstack}{\callstack} :: \actstack'}{\threadpool}{\heap}{\sheap} \Rightarrow \tactconf{\actstack :: \tuactframe{\ell}{s}{\pi'}{\threadstack'}{\callstack'} :: \actstack'}{\threadpool}{\heap'}{\sheap'}} \]

We know that:
\[X  =  \rfastk{\gheap}(\actstack :: \tuactframe{\ell}{s}{\pi}{\threadstack}{\callstack} :: \actstack',\threadpool , (\lheap_l,(\lfilter^{l,j})_{j})_l) \cup \rfheap{\gheap}(\heap) \cup \rfstat{}(\sheap)\]
and that :
\[
  \rfframe{\gheap}(\tuactframe{\ell}{s}{\pi}{\threadstack}{\callstack},\lheap_n,(\lfilter^{n,j})_{j})  \subseteq \rfastk{\gheap}(\actstack :: \tuactframe{\ell}{s}{\pi}{\threadstack}{\callstack} :: \actstack',\threadpool,(\lheap_l,(\lfilter^{l,j})_{j})_l)
\]

Moreover $(\lheapd,\lheap_n,(\lfilter^{n,j})_{j})$ is a local configuration decomposition of $\tmethconf{\callstack}{\pi}{\threadstack}{\heap}{\sheap}{\ell}$. We define $X_{loc}$ as follows:
\begin{align*}
X_{loc} = \quad& \rfframe{\gheap}(\tuactframe{\ell}{s}{\pi}{\threadstack}{\callstack},\lheap_n,(\lfilter^{n,j})_{j}) \cup \rfheap{\gheap}(\heap) \cup \rfstat{}(\sheap)\\
= \quad& \rfcall{\ell}(\callstack,\lheap_n,(\lfilter^{n,j})_{j}) \cup \rfdispatch{\ell}(\pi) \cup \rftdispatch{\gheap}(\threadstack) \cup \rfheap{\gheap}(\heap) \cup \rfstat{}(\sheap)\\
\in \quad& \rflconf(\tmethconf{\callstack}{\pi}{\threadstack}{\heap}{\sheap}{\ell})
\end{align*}

Therefore we know that $X_{loc} \in \rflconf(\tmethconf{\callstack}{\pi}{\threadstack}{\heap}{\sheap}{\ell})$ with local configuration decomposition  $\lheapd,\lheap_n,(\lfilter^{n,j})_{j}$. Besides $X_{loc} \subseteq X$, hence by Lemma~\ref{lem:sub-order} we have $X_{loc} <: \absprog$. By Lemma~\ref{thm:local} we know that there exists $\absprog_{loc}'$ and $X_{loc}' \in \rflconf(\tmethconf{\callstack'}{\pi'}{\threadstack'}{\heap'}{\sheap'}{\ell})$  with local configuration decomposition $\lheapd',\lheap'_n,(\lfilter'^{n,j})_{j}$ such that  $\forall i \ne n, \lheap_i = \lheap_i'$,  $\absprog_{loc}' :> X_{loc}'$ and $\translate{P} \cup \absprog \vdash \absprog_{loc}'$.

For all $j$ and $l \ne n$, let $\lfilter'^{l,j} = \lfilter^{l,j}$. Then it is quite easy to check that \cheapdp is a configuration decomposition of $\Psi'$. We define $X'$ by:
\[X'  =  \rfastk{\gheap'}(\actstack :: \tuactframe{\ell}{s}{\pi'}{\threadstack'}{\callstack'} :: \actstack',\threadpool,(\lheap'_l,(\lfilter'^{l,j})_{j})_l) \cup \rfheap{\gheap'}(\heap') \cup \rfstat{}(\sheap') \]

Let $n$ be such that $\actstack$ is of length $n-1$,  $n'$ be the length of $\actstack'$ and  $m$ be the length of $\threadpool$. We know that:
\begin{align*}
  & \rfastk{\gheap'}(\actstack :: \tuactframe{\ell}{s}{\pi'}{\threadstack'}{\callstack'} :: \actstack',\threadpool,(\lheap'_l,(\lfilter'^{l,j})_{j})_l) \backslash  \rfframe{\gheap'}(\tuactframe{\ell}{s}{\pi'}{\threadstack'}{\callstack'},\lheap'_n,(\lfilter'^{n,j})_{j})\\
 = & \left(\bigcup_{ l = 1}^{ n - 1} \rfframe{\gheap'}(\actstack_l,\lheap'_l,(\lfilter'^{l,j})_{j})\right)  \cup \left(\bigcup_{l = 1}^{ n'} \rfframe{\gheap'}(\actstack'_l,\lheap'_{l+n},(\lfilter'^{l+n,j})_{j})\right) \cup \left(\bigcup_{l = 1}^{ m} \rfframe{\gheap'}(\threadpool_l,\lheap'_{l+n+n'},(\lfilter'^{l+n+n',j})_{j})\right)\\
&\hspace{-2.5em}\text{which by  Proposition~\ref{prop:rfconfiguration} is equal to}\\
 = & \left(\bigcup_{ l = 1}^{n-1} \rfframe{\gheap}(\actstack_l,\lheap_l,(\lfilter^{l,j})_{j})\right)   \cup \left(\bigcup_{l = 1}^{n'} \rfframe{\gheap}(\actstack'_l,\lheap_{l+n},(\lfilter^{l+n,j})_{j})\right) \cup \left(\bigcup_{ l = 1}^{m} \rfframe{\gheap}(\threadpool_l,\lheap_{l+n+n'},(\lfilter^{l+n+n',j})_{j})\right)
\end{align*}
Which implies that:
\begin{equation*}
X' \backslash X \subseteq \rfframe{\gheap'}(\tuactframe{\ell}{s}{\pi'}{\threadstack'}{\callstack'},\lheap'_n,(\lfilter'^{n,j})_{j}) \cup \rfheap{\gheap'}(\heap') \cup \rfstat{}(\sheap')= X_{loc}'
\end{equation*}
We define $\absprog' = \absprog \cup \absprog'_{loc}$.We know that $X <: \absprog$ and $X'_{loc} <: \absprog_{loc}'$, therefore by Lemma~\ref{lem:join-order} we have  $X \cup X'_{loc} <:  \absprog \cup \absprog_{loc}' = \absprog'$. Moreover $X' \subseteq X \cup X_{loc}'$, therefore by Lemma~\ref{lem:sub-order} we have $X' <: \absprog'$. We conclude by observing that since $\translate{P} \cup \absprog \vdash \absprog_{loc}'$, we trivially have $\translate{P} \cup \absprog \vdash \absprog'$.

\item  Rule applied is \irule{A-Deactivate}:  
\[\inferrule[(A-Deactivate)]{ }{\tactconf{\actstack :: \tuactframe{\ell}{s}{\pi}{\threadstack}{\ocallstack} :: \actstack'}{\threadpool}{\heap}{\sheap} \Rightarrow \tactconf{\actstack :: \tactframe{\ell}{s}{\pi}{\threadstack}{\ocallstack} :: \actstack'}{\threadpool}{\heap}{\sheap}}\]

In this case $\rfconf(\tactconf{\actstack :: \tuactframe{\ell}{s}{\pi}{\threadstack}{\ocallstack} :: \actstack'}{\threadpool}{\heap}{\sheap}) = \rfconf(\tactconf{\actstack :: \tactframe{\ell}{s}{\pi}{\threadstack}{\ocallstack} :: \actstack'}{\threadpool}{\heap}{\sheap})$, hence the conclusion immediately follows from the induction hypothesis.

\item  Rule applied is \irule{A-Step}: 
\[\inferrule[(A-Step)]{(s,s') \in \lifecycle \\\pi \neq \varepsilon \Rightarrow (s,s') = (\actstate{running},\actstate{onPause}) \\\heap(\ell).\finished = \true \Rightarrow (s,s') \in \{(\actstate{running},\actstate{onPause}),(\actstate{onPause},\actstate{onStop}),(\actstate{onStop},\actstate{onDestroy})\}}{\tactconf{\tactframe{\ell}{s}{\pi}{\threadstack}{\ocallstack} :: \actstack}{\threadpool}{\heap}{\sheap} \Rightarrow \tactconf{\tuactframe{\ell}{s'}{\pi}{\threadstack}{\getcb{\ell}{s'}} :: \actstack}{\threadpool}{\heap}{\sheap}}\]

We have:
\[X  =  \rfastk{\gheap}(\tuactframe{\ell}{s}{\pi}{\threadstack}{\ocallstack} :: \actstack,\threadpool,(\lheap_l,(\lfilter^{l,j})_{j})_l) \cup \rfheap{\gheap}(\heap) \cup \rfstat{}(\sheap)\]

Since we only focus on well-formed configurations, we have $\heap(\ell) = \obj{c}{(f \mapsto u)^*}$ for some activity class $c$ and $\ell = \pointer{p}{c}$ for some pointer $p$. We then observe that $\getcb{\ell}{s'} = \locstate{c',m,0}{\stm^*}{R}{v^*} :: \varepsilon$, where $(c',\stm^*) = \lookup(c,m)$ for some $m \in \cb(c,s)$, $\sign(c',m) = \methsign{\tau_1,\ldots,\tau_n}{\tau}{\loc}$ and:
\[ R = ((r_i \mapsto \defvalue)^{i \leq \loc}, r_{\loc+1} \mapsto \ell, (r_{\loc+1+j} \mapsto v_j)^{j \leq n})\]
for some values $v_1,\ldots,v_n$ of the correct type $\tau_1,\ldots,\tau_n$. By Assumption~\ref{asm:overriding}, we also have $c \subtype c'$.

Given that $\absprog :> X \in \rfconf(\Psi)$, we have $\absprog :> \rfheap{\gheap}(\heap)$. We know that $\ell = \pointer{p}{c} \in dom(\heap)$, and since local heaps contain only locations whose annotations are program points, we know that $\ell \in dom(\gheap)$. Therefore there exists $\absheap(\absloc,\absblock) \in \absprog$ such that $\absloc = \rflab(\ell) = c$ and $ \rfblock{}(\obj{c}{(f \mapsto u)^*}) \poblk\absblock $. This implies that $\absblock = \absobj{c}{(f \mapsto \absval)^*}$ for some $\absval^*$ such that $\forall i, \rfval{}(u_i) \poval \absval_i$. Hence using the implications $\rulename{Cbk}$ included in $\translate{P}$ we get that:
\begin{equation}
\label{eq:step0}
\translate{P} \cup \absprog \vdash \absreg{\mathsf{c'},\mathsf{m},\mathsf{0}}{(\absg{c},(\top_{\tau_j})^{j \leq n})}{(\adefvalue_k)^{k \leq \loc},\absg{c},(\top_{\tau_j})^{j \leq n}}{(\bot)^*}{0^*}
\end{equation}
Let $\absprog' = \absprog \cup \{\absreg{\mathsf{c'},\mathsf{m},\mathsf{0}}{(\absg{c},(\top_{\tau_j})^{j \leq n})}{(\adefvalue_k)^{k \leq \loc},\absg{c},(\top_{\tau_j})^{j \leq n}}{(\bot)^*}{0^*} \}$. From Equation~\ref{eq:step0} we get that $\translate{P} \cup \absprog \vdash \absprog'$.

 Let $\gheap' = \gheap$, for all $i > 1$ let $\lheap_i' = \lheap_i$ and for all $j > 1, (\lfilter'^{l,j})_j = (\lfilter^{l,j})_j$. Let also $\lheap'_1$ be a fresh empty local heap and $(\lfilter'^{1,j})_j = (\{(\ell \mapsto 0) ~|~ \ell \}) :: \varepsilon$. Using Assumption~\ref{asm:bundle}, it is simple to show that \cheapdp is a configuration decomposition of $\tactconf{\tuactframe{\ell}{s'}{\pi}{\threadstack}{\getcb{\ell}{s'}} :: \actstack}{\threadpool}{\heap}{\sheap}$, and that:
\begin{equation}
\absprog' >: \{\absreg{\mathsf{c'},\mathsf{m},\mathsf{0}}{(\absg{c},(\top_{\tau_j})^{j \leq n})}{(\adefvalue_k)^{k \leq \loc},\absg{c},(\top_{\tau_j})^{j \leq n}}{(\bot)^*}{0^*}\} :> \rfcall{\ell}(\getcb{\ell}{s'},\lheap'_1,(\lfilter'^{1,j})_{j})\label{eq:step1}
\end{equation}
Observe that $\rftdispatch{\gheap}(\threadstack) = \rftdispatch{\gheap'}(\threadstack)$. Besides $\absprog :> \rfconf(\tactconf{\actstack}{\threadpool}{\heap}{\sheap})$ implies that $\rfdispatch{\ell}(\pi)\cup\rftdispatch{\gheap}(\threadstack) <: \absprog$, and we know that since $\absprog \subseteq \absprog'$ we have $\absprog <: \absprog'$. Therefore by transitivity of $<:$ we have :
\begin{equation}
\rfdispatch{\ell}(\pi) \cup \rftdispatch{\gheap'}(\threadstack) <: \absprog'
\label{eq:step2}
\end{equation}

It is easy to check that $X' \in \rfconf(\Psi')$, where $X'$ is the following set of facts:
\[
X'=   \rfastk{\gheap'}(\tuactframe{\ell}{s'}{\pi}{\threadstack}{\getcb{\ell}{s'}} :: \actstack,\threadpool,(\lheap'_l,(\lfilter'^{l,j})_{j})_l)\cup \rfheap{\gheap}(\heap) \cup \rfstat{}(\sheap)
\]

Using Proposition~\ref{prop:rfconfiguration}, one can check that:
\[
X' \backslash X = \rfcall{\ell}(\getcb{\ell}{s'},\lheap'_1,(\lfilter'^{1,j})_{j}) \cup \rfdispatch{\ell}(\pi) \cup \rftdispatch{\gheap'}(\threadstack)
\]
Equation~\ref{eq:step1} and Equation~\ref{eq:step2} give us that $X' \backslash X <: \absprog'$. We conclude by observing that since $X <: \absprog <: \absprog'$ and $X' \subseteq X \cup (X' \backslash X)$, we have $X' <: \absprog'$.

\item  Rule applied is \irule{A-Hidden}:
\[\inferrule[(A-Hidden)]{\varphi = \tactframe{\ell}{s}{\pi}{\threadstack}{\ocallstack} \\s \in \{\actstate{onResume},\actstate{onPause}\} \\(s',s'') \in \{(\actstate{onPause},\actstate{onStop}),(\actstate{onStop},\actstate{onDestroy})\} }{\tactconf{\varphi :: \actstack :: \tactframe{\ell'}{s'}{\pi'}{\threadstack'}{\ocallstack'} :: \actstack'}{\threadpool}{\heap}{\sheap} \Rightarrow \tactconf{\varphi :: \actstack :: \tuactframe{\ell'}{s''}{\pi'}{\threadstack'}{\getcb{\ell'}{s''}} :: \actstack'}{\threadpool}{\heap}{\sheap}}\]

This case is analogous to the case \irule{A-Step}.

\item  Rule applied is \irule{A-Destroy}: 
\[\inferrule[(A-Destroy)]{\heap(\ell).\finished = \true}{\tactconf{\actstack :: \tactframe{\ell}{\actstate{onDestroy}}{\pi}{\threadstack}{\ocallstack} :: \actstack'}{\threadpool}{\heap}{\sheap} \Rightarrow \tactconf{\actstack :: \actstack'}{\threadpool}{\heap}{\sheap}}\]

Let $n$ be the length of $\actstack$. It is easy to check that $(\gheap\cup \lheap_n,(\lheap_l,(\lfilter^{l,j})_j)_{l \ne n})$ is a configuration decomposition of $\tactconf{\actstack :: \actstack'}{\threadpool}{\heap}{\sheap}$, and that $X' \in \rfconf(\Psi')$ where:
\[ X' = \rfastk{\gheap\cup\lheap_n}(\actstack :: \actstack',\threadpool,(\lheap_l,(\lfilter^{l,j})_j)_{l \ne n}) \cup \rfheap{\gheap}(\heap) \cup \rfstat{}(\sheap) \subseteq X\]
Since $X <: \absprog$, this implies that $X' <: \absprog$. We conclude with the trivial observation that $\translate{P} \cup \absprog \vdash \absprog$.

\item  Rule applied is \irule{A-Back}: 
\[\inferrule[(A-Back)]{\heap' = \heap[\ell \mapsto \heap(\ell)[\finished \mapsto \true]]}{\tactconf{\tactframe{\ell}{\actstate{running}}{\varepsilon}{\threadstack}{\ocallstack} :: \actstack}{\threadpool}{\heap}{\sheap} \Rightarrow \tactconf{\tactframe{\ell}{\actstate{running}}{\varepsilon}{\threadstack}{\ocallstack} :: \actstack}{\threadpool}{\heap'}{\sheap}}\]

Let $b = \heap(\ell)$. Since we only focus on well-formed configurations, we have $b = \obj{c}{(f \mapsto u)^*,\finished \mapsto v}$ for some activity class $c$ and some boolean value $v$. Let then $b' = \heap'(\ell) = \obj{c}{(f \mapsto u)^*,\finished \mapsto \true}$ according to the reduction rule. 

Given that $\absprog :> X \in \rfconf(\Psi)$, we have $\absprog :> \rfheap{\gheap}(\heap)$. We know that $\ell = \pointer{p}{c} \in dom(\heap)$, and since local heaps contain only locations whose annotations are program points, we know that $\ell \in dom(\gheap)$. Therefore there exists $\absheap(\absloc,\absblock) \in \absprog$ such that $\absloc = \rflab(\ell) = c$ and $ \rfblock{}(\obj{c}{(f \mapsto u)^*,\finished \mapsto v}) \poblk\absblock $. This implies that $\absblock = \absobj{c}{(f \mapsto \absual)^*,\finished \mapsto \absval}$ for some $\absual^*, \absval$ such that $\forall i, \rfval{}(u_i) \poval \absual_i$ and $\rfval{}(v) \poval \absval$. It is easy to check that:
\[
\rfblock{}(b') = \absobj{c}{(f \mapsto \rfval{}(u))^*, \finished \mapsto \widehat{\true}}
\]

We define $\absprog' = \absprog \cup \{ \absheap(\absloc,\absobj{c}{(f \mapsto \absual)^*,\finished \mapsto \top_{\type{bool}}})\}$. Since $\absheap(\absloc,\absblock) \in \absprog$ we have by using the implication $\rulename{Fin}$ in $\translate{P}$ that:
\[
\translate{P} \cup \absprog \vdash \absheap(\absloc,\absobj{c}{(f \mapsto \absual)^*,\finished \mapsto \top_{\type{bool}}})
\]
Therefore $\translate{P} \cup \absprog \vdash \absprog'$. We then observe that:
\begin{alignat*}{2}
 \absheap(\rflab(\ell), \rfblock{}(b')) \quad&\poblk\quad&& \absheap(\absloc,\absobj{c}{(f \mapsto \absual)^*,\finished \mapsto \widehat{\true}}) \\
 & \poblk &&\absheap(\absloc,\absobj{c}{(f \mapsto \absual)^*,\finished \mapsto \top_{\type{bool}}})
\end{alignat*}

Hence $\rfheap{\gheap}(\heap') <: \absprog'$. It is then easy to conclude this case.

\item Rule applied is \irule{A-Swap}: 
\[\inferrule[(A-Swap)]{\varphi' = \tactframe{\ell'}{\actstate{onPause}}{\varepsilon}{\threadstack'}{\ocallstack'} \\\heap(\ell').\finished = \true \\\varphi = \tactframe{\ell}{s}{i :: \pi}{\threadstack}{\ocallstack} \\s \in \{\actstate{onPause},\actstate{onStop}\} \\\heap(\ell').\parent = \ell}{\tactconf{\varphi' :: \varphi :: \actstack}{\threadpool}{\heap}{\sheap} \Rightarrow \tactconf{\varphi :: \varphi' :: \actstack}{\threadpool}{\heap}{\sheap}}\]

Just take $\gheap' = \gheap, \lheap_1' = \lheap_2, \lheap_2' = \lheap_1$, for all $j, \lfilter'^{1,j} = \lfilter^{2,j}$, $\lfilter'^{2,j} = \lfilter^{1,j}$ (we simply exchange the first local heap and filters with the second local heap and filters). The rest is kept unchanged: for all $ l > 2$, for all $j$, $\lheap'_i = \lheap_i$ and $\lfilter'^{l,j}= \lfilter^{l,j}$.

It is quite simple to check that \cheapd is a configuration decomposition and that the corresponding set of abstract facts are the same.

Therefore $\rfconf(\Psi) = \rfconf(\Psi')$, which concludes this case.

\item Rule applied is \irule{A-Start}: 
\[\inferrule[(A-Start)]{s \in \{\actstate{onPause},\actstate{onStop}\} \\i = \intent{c}{(k \mapsto v)^*} \\\emptyset \vdash \serblock{\heap}(i) = (i',\heap') \\\pointer{p}{c},\pointer{p'}{\astart{c}} \not\in \dom(\heap,\heap') \\o = \obj{c}{(f_{\tau} \mapsto \defvalue_{\tau})^*,\finished \mapsto \false, \fintent \mapsto \pointer{p'}{\astart{c}}, \parent \mapsto \ell} \\\heap'' = \heap,\heap',\pointer{p}{c} \mapsto o, \pointer{p'}{\astart{c}} \mapsto i'}{\tactconf{\tactframe{\ell}{s}{i :: \pi}{\threadstack}{\ocallstack} :: \actstack}{\threadpool}{\heap}{\sheap} \Rightarrow \tactconf{\tuactframe{\pointer{p}{c}}{\actstate{constructor}}{\varepsilon}{\varepsilon}{\getcb{\pointer{p}{c}}{\actstate{constructor}}} :: \tactframe{\ell}{s}{\pi}{\threadstack}{\ocallstack} :: \actstack}{\threadpool}{\heap''}{\sheap}}\]

Since we only focus on well-formed configurations, we know that $\ell = \pointer{p''}{c''}$ for some pointer $p''$ and some activity class $c''$. We then observe that $\getcb{\pointer{p}{c}}{\actstate{constructor}} = \locstate{c',m,0}{\stm^*}{R}{v^*} :: \varepsilon$, where $(c',\stm^*) = \lookup(c,\actstate{constructor})$, $\sign(c',\actstate{constructor}) = \methsign{\tau_1,\ldots,\tau_n}{\tau}{\loc}$ and:
\[ 
R = ((r_i \mapsto \defvalue)^{i \leq \loc}, r_{\loc+1} \mapsto \pointer{p}{c}, (r_{\loc+1+j} \mapsto v_j')^{j \leq n}),
\]
for some values $v_1',\ldots,v_n'$ of the correct type $\tau_1,\ldots,\tau_n$. By Assumption~\ref{asm:overriding}, we also have $c \subtype c'$.

Given that $X <: \absprog$, we have $\absprog :> \rfdispatch{\ell}(i :: \pi)$, which implies that there exists $\absdispatch_\ann(\absblock) \in \absprog$ such that $\ann = \rflab(\ell) = c'$ and $ \rfblock{}(i) \poblk \absblock$. This implies that $\absblock = \absintent{c}{\absval}$ for some $\absval$ such that $ \sqcup_i\, \rfval{}(v_i) \poval \absval $. Using the implications $\rulename{Act}$  in $\translate{P}$ we get:
\begin{alignat}{2}
\translate{P} \cup \absprog &\;\;\vdash \;\;&& \absheap(\astart{c},\absintent{c}{\absval})\label{eq:start0}\\
\translate{P} \cup \absprog &\;\;\vdash&& \absheap(c, \absobj{c}{(f \mapsto \adefvalue_{\tau})^*, \finished \mapsto \widehat{\false}, \parent \mapsto c', \fintent \mapsto \astart{c}})\label{eq:start1}
\end{alignat}
Hence using the implications $\rulename{Cbk}$ included in $\translate{P}$ we get that:
\begin{multline}
\translate{P} \cup \{\absheap(c, \absobj{c}{(f \mapsto \adefvalue_{\tau})^*, \finished \mapsto \widehat{\false}, \parent \mapsto c', \fintent \mapsto \astart{c}})\}\\ \vdash \absreg{\mathsf{c'},\mathsf{m},\mathsf{0}}{(\absg{c},(\top_{\tau_j})^{j \leq n})}{(\adefvalue_k)^{k \leq \loc},\absg{c},(\top_{\tau_j})^{j \leq n}}{(\bot)^*}{0^*}\label{eq:start2}
\end{multline}
We define the set of abstract fact:
\begin{alignat*}{2}
\absprog' \;\;&=\;\;&& \absprog \cup \{\absreg{\mathsf{c'},\mathsf{m},\mathsf{0}}{(\absg{c},(\top_{\tau_j})^{j \leq n})}{(\adefvalue_k)^{k \leq \loc},\absg{c},(\top_{\tau_j})^{j \leq n}}{(\bot)^*}{0^*} \} \cup\{\absheap(\astart{c},\absintent{c}{\absval})\}\\
&&& \cup \{\absheap(c, \absobj{c}{(f \mapsto \adefvalue_{\tau})^*, \finished \mapsto \widehat{\false}, \parent \mapsto c', \fintent \mapsto \astart{c}})\}
\end{alignat*}
From Equation~\ref{eq:start0}, Equation~\ref{eq:start1} and Equation~\ref{eq:start2} we get that $\translate{P} \cup \absprog \vdash \absprog'$.

\paragraph{Configuration Decomposition} Let $\lheap'_0$ be an fresh empty local heap. We take $\gheap'=\gheap \cup \heap' \cup \{\pointer{p}{c},\pointer{p'}{\astart{c}} \}$,  $(\lheap_l')_l = \lheap_0' :: (\lheap)_l$ and $(\lfilter'^{l,j})_{l,j} = ((\{(\ell \mapsto 0) ~|~ \ell \}) :: \varepsilon) :: (\lfilter^{l,j})_{l,j}$. 

Since $(\gheap,(\lheap_i), \lheap_1,(\lfilter^{1,j})_j)$ is a local configuration decomposition of $\tmethconf{\ocallstack}{(i :: \pi)}{\threadstack}{\heap}{\sheap}{\ell}$, we know that there exists $\ell'$ such that $(\ell' \mapsto i) \in \gheap$. Moreover $\absprog :> \rfheap{\gheap}(\heap)$ and $\serblock{\heap}(i) = (i',\heap')$, therefore by applying  Lemma~\ref{lem:heap-serialization}  we know that $\absprog :> \rfheap{\gheap}(\heap')$ and that $\gheap \cup \heap' , (\lheap_i)_i$ is a heap decomposition of $\heap \cup \heap'\cdot\sheap$.

Since $\ell = \pointer{p''}{c}$ we know that $\ell \in \gheap$, hence for all $i$, $o \not \heapto \lheap_i$. By Lemma~\ref{lem:serialization-prop2} we know that for all $i$, $i \not \heapto \lheap_i$. Moreover $\pointer{p}{c}$ and $\pointer{p'}{\astart{c}}$ are fresh locations, therefore $\gheap', (\lheap_i)_i$ is a heap decomposition of $\heap''\cdot\sheap$. Since $\lheap'_0$ is a fresh empty local heap we easily get from this that $\gheap', (\lheap'_i)_i$ is a heap decomposition of $\heap''\cdot\sheap$. 

Using Assumption~\ref{asm:bundle}, it is simple to check that \cheapdp is a configuration decomposition of $\Psi'$.

Let $X'$ be the corresponding set of facts:
\[
 \rfastk{\gheap'}(\tuactframe{\pointer{p}{c}}{\actstate{constructor}}{\varepsilon}{\varepsilon}{\getcb{\pointer{p}{c}}{\actstate{constructor}}} :: \tactframe{\ell}{s}{\pi}{\threadstack}{\ocallstack} :: \actstack,\threadpool,(\lheap'_l,(\lfilter'^{l,j})_{j})_l)\cup \rfheap{\gheap'}(\heap'') \cup \rfstat{}(\sheap)
\]
We are going to prove that $X'$ is over-approximated by the set of abstract facts $\absprog'$.

\paragraph{Heap} We already saw that  $\absprog :> \rfheap{\gheap}(\heap')$, and by applying Lemma~\ref{lem:serialization} we know that $\rfblock(i) = \rfblock(i')$. We then observe that:
\[
\begin{array}{lcll}
\{\absheap(\astart{c},\absintent{c}{\absval})\} & :> & \{\absheap(\astart{c},\rfblock{}(i)\} & \text{since } \rfblock{}(i) \poblk \absblock = \absintent{c}{\absval})   \numberthis\label{eq:starth1}\\
& = & \{\absheap(\astart{c},\rfblock{}(i')\} & \text{since } \rfblock{}(i) = \rfblock{}(i')\\
& = & \{\absheap(\rflab(\pointer{p'}{\astart{c}}),\rfblock{}(i')\} \qquad\null& \text{by definition}\\
\end{array}
\]
Also notice that:
\begin{equation}
\{\absheap(c, \absobj{c}{(f \mapsto \adefvalue_{\tau})^*, \finished \mapsto \widehat{\false}, \parent \mapsto c', \fintent \mapsto \astart{c}})\} = \absheap(\rflab(\pointer{p}{c}), \rfblock{}(o))\label{eq:starth2}
\end{equation}

Moreover it is simple to see that we have:
\[\rfheap{\gheap'}(\heap'') = \rfheap{\gheap}(\heap) \cup \rfheap{\gheap \cup \absheap'}(\heap') \cup \{\absheap(\rflab(\pointer{p}{c}),\rfblock{}(o))\} \cup \{ \{\absheap(\rflab(\pointer{p'}{\astart{c}}),\rfblock{}(i')\}\}\]

We already saw that $\rfheap{\gheap \cup \absheap'}(\heap') <: \absprog <: \absprog'$. This together with Equation~\ref{eq:starth1} and Equation~\ref{eq:starth2} shows that $\rfheap{\gheap'}(\heap'') <: \absprog'$.

\paragraph{Activity Stack}
Let $n$ be the length of $\actstack$, and let $m$ be the length of $\threadpool$.
\begin{align*}
 & \rfastk{\gheap'}(\tuactframe{\pointer{p}{c}}{\actstate{constructor}}{\varepsilon}{\varepsilon}{\getcb{\pointer{p}{c}}{\actstate{constructor}}} :: \tactframe{\ell}{s}{\pi}{\threadstack}{\ocallstack} :: \actstack,\threadpool,(\lheap'_l,(\lfilter'^{l,j})_{j})_l)&\\
 =\quad &  \rfframe{\gheap'}(\tuactframe{\pointer{p}{c}}{\actstate{constructor}}{\varepsilon}{\varepsilon}{\getcb{\pointer{p}{c}}{\actstate{constructor}}},\lheap'_0,(\lfilter'^{0,j})_{j}) 
\cup\rfframe{\gheap'}(\tactframe{\ell}{s}{\pi}{\threadstack}{\ocallstack},\lheap'_1,(\lfilter'^{1,j})_{j})\\
&\cup \left( \bigcup_{1 \le l \le n} \rfframe{\gheap'}(\actstack_l,\lheap'_{l+1},(\lfilter'^{l+1,j})_{j})\right)
\cup \left( \bigcup_{1 \le l \le m} \rfframe{\gheap'}(\threadpool_l,\lheap'_{l+n+1},(\lfilter'^{l+n+1,j})_{j})\right)
\end{align*}
By Proposition~\ref{prop:rfconfiguration} this is equal to:
\begin{align*}
&\rfframe{\gheap'}(\tuactframe{\pointer{p}{c}}{\actstate{constructor}}{\varepsilon}{\varepsilon}{\getcb{\pointer{p}{c}}{\actstate{constructor}}},\lheap'_0,(\lfilter'^{0,j})_{j}) 
\cup\rfframe{\gheap}(\tactframe{\ell}{s}{\pi}{\threadstack}{\ocallstack},\lheap_1,(\lfilter^{1,j})_{j})\\
&\cup \left( \bigcup_{1 \le l \le n} \rfframe{\gheap}(\actstack_l,\lheap_{l+1},(\lfilter^{l+1,j})_{j})\right)
\cup \left( \bigcup_{1 \le l \le m} \rfframe{\gheap}(\threadpool_l,\lheap_{l+n+1},(\lfilter^{l+n+1,j})_{j})\right)
\end{align*}

We then observe that:
\begin{alignat*}{2}
\absprog' \;\;&:>\;\;&& \{\absreg{\mathsf{c'},\mathsf{m},\mathsf{0}}{(\absg{c},(\top_{\tau_j})^{j \leq n})}{(\adefvalue_k)^{k \leq \loc},\absg{c},(\top_{\tau_j})^{j \leq n}}{(\bot)^*}{0^*}\} \\
&:>&& \rfframe{\gheap'}(\tuactframe{\pointer{p}{c}}{\actstate{constructor}}{\varepsilon}{\varepsilon}{\getcb{\pointer{p}{c}}{\actstate{constructor}}},\lheap'_0,(\lfilter'^{0,j})_{0,j})
\end{alignat*}
This proves that the changes to the activity stack are over-approximated by $\absprog'$.

\item  Rule applied is \irule{A-Replace}:
\[\inferrule[(A-Replace)]{\heap(\ell) = \obj{c}{(f_{\tau} \mapsto v)^*,\finished \mapsto u} \\ \pointer{p}{c} \not \in dom(\heap) \\ o = \obj{c}{(f_{\tau} \mapsto \defvalue_{\tau})^*,\finished \mapsto \false} \\ \heap' = \heap, \pointer{p}{c} \mapsto o}{\tactconf{\tactframe{\ell}{\actstate{onDestroy}}{\pi}{\threadstack}{\ocallstack} :: \actstack}{\threadpool}{\heap}{\sheap} \Rightarrow \tactconf{\tuactframe{\pointer{p}{c}}{\actstate{constructor}}{\pi}{\threadstack}{\getcb{\pointer{p}{c}}{\actstate{constructor}}} :: \actstack}{\threadpool}{\heap'}{\sheap}}\]

Since we only focus on well-formed configurations, we know that $c$ is an activity class and $\ell = \pointer{p'}{c}$ for some pointer $p'$.

We then observe that $\getcb{\pointer{p}{c}}{\actstate{constructor}} = \locstate{c',m,0}{\stm^*}{R}{v^*} :: \varepsilon$, where $(c',\stm^*) = \lookup(c,\actstate{constructor})$, $\sign(c',\actstate{constructor}) = \methsign{\tau_1,\ldots,\tau_n}{\tau}{\loc}$ and:
\[ 
R = ((r_i \mapsto \defvalue)^{i \leq \loc}, r_{\loc+1} \mapsto \pointer{p}{c}, (r_{\loc+1+j} \mapsto v_j')^{j \leq n}),
\]
for some values $v_1',\ldots,v_n'$ of the correct type $\tau_1,\ldots,\tau_n$. By Assumption~\ref{asm:overriding}, we also have $c \subtype c'$.

Given that $\absprog :> X \in \rfconf(\Psi)$, we have $\absprog :> \rfheap{\gheap}(\heap)$. We know that $\ell = \pointer{p'}{c} \in dom(\heap)$, and since local heaps contain only locations whose annotations are program points, we know that $\ell \in dom(\gheap)$. Therefore there exists $\absheap(\absloc,\absblock) \in \absprog$ such that $\absloc = \rflab(\ell) = c$ and $ \rfblock{}(\obj{c}{(f \mapsto v)^*,\finished \mapsto u}) \poblk\absblock $. This implies that $\absblock = \absobj{c}{(f \mapsto \absval)^*,\finished \mapsto \absual}$ for some $\absval^*, \absual$ such that $\forall i, \rfval{}(v_i) \poval \absval_i$ and $\rfval{}(u) \poval \absual$. Hence using the implications $\rulename{Cbk}$ and $\rulename{Rep}$\footnote{We assume here that boolean fields are initialized to $\false$. The proof can be adapted to the case where they are initialized to $\true$ by using the implication in rule $\rulename{Fin}$.} included in $\translate{P}$ we get that:
\begin{gather}
\translate{P} \cup \absprog \vdash \absreg{\mathsf{c'},\mathsf{m},\mathsf{0}}{(\absg{c},(\top_{\tau_j})^{j \leq n})}{(\adefvalue_k)^{k \leq \loc},\absg{c},(\top_{\tau_j})^{j \leq n}}{(\bot)^*}{0^*}\label{eq:replace0}\\
\translate{P} \cup \absprog \vdash \absheap(c,\absobj{c}{(f \mapsto \adefvalue_{\tau})^*,\finished \mapsto \widehat{\false}}))\label{eq:replace1}
\end{gather}

We define the set of abstract $\absprog'$ by:
\begin{alignat*}{2}
\absprog' = \absprog\quad& \cup\quad&& \left\{ \absreg{\mathsf{c'},\mathsf{m},\mathsf{0}}{(\absg{c},(\top_{\tau_j})^{j \leq n})}{(\adefvalue_k)^{k \leq \loc},\absg{c},(\top_{\tau_j})^{j \leq n}}{(\bot)^*}{0^*} \right\}\\
& \cup&& \left\{ \absheap(c,\absobj{c}{(f \mapsto \adefvalue_{\tau})^*,\finished \mapsto \widehat{\false}}))\right\}
\end{alignat*}

 Let $\gheap' = \gheap \cup \{\pointer{p}{c}\}$, for all $i > 1$ let $\lheap_i' = \lheap_i$ and for all $j > 1, (\lfilter'^{l,j})_j = (\lfilter^{l,j})_j$. Let also $\lheap'_1$ be a fresh empty local heap and $(\lfilter'^{1,j})_j = (\{(\ell \mapsto 0) ~|~ \ell \}) :: \varepsilon$. Using Assumption~\ref{asm:bundle}, it is simple to show that \cheapdp is a configuration decomposition of $\tactconf{\tuactframe{\ell}{s'}{\pi}{\threadstack}{\getcb{\pointer{p}{c}}{\actstate{constructor}}} :: \actstack}{\threadpool}{\heap'}{\sheap}$ and that:
\begin{equation}
\rfcall{\ell}(\getcb{\pointer{p}{c}}{constructor},\lheap'_1,(\lfilter'^{1,j})_{j}) <: \{\absreg{\mathsf{c'},\mathsf{m},\mathsf{0}}{(\absg{c},(\top_{\tau_j})^{j \leq n})}{(\adefvalue_k)^{k \leq \loc},\absg{c},(\top_{\tau_j})^{j \leq n}}{(\bot)^*}{0^*}\} <:\absprog'
\label{eq:replace2}
\end{equation}

Observe that $\rftdispatch{\gheap}(\threadstack) = \rftdispatch{\gheap'}(\threadstack)$. Besides $\absprog :> \rfconf(\tactconf{\actstack}{\threadpool}{\heap}{\sheap})$ implies that $\rfdispatch{\ell}(\pi)\cup\rftdispatch{\gheap}(\threadstack) <: \absprog$, and we know that since $\absprog \subseteq \absprog'$ we have $\absprog <: \absprog'$. Therefore by transitivity of $<:$ we have :
\begin{equation}
\rfdispatch{\ell}(\pi)\cup \rftdispatch{\gheap'}(\threadstack) <: \absprog'
\label{eq:replace3}
\end{equation}

Moreover:
\begin{alignat*}{4}
  \rfheap{\gheap'}(\heap') \;\;&=\;\;&&  \rfheap{\gheap}(\heap) \;\;&\cup\;\;& \absheap(\rflab(\pointer{p}{c}),\rfblock(o))\\
&=&&  \rfheap{\gheap}(\heap) &\cup\;\;& \absheap(c,\rfblock(\obj{c}{(f_{\tau} \mapsto \defvalue_{\tau})^*,\finished \mapsto \false}))\\
&<:&& \quad\;\;\absprog &\cup\;\;& \absheap(c,\absobj{c}{(f \mapsto \adefvalue_{\tau})^*,\finished \mapsto \widehat{\false}}))\\
&<:&& \quad\;\;\absprog'\numberthis\label{eq:replace4}
\end{alignat*}

It is easy to check that $X' \in \rfconf(\Psi')$, where $X'$ is the following set of facts:
\[
X'=  \rfastk{\gheap'}(\tuactframe{\ell}{s'}{\pi}{\threadstack}{\getcb{\pointer{p}{c}}{\actstate{constructor}}} :: \actstack, \threadpool,(\lheap'_l,(\lfilter'^{l,j})_{j})_l)\cup \rfheap{\gheap'}(\heap') \cup \rfstat{}(\sheap)
\]
Using Proposition~\ref{prop:rfconfiguration} one can check that:
\[
X' \backslash X = \rfcall{\ell}(\getcb{\ell}{s'},\lheap'_1,(\lfilter'^{1,j})_{j}) \cup \rfdispatch{\ell}(\pi)\cup \rftdispatch{\gheap'}(\threadstack) \cup   \rfheap{\gheap'}(\heap')
\]

Equation~\ref{eq:replace2}, Equation~\ref{eq:replace3} and Equation~\ref{eq:replace4} give us that $X' \backslash X <: \absprog'$. We conclude by observing that since $X <: \absprog <: \absprog'$ and $X' \subseteq X \cup (X' \backslash X)$ we have $X' <: \absprog'$.

\item Rule applied is \irule{A-Result}: 
\[\inferrule[(A-Result)]{\varphi' = \tactframe{\ell'}{\actstate{onPause}}{\varepsilon}{\threadstack'}{\ocallstack'} \\\heap(\ell').\finished = \true \\\varphi = \tactframe{\ell}{s}{\varepsilon}{\threadstack}{\ocallstack} \\s \in \{\actstate{onPause},\actstate{onStop}\} \\\heap(\ell').\parent = \ell \\\emptyset \vdash \serval{\heap}(\heap(\ell').result) = (w',\heap') \\\heap'' = (\heap,\heap')[\ell \mapsto \heap(\ell)[\result \mapsto w']]}{\tactconf{\varphi' :: \varphi :: \actstack}{\threadpool}{\heap}{\sheap} \Rightarrow \tactconf{\tuactframe{\ell}{s}{\varepsilon}{\threadstack}{\getcb{\ell}{\actstate{onActivityResult}}} :: \varphi' :: \actstack}{\threadpool}{\heap''}{\sheap}}\]

Since we focus only on well-formed configurations, we have $\ell = \pointer{p}{c}$ and $\ell' = \pointer{p'}{c'}$ for some pointers $p,p'$ and some activity classes $c,c'$. Also, let $\heap(\ell) = \obj{c}{(f \mapsto \absval)^*}$ and $\heap(\ell') = \obj{c'}{(f' \mapsto \absval')^*, \parent \mapsto \ell, result \mapsto w}$. We then observe that $\getcb{\pointer{p}{c}}{\actstate{onActivityResult}} = \locstate{c'',m,0}{\stm^*}{R}{v^*} :: \varepsilon$, where $(c'',\stm^*) = \lookup(c,\actstate{onActivityResult})$, $\sign(c'',\actstate{onActivityResult}) = \methsign{\tau_1,\ldots,\tau_n}{\tau}{\loc}$ and:
\[ 
R = ((r_i \mapsto \defvalue)^{i \leq \loc}, r_{\loc+1} \mapsto \pointer{p}{c}, (r_{\loc+1+j} \mapsto v_j')^{j \leq n}),
\]
for some values $v_1',\ldots,v_n'$ of the correct type $\tau_1,\ldots,\tau_n$. By Assumption~\ref{asm:overriding}, we also have $c \subtype c''$.

Given that $\absprog :> X \in \rfconf(\Psi)$, we have $\absprog :> \rfheap{\gheap}(\heap)$. We know that $\ell = \pointer{p}{c} \in dom(\heap)$, and since local heaps contain only locations whose annotations are program points, we know that $\ell \in dom(\gheap)$. Therefore there exists $\absheap(\absloc,\absblock) \in \absprog$ such that $\absloc = \rflab(\ell) = c$ and $ \rfblock{}(\obj{c}{(f \mapsto v)^*}) \poblk\absblock $. This implies that $\absblock = \absobj{c}{(f \mapsto \absval)^*}$ for some $\absval^*$ such that $\forall i, \rfval{}(v_i) \poval \absval_i$. Hence using the implications $\rulename{Cbk}$ included in $\translate{P}$ we get that:
\begin{gather}
\translate{P} \cup \absprog \vdash \absreg{\mathsf{c''},\mathsf{m},\mathsf{0}}{(\absg{c},(\top_{\tau_j})^{j \leq n})}{(\adefvalue_k)^{k \leq \loc},\absg{c},(\top_{\tau_j})^{j \leq n}}{(\bot)^*}{0^*}\label{eq:result0}
\end{gather}

Similarly,   there exists $\absheap(\absloc',\absblock') \in \absprog$ such that $\absloc' = \rflab(\ell') = c'$ and $\rfblock(\heap(\ell')) \poblk \absblock'$, which implies that $\absblock' = \absobj{c'}{(f' \mapsto \absval')^*,\parent \mapsto c,\result \mapsto \abswal}$ for some $\absval'^*,\ann''$ such that $\forall i.   \rfval(v_i') \poval \absval_i'$ and $\rfval(w) \poval \abswal$. Hence by using the implication $\rulename{Res}$ we get
\begin{equation}
\translate{P} \cup \absprog \vdash \absheap(c,\absobj{c}{(f \mapsto \absval)^*[\result \mapsto \abswal]})\label{eq:result1}
\end{equation}

We define the following set of facts:
\[\absprog' = \absprog \cup \{\absreg{\mathsf{c''},\mathsf{m},\mathsf{0}}{(\absg{c},(\top_{\tau_j})^{j \leq n})}{(\adefvalue_k)^{k \leq \loc},\absg{c},(\top_{\tau_j})^{j \leq n}}{(\bot)^*}{0^*}\} \cup \{\absheap(c,\absobj{c}{(f \mapsto \absval)^*[\result \mapsto \abswal]})\}\]
Equation~\ref{eq:result0} and Equation~\ref{eq:result1} prove that $\translate{P} \cup \absprog \vdash \absprog'$.

 Let $\lheap'_1$ be an fresh empty local heap. We take $\gheap'=\gheap[\ell \mapsto \heap(\ell)[result \mapsto w']]] \cup \heap'$,  $(\lheap_l')_l = \lheap_1' :: \lheap_1 :: (\lheap)_{l > 3}$ and $(\lfilter'^{l,j})_{l,j} = ((\{(\ell \mapsto 0) ~|~ \ell \}) :: \varepsilon) :: (\lfilter^{1,j})_{j} :: (\lfilter^{l,j})_{l > 3,j}$. 

Recall that $\ell \in \gheap$, therefore $w = \heap(\ell).result$ is either a primitive value or in $dom(\gheap)$. Besides  $\absprog :> \rfheap{\gheap}(\heap)$ and $\serval{\heap}(w) = (w',\heap')$, therefore by applying  Lemma~\ref{lem:heap-serialization}  we know that $\absprog :> \rfheap{\gheap\cup \heap'}(\heap')$ and that $\gheap \cup \heap' , (\lheap_i)_i$ is a heap decomposition of $\heap \cup \heap'\cdot\sheap$.

By Lemma~\ref{lem:serialization-prop2} we know that for all $i$, $w' \not\in dom(\lheap_i)$, therefore $\gheap', (\lheap_i)_i$ is a heap decomposition of $\heap''\cdot\sheap$. Since $\lheap'_0$ is a fresh empty local heap we get from this that $\gheap', (\lheap'_i)_i$ is a heap decomposition of $\heap''\cdot\sheap$. 

Using Assumption~\ref{asm:bundle}, it is simple to check that \cheapdp is a configuration decomposition of $\Psi'$.

Let $X'$ be the corresponding set of facts in $\rfconf(\Psi')$:
\[
 X' = \rfastk{\gheap'}(\tuactframe{\ell}{s}{\varepsilon}{\threadstack}{\getcb{\ell}{\actstate{onActivityResult}}} :: \varphi' :: \actstack,\threadpool,(\lheap'_l,(\lfilter'^{l,j})_{j})_l)\cup \rfheap{\gheap'}(\heap'') \cup \rfstat{}(\sheap)
\]
We are going to prove that $X'$ is over-approximated by the set of abstract facts $\absprog'$. Similarly to what we did in the previous cases, one can check that:
\[ X' \backslash X = \rfframe{\gheap'}(\tuactframe{\ell}{s}{\varepsilon}{\threadstack}{\getcb{\ell}{\actstate{onActivityResult}}},\lheap'_1,(\lfilter'^{1,j})_{j}) \cup \rfheap{\gheap'}(\heap'')\]
And besides:
\[ \rfheap{\gheap'}(\heap'') = \rfheap{\gheap}(\heap_{|dom(\heap)\backslash \ell}) \cup \rfheap{\gheap \cup \absheap'}(\heap') \cup \absheap(c,\rfblock(\heap(\ell)[result \mapsto w']])) \]

\begin{alignat*}{3}
\absheap(c,\rfblock(\heap(\ell)[result \mapsto w']])) \;\;&=&\;\;& \absheap(c,\rfblock(\heap(\ell))[result \mapsto \rfval(w')]]))&&\\
&=&& \absheap(c,\rfblock(\heap(\ell))[result \mapsto \rfval(w)]]))&\;\;\qquad & \text{(by lemma \ref{lem:serialization})}\\
&<:&&\absheap(c,\absblock[result \mapsto \abswal]]))&&\text{(by Proposition~\ref{prop:fieldr})}\\
&<:&&\absprog'&&\numberthis\label{eq:result2}
\end{alignat*}

We already saw that $\rfheap{\gheap \cup \absheap'}(\heap') <: \absprog <: \absprog'$. Moreover $\rfheap{\gheap}(\heap_{|dom(\heap)\backslash \ell}) \subseteq \rfheap{\gheap}(\heap) <: \absprog <: \absprog'$. These two fact and Equation~\ref{eq:result2} show that $\rfheap{\gheap'}(\heap'') <: \absprog'$. We can also check that:
\begin{multline*}
\rfframe{\gheap'}(\tuactframe{\ell}{s}{\varepsilon}{\threadstack}{\getcb{\ell}{\actstate{onActivityResult}}},\lheap'_1,(\lfilter'^{1,j})_{j}) \\
<:\absreg{\mathsf{c''},\mathsf{m},\mathsf{0}}{(\absg{c},(\top_{\tau_j})^{j \leq n})}{(\adefvalue_k)^{k \leq \loc},\absg{c},(\top_{\tau_j})^{j \leq n}}{(\bot)^*}{0^*} <: \absprog'
\end{multline*}

Hence $X' \backslash X <: \absprog'$. We conclude by observing that since $X <: \absprog <: \absprog'$ and $X' \subseteq X \cup (X' \backslash X)$ we have $X' <: \absprog'$.

\item  Rule applied is \irule{A-ThreadStart}:
\[\inferrule[(A-ThreadStart)]
{\varphi = \tuactframe{\ell}{s}{\pi}{\ell'' :: \threadstack}{\callstack}\\
\varphi' =\tuactframe{\ell}{s}{\pi}{\threadstack}{\callstack}\\
\psi =\threadframe{\ell}{\ell''}{\varepsilon}{\varepsilon}{\callstack'}\\
\heap(\ell'') = \threadobj{c'}{(f\mapsto v)^*}\\
\lookup(c',\threadrun) = (c'',\stm^*)\\
\sign(c'',\threadrun) = \methsign{\tau}{\tau'}{\loc}\\
\callstack' = \locstate{c'',\threadrun,0}{\stm^*}{(r_k \mapsto \defvalue)^{k \leq loc},r_{loc + 1} \mapsto \ell''}{\ell''}}
{\tactconf{\actstack :: \varphi:: \actstack'}{\threadpool}{\heap}{\sheap} \Rightarrow \tactconf{\actstack :: \varphi' :: \actstack'}{\psi :: \threadpool}{\heap}{\sheap}}\]

Given that $X <: \absprog$, we have $\absprog :> \rftdispatch{\gheap}(\ell'' :: \threadstack)$. Moreover $\heap(\ell'') = \threadobj{c'}{(f\mapsto v)^*}$, therefore there exists $\abstdispatch(\ann,\absblock) \in \absprog$ such that $\ann = \rflab(\ell'')$ and $ \rfblock{}(\threadobj{c'}{(f\mapsto v)^*}) \poblk \absblock$. This implies that $\absblock = \absthreadobj{c'}{\absval^*}$ for some $\absval^*$ such that $\forall i,  \rfval{}(v_i) \poval \absval_i $.

By well-formedness we get that $c' \le \thread$, and by  Assumption~\ref{asm:overriding} we know that  $\lookup(c',\threadrun) = (c'',\stm^*)$ implies that $c' \le c''$. Moreover since $\lookup(c',\threadrun) = (c'',\stm^*)$ we know that $c'' \in \abslookup(\threadrun)$, hence we can use the rule $\rulename{Tstart}$ included in $\translate{P}$:
\begin{equation}
\abstdispatch(\absloc,\absobj{c'}{(f \mapsto \_)^*}) \wedge c' \subtype \const{c''} \wedge c' \le \thread \implies \absreg{\const{c''},\threadrun,\mathsf{0}}{(\absg{\absloc},\absg{\absloc})}{(\adefvalue_k)^{k \leq \loc},\absg{\absloc}}{(\bot)^*}{0^*}\label{tstart0}
\end{equation}

We define the set of abstract fact:
\begin{alignat*}{2}
\absprog' \;\;&=\;\;&& \absprog \cup \{\absreg{\const{c''},\threadrun,\mathsf{0}}{(\absg{\absloc},\absg{\absloc})}{(\adefvalue_k)^{k \leq \loc},\absg{\absloc}}{(\bot)^*}{0^*}\}
\end{alignat*}
From Equation~\ref{tstart0} we get that $\translate{P} \cup \absprog \vdash \absprog'$.

Let $n$ be the length of $\actstack :: \varphi:: \actstack'$, and $m$ the length of $\threadpool$. Let $\lheap'_{t}$ be an fresh empty local heap. We take $\gheap'=\gheap$ and :
\[(\lheap_l',(\lfilter'^{l,j})_{j})_{l \le n+m+1} = (\lheap_l,(\lfilter^{l,j})_{j})_{l \le n} :: (\lheap'_{t},((\{(\ell \mapsto 0) ~|~ \ell \}) :: \varepsilon)) :: (\lheap_l,(\lfilter^{l,j})_{j})_{n + 1 \le l \le n+m}\]

Since \cheapd is a configuration decomposition of $\Psi$ we know that $\ell'' \in dom(\gheap)$. With this one can check that \cheapdp is a configuration decomposition of $\Psi'$.

Let $X' \in \rfconf(\Psi')$ be the corresponding set of facts:
\[
 \rfastk{\gheap'}(\actstack :: \varphi' :: \actstack',\psi :: \threadpool,(\lheap'_l,(\lfilter'^{l,j})_{j})_l)\cup \rfheap{\gheap'}(\heap) \cup \rfstat{}(\sheap)
\]
Let $n_0$ be such that  $\actstack$ is of length $n_0-1$. It is quite easy to check that:
\[
X' \backslash X \subseteq  \rfframe{\gheap'}(\tuactframe{\ell}{s}{\pi}{\threadstack}{\callstack},\lheap'_{n_0},(\lfilter'^{{n_0},j})_{j}) 
\cup  \rfframe{\gheap'}(\threadframe{\ell}{\ell''}{\varepsilon}{\varepsilon}{\callstack'},\lheap'_{n+1},(\lfilter'^{{n+1},j})_{j})
\]

Since $\ell'' \in dom(\gheap)$, we have that:
\begin{alignat*}{2}
\absprog' \;\;&:>\;\;&& \{\absreg{\const{c''},\threadrun,\mathsf{0}}{(\absg{\absloc},\absg{\absloc})}{(\adefvalue_k)^{k \leq \loc},\absg{\absloc}}{(\bot)^*}{0^*}\} \\
&:>&&  \rfframe{\gheap'}(\threadframe{\ell}{\ell''}{\varepsilon}{\varepsilon}{\callstack'},\lheap'_{n+1},(\lfilter'^{{n+1},j})_{j})
\end{alignat*}

Moreover since $\phi'$ only differ from $\phi$ in the fact that it has a smaller thread stack, we have:
\[
\rfframe{\gheap'}(\tuactframe{\ell}{s}{\pi}{\threadstack}{\callstack},\lheap'_{n_0},(\lfilter'^{{n_0},j})_{j}) \subseteq \rfframe{\gheap}(\tuactframe{\ell}{s}{\pi}{\ell'' :: \threadstack}{\callstack},\lheap_{n_0},(\lfilter^{{n_0},j})_{j}) <: \Delta
\]
This proves that $X' :> \absprog'$.

\item  Rule applied is \irule{T-Reduce}:
\[\inferrule[(T-Reduce)]
{\tmethconf{\callstack}{\pi}{\threadstack}{\heap}{\sheap}{\ell'} \rightsquigarrow \tmethconf{\callstack'}{\pi'}{\threadstack'}{\heap'}{\sheap'}{\ell'}}
{\tactconf{\actstack}{\threadpool :: \threadframe{\ell}{\ell'}{\pi}{\threadstack}{\callstack} :: \threadpool'}{\heap}{\sheap} \Rightarrow \tactconf{\actstack}{\threadpool :: \threadframe{\ell}{\ell'}{\pi'}{\threadstack'}{\callstack'} :: \threadpool'}{\heap'}{\sheap'}}\]

Exactly like the  \irule{A-Reduce} case.

\item  Rule applied is \irule{T-Kill}:
\[\inferrule[(T-Kill)]
{ \heap(\ell') = \obj{c}{(f\mapsto v)^*,\finished \mapsto \_}\\
\heap'= \ \heap[\ell' \mapsto \obj{c}{(f\mapsto v)^*,\finished \mapsto \true}]
}
{\tactconf{\actstack}{\threadpool :: \threadframe{\ell}{\ell'}{\varepsilon}{\varepsilon}{\ocallstack} :: \threadpool'}{\heap}{\sheap} \Rightarrow \tactconf{\actstack}{\threadpool :: \threadpool'}{\heap'}{\sheap} }\]

Exactly like the  \irule{A-Destroy} case.

\item  Rule applied is \irule{T-Intent}:
\[\inferrule[(T-Intent)]
{(\varphi,\varphi') \in \{(\tactframe{\ell}{s}{\pi}{\threadstack}{\callstack},\tactframe{\ell}{s}{i :: \pi}{\threadstack}{\callstack}),(\tuactframe{\ell}{s}{\pi}{\threadstack}{\callstack},\tuactframe{\ell}{s}{i :: \pi}{\threadstack}{\callstack})\}}
{\tactconf{\actstack :: \varphi :: \actstack'}{\threadpool :: \threadframe{\ell}{\ell'}{i :: \pi'}{\threadstack'}{\callstack'} :: \threadpool'}{\heap}{\sheap} \Rightarrow \tactconf{\actstack :: \varphi' :: \actstack'}{\threadpool :: \threadframe{\ell}{\ell'}{\pi'}{\threadstack'}{\callstack'} :: \threadpool'}{\heap}{\sheap}}\]

Trivial since there are no changes to the abstraction: $\rfconf(\Psi) = \rfconf(\Psi')$.

\item  Rule applied is \irule{T-Thread}:
\[\inferrule[(T-Thread)]
{(\varphi,\varphi') \in \{(\tactframe{\ell}{s}{\pi}{\threadstack}{\callstack},\tactframe{\ell}{s}{\pi}{\ell_t :: \threadstack}{\callstack}),(\tuactframe{\ell}{s}{\pi}{\threadstack}{\callstack},\tuactframe{\ell}{s}{\pi}{\ell_t :: \threadstack}{\callstack})\}}
{\tactconf{\actstack :: \varphi :: \actstack'}{\threadpool :: \threadframe{\ell}{\ell'}{\pi'}{\ell_t :: \threadstack'}{\callstack'} :: \threadpool'}{\heap}{\sheap} \Rightarrow \tactconf{\actstack :: \varphi' :: \actstack'}{\threadpool :: \threadframe{\ell}{\ell'}{\pi'}{\threadstack'}{\callstack'} :: \threadpool}{\heap}{\sheap}}\]

Trivial since there are no changes to the abstraction: $\rfconf(\Psi) = \rfconf(\Psi')$.

\end{itemize}
\end{IEEEproof}



\fi

\end{document}